\documentclass[acmtog]{acmart}
\AtBeginDocument{}

\acmSubmissionID{298}

\usepackage[nomessages]{fp}
\usepackage{subcaption}
\usepackage{multirow}
\usepackage{enumitem}
\usepackage{bm}
\usepackage{makecell}
\usepackage{pifont}

\usepackage{algorithm}
\usepackage{algpseudocode}

\let\REQUIRE\Require
\let\ENSURE\Ensure
\let\STATE\State
\let\FOR\For
\let\ENDFOR\EndFor
\let\WHILE\While
\let\ENDWHILE\EndWhile

\let\RETURN\Return
\let\COMMENT\Comment
\let\CALL\Call

\setlist[itemize]{align=parleft,left=0pt..1em,noitemsep,topsep=0pt}
\AtBeginDocument{
\setlength{\abovedisplayskip}{4pt plus 0pt minus 1pt}
\setlength{\belowdisplayskip}{5pt plus 0pt minus 1pt}
\setlength{\abovedisplayshortskip}{4pt plus 0pt minus 1pt}
\setlength{\belowdisplayshortskip}{5pt plus 0pt minus 1pt}
}

\copyrightyear{2026}
\acmYear{2026}
\setcopyright{cc}
\setcctype{by}
\acmConference[SIGGRAPH Conference Papers '26]{Special Interest Group on Computer Graphics and Interactive Techniques Conference Conference Papers}{July 19--23, 2026}{Los Angeles, CA, USA}
\acmBooktitle{Special Interest Group on Computer Graphics and Interactive Techniques Conference Conference Papers (SIGGRAPH Conference Papers '26), July 19--23, 2026, Los Angeles, CA, USA}
\acmDOI{10.1145/3799902.3811058}
\acmISBN{979-8-4007-2554-8/2026/07} 
\begin{document}

\title{CAGS: Color-Adaptive Volumetric Video Streaming with Dynamic 3D Gaussian Splatting}

\author{Daheng Yin}
\orcid{0000-0002-9431-4240}
\affiliation{\institution{Simon Fraser University}
 \city{Burnaby}
 \country{Canada}}
\affiliation{\institution{Jiangxing Intelligence Inc.}
\city{Shenzhen}
\country{China}}
\email{dya64@sfu.ca}

\author{Yili Jin}
\orcid{0000-0002-7127-8902}
\affiliation{\institution{McGill University}
\city{Montreal}
\country{Canada}}
\affiliation{\institution{Simon Fraser University}
 \city{Burnaby}
 \country{Canada}}
\email{yili.jin@mail.mcgill.ca}

\author{Jianxin Shi}
\orcid{0000-0002-7687-8480}
\affiliation{\institution{Nankai University}
\city{Tianjin}
\country{China}}
\affiliation{\institution{Simon Fraser University}
 \city{Burnaby}
 \country{Canada}}
\email{jxshi@nankai.edu.cn}

\author{Isaac Ding}
\orcid{0009-0004-1117-3860}
\affiliation{\institution{Simon Fraser University}
 \city{Burnaby}
 \country{Canada}}
\email{isaac_ding@sfu.ca}

\author{Miao Zhang}
\orcid{0000-0001-6126-6142}
\affiliation{\institution{Simon Fraser University}
 \city{Burnaby}
 \country{Canada}}
\email{mza94@sfu.ca}

\author{Fangxin Wang}
\orcid{0000-0003-2559-045X}
\affiliation{\institution{The Chinese University of Hong Kong, Shenzhen}
 \city{Shenzhen}
 \country{China}}
\email{wangfangxin@cuhk.edu.cn}

\author{Zhaowu Huang}
\orcid{0000-0002-3941-6412}
\affiliation{\institution{Fuzhou University}
 \city{Fuzhou}
 \country{China}}
 \affiliation{\institution{Southeast University}
  \city{Nanjing}
  \country{China}}
\email{zwh@fzu.edu.cn}

\author{Cong Zhang}
\orcid{0000-0002-9439-6725}
\affiliation{\institution{Simon Fraser University}
 \city{Burnaby}
 \country{Canada}}
\affiliation{\institution{Jiangxing Intelligence Inc.}
\city{Shenzhen}
\country{China}}
\email{congz@ieee.org}

\author{Jiangchuan Liu}
\authornote{Corresponding authors.}
\orcid{0000-0001-6592-1984}
\affiliation{\institution{Simon Fraser University}
\city{Burnaby}
\country{Canada}}
\email{jcliu@cs.sfu.ca}

\author{Fang Dong}
\orcid{0000-0001-6770-326X}
\affiliation{\institution{Southeast University}
\city{Nanjing}
\country{China}}
\email{fdong@seu.edu.cn}

\begin{abstract}
  Volumetric video (VV) streaming delivers truly immersive viewing experiences over the Internet, serving as a critical foundation for next-generation applications, including immersive telepresence in the metaverse, the surveillance of remote ecological systems, and robotic teleoperation for embodied AI, and beyond.
Beyond immersive viewing, these applications turn VV streaming into a real-time interface to remote physical environments, imposing new system-level demands for photorealistic scene representation, low-latency interaction, and robust performance under heterogeneous network conditions.
3D Gaussian Splatting (3DGS) has been widely used for real-time photorealistic rendering, offering superior visual quality and rendering performance, but it faces challenges due to bandwidth consumption.
Furthermore, as the foundation of adaptive VV streaming, existing Levels of Detail (LoD) methods based on density are not well-suited to Gaussian representations, leading to visible gaps and severe quality degradation.
Recent studies have also explored attribute compression techniques to reduce bandwidth consumption.
Our preliminary studies reveal that aggressive attribute compression primarily causes color distortion, which can be effectively corrected in the rendered image using a reference image.
Motivated by these findings, we propose a novel Color-Adaptive scheme for adaptive VV streaming that uses vector quantization (VQ) to establish LoDs and correct color distortions with low-resolution reference images.
We further present CAGS, an adaptive VV streaming system compatible with diverse Gaussian representations, which integrates the Color-Adaptive scheme by rendering reference images on the streaming server and performing color restoration on the client.
Extensive experiments on our prototype system demonstrate that CAGS outperforms the existing adaptive streaming systems in PSNR by 5$\sim$20 dB under fluctuating bandwidth, operates significantly faster than existing scalable Gaussian compression methods, and generalizes across different Gaussian representations.
The code is available at \href{https://github.com/yindaheng98/ColorAdaptiveGaussianSplatting}{https://github.com/yindaheng98/ColorAdaptiveGaussianSplatting}. \end{abstract}



\maketitle

\section{Introduction}

Modern Internet infrastructure and real-time streaming technologies are reshaping online media, evolving it from delivering static images and videos toward live, interactive, photorealistic spatial experiences.
Volumetric Video (VV) streaming is central to this evolution, which captures dynamic scenes as evolving 3D content, allowing interactive exploration from arbitrary viewpoints with six degrees of freedom (6DoF).
Built on this infrastructure, emerging applications such as immersive communication, live digital twins, and embodied intelligence are turning VV streaming from an immersive viewing medium into a real-time interface to the physical world.
This transition introduces new system-level demands: VV streaming must deliver high visual fidelity for accurate interpretation, minimal latency for responsive interaction, and robust performance across heterogeneous network conditions for reliable deployment.

Prior studies have established VV streaming as an effective 6DoF medium for real-world scenes, supporting more interactive experiences than conventional video in telepresence, collaboration, education, training, and immersive media \cite{hanViVoVisibilityawareMobile2020,guanMetaStreamLiveVolumetric2023}.
These capabilities naturally extend to a wide range of domains where remote perception, coordination, and action are critical, from healthcare~\cite{gasques2021artemis} and medical training~\cite{rojas2019surgical} to digital twins, robotics, autonomous systems~\cite{zimmer2024tumtraf,tao2018digital}, and environmental or industrial monitoring~\cite{hazeleger2024digital,de2023digital}.
In this sense, VV streaming expands the Internet from a medium for transmitting images, videos, and messages into an infrastructure for delivering dynamic, interactive representations of the physical world.

\begin{figure}[t]
	\centering
	\includegraphics[width=\linewidth]{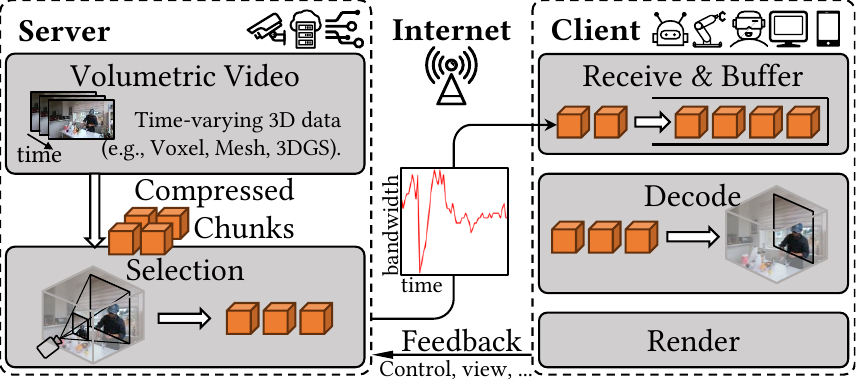}
	\caption{
		Overview of a typical adaptive VV streaming pipeline.
		The server maintains time-varying 3D data, selects an appropriate quality-size version of each scene tile according to viewport and network feedback, and streams it to the client for decoding and rendering.
	}\label{figure:stream}
\end{figure} 
To support such applications, VV streaming systems must deliver interactive 6DoF content with stable geometry for navigation, accurate visuals for interpretation, low latency for responsiveness, and consistent performance across varying network bandwidths.
Adaptive VV streaming systems address these requirements by encoding scene tiles at multiple quality-size levels and dynamically selecting optimal versions based on viewport feedback, user interactions, or network status (Figure~\ref{figure:stream}) \cite{zhangYuZuNeuralEnhancedVolumetric2022}.

As a spatially decomposable and rasterization-friendly technique, 3D Gaussian Splatting (3DGS)~\cite{kerbl3DGaussianSplatting2023} is widely used for real-time photorealistic 3D rendering.
It represents scenes using semi-transparent ellipsoids (``Gaussians'') with Spherical Harmonics (SH) for view-dependent photorealism, while enabling real-time rendering via rasterization.
Despite these advantages, Gaussians carry complex attributes that produce large data volumes, posing significant challenges for streaming applications~\cite{zhuSGSSStreaming6DoF2025,sunLTSDASHStreaming2025,huangACPGSBandwidthEfficientDelivery2026,kimVegaFullyImmersive2025}.
Data reduction typically follows two orthogonal strategies: decreasing the number of Gaussians or compressing their attributes.

\begin{figure}[t]
	\centering
	\newcommand{\includegraphicstrimcoffee}[1]{\includegraphics[width=\linewidth,trim={250mm 50mm 250mm 240mm},clip]{figures/LoDinIntro/#1/00004.png}}
	\newcommand{\includegraphicstrimrobot}[1]{\includegraphics[angle=270,origin=c,width=\linewidth,trim={150mm 0 150mm 0},clip]{figures/LoDinIntro/Robo360/#1/ours_30000/renders/00072.png}}
	\newcommand{\includegraphicstrimlilyrun}[1]{\includegraphics[width=\linewidth,trim={380mm 240mm 400mm 160mm},clip]{figures/LoDinIntro/PARIS_dataset/#1/ours_30000/renders/00001.png}}
	\newcommand{\includegraphicstrimmvpsp}[1]{\includegraphics[width=\linewidth]{figures/LoDinIntro/mvpsp/#1/ours_30000/renders/00004.png}}

	\begin{subfigure}[b]{0.32\linewidth}
		\centering
        \includegraphicstrimcoffee{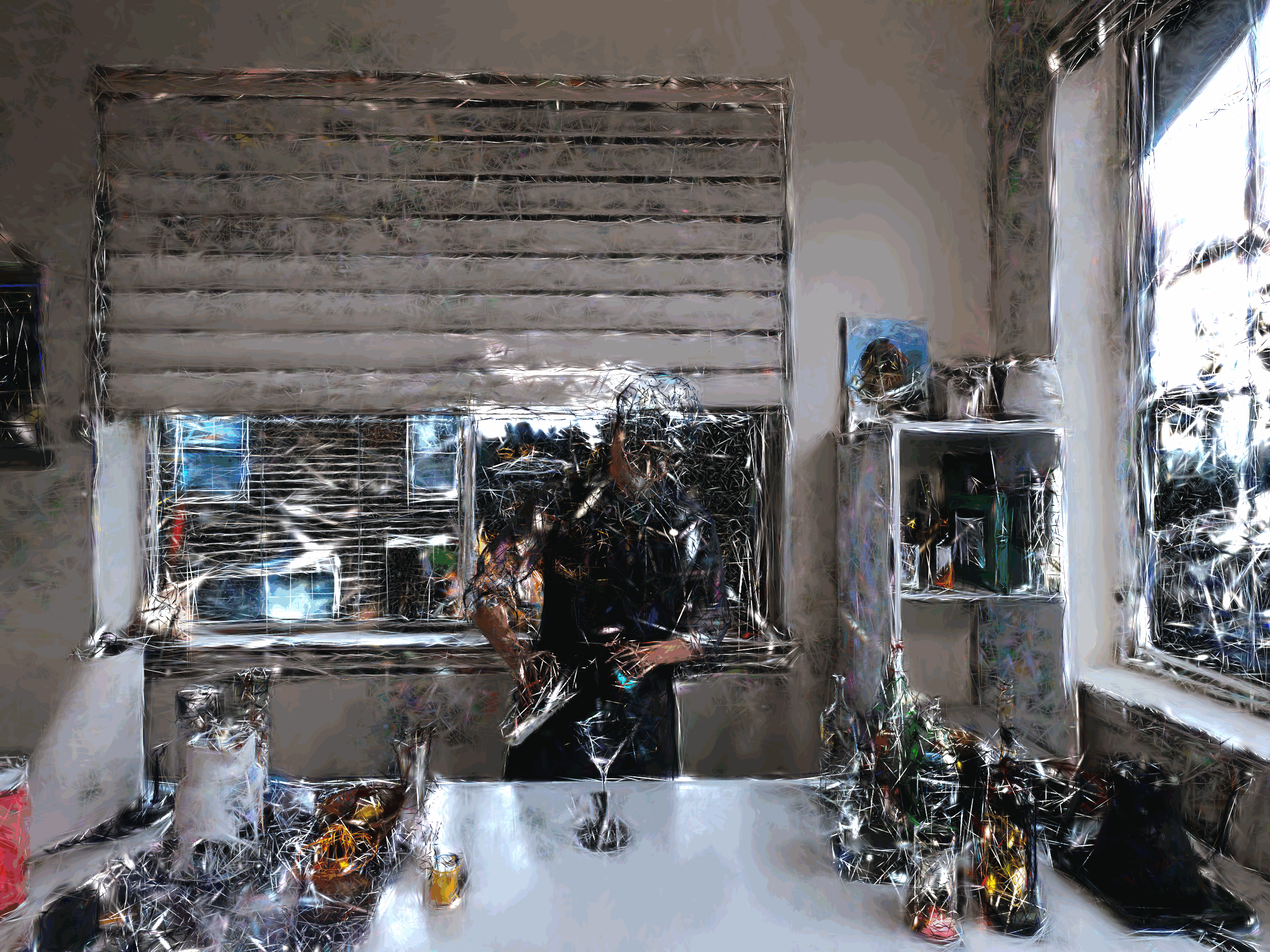}
        \includegraphicstrimrobot{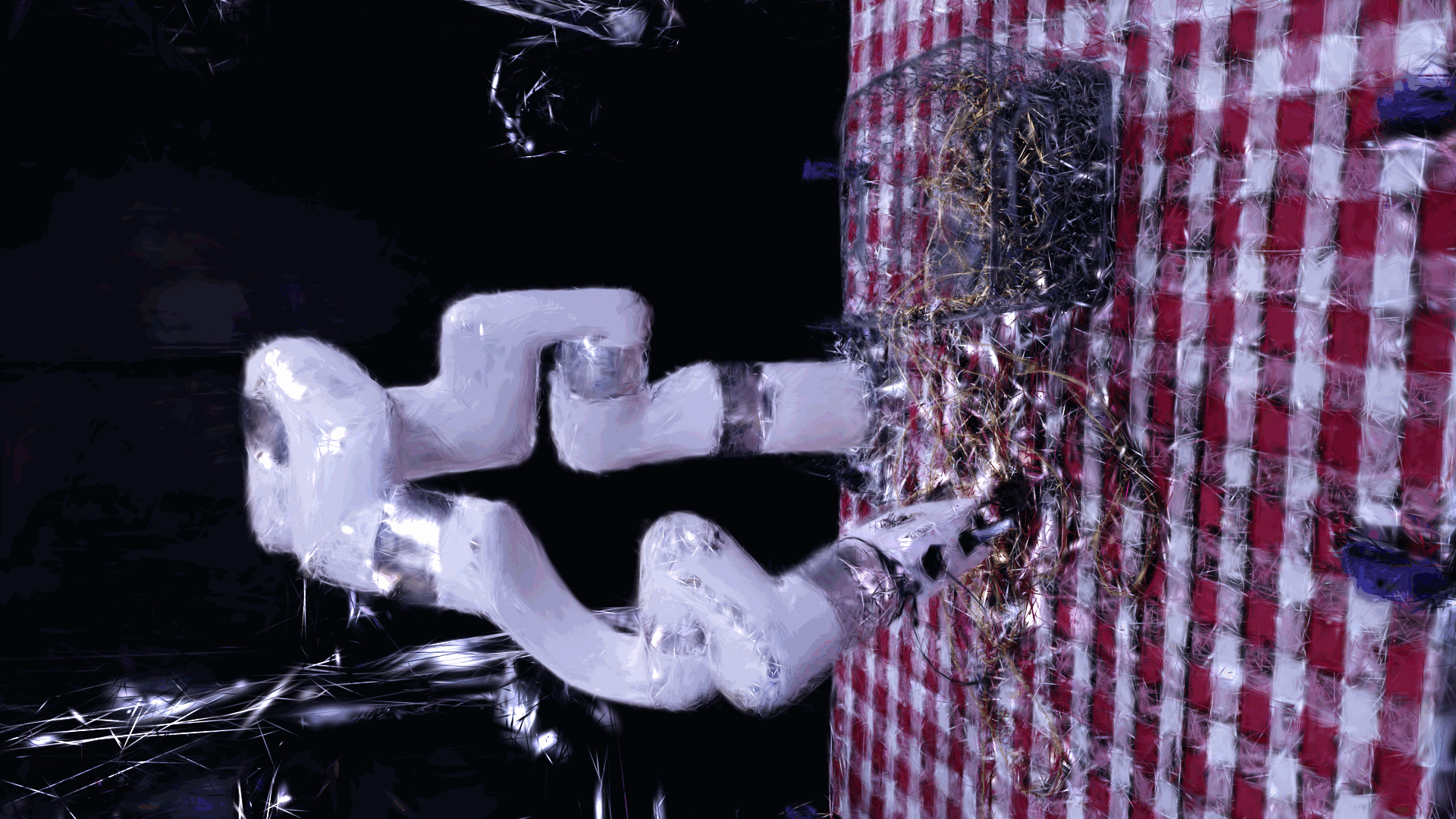}
		\includegraphicstrimlilyrun{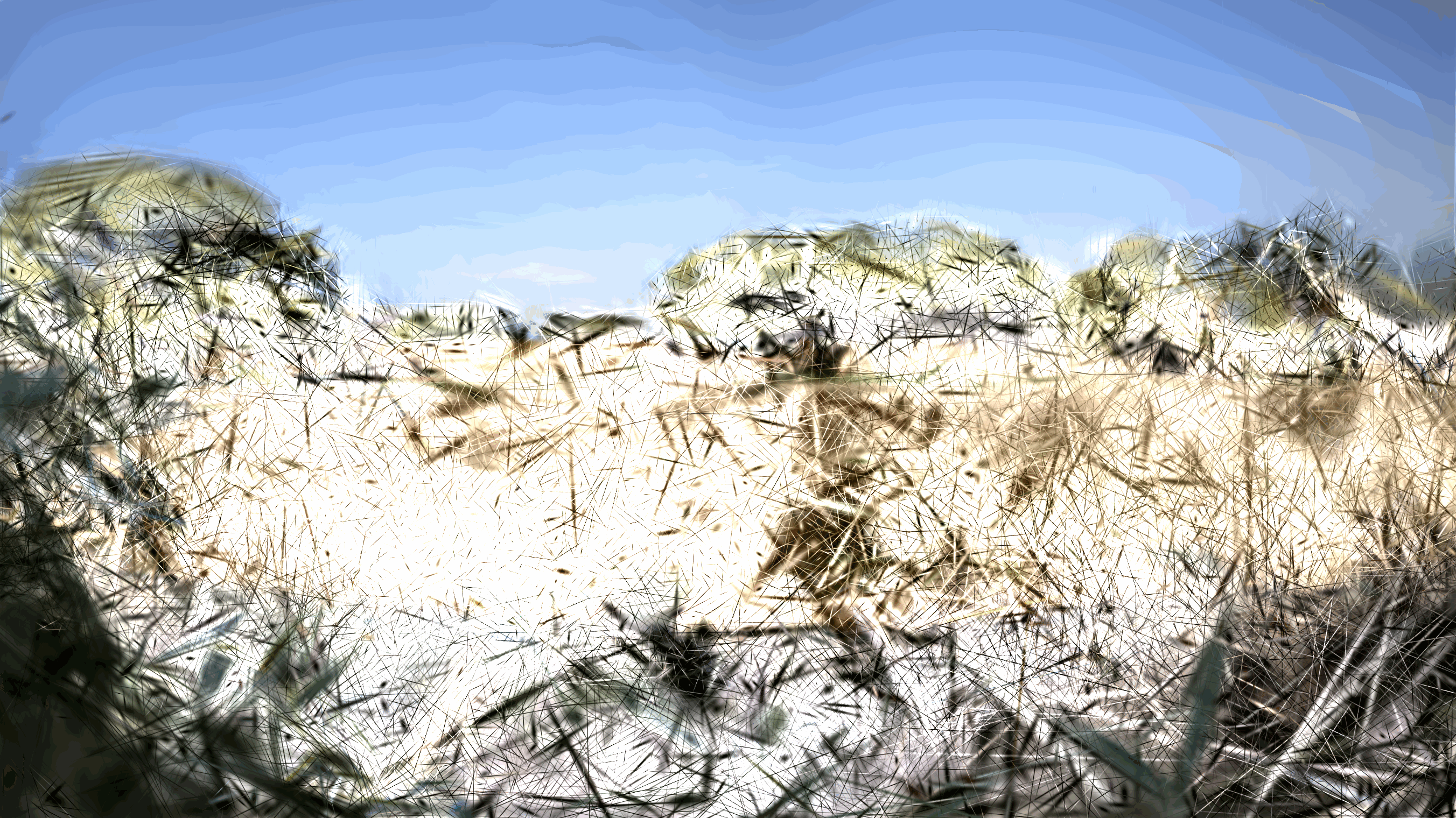}
		\includegraphicstrimmvpsp{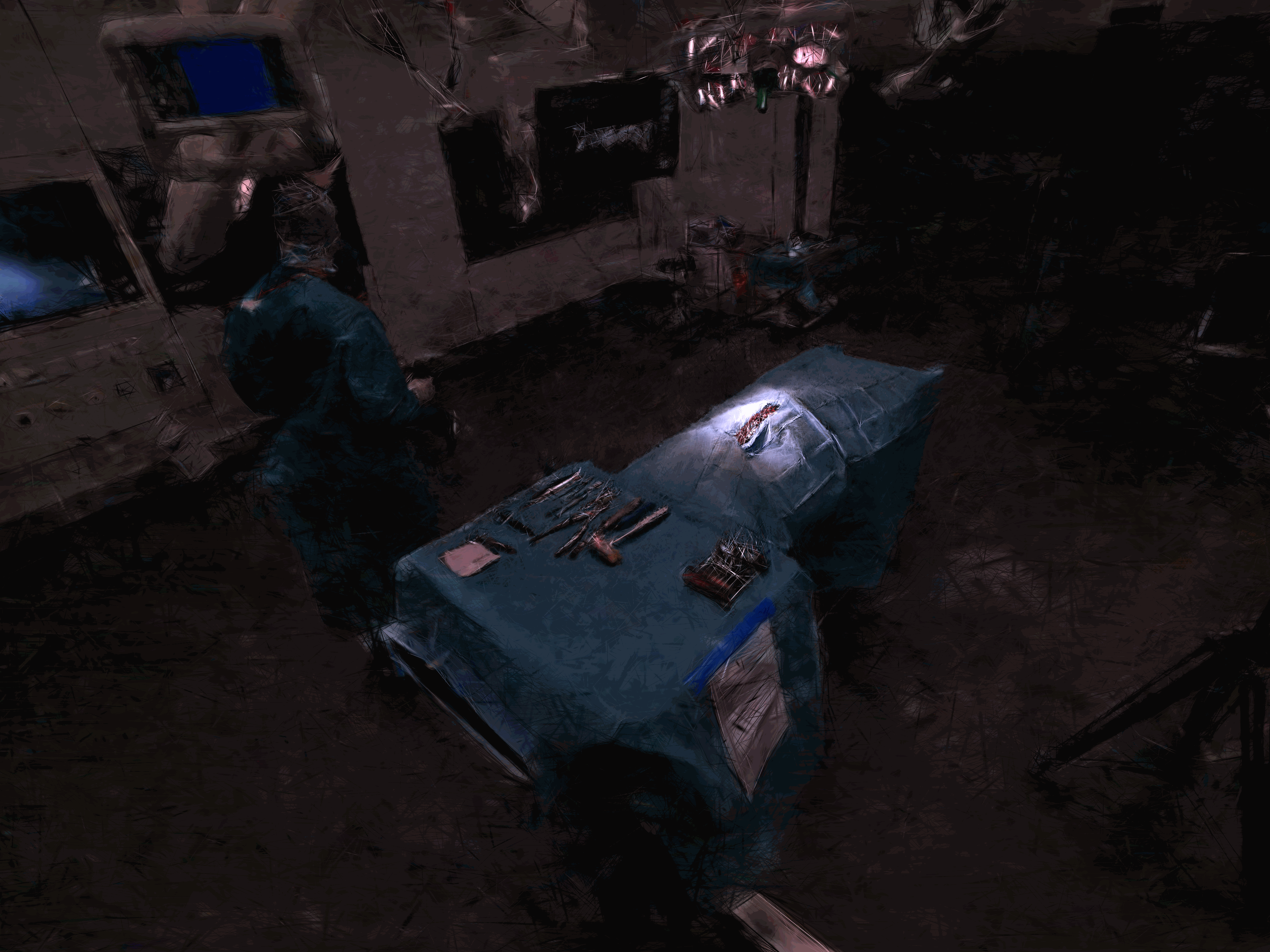}
		\caption{LoD 0 (\textasciitilde 21 MB)}\label{figure:Octree-lod}
	\end{subfigure}
	\begin{subfigure}[b]{0.32\linewidth}
		\centering
        \includegraphicstrimcoffee{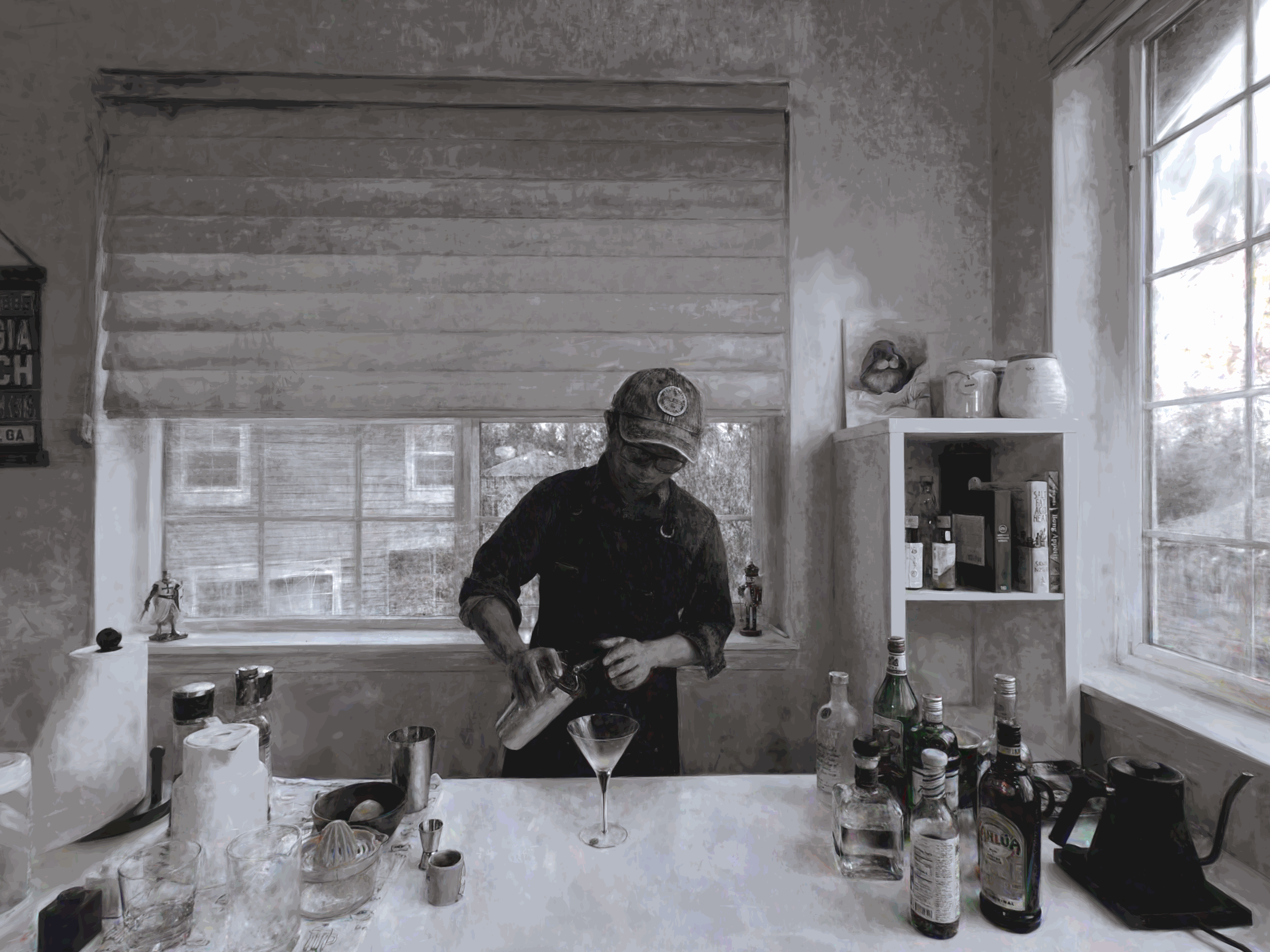}
        \includegraphicstrimrobot{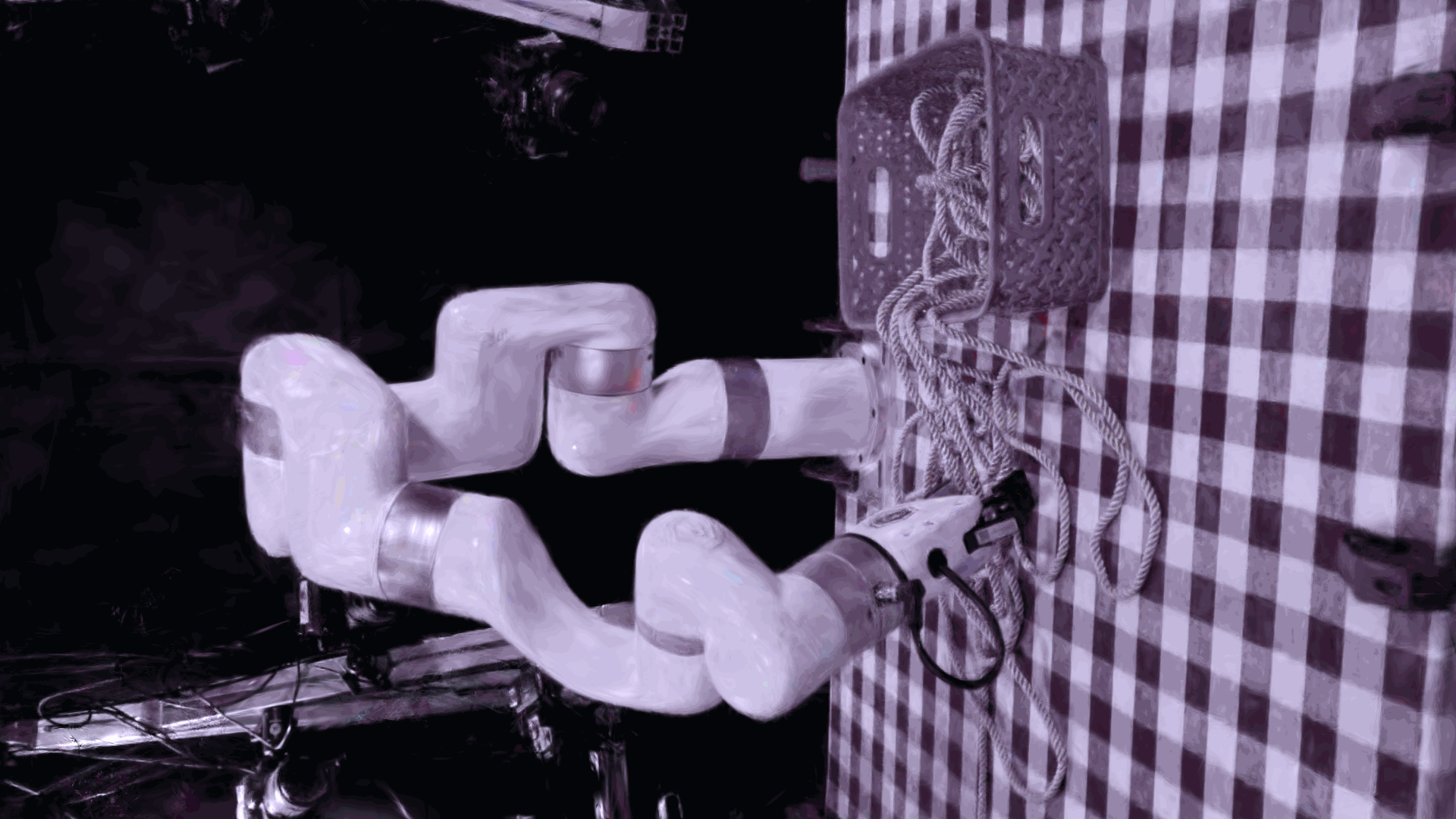}
		\includegraphicstrimlilyrun{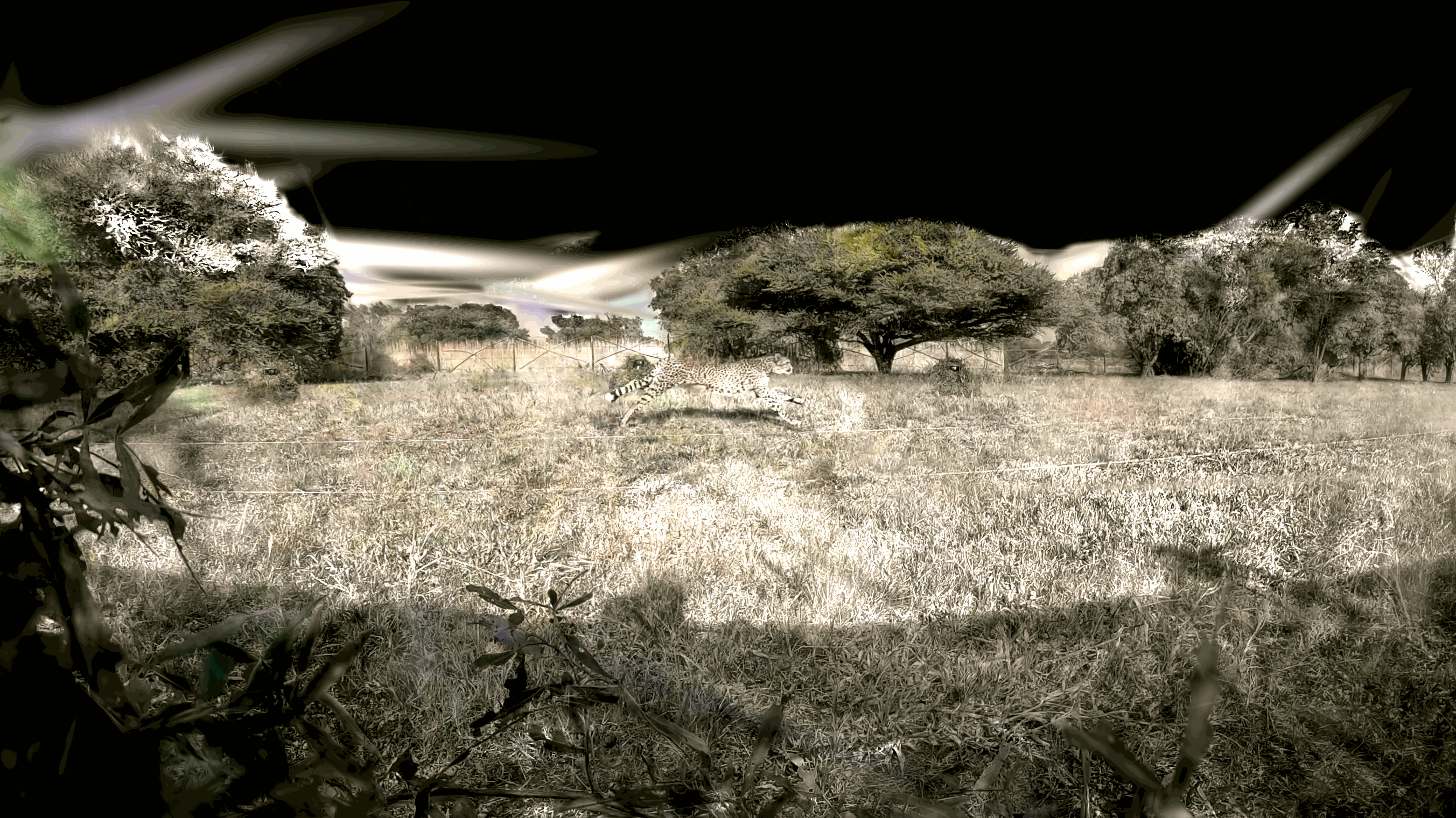}
		\includegraphicstrimmvpsp{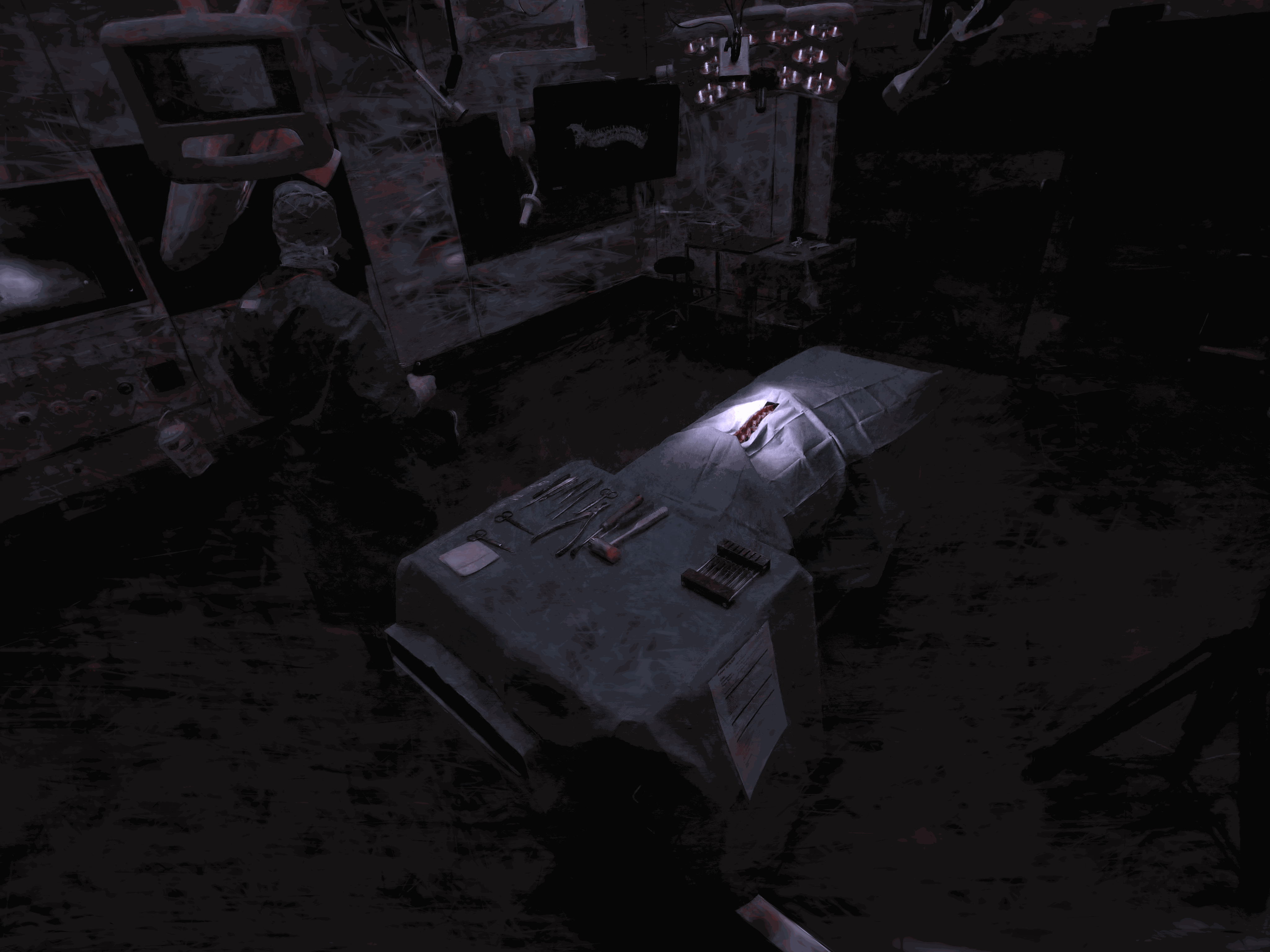}
		\caption{40-bit VQ (\textasciitilde 14 MB)}\label{figure:Octree-vq}
	\end{subfigure}
	\begin{subfigure}[b]{0.32\linewidth}
		\centering
        \includegraphicstrimcoffee{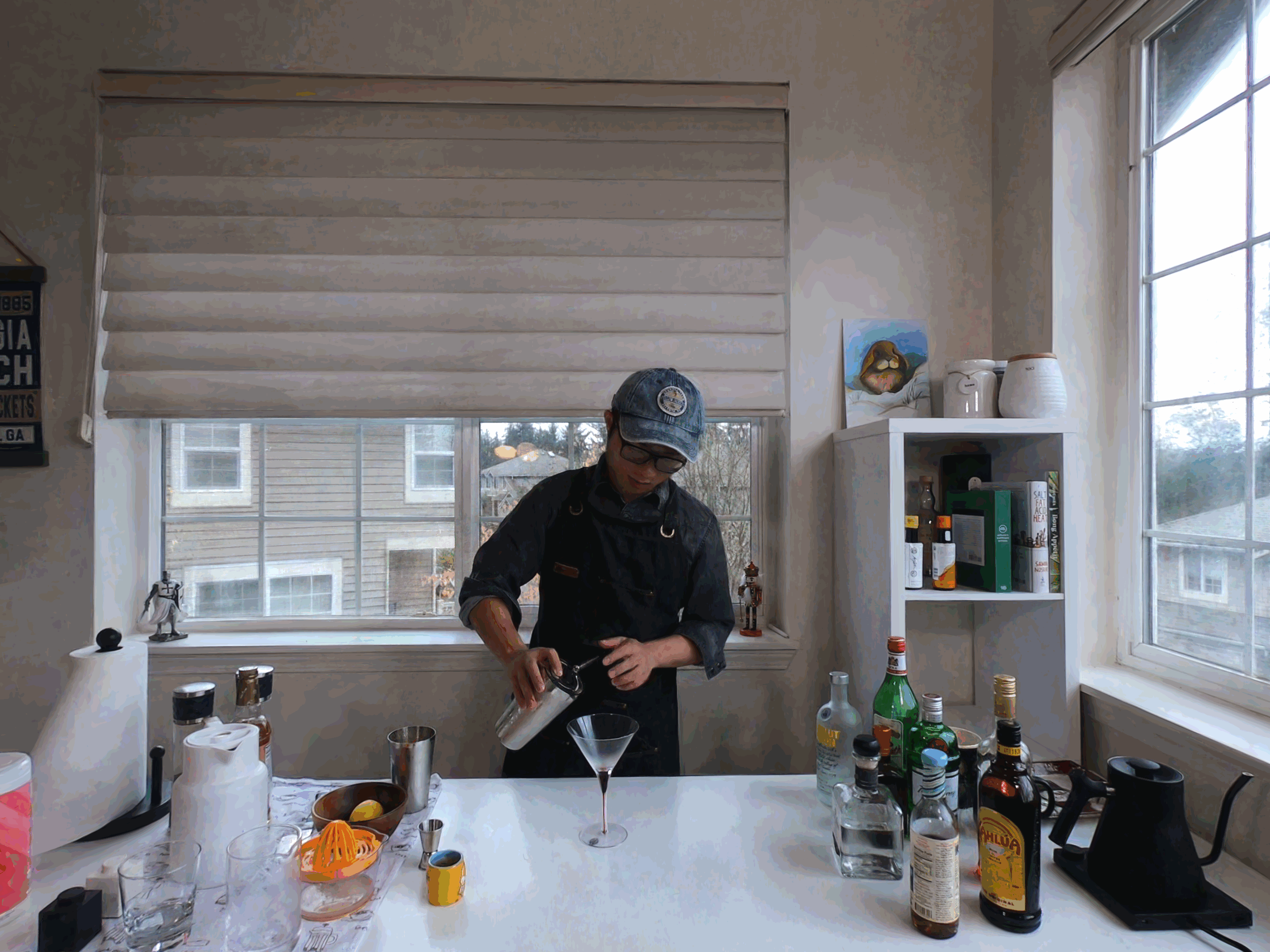}
        \includegraphicstrimrobot{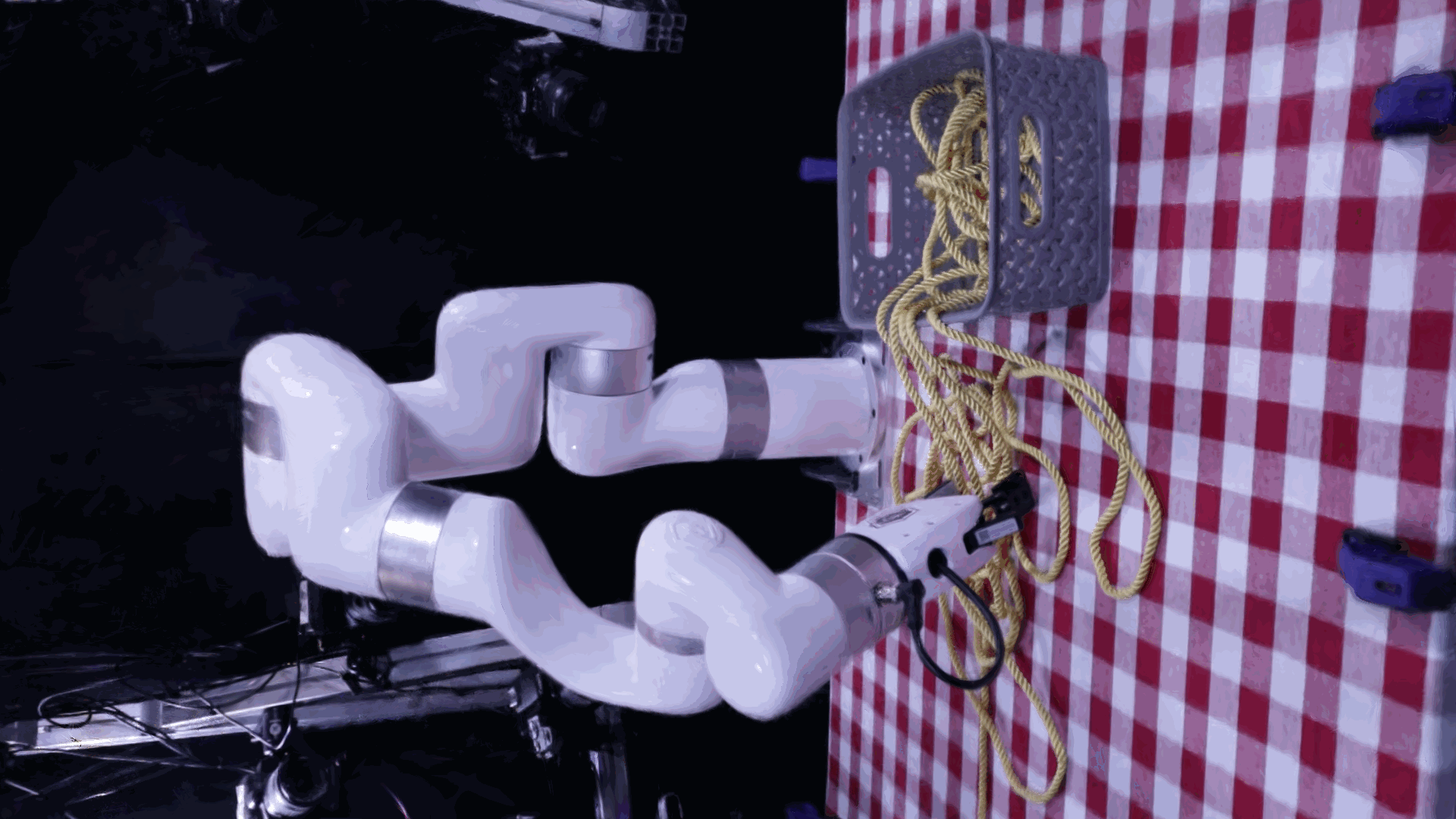}
		\includegraphicstrimlilyrun{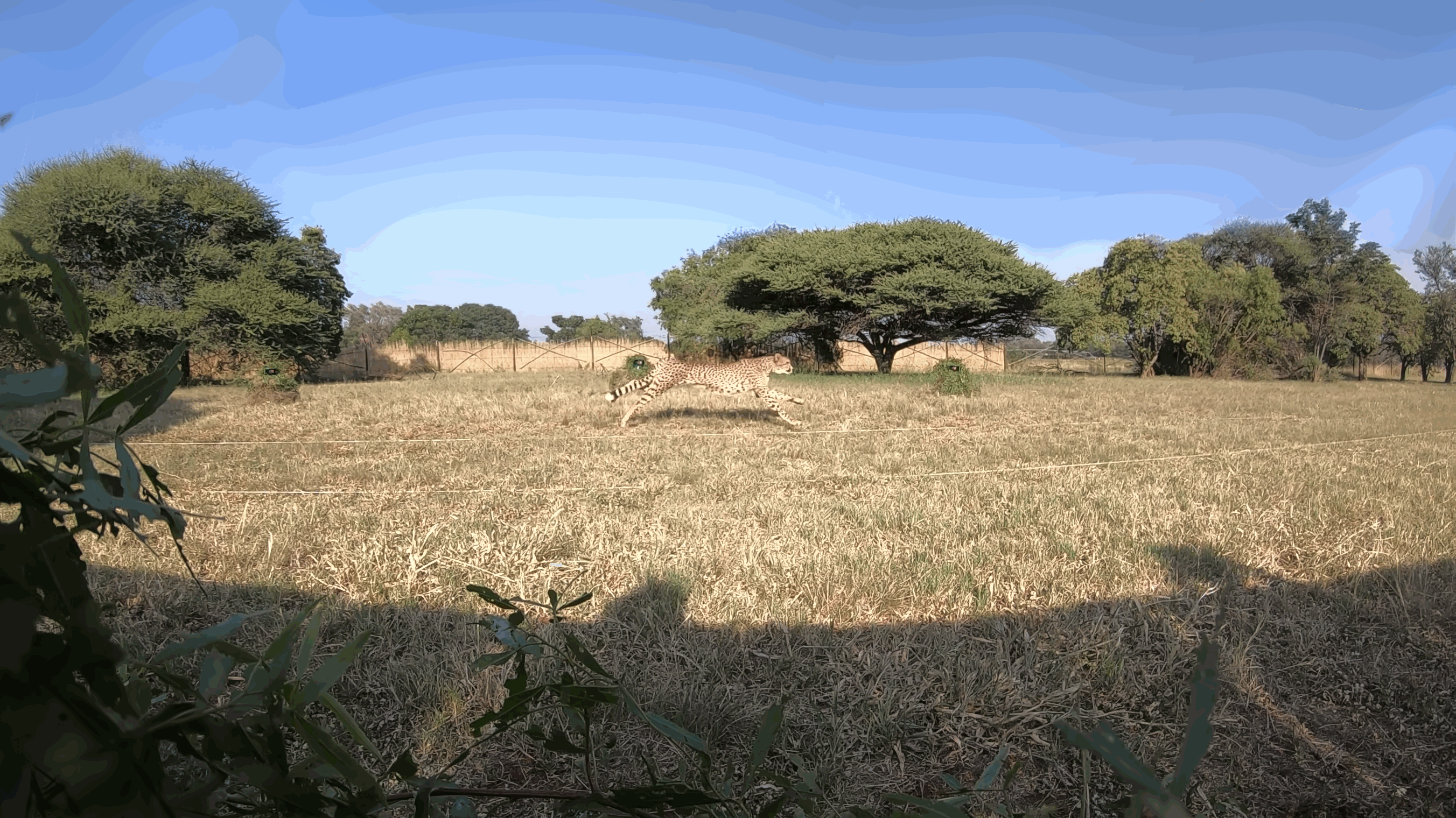}
		\includegraphicstrimmvpsp{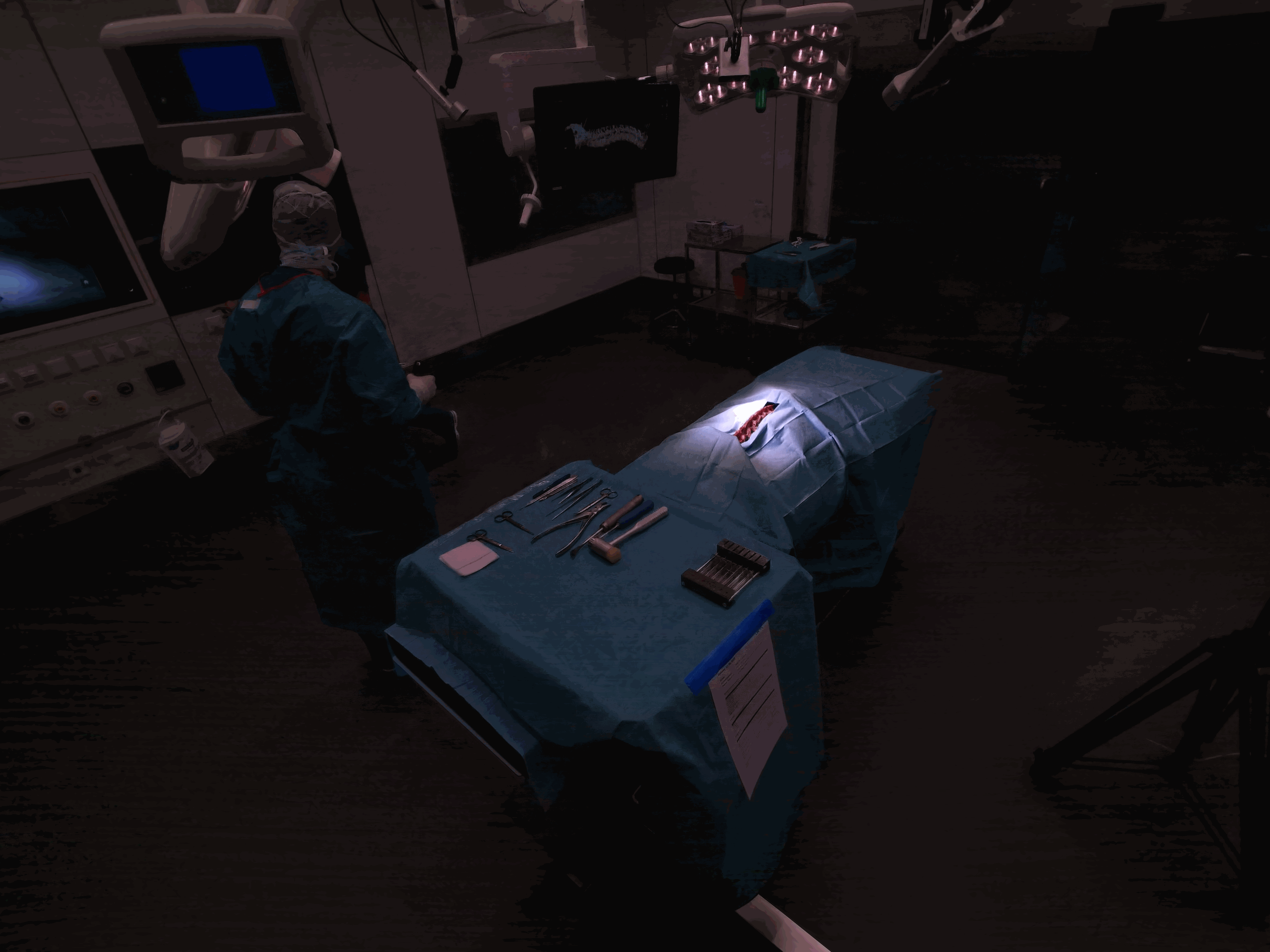}
		\caption{100-bit VQ (\textasciitilde 34 MB)}
	\end{subfigure}
	\caption{
		Visual comparison of density-based LoD used in LTS~\cite{sunLTSDASHStreaming2025} and attribute compression by Vector Quantization (VQ) (both compressed further by Draco~\cite{draco}).
		Removing Gaussians causes structural gaps at low bitrates (a), while attribute compression better preserves structure with color distortion (b).
		Scenes are selected from the N3DV~\cite{liNeural3DVideo2022}, Robo360~\cite{liang2023robo360}, AcinoSet~\cite{joska2021acinoset} and MVSPS~\cite{HEIN2025103613} datasets.}\label{figure:Octree}
\end{figure}

Existing systems typically support network adaptation through \textit{density-based Level of Detail (LoD)}~\cite{liuCaV3CacheassistedViewport2023,shiLapisGSLayeredProgressive2025}, which provides progressive scene representations with different numbers of primitives.
Since each Gaussian captures fine-grained details within a volumetric region, removing Gaussians can lead to insufficient scene coverage and introduce structural gaps (Figure~\ref{figure:Octree-lod}).

Alternatively, attribute compression reduces data by compressing Gaussian attributes.
A representative method is Vector Quantization (VQ), which maps high-dimensional Gaussian attributes to compact codebook indices and enables fast index-based decoding~\cite{leeCompact3DGaussian2024}.
Nonetheless, existing attribute compression techniques primarily focus on static compression and lack the quality scalability required for adaptive VV streaming under fluctuating bandwidth.

In this paper, we explore \emph{scalable attribute compression} for adaptive, photorealistic VV streaming based on Gaussian representations.
Our preliminary studies (Sec.~\ref{sec:motivate}) highlight that attribute compression largely avoids structural gaps but introduces color distortion as the dominant visual artifact.
Accurate color representation is critical in streamed 3D scenes, as colors indicate object types and materials in robotics, liquid states and interactions in human contexts, tissue and instrument conditions in clinical settings, and species or environmental specifics in ecological observations (Figure~\ref{figure:Octree-vq}).
Distorted color streams may obscure visual information required for accurate human interpretation and effective machine perception.

Encouragingly, our preliminary studies further reveal that such \textit{color distortion caused by VQ can be effectively corrected in rendered images using a low-resolution reference image with accurate colors}.
Motivated by this finding, we advocate a \textbf{Color Adaptation scheme} that \textit{scalably compresses Gaussian attributes by VQ to establish LoDs, and restores colors using low-resolution reference images}.
In streaming systems, the server renders these reference images from the high-fidelity Gaussians and streams them to clients to enable color restoration.

Implementing this concept in practical streaming systems introduces three challenges.
First, adaptive streaming requires scalable VQ rather than traditional flat VQ.
We therefore design \textbf{Scalable Vector Quantization (SVQ)} that organizes Gaussian attributes into a base layer and multiple enhancement layers, establishing LoDs with different levels of quantization error.
Second, latency constraints require the streaming server to predict the client's viewport and render reference images in advance.
Prediction errors inevitably result in mismatches between reference images and actual client viewports, severely degrading color restoration quality.
We address this with \textbf{Post-Render Perspective Alignment (PRPA)}, which realigns reference images on the client side using locally rendered depth.
Third, prediction errors may leave visible regions uncovered by reference images, creating gaps during PRPA.
We introduce an \textbf{Adaptive Field of View} strategy that uses server-side LSTM-based prediction to dynamically adjust the reference field of view (FoV), balancing visible-region coverage and reference quality.

Based on these designs, we introduce \textbf{Color-Adaptive Gaussian Streaming (CAGS)}, an adaptive VV streaming system compatible with Gaussian representations organized as differential volumetric frames (Sec.~\ref{sec:system}).
This design facilitates integration with diverse Gaussian representations and can benefit from ongoing advances in 3D/4D Gaussian compression.
The contributions of this paper are:

\begin{itemize}
    \item We analyze the interaction between VQ and color restoration, revealing an opportunity to build adaptive VV streaming around a novel Color Adaptation scheme.
    \item We identify key system challenges of bringing Color Adaptation into adaptive VV streaming, and address them with SVQ, PRPA, and Adaptive FoV.
    \item We implement and evaluate a CAGS prototype on volumetric video datasets and real-world network traces, demonstrating 5$\sim$20 dB PSNR gains over existing LoD methods and broad generalizability across Gaussian representations.
\end{itemize} 
\section{Related Work}

\subsection{3D Gaussian Splatting for Volumetric Video Streaming}
As the most popular representation for VV streaming systems~\cite{guanMetaStreamLiveVolumetric2023,liuCaV3CacheassistedViewport2023,zhangHabitusBoostingMobile2024,wangBandwidthEfficientMobileVolumetric2024}, point clouds require high-density data and significant bandwidth (300--800 Mbps) to ensure good visual quality~\cite{hanViVoVisibilityawareMobile2020} due to their discrete nature~\cite{hanViVoVisibilityawareMobile2020} and lack view-dependent effects that are critical for photorealism (e.g., specular highlights)~\cite{wegenSurveyNonphotorealisticRendering2024}.
While Neural Radiance Fields (NeRF)~\cite{mildenhallNeRFRepresentingScenes2021} offer superior fidelity with view-dependent effects, they suffer from severe rendering latency~\cite{shiFullsceneVolumetricVideo2024,yinFSVFGImmersiveFullScene2024}.
Alternatively, 3D Gaussian Splatting (3DGS)~\cite{kerbl3DGaussianSplatting2023} enables real-time photorealistic rendering~\cite{leeCompact3DGaussian2024} and has been extended to dynamic scenes~\cite{luiten2023dynamic,liLoopGaussianCreating3D2024,yan4DGaussianSplatting2024}.
Despite its advantages, the large data footprint of high-dimensional Gaussian attributes~\cite{papantonakisReducingMemoryFootprint2024} poses significant challenges for bandwidth-constrained streaming.

Vector Quantization (VQ) has become popular in recent Gaussian compression and streaming systems~\cite{wangV^3ViewingVolumetric2024,girishQUEENQUantizedEfficient2024,liGIFStream4DGaussianbased2025}, enhancing compression efficiency through advanced codebook designs~\cite{xuGrid4D4DDecomposed2024,wangV^3ViewingVolumetric2024} and clustering techniques~\cite{xuRepresentingLongVolumetric2024,xieSizeGSSizeawareCompression2025}.
Critically, most existing VQ-based methods prioritize compression efficiency for storage, overlooking the scalability indispensable for adaptive streaming~\cite{hanViVoVisibilityawareMobile2020,guanMetaStreamLiveVolumetric2023}.
While other approaches explore scalable compression for 3DGS~\cite{liuCompGSEfficient3D2024,chenHAC100XCompression2025}, they typically involve complex structures that introduce high decoding latency (Table~\ref{tab:sota}), thus hindering real-time streaming capabilities.
To address this gap, we propose a scalable VQ framework that retains the fast decoding speed of VQ while enabling bitrate adaptation for robust VV streaming.

\subsection{Level of Detail for Adaptive Streaming}
Level of Detail (LoD) has long been used to reduce rendering cost~\cite{kerblHierarchical3DGaussian2024,cuiLetsGoLargeScaleGarage2024,yanMultiScale3DGaussian2024}.
In adaptive streaming, LoD also facilitates bitrate adaptation by allowing clients to switch between quality levels according to real-time network conditions.
To maximize visual quality without causing playback stalls, existing adaptive VV streaming systems typically combine visibility-aware transmission~\cite{zhuSGSSStreaming6DoF2025,10.1145/3355089.3356530} with LoD selection~\cite{hanViVoVisibilityawareMobile2020,sunLTSDASHStreaming2025,liuCaV3CacheassistedViewport2023}, prioritizing content that is visible or likely to become visible.
For Gaussian-based representations, most LoD designs are density-based.
Representative approaches include progressive training~\cite{kerblHierarchical3DGaussian2024,scaffoldgs,shiLapisGSLayeredProgressive2025} and importance evaluation based on geometric properties~\cite{fanLightGaussianUnbounded3D2024,papantonakisReducingMemoryFootprint2024}.
However, since each Gaussian occupies considerable space, low-density layers lack sufficient Gaussians to preserve fine details, resulting in significant quality degradation at lower LoDs.
Alternatively, recent scalable neural codecs~\cite{liuCompGSEfficient3D2024,chenHACHashGridAssisted2025} offer LoD-like quality-size scalability, but their complex decoding pipelines incur substantial latency (Table~\ref{tab:sota}), limiting their suitability for real-time interaction.
Other latency-resilient systems~\cite{hladkyQuadStream2022,luQUASARQuadbasedAdaptive2025} approximate views using geometry proxies, relying on server-side rendering, which increases computational load proportionally to client display resolution.
In contrast, our method keeps Gaussian rendering on the client and uses only a fixed low-resolution reference image for color restoration, decoupling server-side rendering cost from the final display resolution.

\subsection{Image Restoration}
Learning-based image restoration has been widely studied for tasks such as super-resolution~\cite{luoLearningDegradationDistribution2022} and color restoration~\cite{bozicVersatileVisionFoundation2024}.
While related techniques have been combined with neural scene representations to accelerate rendering~\cite{wangNeRFSRHighQuality2022,huangRefSRNeRFHighFidelity2023}, their potential for mitigating compression artifacts in 3DGS-based VV streaming remains largely unexplored.
In this paper, we explore example-based color restoration~\cite{congAutomaticControllableColorization2024,zhangDeepExemplarBasedVideo2019} to address the color distortion caused by VQ on Gaussians for VV streaming systems.

\begin{figure*}[t]
	\centering
	\includegraphics[width=\linewidth]{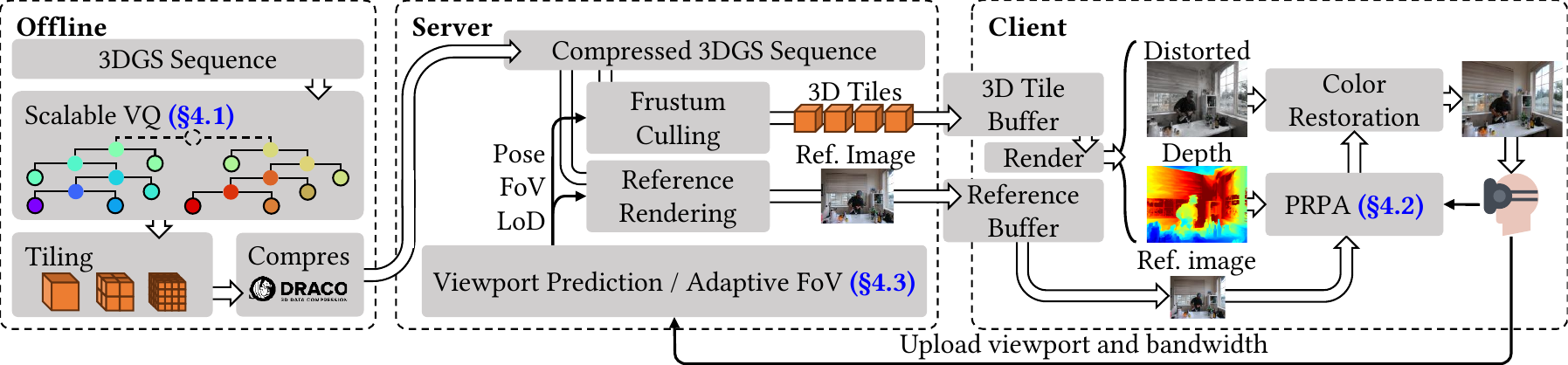}
	\caption{Overview of CAGS.
		The server predicts the viewport, selects tiles and LoDs, renders a low-resolution reference image from the highest-quality layer of the compressed 3DGS, and streams it with Gaussian data.
		The client renders the tiles, aligns the reference image using PRPA, and restores colors for display.
		}\label{figure:overview}\end{figure*}
 \section{Measurement and Motivation}\label{sec:motivate}

To motivate our proposed Color Adaptation scheme, we conduct preliminary experiments to evaluate the effectiveness of example-based color restoration in addressing color distortions introduced by attribute compression.

\subsection{Experimental Setup}\label{sec:motivate-model}
We select three videos from the N3DV dataset~\cite{liNeural3DVideo2022} and train a 3DGS scene on the first frame of each video.
We render each scene from interpolated training viewports at two resolutions: 1600$\times$1200 as ground truth and 400$\times$300 as reference images.
We compress each 3DGS scene using KMeans VQ at five quality levels, followed by lossless compression with Draco~\cite{draco}.
After decompression and dequantization, we render the dequantized scenes from the same viewports to obtain color-distorted images.
We adapt a lightweight SRResNet~\cite{ledigPhotoRealisticSingleImage2017} for restoration by modifying its input layer to support three settings: 1) color restoration from the distorted image, 2) super-resolution from the reference image, and 3) example-based restoration from both images.

\subsection{Measurement Insights}
\begin{figure}[t]
	\centering
	\includegraphics[width=\linewidth]{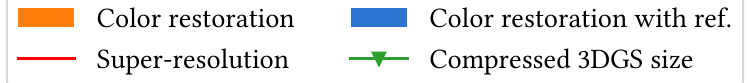}
	\includegraphics[width=\linewidth]{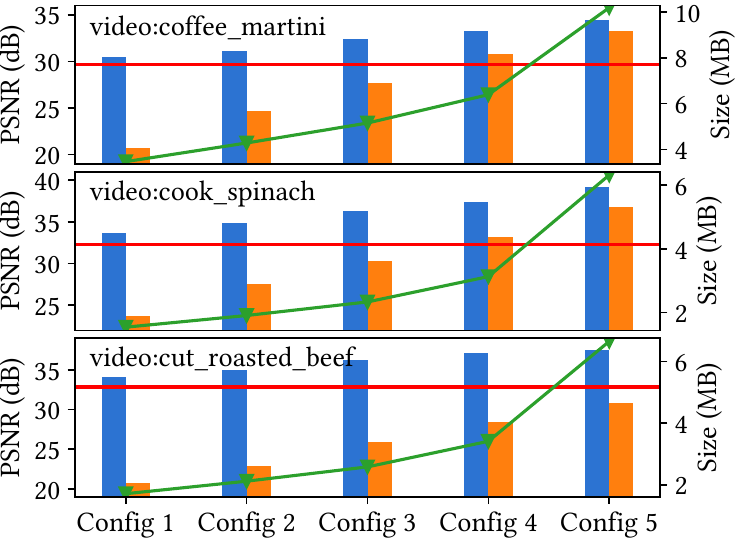}
	\caption{
Comparison of compressed frame size and PSNR for super-resolution, single-image color restoration, and example-based color restoration.
		Configs 1--5 correspond to KMeans VQ settings where scales are quantized to 8/10/12/14/16 bits; rotation and SH are quantized to 4/7/10/13/16 bits; and opacity is fixed at 4 bits.
		}\label{figure:SRMotivation}
\end{figure}

Figure~\ref{figure:SRMotivation} shows that example-based color restoration consistently achieves better visual quality than single-image color restoration and super-resolution.
The results indicate that aggressive VQ mainly damages color while preserving much of the scene structure.
Hence, example-based color restoration leverages preserved structural details to achieve more accurate and visually appealing results.
Moreover, when severe distortion makes the distorted image less trustworthy, example-based restoration relies more on the reference image and resembles super-resolution results.
Thus, super-resolution defines the lower bound of example-based restoration.
These findings highlight the potential benefit of integrating VQ and example-based color restoration into adaptive VV streaming systems.

\section{System Design}
\newcommand{\abswidthtree}{9.53}
\newcommand{\abswidthcodebook}{10.93}
\newcommand{\abswidthcode}{8.09}
\newcommand{\vqfigurescale}{0.94}
\FPeval\widthall{\abswidthtree+\abswidthcodebook+\abswidthcode}
\FPeval\widthtree{\abswidthtree*\vqfigurescale/\widthall}
\FPeval\widthcodebook{\abswidthcodebook*\vqfigurescale/\widthall}
\FPeval\widthcode{\abswidthcode*\vqfigurescale/\widthall}
\newcommand{\vqcaptionskip}{0pt}
\begin{figure*}[t]
	\centering
	\begin{subfigure}[t]{\widthtree\linewidth}
		\captionsetup{skip=\vqcaptionskip}
        \includegraphics[width=\linewidth]{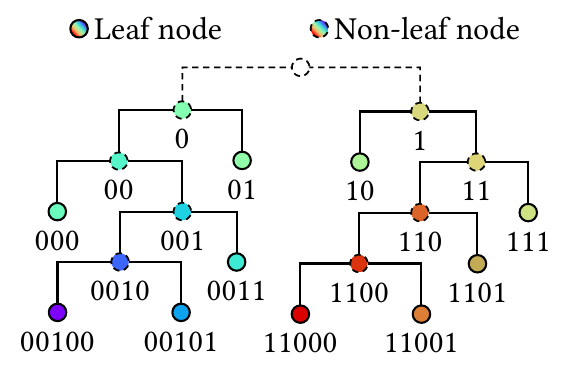}
        \caption{Cluster Tree}\label{fig:vq-tree}
    \end{subfigure}
	\hfill
	\begin{subfigure}[t]{\widthcodebook\linewidth}
		\captionsetup{skip=\vqcaptionskip}
        \includegraphics[width=\linewidth]{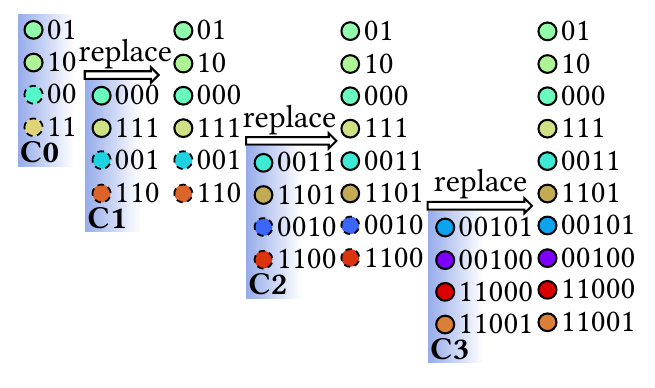}
        \caption{Structure of Codebook}\label{fig:vq-codebook}
    \end{subfigure}
	\hfill
	\begin{subfigure}[t]{\widthcode\linewidth}
		\captionsetup{skip=\vqcaptionskip}
        \includegraphics[width=\linewidth]{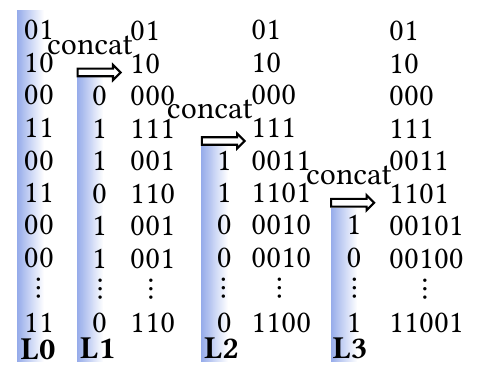}
        \caption{Structure of Quantized Data}\label{fig:vq-code}
    \end{subfigure}\\
	\caption{Illustration of Scalable Vector Quantization (L0: the base layer quantized data; L1--L3: enhancement layers quantized data; C0--C3: corresponding codebook layers; colors indicate cluster centers; each row in (c) represents a Gaussian).}\label{figure:vq}
\end{figure*}
 \label{sec:system}
Figure~\ref{figure:overview} provides an overview of the CAGS pipeline.
CAGS is compatible with Gaussian representations organized as differential frames, where each frame stores only the Gaussians that differ from the previous frame.
In the offline phase, CAGS establishes LoDs by \textbf{Scalable Vector Quantization} (Sec.~\ref{sec:SVQ}).
Each quantized frame is then tiled and compressed with Draco~\cite{draco}.
In the online streaming phase, the server predicts the client viewport from the uploaded viewport history and determines an \textbf{Adaptive Field of View} (Sec.~\ref{sec:adaptive fov}).
Based on the available bandwidth and predicted viewport, the server then renders low-resolution reference images and selects suitable tiles and LoDs.
Low-resolution reference images are encoded as a video stream and streamed together with the selected tiles.
For each frame, the client renders the received tiles to obtain a color-distorted image and a depth map, aligns the reference image using \textbf{Post-Render Perspective Alignment} (Sec.~\ref{sec:PRPA}), and finally applies a lightweight network to restore colors.

\subsection{Scalable Vector Quantization}\label{sec:SVQ}

While KMeans VQ effectively compresses Gaussian attributes~\cite{fanLightGaussianUnbounded3D2024,leeCompact3DGaussian2024}, its flat clustering structure lacks the scalability required for adaptive streaming.
Hierarchical vector quantization has been studied extensively for organizing codebooks across quality levels~\cite{gershoShohamHierarchicalVQ1984}.
A representative method is Agglomerative Hierarchical Clustering (AHC), which builds a cluster tree by repeatedly merging the closest clusters.
Unfortunately, AHC involves computing pairwise distances in each iteration, which is computationally prohibitive for Gaussian representations that typically contain >100k Gaussians per frame.
To bridge this gap, we propose Scalable Vector Quantization (SVQ) that integrates the efficiency of KMeans clustering with the hierarchical structure of AHC, with an index assignment strategy to build a scalable codebook.

\subsubsection{Hierarchical Codebook Construction}\label{sec:initialization}\label{sec:merging}
SVQ first runs KMeans on a random subset of Gaussians, then iteratively merges these clusters into a binary tree, similar to AHC.
To preserve fidelity, we define a cluster distance metric based on the quantization error introduced by merging two clusters $d(C_1, C_2)$:
$$d(C_1,C_2)=\text{mean}(\{c-\text{mean}(C_1\cup C_2)|c\in C_1\cup C_2\})$$
At each iteration, SVQ merges the pair of clusters with the smallest distance, progressively forming a binary tree (Figure~\ref{fig:vq-tree}), until the number of clusters matches the target codebook size.
By starting from KMeans clusters rather than individual Gaussians, SVQ significantly reduces computational complexity compared to AHC.

\subsubsection{Scalable Indexing}
After building the tree, SVQ assigns indices to clusters to create a scalable codebook.
Inspired by Huffman coding, indices are assigned based on cluster positions in the binary tree (Figure~\ref{fig:vq-tree}).
This design provides inherent scalability: truncating lower-order bits naturally maps to its parent cluster at a coarser LoD (Figure~\ref{fig:vq-codebook}).
Consequently, the codebook is organized into a base layer to store higher-order index bits and subsequent enhancement layers to store lower-order bits (Figure~\ref{fig:vq-codebook}).

\begin{figure}[t]
	\centering
	\includegraphics[width=\linewidth]{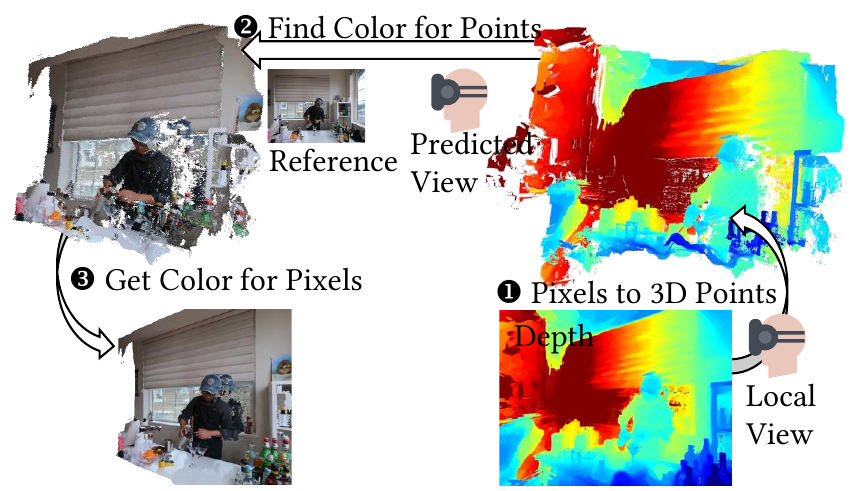}
	\caption{
		PRPA aligns the reference image in three steps:
		\ding{182} unprojects client-view pixels using the depth map,
		\ding{183} reprojects them into the reference view, and
		\ding{184} samples the corresponding colors to produce an aligned image.}\label{figure:aligning}
\end{figure} 
\subsubsection{Dequantization}
Dequantization involves only concatenating the received bits (Figure~\ref{fig:vq-code}) and indexing into the codebook (Figure~\ref{fig:vq-codebook}), and is therefore computationally efficient without becoming a performance bottleneck (Table~\ref{tab:performance}).
 
\subsection{Post-Render Perspective Alignment}\label{sec:PRPA}
To support real-time color restoration at high resolutions, CAGS relies on lightweight models such as SRResNet~\cite{ledigPhotoRealisticSingleImage2017}.
Our evaluation shows that restoration quality depends on accurate alignment between the reference and distorted images (Sec.~\ref{sec:ablation}).

Unfortunately, misalignment is inevitable in server-side rendering.
Due to inherent latency on the Internet (typically 40--90 ms), uploading the client viewport and waiting for the corresponding reference image for every frame would exceed the latency budget for interactive VV streaming (33.3 ms per frame at 30 FPS)~\cite{gulLatencyCompensationImage2022}.
Han et al.~\cite{hanViVoVisibilityawareMobile2020} have shown that practical systems must predict future viewports to satisfy this constraint.
Accordingly, the CAGS server renders low-resolution reference images in advance from the highest-quality level of the compressed Gaussian sequence.
The prediction errors inevitably introduce misalignment between the reference and distorted images.

To mitigate this issue, we propose Post-Render Perspective Alignment (PRPA).
PRPA takes the server-rendered reference image, its rendering viewport, and the client-side depth map produced by the 3DGS rasterizer.
It aligns the reference image to the actual client viewport before color restoration.

\subsubsection{Reference Image Alignment}

As shown in Figure~\ref{figure:aligning}, PRPA reprojects pixels from the client-side depth map to the server-rendered reference and samples the corresponding colors to produce an aligned image.
Note that PRPA fundamentally differs from Depth Image Based Rendering (DIBR)~\cite{wuZGamingZeroLatency3D2023,10.1117/12.524762}.
DIBR warps a reference image using its \textit{own} depth to a target view.
In contrast, PRPA aligns the reference image using the target depth.

\subsubsection{Error Erosion on Occluded Regions}
\begin{figure}[t]
	\newcommand{\includeTrimAligning}[1]{\includegraphics[width=\linewidth,trim={600 0 0 450},clip]{#1}}
	\centering
	\begin{subfigure}[b]{0.493\linewidth}
		\centering
        \includegraphics[width=\linewidth]{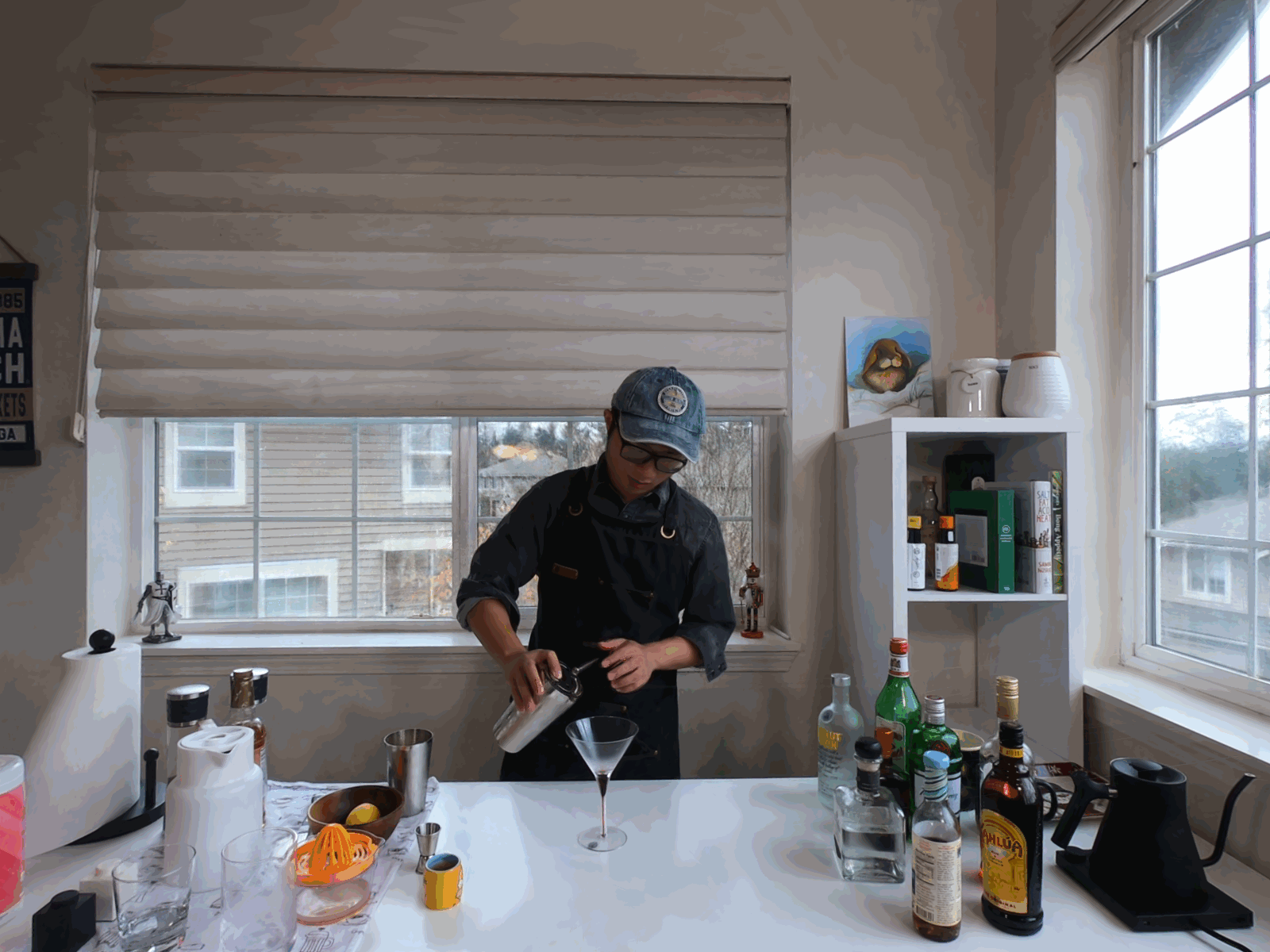}
		\caption{Reference image}\label{figure:reference image}
	\end{subfigure}
\begin{subfigure}[b]{0.493\linewidth}
		\centering
        \includeTrimAligning{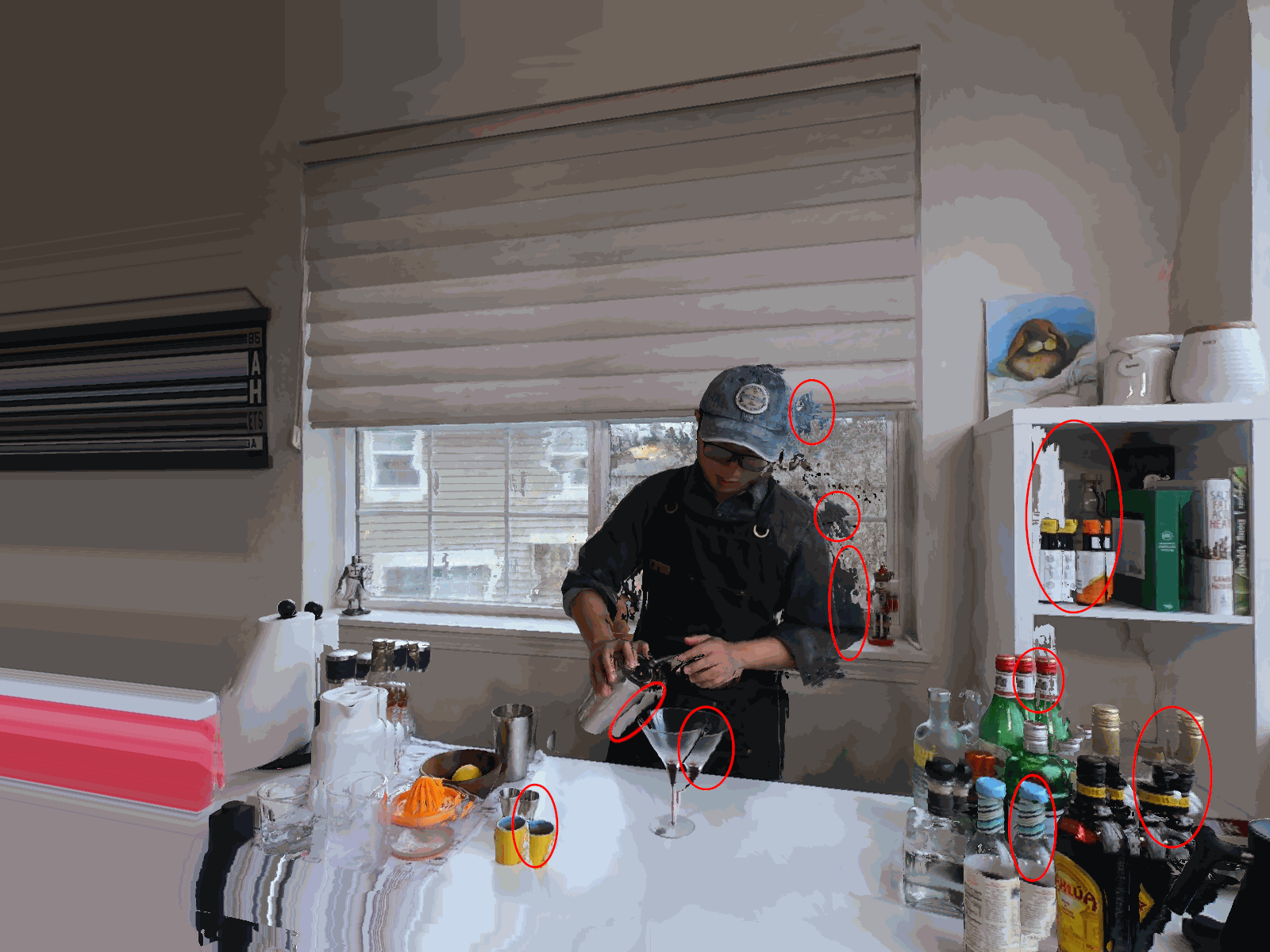}
		\caption{Aligned reference image}\label{figure:alignment error}
	\end{subfigure}
\newcommand{\includEerosion}[1]{\includegraphics[width=\linewidth,trim={2.6 0 2.6 0},clip]{#1}}
	\begin{subfigure}[b]{0.327\linewidth}
		\includEerosion{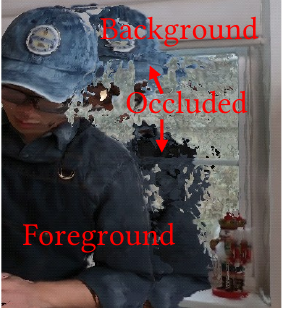}
		\caption{Before erosion}
	\end{subfigure}
	\begin{subfigure}[b]{0.327\linewidth}
		\includEerosion{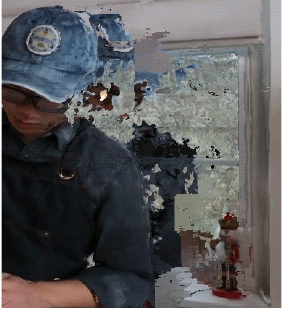}
		\caption{Erosion Step 3}
	\end{subfigure}
	\begin{subfigure}[b]{0.327\linewidth}
		\includEerosion{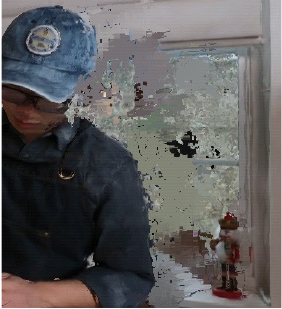}
		\caption{Erosion Step 30}\label{figure:after erosion}
	\end{subfigure}
	\caption{
		Illustration of alignment errors and error erosion in PRPA.
		(b) Aligned reference image before error erosion, with red circles highlighting occlusion artifacts.
		(c)-(e) Iterative error erosion reduces these artifacts.
	}
\end{figure}

Naive alignment may map pixels to occluded regions in the reference view, causing visual artifacts (Figure~\ref{figure:alignment error}).
PRPA reduces these artifacts through error erosion.
During reprojection into the reference view, PRPA records the projected depth of each pixel.
When multiple pixels map to the same reference pixel, only the pixel with the smallest depth is treated as visible, while the others are marked as occluded.
PRPA then iteratively replaces each occluded pixel with the average color of its non-occluded neighbors, yielding a cleaner aligned reference image for color restoration (Figure~\ref{figure:after erosion}).

Implemented with optimized GPU matrix operations, PRPA introduces negligible overhead that does not bottleneck real-time applications (Table~\ref{tab:performance}).
Pseudocode is provided in supplementary.

\subsection{Adaptive Field of View}\label{sec:adaptive fov}

\begin{figure}[t]
	\centering
	\begin{subfigure}[b]{0.493\linewidth}
		\centering
        \includegraphics[width=\linewidth]{figures/aligning/00000.png}
		\caption{Small FoV ref. image}\label{fig:smallfovref}
	\end{subfigure}
	\begin{subfigure}[b]{0.493\linewidth}
		\centering
        \includegraphics[width=\linewidth]{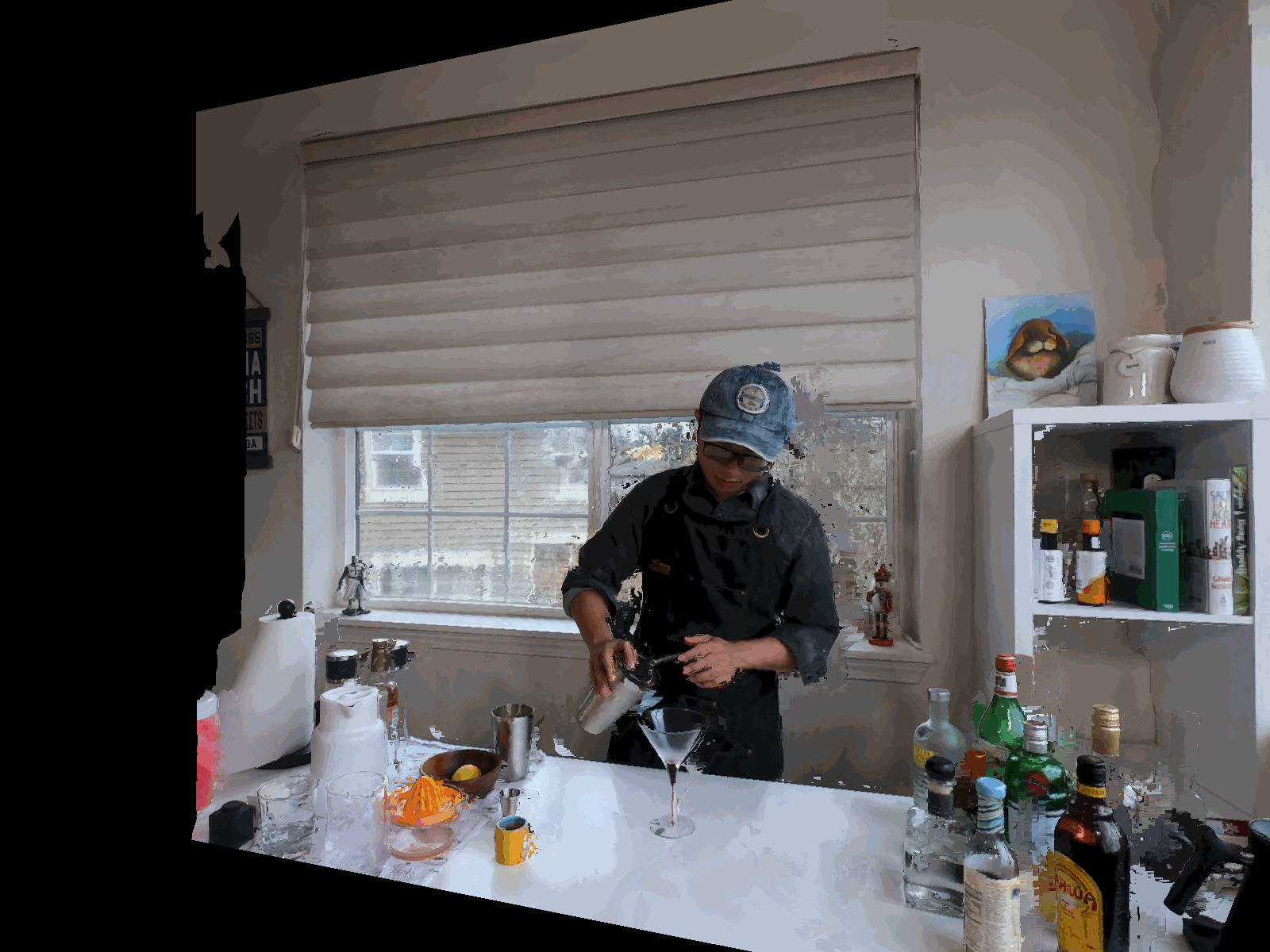}
		\caption{Aligned small FoV ref. image}\label{fig:smallfovrefaligned}
	\end{subfigure}
	\centering
	\begin{subfigure}[b]{0.493\linewidth}
		\centering
        \includegraphics[width=\linewidth]{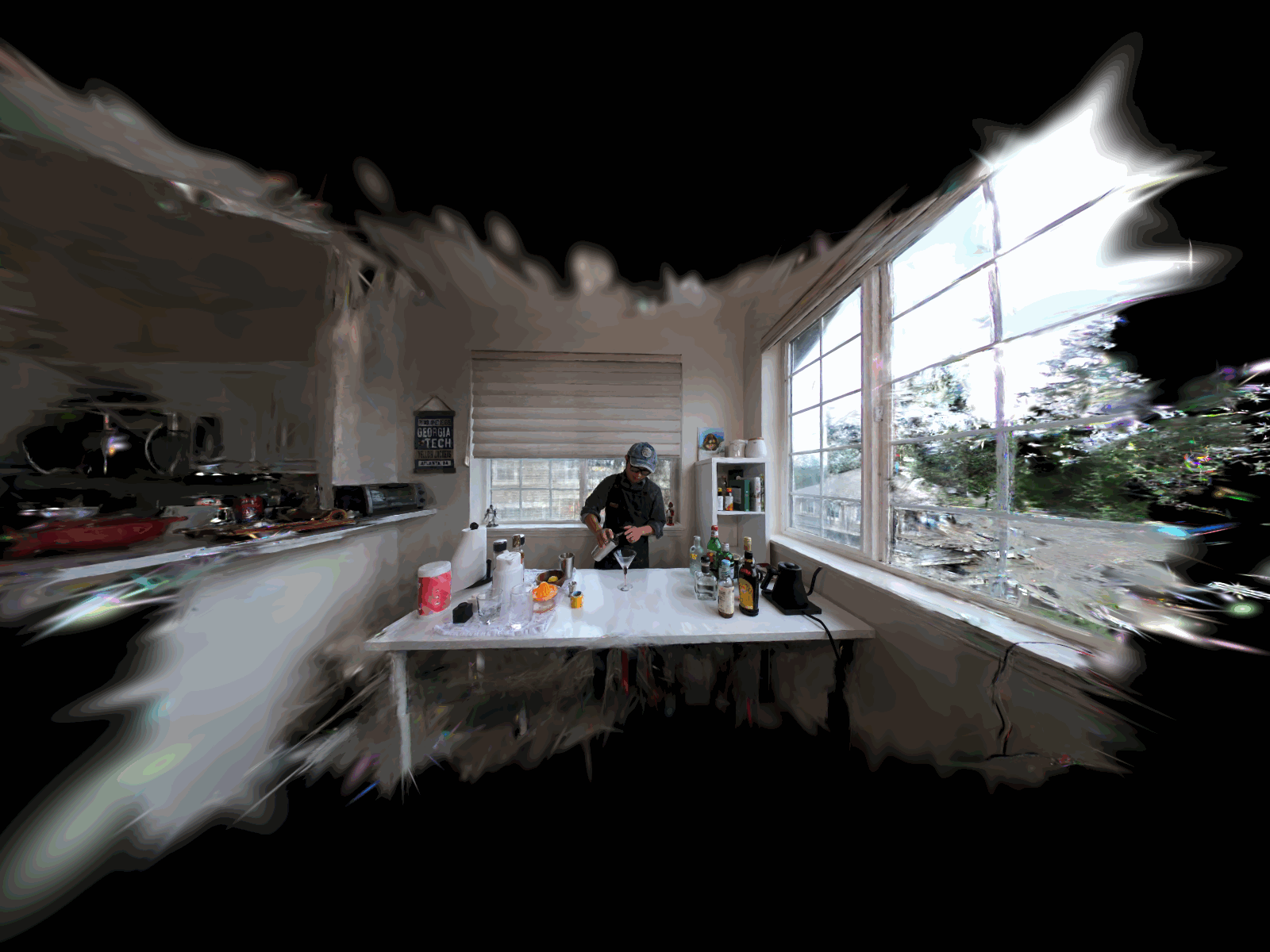}
		\caption{Large FoV ref. image}\label{fig:largefovref}
	\end{subfigure}
	\begin{subfigure}[b]{0.493\linewidth}
		\centering
        \includegraphics[width=\linewidth]{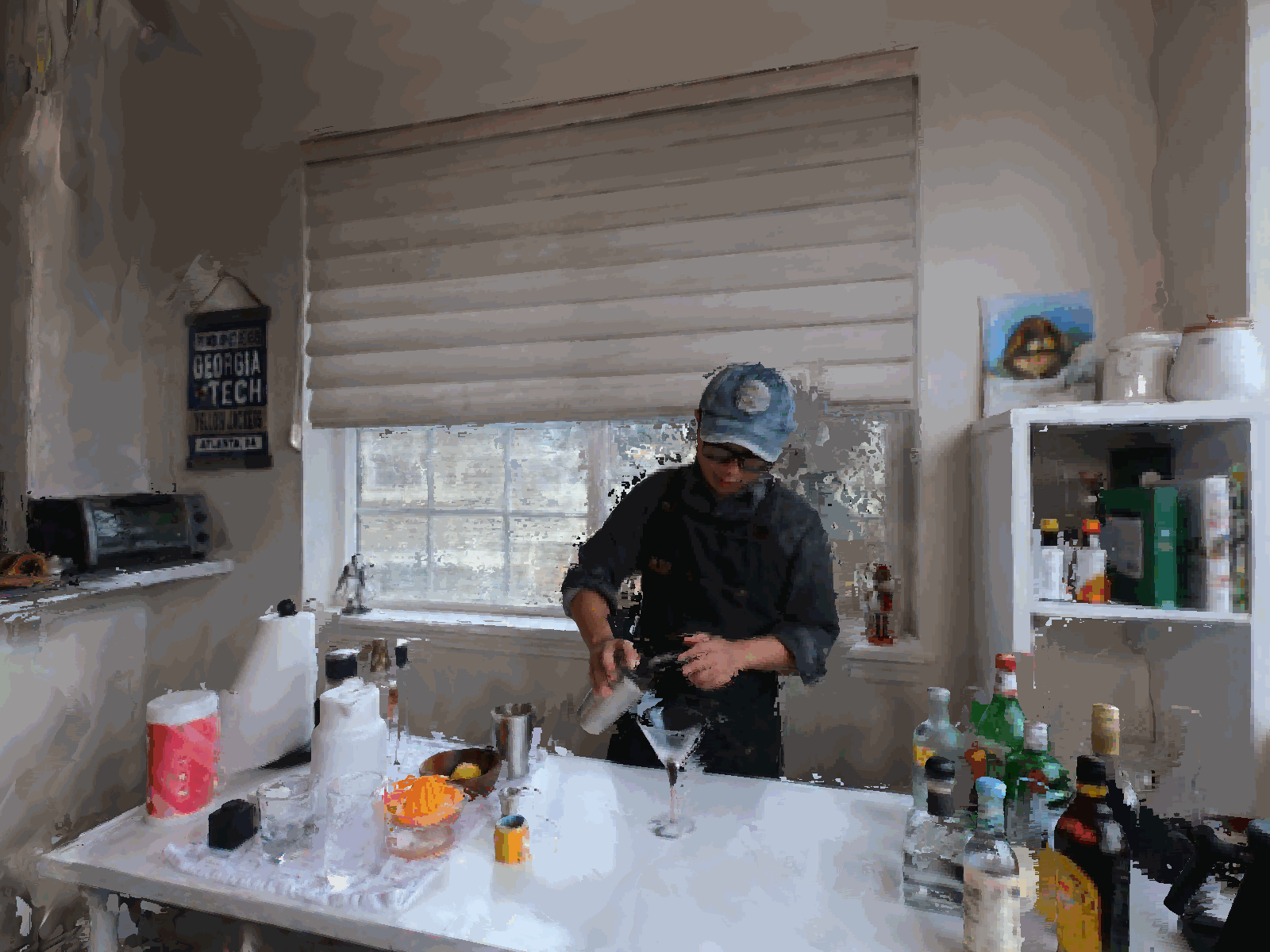}
		\caption{Aligned large FoV ref. image}\label{fig:largefovrefaligned}
	\end{subfigure}
	\caption{
		Visualization of PRPA results with different reference FoVs.
		A small FoV (a) misses boundary content (b), whereas a large FoV (c) preserves coverage but reduces pixel density in the aligned reference image (d).
	}
\end{figure}

Viewport prediction errors may cause the reference image to cover only part of the client viewport, leaving missing regions after PRPA (Figure~\ref{fig:smallfovrefaligned}) and degrading restoration quality.
While enlarging the reference FoV improves coverage (Figure~\ref{fig:largefovref}), it reduces pixel density in target regions (Figure~\ref{fig:largefovrefaligned}), which also hurts quality.

To balance coverage and density, we propose an Adaptive FoV strategy driven by a lightweight LSTM model.
We choose LSTM because it is efficient and effective for sequential, latency-sensitive prediction.
Though complex backbones like Transformers could achieve similar accuracy, they usually require additional optimization for real-time use.
Notably, our strategy is decoupled from the viewport prediction, providing a plug-and-play solution compatible with various prediction methods~\cite{hanViVoVisibilityawareMobile2020,liuCaV3CacheassistedViewport2023}.
In our evaluation, we use autoregression for viewport prediction.

\subsubsection{LSTM-based Adaptive FoV Model}
The LSTM predicts a scaling factor $s_i=(s_i^{x},s_i^{y})$ for frame $i$, which adjusts the reference FoV $(F^{x},F^{y})$ relative to the fixed client viewport FoV $(F_0^{x},F_0^{y})$, such that $F^{x}=(1+s_i^{x})F_0^{x}$ and $F^{y}=(1+s_i^{y})F_0^{y}$.
The model takes the client viewport (rotation $q_i$ and position $p_i$), their temporal changes, and historical states as inputs.
Formally, the predicted scaling factor $\hat s_i$ and hidden state $h_i$ are computed as:
$$(\hat s_i,h_i)=\text{LSTM}(q_i,q_i-q_{i-1},p_i,p_i-p_{i-1},s_{i-1}^a,h_{i-1})$$
where $s_{i-1}^a$ denotes the approximate ground-truth FoV scaling factor from the previous frame (Sec.~\ref{sec:approximate fov}).

In a real streaming system, viewport predictions typically span multiple future frames~\cite{liuCaV3CacheassistedViewport2023}, where actual client viewports are unavailable.
In this case, the model uses predicted viewports as $q_i$ and $p_i$, and uses the previously predicted FoV scaling factor $\hat s_{i-1}$ as $s_{i-1}^a$.
To limit error accumulation, we refresh the LSTM hidden state whenever actual client viewport updates arrive.

\subsubsection{Fast Approximate Ground-truth FoV}\label{sec:approximate fov}
Computing the exact ground-truth FoV requires projecting all pixels from the reference image into the client viewport and then finding the smallest FoV that covers them, which is computationally prohibitive.
Instead, we approximate the ground-truth FoV by projecting only the four corner pixels of the reference image into the client viewport at a fixed depth (e.g., 10 m).
Given that viewport prediction errors are usually small over short intervals~\cite{hanViVoVisibilityawareMobile2020}, this approximation can greatly reduce computation while maintaining sufficient accuracy.
Since the server cannot access the actual client viewport in real-time, this approximate FoV is only used for offline supervision and to update $s_i^a$ once client viewport updates arrive.
The FoV for reference rendering is predicted by the Adaptive FoV model.

\subsubsection{LSTM Model Training}\label{sec:fovtraining}
We train the LSTM model offline using collected viewport datasets.
First, we run the viewport prediction on ground-truth viewports to obtain predicted viewports.
Then we compute the ground-truth FoV scaling factors $s_i$ and their approximations $s_i^a$ using the method above.
The training loss $\mathcal L$ is computed between $\hat s_i$ and $s_i$ over all $N$ frames:

$$\mathcal L=\frac{1}{N}\sum_{i=1}^{N}|\hat{s}_i - s_i|$$

\section{Evaluation}
\subsection{Prototype Implementation}
\label{sec:implement}
\label{sec:prepare}
\label{sec:abr}

We implemented a CAGS prototype to evaluate its effectiveness, following prior VV streaming practice~\cite{sunLTSDASHStreaming2025,wangV^3ViewingVolumetric2024}.
We prepare volumetric videos using TrackerSplat~\cite{yinTrackerSplatExploitingPoint2025} and apply importance-based pruning~\cite{papantonakisReducingMemoryFootprint2024,zhouGaussianSplattingNeural2024} to remove redundant Gaussians.
To enable adaptive streaming, we build a linear LoD hierarchy by interleaving SVQ layers according to the empirically measured visual impact of Gaussian attributes, prioritizing high-impact layers at lower LoDs.
We tile Gaussians via Morton sorting~\cite{jiangTopologyAwareOptimizationGaussian2025} to balance the number of Gaussians per tile.
Spatial data and base LoD are compressed with Draco~\cite{draco}, while enhancement layers and the codebook are compressed with Gzip.
During streaming, the server applies a bandwidth-aware adaptation strategy that prioritizes tiles covering visible Gaussians (identified via reference rendering) and progressively raises their LoDs until reaching the bandwidth limit.
Our prototype is deployed on consumer-grade GPUs (RTX 3080) for both server and client, streaming at 30 FPS while rendering at 60 FPS on the client side.
Additional engineering details and underlying rationales are provided in the supplementary. 

\subsection{Evaluation Setup}

\subsubsection{Dataset}\label{sec:dataset}

We evaluate CAGS on four datasets with diverse motion patterns, scene scales, and capture resolutions: Neural 3D Video Synthesis (N3DV)~\cite{liNeural3DVideo2022}, ST-NeRF~\cite{zhangEditableFreeviewpointVideo2021}, Meeting Room~\cite{liStreamingRadianceFields2022}, and Dynamic 3DGS~\cite{luiten2023dynamic}.
We adapt UnityGS~\cite{UnityGS} and develop an application to record viewport traces using a Meta Quest 3.
The traces used to train color restoration and FoV prediction are collected independently from those used for evaluation.

\subsubsection{Metrics}
We render the original uncompressed Gaussians at the corresponding viewports as ground truth.
We report PSNR, SSIM, and LPIPS~\cite{zhang2018perceptual}.

\subsubsection{Network Settings}
\begin{figure}[htbp]
	\centering
	\includegraphics[width=\linewidth]{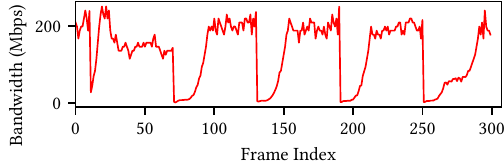}
	\caption{Throughput of the selected network trace from the 5Gophers dataset (lines 3249 to 3549, 1.47--251.7 Mbps, average 133.7 Mbps).}\label{figure:bandwidth}
\end{figure}

We evaluate both fixed and dynamic bandwidth settings.
For fixed bandwidth, we set the bandwidth limit (Sec.~\ref{sec:abr}) to 30, 60, 90, 120, and 150 Mbps.
For dynamic bandwidth, we use a representative segment from 5Gophers~\cite{narayananFirstLookCommercial2020}, as shown in Figure~\ref{figure:bandwidth}.

\subsubsection{Baselines}
We compare CAGS with two state-of-the-art Gaussian-based VV streaming systems and two restoration variants, specifically focusing on methods that are compatible with the real-time client loop required for interactive streaming.

\noindent\textbf{LTS-F}:
LTS~\cite{sunLTSDASHStreaming2025} is an adaptive VV streaming system based on density-based LoD.
We use its released LoD component and implement its corresponding network adaptation strategy as the baseline LTS-F.
For consistency with Sec.~\ref{sec:implement}, we encode the sequence into independently decodable Groups of Frames (GoF) and select the LoD for each frame during streaming.

\noindent\textbf{V$^3$-A}:
V$^3$~\cite{wangV^3ViewingVolumetric2024} is a VV streaming system based on VQ (hash grids trained with entropy loss).
It does not support LoD and thus lacks a built-in network adaptation strategy.
For the V$^3$-A baseline, we integrate its released code as a compression module into our pipeline (including PRPA and adaptive FoV for fairness), and implement network adaptation in Sec.~\ref{sec:abr}, except that already-transmitted tiles are resent at higher quality until the bandwidth limit is reached.

\noindent\textbf{SR align ref.}: Uses SRResNet~\cite{ledigPhotoRealisticSingleImage2017} for super-resolution on the reference image aligned by PRPA.

\noindent\textbf{CR w/o ref.}: Performs color restoration (Sec.~\ref{sec:motivate-model}) directly on the distorted image without a reference.

\subsection{Evaluation Results}

\subsubsection{Evaluation of Scalable Vector Quantization}

Before evaluating the full system, we benchmark SVQ against state-of-the-art scalable compression methods: SPZ~\cite{spz}, CompGS~\cite{liuCompGSEfficient3D2024}, HAC~\cite{chenHACHashGridAssisted2025}, and HAC++~\cite{chenHAC100XCompression2025}.
As shown in Table~\ref{tab:sota}, SVQ achieves comparable rate-distortion performance to these methods and outperforms them in decoding latency, which is uniquely suited for real-time streaming.

\subsubsection{Evaluation under Fixed Bandwidth}
Figure~\ref{fig:FixedBandwidth} reports the average PSNR and SSIM under fixed bandwidth constraints.
CAGS achieves higher PSNR than the baselines in most cases, indicating better color fidelity.
LTS-F is limited by density-based LoD, which provides a weaker rate--quality trade-off and therefore selects lower-quality LoDs under the same bandwidth.
V$^3$-A lacks scalability and causes redundant transmissions, resulting in lower quality.

In some cases such as the ``walking'' (Figure~\ref{fig:walking}), LTS-F achieves higher SSIM at high bandwidth.
Analysis shows this sequence contains fewer Gaussians, allowing density-based LoD to approach uncompressed quality when bandwidth is sufficient, while CAGS and V$^3$-A remain bounded by lossy VQ compression.

\subsubsection{Evaluation under Fluctuating Bandwidth}\label{sec:dynamicbandwidth}
Figure~\ref{fig:DynamicBandwidth} reports per-frame PSNR/SSIM under the fluctuating 5G trace.
CAGS achieves higher and more stable quality over time, demonstrating robustness to bandwidth variations.
Similar to fixed bandwidth results, the ``walking'' sequence favors the baseline due to its fewer Gaussians.

Figure~\ref{fig:DynamicBandwidth} also reveals occasional quality drops (e.g., frame 32 of ``basketball'' and frame 240 of ``discussion'').
These drops are caused by rapid head movements, which increase viewport prediction errors and lead to either large missing regions in the PRPA output or overly large FoV predictions.
While noticeable in measurements, these degradations have minimal perceptual impact because: 1) they are short-lived, since humans cannot move rapidly for long, and once movement stabilizes, prediction errors drop and quality quickly recovers; and 2) rapid movements naturally limit human visual perception, making such temporary degradations less noticeable.

\begin{table}[t]
\caption{
    Performance and quality comparison of our SVQ method (with 66-bit and 76-bit initialization) against state-of-the-art \textbf{scalable} compression methods at the highest quality level.
Best and second-best results are highlighted in \textbf{bold} and \underline{underline}, respectively.
    Size is in MB and decoding time ("dec.t") in seconds.
    Full results including SSIM and LPIPS are provided in the supplementary material.
}
\begin{center}
\setlength{\tabcolsep}{1.5pt}
\resizebox{\columnwidth}{!}{\begin{tabular}{l|crr|crr|crr}
\hline
Datasets & 
\multicolumn{3}{c|}{\textbf{Neural3DV}} & 
\multicolumn{3}{c|}{\textbf{Meet.Room}} &
\multicolumn{3}{c}{\textbf{Dyn.3DGS}} \\
Methods & 
psnr & size & dec.t & 
psnr & size & dec.t & 
psnr & size & dec.t \\
\hline
SPZ low & 23.3 & 3.33 & 0.076 & 21.8 & 2.54 & 0.062 & 21.1 & 2.26 & 0.052 \\
SPZ high & 23.1 & 4.53 & 0.075 & 21.8 & 3.41 & 0.057 & 21.0 & 3.32 & 0.057 \\
CompGS & \underline{24.9} & 22.20 & 6.213 & 25.0 & 5.58 & 0.799 & 19.1 & 4.27 & 2.772 \\
HAC & 22.2 & 25.69 & 24.407 & 25.2 & 8.69 & 9.135 & 20.5 & 2.84 & 1.571 \\
HAC++ & 21.5 & 17.98 & 51.708 & 25.2 & 6.78 & 14.706 & 20.9 & \textbf{2.00} & 6.169 \\
\hline
SVQ 66bit & 24.5 & \textbf{2.00} & \textbf{0.015} & \underline{26.0} & \textbf{1.53} & \textbf{0.018} & \textbf{25.3} & \underline{2.25} & \textbf{0.028} \\
SVQ 76bit & \textbf{25.1} & \underline{2.14} & \underline{0.015} & \textbf{26.4} & \underline{1.69} & \underline{0.019} & \underline{23.3} & 2.42 & \underline{0.029} \\
\hline
\end{tabular}}
\label{tab:sota}
\end{center}
\end{table} 
\subsubsection{Visualization Results}

Figure~\ref{fig:visual} shows a representative example of color restoration.
The color distortion in Figure~\ref{fig:visualdistorted} is effectively corrected in Figure~\ref{fig:visualrestored}, demonstrating the effectiveness of our restoration method.

\subsubsection{System Performance}
We profile key components on both server and client.
Table~\ref{tab:performance} summarizes the profiling results, demonstrating that CAGS supports real-time streaming and rendering.

\subsection{Ablation Studies}\label{sec:ablation}
We conduct two ablation studies to quantify the contributions of key components:
1) \textbf{w/o PRPA}: feeds the misaligned reference image directly to restoration.
2) \textbf{w/o Adaptive FoV}: fixes the reference rendering FoV to 10\% larger than the client viewport.

Figures~\ref{fig:FixedBandwidth} and~\ref{fig:DynamicBandwidth} show that removing PRPA leads to a significant quality drop, confirming that accurate reference--target alignment is essential for effective restoration.
Adaptive FoV also provides meaningful improvements, indicating that sufficient boundary coverage is important for reducing the impact of viewport prediction errors.
These components effectively handle viewport prediction errors:
PRPA aligns reference images rendered from predicted viewpoints with the actual viewport, enabling accurate color restoration.
Adaptive FoV expands the reference FoV according to head motion, ensuring complete coverage.
Together, they ensure the quality stability observed in Sec.~\ref{sec:dynamicbandwidth}.

\subsection{Generality to Other Representations}
CAGS is designed for Gaussian representations where each frame stores the Gaussians that differ from the previous frame.
To validate its generality, we integrate CAGS with three representative methods for preparing volumetric videos: Dynamic 3DGS~\cite{luiten2023dynamic}, 4DGS~\cite{wu4DGaussianSplatting2024} and HiCoM~\cite{gaoHiCoMHierarchicalCoherent2024}, and evaluate them under the same bandwidth trace using the baselines (SR align ref., and CR w/o ref.) and ablations (w/o PRPA, w/o Adaptive FoV).

Figures~\ref{fig:DynamicBandwidth-dynamic3dgs} and~\ref{fig:DynamicBandwidth-hicom} show that CAGS consistently improves quality across different preparation methods.
These results suggest that CAGS can extend beyond the tested methods, potentially supporting a wide range of 3D/4D Gaussian representations and benefiting from ongoing evolutions in 3D/4D Gaussian compression.

\begin{table}[t]
\caption{Performance of system components.}
\begin{center}
\begin{tabular}{llr}
\hline
& \textbf{Component} & \textbf{Time}\\
\hline
\multirow{2}{*}{\makecell[l]{\textbf{Encoding (Offline)}}} 
& SVQ Codebook (per-video)  & 36.8 s \\
& SVQ \& Draco Encoding & 370 ms \\
\hline
\multirow{3}{*}{\textbf{Server}} 
& Viewport Prediction & 1 ms \\
& Dynamic FoV & 1 ms \\
& Rendering (400x300) & 1.7 ms \\
\hline
\multirow{2}{*}{\textbf{Client Decoding}} 
& Draco Decoding & 9 ms \\
& SVQ Decoding & 2.58 ms \\
\hline
\multirow{3}{*}{\makecell[l]{\textbf{Client Rendering}\\(1600x1200)}} 
 & Render Distorted+Depth & 9.32 ms \\
& PRPA & 2.11 ms \\
& Color Restoration & 6.5 ms \\ \hline
\multirow{3}{*}{\makecell[l]{\textbf{Client Rendering}\\(1600x1200, RTX 4090)}}
 & Render Distorted+Depth & 1.21 ms \\
& PRPA & 1 ms \\
& Color Restoration & 3.3 ms \\ \hline
\end{tabular}
\label{tab:performance}
\end{center}
\end{table}

\section{Limitation and Future Work}

CAGS has certain limitations and leaves room for improvement.

Error erosion reduces color distortion in occluded regions but still leaves artifacts in the PRPA output.
To ensure real-time performance, we did not complicate our restoration model to specifically handle these artifacts.
Although our lightweight restoration model can learn to suppress many of them during training, small artifacts may still remain (can be observed in our provided video results).
Future work can explore artifact handling without sacrificing speed.

Although our VQ design is developed for Gaussian representations, VQ itself is more general and can also be applied to other 3D representations.
Prior works have shown its effectiveness for NeRFs~\cite{zhongVQNeRFNeuralReflectance2024} and also report color distortion as a side effect~\cite{takikawaVariableBitrateNeural2022}, suggesting that such distortion may be a common issue inherent to VQ across different 3D representations.
This highlights the broader potential of the Color Adaptation scheme beyond Gaussian-based streaming: it could serve as a general solution for streaming 3D content across diverse scene representations.
Future work can explore integrating the color-adaptive scheme with a wider range of 3D representations. 
\section{Conclusion}

In this paper, we present a Color Adaptation scheme for volumetric video streaming based on dynamic 3D Gaussian Splatting.
We first identify limitations of existing LoD methods for Gaussian representations.
We reveal that vector quantization primarily causes color distortion, which can be effectively corrected using reference images.
We also identify the challenges of implementing Color Adaptation in real systems and address them with the CAGS system.
Extensive evaluation of our prototype demonstrates the effectiveness and efficiency of our design.

\balance \bibliographystyle{ACM-Reference-Format}

\begin{thebibliography}{73}



\ifx \showCODEN    \undefined \def \showCODEN     #1{\unskip}     \fi
\ifx \showISBNx    \undefined \def \showISBNx     #1{\unskip}     \fi
\ifx \showISBNxiii \undefined \def \showISBNxiii  #1{\unskip}     \fi
\ifx \showISSN     \undefined \def \showISSN      #1{\unskip}     \fi
\ifx \showLCCN     \undefined \def \showLCCN      #1{\unskip}     \fi
\ifx \shownote     \undefined \def \shownote      #1{#1}          \fi
\ifx \showarticletitle \undefined \def \showarticletitle #1{#1}   \fi
\ifx \showURL      \undefined \def \showURL       {\relax}        \fi
\providecommand\bibfield[2]{#2}
\providecommand\bibinfo[2]{#2}
\providecommand\natexlab[1]{#1}
\providecommand\showeprint[2][]{arXiv:#2}

\bibitem[Bozic et~al\mbox{.}(2024)]{bozicVersatileVisionFoundation2024}
\bibfield{author}{\bibinfo{person}{Vukasin Bozic}, \bibinfo{person}{Abdelaziz Djelouah}, \bibinfo{person}{Yang Zhang}, \bibinfo{person}{Radu Timofte}, \bibinfo{person}{Markus Gross}, {and} \bibinfo{person}{Christopher Schroers}.} \bibinfo{year}{2024}\natexlab{}.
\newblock \showarticletitle{Versatile {{Vision Foundation Model}} for {{Image}} and {{Video Colorization}}}. In \bibinfo{booktitle}{\emph{{{ACM SIGGRAPH}} 2024 {{Conference Papers}}}} \emph{(\bibinfo{series}{{{SIGGRAPH}} '24})}. \bibinfo{pages}{1--11}.
\newblock


\bibitem[Chen et~al\mbox{.}(2025a)]{chenHACHashGridAssisted2025}
\bibfield{author}{\bibinfo{person}{Yihang Chen}, \bibinfo{person}{Qianyi Wu}, \bibinfo{person}{Weiyao Lin}, \bibinfo{person}{Mehrtash Harandi}, {and} \bibinfo{person}{Jianfei Cai}.} \bibinfo{year}{2025}\natexlab{a}.
\newblock \showarticletitle{{{HAC}}: {{Hash-Grid Assisted Context}} for~{{3D Gaussian Splatting Compression}}}. In \bibinfo{booktitle}{\emph{Computer {{Vision}} -- {{ECCV}} 2024}}, \bibfield{editor}{\bibinfo{person}{Ale{\v s} Leonardis}, \bibinfo{person}{Elisa Ricci}, \bibinfo{person}{Stefan Roth}, \bibinfo{person}{Olga Russakovsky}, \bibinfo{person}{Torsten Sattler}, {and} \bibinfo{person}{G{\"u}l Varol}} (Eds.). \bibinfo{pages}{422--438}.
\newblock


\bibitem[Chen et~al\mbox{.}(2025b)]{chenHAC100XCompression2025}
\bibfield{author}{\bibinfo{person}{Yihang Chen}, \bibinfo{person}{Qianyi Wu}, \bibinfo{person}{Weiyao Lin}, \bibinfo{person}{Mehrtash Harandi}, {and} \bibinfo{person}{Jianfei Cai}.} \bibinfo{year}{2025}\natexlab{b}.
\newblock \bibinfo{title}{{{HAC}}++: {{Towards 100X Compression}} of {{3D Gaussian Splatting}}}.
\newblock
\href{https://doi.org/10.48550/arXiv.2501.12255}{doi:\nolinkurl{10.48550/arXiv.2501.12255}}


\bibitem[Cong et~al\mbox{.}(2024)]{congAutomaticControllableColorization2024}
\bibfield{author}{\bibinfo{person}{Xiaoyan Cong}, \bibinfo{person}{Yue Wu}, \bibinfo{person}{Qifeng Chen}, {and} \bibinfo{person}{Chenyang Lei}.} \bibinfo{year}{2024}\natexlab{}.
\newblock \showarticletitle{Automatic {{Controllable Colorization}} via {{Imagination}}}. In \bibinfo{booktitle}{\emph{Proceedings of the {{IEEE}}/{{CVF Conference}} on {{Computer Vision}} and {{Pattern Recognition}}}}. \bibinfo{pages}{2609--2619}.
\newblock


\bibitem[Cui et~al\mbox{.}(2024)]{cuiLetsGoLargeScaleGarage2024}
\bibfield{author}{\bibinfo{person}{Jiadi Cui}, \bibinfo{person}{Junming Cao}, \bibinfo{person}{Fuqiang Zhao}, \bibinfo{person}{Zhipeng He}, \bibinfo{person}{Yifan Chen}, \bibinfo{person}{Yuhui Zhong}, \bibinfo{person}{Lan Xu}, \bibinfo{person}{Yujiao Shi}, \bibinfo{person}{Yingliang Zhang}, {and} \bibinfo{person}{Jingyi Yu}.} \bibinfo{year}{2024}\natexlab{}.
\newblock \showarticletitle{{{LetsGo}}: {{Large-Scale Garage Modeling}} and {{Rendering}} via {{LiDAR-Assisted Gaussian Primitives}}}. In \bibinfo{booktitle}{\emph{{{SIGGRAPH Asia}} 2024 {{Conference Papers}}}} \emph{(\bibinfo{series}{{{SA}} '24})}.
\newblock


\bibitem[de~Koning et~al\mbox{.}(2023)]{de2023digital}
\bibfield{author}{\bibinfo{person}{Koen de Koning}, \bibinfo{person}{Jeroen Broekhuijsen}, \bibinfo{person}{Ingolf K{\"u}hn}, \bibinfo{person}{Otso Ovaskainen}, \bibinfo{person}{Franziska Taubert}, \bibinfo{person}{Dag Endresen}, \bibinfo{person}{Dmitry Schigel}, {and} \bibinfo{person}{Volker Grimm}.} \bibinfo{year}{2023}\natexlab{}.
\newblock \showarticletitle{Digital twins: dynamic model-data fusion for ecology}.
\newblock \bibinfo{journal}{\emph{Trends in ecology \& evolution}} \bibinfo{volume}{38}, \bibinfo{number}{10} (\bibinfo{year}{2023}), \bibinfo{pages}{916--926}.
\newblock


\bibitem[Fan et~al\mbox{.}(2024)]{fanLightGaussianUnbounded3D2024}
\bibfield{author}{\bibinfo{person}{Zhiwen Fan}, \bibinfo{person}{Kevin Wang}, \bibinfo{person}{Kairun Wen}, \bibinfo{person}{Zehao Zhu}, \bibinfo{person}{Dejia Xu}, {and} \bibinfo{person}{Zhangyang Wang}.} \bibinfo{year}{2024}\natexlab{}.
\newblock \showarticletitle{{{LightGaussian}}: {{Unbounded 3D Gaussian Compression}} with 15x {{Reduction}} and 200+ {{FPS}}}.
\newblock \bibinfo{journal}{\emph{Advances in Neural Information Processing Systems}}  \bibinfo{volume}{37} (\bibinfo{year}{2024}), \bibinfo{pages}{140138--140158}.
\newblock


\bibitem[Fehn(2004)]{10.1117/12.524762}
\bibfield{author}{\bibinfo{person}{Christoph Fehn}.} \bibinfo{year}{2004}\natexlab{}.
\newblock \showarticletitle{{Depth-image-based rendering (DIBR), compression, and transmission for a new approach on 3D-TV}}. In \bibinfo{booktitle}{\emph{Stereoscopic Displays and Virtual Reality Systems XI}}, Vol.~\bibinfo{volume}{5291}. International Society for Optics and Photonics, \bibinfo{publisher}{SPIE}, \bibinfo{pages}{93 -- 104}.
\newblock


\bibitem[Gao et~al\mbox{.}(2024)]{gaoHiCoMHierarchicalCoherent2024}
\bibfield{author}{\bibinfo{person}{Qiankun Gao}, \bibinfo{person}{Jiarui Meng}, \bibinfo{person}{Chengxiang Wen}, \bibinfo{person}{Jie Chen}, {and} \bibinfo{person}{Jian Zhang}.} \bibinfo{year}{2024}\natexlab{}.
\newblock \showarticletitle{{{HiCoM}}: {{Hierarchical Coherent Motion}} for {{Dynamic Streamable Scenes}} with {{3D Gaussian Splatting}}}. In \bibinfo{booktitle}{\emph{The {{Thirty-eighth Annual Conference}} on {{Neural Information Processing Systems}}}}.
\newblock


\bibitem[Gasques et~al\mbox{.}(2021)]{gasques2021artemis}
\bibfield{author}{\bibinfo{person}{Danilo Gasques}, \bibinfo{person}{Janet~G Johnson}, \bibinfo{person}{Tommy Sharkey}, \bibinfo{person}{Yuanyuan Feng}, \bibinfo{person}{Ru Wang}, \bibinfo{person}{Zhuoqun~Robin Xu}, \bibinfo{person}{Enrique Zavala}, \bibinfo{person}{Yifei Zhang}, \bibinfo{person}{Wanze Xie}, \bibinfo{person}{Xinming Zhang}, {et~al\mbox{.}}} \bibinfo{year}{2021}\natexlab{}.
\newblock \showarticletitle{Artemis: A collaborative mixed-reality system for immersive surgical telementoring}. In \bibinfo{booktitle}{\emph{Proceedings of the 2021 CHI conference on human factors in computing systems}}. \bibinfo{pages}{1--14}.
\newblock


\bibitem[Gersho and Shoham(1984)]{gershoShohamHierarchicalVQ1984}
\bibfield{author}{\bibinfo{person}{A. Gersho} {and} \bibinfo{person}{Y. Shoham}.} \bibinfo{year}{1984}\natexlab{}.
\newblock \showarticletitle{Hierarchical vector quantization of speech with dynamic codebook allocation}. In \bibinfo{booktitle}{\emph{ICASSP '84. IEEE International Conference on Acoustics, Speech, and Signal Processing}}, Vol.~\bibinfo{volume}{9}. \bibinfo{pages}{416--419}.
\newblock


\bibitem[Girish et~al\mbox{.}(2024)]{girishQUEENQUantizedEfficient2024}
\bibfield{author}{\bibinfo{person}{Sharath Girish}, \bibinfo{person}{Tianye Li}, \bibinfo{person}{Amrita Mazumdar}, \bibinfo{person}{Abhinav Shrivastava}, \bibinfo{person}{David Luebke}, {and} \bibinfo{person}{Shalini~De Mello}.} \bibinfo{year}{2024}\natexlab{}.
\newblock \showarticletitle{{{QUEEN}}: {{QUantized Efficient ENcoding}} of {{Dynamic Gaussians}} for {{Streaming Free-viewpoint Videos}}}. In \bibinfo{booktitle}{\emph{The {{Thirty-eighth Annual Conference}} on {{Neural Information Processing Systems}}}}.
\newblock


\bibitem[Google(2017)]{draco}
\bibfield{author}{\bibinfo{person}{Google}.} \bibinfo{year}{2017}\natexlab{}.
\newblock \bibinfo{title}{Draco 3D Graphics Compression}.
\newblock \bibinfo{howpublished}{https://github.com/google/draco}.
\newblock


\bibitem[Guan et~al\mbox{.}(2023)]{guanMetaStreamLiveVolumetric2023}
\bibfield{author}{\bibinfo{person}{Yongjie Guan}, \bibinfo{person}{Xueyu Hou}, \bibinfo{person}{Nan Wu}, \bibinfo{person}{Bo Han}, {and} \bibinfo{person}{Tao Han}.} \bibinfo{year}{2023}\natexlab{}.
\newblock \showarticletitle{{{MetaStream}}: {{Live Volumetric Content Capture}}, {{Creation}}, {{Delivery}}, and {{Rendering}} in {{Real Time}}}.
\newblock In \bibinfo{booktitle}{\emph{Proceedings of the 29th {{Annual International Conference}} on {{Mobile Computing}} and {{Networking}}}}. Number~29. \bibinfo{pages}{1--15}.
\newblock


\bibitem[G{\"u}l et~al\mbox{.}(2022)]{gulLatencyCompensationImage2022}
\bibfield{author}{\bibinfo{person}{Serhan G{\"u}l}, \bibinfo{person}{Cornelius Hellge}, {and} \bibinfo{person}{Peter Eisert}.} \bibinfo{year}{2022}\natexlab{}.
\newblock \showarticletitle{Latency {{Compensation Through Image Warping For Remote Rendering-Based Volumetric Video Streaming}}}. In \bibinfo{booktitle}{\emph{2022 {{IEEE International Conference}} on {{Image Processing}} ({{ICIP}})}}. \bibinfo{pages}{2026--2030}.
\newblock


\bibitem[Han et~al\mbox{.}(2020)]{hanViVoVisibilityawareMobile2020}
\bibfield{author}{\bibinfo{person}{Bo Han}, \bibinfo{person}{Yu Liu}, {and} \bibinfo{person}{Feng Qian}.} \bibinfo{year}{2020}\natexlab{}.
\newblock \showarticletitle{{{ViVo}}: Visibility-Aware Mobile Volumetric Video Streaming}. In \bibinfo{booktitle}{\emph{Proceedings of the 26th {{Annual International Conference}} on {{Mobile Computing}} and {{Networking}}}}. \bibinfo{pages}{1--13}.
\newblock


\bibitem[Hazeleger et~al\mbox{.}(2024)]{hazeleger2024digital}
\bibfield{author}{\bibinfo{person}{W Hazeleger}, \bibinfo{person}{JPM Aerts}, \bibinfo{person}{Peter Bauer}, \bibinfo{person}{MFP Bierkens}, \bibinfo{person}{Gustau Camps-Valls}, \bibinfo{person}{MM Dekker}, \bibinfo{person}{FJ Doblas-Reyes}, \bibinfo{person}{Veronika Eyring}, \bibinfo{person}{C Finkenauer}, \bibinfo{person}{Arthur Grundner}, {et~al\mbox{.}}} \bibinfo{year}{2024}\natexlab{}.
\newblock \showarticletitle{Digital twins of the Earth with and for humans}.
\newblock \bibinfo{journal}{\emph{Communications earth \& environment}} \bibinfo{volume}{5}, \bibinfo{number}{1} (\bibinfo{year}{2024}), \bibinfo{pages}{463}.
\newblock


\bibitem[Hein et~al\mbox{.}(2025)]{HEIN2025103613}
\bibfield{author}{\bibinfo{person}{Jonas Hein}, \bibinfo{person}{Nicola Cavalcanti}, \bibinfo{person}{Daniel Suter}, \bibinfo{person}{Lukas Zingg}, \bibinfo{person}{Fabio Carrillo}, \bibinfo{person}{Lilian Calvet}, \bibinfo{person}{Mazda Farshad}, \bibinfo{person}{Nassir Navab}, \bibinfo{person}{Marc Pollefeys}, {and} \bibinfo{person}{Philipp Fürnstahl}.} \bibinfo{year}{2025}\natexlab{}.
\newblock \showarticletitle{Next-generation surgical navigation: Marker-less multi-view 6DoF pose estimation of surgical instruments}.
\newblock \bibinfo{journal}{\emph{Medical Image Analysis}}  \bibinfo{volume}{103} (\bibinfo{year}{2025}), \bibinfo{pages}{103613}.
\newblock
\showISSN{1361-8415}


\bibitem[Hladky et~al\mbox{.}(2019)]{10.1145/3355089.3356530}
\bibfield{author}{\bibinfo{person}{Jozef Hladky}, \bibinfo{person}{Hans-Peter Seidel}, {and} \bibinfo{person}{Markus Steinberger}.} \bibinfo{year}{2019}\natexlab{}.
\newblock \showarticletitle{The camera offset space: real-time potentially visible set computations for streaming rendering}.
\newblock \bibinfo{journal}{\emph{ACM Trans. Graph.}} \bibinfo{volume}{38}, \bibinfo{number}{6} (\bibinfo{year}{2019}), \bibinfo{numpages}{14}~pages.
\newblock
\showISSN{0730-0301}


\bibitem[Hladky et~al\mbox{.}(2022)]{hladkyQuadStream2022}
\bibfield{author}{\bibinfo{person}{Jozef Hladky}, \bibinfo{person}{Michael Stengel}, \bibinfo{person}{Nicholas Vining}, \bibinfo{person}{Bernhard Kerbl}, \bibinfo{person}{Hans-Peter Seidel}, {and} \bibinfo{person}{Markus Steinberger}.} \bibinfo{year}{2022}\natexlab{}.
\newblock \showarticletitle{QuadStream: A Quad-Based Scene Streaming Architecture for Novel Viewpoint Reconstruction}.
\newblock \bibinfo{journal}{\emph{ACM Trans. Graph.}} \bibinfo{volume}{41}, \bibinfo{number}{6} (\bibinfo{year}{2022}), \bibinfo{numpages}{13}~pages.
\newblock
\showISSN{0730-0301}


\bibitem[Huang et~al\mbox{.}(2023)]{huangRefSRNeRFHighFidelity2023}
\bibfield{author}{\bibinfo{person}{Xudong Huang}, \bibinfo{person}{Wei Li}, \bibinfo{person}{Jie Hu}, \bibinfo{person}{Hanting Chen}, {and} \bibinfo{person}{Yunhe Wang}.} \bibinfo{year}{2023}\natexlab{}.
\newblock \showarticletitle{RefSR-NeRF: Towards High Fidelity and Super Resolution View Synthesis}. In \bibinfo{booktitle}{\emph{Proceedings of the IEEE/CVF Conference on Computer Vision and Pattern Recognition (CVPR)}}. \bibinfo{pages}{8244--8253}.
\newblock


\bibitem[Huang et~al\mbox{.}(2026)]{huangACPGSBandwidthEfficientDelivery2026}
\bibfield{author}{\bibinfo{person}{Zhaohui Huang}, \bibinfo{person}{Cong Zhang}, \bibinfo{person}{Jianxin Shi}, \bibinfo{person}{Xiaoyi Fan}, \bibinfo{person}{Laizhong Cui}, {and} \bibinfo{person}{Jiangchuan Liu}.} \bibinfo{year}{2026}\natexlab{}.
\newblock \showarticletitle{{{ACPGS}}: {{Towards Bandwidth-Efficient Delivery}} of {{3D Gaussian Splatting}}}. In \bibinfo{booktitle}{\emph{Proceedings of the 36th {{Workshop}} on {{Network}} and {{Operating System Support}} for {{Digital Audio}} and {{Video}}}} \emph{(\bibinfo{series}{{{NOSSDAV}} '26})}. \bibinfo{pages}{127--133}.
\newblock


\bibitem[Jiang et~al\mbox{.}(2025)]{jiangTopologyAwareOptimizationGaussian2025}
\bibfield{author}{\bibinfo{person}{Yuheng Jiang}, \bibinfo{person}{Chengcheng Guo}, \bibinfo{person}{Yize Wu}, \bibinfo{person}{Yu Hong}, \bibinfo{person}{Shengkun Zhu}, \bibinfo{person}{Zhehao Shen}, \bibinfo{person}{Yingliang Zhang}, \bibinfo{person}{Shaohui Jiao}, \bibinfo{person}{Zhuo Su}, \bibinfo{person}{Lan Xu}, \bibinfo{person}{Marc Habermann}, {and} \bibinfo{person}{Christian Theobalt}.} \bibinfo{year}{2025}\natexlab{}.
\newblock \showarticletitle{Topology-{{Aware Optimization}} of {{Gaussian Primitives}} for {{Human-Centric Volumetric Videos}}}. In \bibinfo{booktitle}{\emph{Proceedings of the {{SIGGRAPH Asia}} 2025 {{Conference Papers}}}} \emph{(\bibinfo{series}{{{SA Conference Papers}} '25})}. \bibinfo{pages}{1--12}.
\newblock
\showISBNx{979-8-4007-2137-3}


\bibitem[Joska et~al\mbox{.}(2021)]{joska2021acinoset}
\bibfield{author}{\bibinfo{person}{Daniel Joska}, \bibinfo{person}{Liam Clark}, \bibinfo{person}{Naoya Muramatsu}, \bibinfo{person}{Ricardo Jericevich}, \bibinfo{person}{Fred Nicolls}, \bibinfo{person}{Alexander Mathis}, \bibinfo{person}{Mackenzie~W. Mathis}, {and} \bibinfo{person}{Amir Patel}.} \bibinfo{year}{2021}\natexlab{}.
\newblock \bibinfo{title}{AcinoSet: A 3D Pose Estimation Dataset and Baseline Models for Cheetahs in the Wild}.
\newblock
\showeprint[arxiv]{2103.13282}~[cs.CV]


\bibitem[Kerbl et~al\mbox{.}(2023)]{kerbl3DGaussianSplatting2023}
\bibfield{author}{\bibinfo{person}{Bernhard Kerbl}, \bibinfo{person}{Georgios Kopanas}, \bibinfo{person}{Thomas Leimkuehler}, {and} \bibinfo{person}{George Drettakis}.} \bibinfo{year}{2023}\natexlab{}.
\newblock \showarticletitle{{{3D Gaussian Splatting}} for {{Real-Time Radiance Field Rendering}}}.
\newblock \bibinfo{journal}{\emph{ACM Transactions on Graphics}} \bibinfo{volume}{42}, \bibinfo{number}{4} (\bibinfo{year}{2023}), \bibinfo{pages}{139:1--139:14}.
\newblock


\bibitem[Kerbl et~al\mbox{.}(2024)]{kerblHierarchical3DGaussian2024}
\bibfield{author}{\bibinfo{person}{Bernhard Kerbl}, \bibinfo{person}{Andreas Meuleman}, \bibinfo{person}{Georgios Kopanas}, \bibinfo{person}{Michael Wimmer}, \bibinfo{person}{Alexandre Lanvin}, {and} \bibinfo{person}{George Drettakis}.} \bibinfo{year}{2024}\natexlab{}.
\newblock \showarticletitle{A {{Hierarchical 3D Gaussian Representation}} for {{Real-Time Rendering}} of {{Very Large Datasets}}}.
\newblock \bibinfo{journal}{\emph{ACM Transactions on Graphics}} \bibinfo{volume}{44}, \bibinfo{number}{3} (\bibinfo{year}{2024}).
\newblock


\bibitem[Kim et~al\mbox{.}(2025)]{kimVegaFullyImmersive2025}
\bibfield{author}{\bibinfo{person}{Gunjoong Kim}, \bibinfo{person}{Seonghoon Park}, \bibinfo{person}{Jeho Lee}, \bibinfo{person}{Chanyoung Jung}, \bibinfo{person}{Hyungchol Jun}, {and} \bibinfo{person}{Hojung Cha}.} \bibinfo{year}{2025}\natexlab{}.
\newblock \showarticletitle{Vega: {{Fully Immersive Mobile Volumetric Video Streaming}} with {{3D Gaussian Splatting}}}. In \bibinfo{booktitle}{\emph{Proceedings of the 31st {{Annual International Conference}} on {{Mobile Computing}} and {{Networking}}}}. \bibinfo{pages}{1106--1120}.
\newblock


\bibitem[Ledig et~al\mbox{.}(2017)]{ledigPhotoRealisticSingleImage2017}
\bibfield{author}{\bibinfo{person}{Christian Ledig}, \bibinfo{person}{Lucas Theis}, \bibinfo{person}{Ferenc Huszar}, \bibinfo{person}{Jose Caballero}, \bibinfo{person}{Andrew Cunningham}, \bibinfo{person}{Alejandro Acosta}, \bibinfo{person}{Andrew Aitken}, \bibinfo{person}{Alykhan Tejani}, \bibinfo{person}{Johannes Totz}, \bibinfo{person}{Zehan Wang}, {and} \bibinfo{person}{Wenzhe Shi}.} \bibinfo{year}{2017}\natexlab{}.
\newblock \showarticletitle{Photo-{{Realistic Single Image Super-Resolution Using}} a {{Generative Adversarial Network}}}. In \bibinfo{booktitle}{\emph{Proceedings of the {{IEEE Conference}} on {{Computer Vision}} and {{Pattern Recognition}}}}. \bibinfo{pages}{4681--4690}.
\newblock


\bibitem[Lee et~al\mbox{.}(2024)]{leeCompact3DGaussian2024}
\bibfield{author}{\bibinfo{person}{Joo~Chan Lee}, \bibinfo{person}{Daniel Rho}, \bibinfo{person}{Xiangyu Sun}, \bibinfo{person}{Jong~Hwan Ko}, {and} \bibinfo{person}{Eunbyung Park}.} \bibinfo{year}{2024}\natexlab{}.
\newblock \showarticletitle{Compact {{3D Gaussian Representation}} for {{Radiance Field}}}. In \bibinfo{booktitle}{\emph{Proceedings of the {{IEEE}}/{{CVF Conference}} on {{Computer Vision}} and {{Pattern Recognition}}}}. \bibinfo{pages}{21719--21728}.
\newblock


\bibitem[Li et~al\mbox{.}(2025)]{liGIFStream4DGaussianbased2025}
\bibfield{author}{\bibinfo{person}{Hao Li}, \bibinfo{person}{Sicheng Li}, \bibinfo{person}{Xiang Gao}, \bibinfo{person}{Abudouaihati Batuer}, \bibinfo{person}{Lu Yu}, {and} \bibinfo{person}{Yiyi Liao}.} \bibinfo{year}{2025}\natexlab{}.
\newblock \showarticletitle{{{GIFStream}}: {{4D Gaussian-based Immersive Video}} with {{Feature Stream}}}. In \bibinfo{booktitle}{\emph{Proceedings of the {{IEEE}}/{{CVF Conference}} on {{Computer Vision}} and {{Pattern Recognition}}}}.
\newblock


\bibitem[Li et~al\mbox{.}(2024)]{liLoopGaussianCreating3D2024}
\bibfield{author}{\bibinfo{person}{Jiyang Li}, \bibinfo{person}{Lechao Cheng}, \bibinfo{person}{Zhangye Wang}, \bibinfo{person}{Tingting Mu}, {and} \bibinfo{person}{Jingxuan He}.} \bibinfo{year}{2024}\natexlab{}.
\newblock \showarticletitle{{{LoopGaussian}}: {{Creating 3D Cinemagraph}} with {{Multi-view Images}} via {{Eulerian Motion Field}}}. In \bibinfo{booktitle}{\emph{Proceedings of the 32nd {{ACM International Conference}} on {{Multimedia}}}} \emph{(\bibinfo{series}{{{MM}} '24})}. \bibinfo{pages}{476--485}.
\newblock


\bibitem[Li et~al\mbox{.}(2022a)]{liStreamingRadianceFields2022}
\bibfield{author}{\bibinfo{person}{Lingzhi Li}, \bibinfo{person}{Zhen Shen}, \bibinfo{person}{Zhongshu Wang}, \bibinfo{person}{Li Shen}, {and} \bibinfo{person}{Ping Tan}.} \bibinfo{year}{2022}\natexlab{a}.
\newblock \showarticletitle{Streaming {{Radiance Fields}} for {{3D Video Synthesis}}}.
\newblock \bibinfo{journal}{\emph{Advances in Neural Information Processing Systems}}  \bibinfo{volume}{35} (\bibinfo{year}{2022}), \bibinfo{pages}{13485--13498}.
\newblock


\bibitem[Li et~al\mbox{.}(2022b)]{liNeural3DVideo2022}
\bibfield{author}{\bibinfo{person}{Tianye Li}, \bibinfo{person}{Mira Slavcheva}, \bibinfo{person}{Michael Zollh{\"o}fer}, \bibinfo{person}{Simon Green}, \bibinfo{person}{Christoph Lassner}, \bibinfo{person}{Changil Kim}, \bibinfo{person}{Tanner Schmidt}, \bibinfo{person}{Steven Lovegrove}, \bibinfo{person}{Michael Goesele}, \bibinfo{person}{Richard Newcombe}, {and} \bibinfo{person}{Zhaoyang Lv}.} \bibinfo{year}{2022}\natexlab{b}.
\newblock \showarticletitle{Neural {{3D Video Synthesis From Multi-View Video}}}. In \bibinfo{booktitle}{\emph{Proceedings of the {{IEEE}}/{{CVF Conference}} on {{Computer Vision}} and {{Pattern Recognition}}}}. \bibinfo{pages}{5521--5531}.
\newblock


\bibitem[Liang et~al\mbox{.}(2023)]{liang2023robo360}
\bibfield{author}{\bibinfo{person}{Litian Liang}, \bibinfo{person}{Liuyu Bian}, \bibinfo{person}{Caiwei Xiao}, \bibinfo{person}{Jialin Zhang}, \bibinfo{person}{Linghao Chen}, \bibinfo{person}{Isabella Liu}, \bibinfo{person}{Fanbo Xiang}, \bibinfo{person}{Zhiao Huang}, {and} \bibinfo{person}{Hao Su}.} \bibinfo{year}{2023}\natexlab{}.
\newblock \showarticletitle{Robo360: a 3D omnispective multi-material robotic manipulation dataset}.
\newblock \bibinfo{journal}{\emph{arXiv preprint arXiv:2312.06686}} (\bibinfo{year}{2023}).
\newblock


\bibitem[Liu et~al\mbox{.}(2023)]{liuCaV3CacheassistedViewport2023}
\bibfield{author}{\bibinfo{person}{Junhua Liu}, \bibinfo{person}{Boxiang Zhu}, \bibinfo{person}{Fangxin Wang}, \bibinfo{person}{Yili Jin}, \bibinfo{person}{Wenyi Zhang}, \bibinfo{person}{Zihan Xu}, {and} \bibinfo{person}{Shuguang Cui}.} \bibinfo{year}{2023}\natexlab{}.
\newblock \showarticletitle{{{CaV3}}: {{Cache-assisted Viewport Adaptive Volumetric Video Streaming}}}. In \bibinfo{booktitle}{\emph{2023 {{IEEE Conference Virtual Reality}} and {{3D User Interfaces}} ({{VR}})}}. \bibinfo{pages}{173--183}.
\newblock


\bibitem[Liu et~al\mbox{.}(2024)]{liuCompGSEfficient3D2024}
\bibfield{author}{\bibinfo{person}{Xiangrui Liu}, \bibinfo{person}{Xinju Wu}, \bibinfo{person}{Pingping Zhang}, \bibinfo{person}{Shiqi Wang}, \bibinfo{person}{Zhu Li}, {and} \bibinfo{person}{Sam Kwong}.} \bibinfo{year}{2024}\natexlab{}.
\newblock \showarticletitle{{{CompGS}}: {{Efficient 3D Scene Representation}} via {{Compressed Gaussian Splatting}}}. In \bibinfo{booktitle}{\emph{Proceedings of the 32nd {{ACM International Conference}} on {{Multimedia}}}} \emph{(\bibinfo{series}{{{MM}} '24})}. \bibinfo{pages}{2936--2944}.
\newblock


\bibitem[Lu and Rowe(2025)]{luQUASARQuadbasedAdaptive2025}
\bibfield{author}{\bibinfo{person}{Edward Lu} {and} \bibinfo{person}{Anthony Rowe}.} \bibinfo{year}{2025}\natexlab{}.
\newblock \showarticletitle{QUASAR: Quad-based Adaptive Streaming And Rendering}.
\newblock \bibinfo{journal}{\emph{ACM Trans. Graph.}} \bibinfo{volume}{44}, \bibinfo{number}{4} (\bibinfo{year}{2025}), \bibinfo{numpages}{18}~pages.
\newblock
\showISSN{0730-0301}


\bibitem[Lu et~al\mbox{.}(2024)]{scaffoldgs}
\bibfield{author}{\bibinfo{person}{Tao Lu}, \bibinfo{person}{Mulin Yu}, \bibinfo{person}{Linning Xu}, \bibinfo{person}{Yuanbo Xiangli}, \bibinfo{person}{Limin Wang}, \bibinfo{person}{Dahua Lin}, {and} \bibinfo{person}{Bo Dai}.} \bibinfo{year}{2024}\natexlab{}.
\newblock \showarticletitle{Scaffold-Gs: {{Structured}} 3d Gaussians for View-Adaptive Rendering}. In \bibinfo{booktitle}{\emph{Proceedings of the {{IEEE}}/{{CVF}} Conference on Computer Vision and Pattern Recognition}}. \bibinfo{pages}{20654--20664}.
\newblock


\bibitem[Luiten et~al\mbox{.}(2024)]{luiten2023dynamic}
\bibfield{author}{\bibinfo{person}{Jonathon Luiten}, \bibinfo{person}{Georgios Kopanas}, \bibinfo{person}{Bastian Leibe}, {and} \bibinfo{person}{Deva Ramanan}.} \bibinfo{year}{2024}\natexlab{}.
\newblock \showarticletitle{Dynamic {{3D}} Gaussians: {{Tracking}} by Persistent Dynamic View Synthesis}. In \bibinfo{booktitle}{\emph{{{3DV}}}}.
\newblock


\bibitem[Luo et~al\mbox{.}(2022)]{luoLearningDegradationDistribution2022}
\bibfield{author}{\bibinfo{person}{Zhengxiong Luo}, \bibinfo{person}{Yan Huang}, \bibinfo{person}{Shang Li}, \bibinfo{person}{Liang Wang}, {and} \bibinfo{person}{Tieniu Tan}.} \bibinfo{year}{2022}\natexlab{}.
\newblock \showarticletitle{Learning the {{Degradation Distribution}} for {{Blind Image Super-Resolution}}}. In \bibinfo{booktitle}{\emph{Proceedings of the {{IEEE}}/{{CVF Conference}} on {{Computer Vision}} and {{Pattern Recognition}}}}. \bibinfo{pages}{6063--6072}.
\newblock


\bibitem[Mildenhall et~al\mbox{.}(2021)]{mildenhallNeRFRepresentingScenes2021}
\bibfield{author}{\bibinfo{person}{Ben Mildenhall}, \bibinfo{person}{Pratul~P. Srinivasan}, \bibinfo{person}{Matthew Tancik}, \bibinfo{person}{Jonathan~T. Barron}, \bibinfo{person}{Ravi Ramamoorthi}, {and} \bibinfo{person}{Ren Ng}.} \bibinfo{year}{2021}\natexlab{}.
\newblock \showarticletitle{{{NeRF}}: Representing Scenes as Neural Radiance Fields for View Synthesis}.
\newblock \bibinfo{journal}{\emph{Commun. ACM}} \bibinfo{volume}{65}, \bibinfo{number}{1} (\bibinfo{year}{2021}), \bibinfo{pages}{99--106}.
\newblock


\bibitem[Narayanan et~al\mbox{.}(2020)]{narayananFirstLookCommercial2020}
\bibfield{author}{\bibinfo{person}{Arvind Narayanan}, \bibinfo{person}{Eman Ramadan}, \bibinfo{person}{Jason Carpenter}, \bibinfo{person}{Qingxu Liu}, \bibinfo{person}{Yu Liu}, \bibinfo{person}{Feng Qian}, {and} \bibinfo{person}{Zhi-Li Zhang}.} \bibinfo{year}{2020}\natexlab{}.
\newblock \showarticletitle{A {{First Look}} at {{Commercial 5G Performance}} on {{Smartphones}}}. In \bibinfo{booktitle}{\emph{Proceedings of {{The Web Conference}} 2020}} \emph{(\bibinfo{series}{{{WWW}} '20})}. \bibinfo{pages}{894--905}.
\newblock


\bibitem[{Niantic Labs}(2025)]{spz}
\bibfield{author}{\bibinfo{person}{{Niantic Labs}}.} \bibinfo{year}{2025}\natexlab{}.
\newblock \bibinfo{title}{spz: File Format for 3D Gaussian Splats}.
\newblock \bibinfo{howpublished}{https://github.com/nianticlabs/spz}.
\newblock


\bibitem[Papantonakis et~al\mbox{.}(2024)]{papantonakisReducingMemoryFootprint2024}
\bibfield{author}{\bibinfo{person}{Panagiotis Papantonakis}, \bibinfo{person}{Georgios Kopanas}, \bibinfo{person}{Bernhard Kerbl}, \bibinfo{person}{Alexandre Lanvin}, {and} \bibinfo{person}{George Drettakis}.} \bibinfo{year}{2024}\natexlab{}.
\newblock \showarticletitle{Reducing the {{Memory Footprint}} of {{3D Gaussian Splatting}}}.
\newblock \bibinfo{journal}{\emph{Proceedings of the ACM on Computer Graphics and Interactive Techniques}} \bibinfo{volume}{7}, \bibinfo{number}{1} (\bibinfo{year}{2024}), \bibinfo{pages}{16:1--16:17}.
\newblock


\bibitem[Pranckevičius(2023)]{UnityGS}
\bibfield{author}{\bibinfo{person}{Aras Pranckevičius}.} \bibinfo{year}{2023}\natexlab{}.
\newblock \bibinfo{title}{Gaussian Splatting playground in Unity}.
\newblock \bibinfo{howpublished}{\url{https://github.com/aras-p/UnityGaussianSplatting}}.
\newblock


\bibitem[Rojas-Mu{\~n}oz et~al\mbox{.}(2019)]{rojas2019surgical}
\bibfield{author}{\bibinfo{person}{Edgar Rojas-Mu{\~n}oz}, \bibinfo{person}{Maria~Eugenia Cabrera}, \bibinfo{person}{Daniel Andersen}, \bibinfo{person}{Voicu Popescu}, \bibinfo{person}{Sherri Marley}, \bibinfo{person}{Brian Mullis}, \bibinfo{person}{Ben Zarzaur}, {and} \bibinfo{person}{Juan Wachs}.} \bibinfo{year}{2019}\natexlab{}.
\newblock \showarticletitle{Surgical telementoring without encumbrance: a comparative study of see-through augmented reality-based approaches}.
\newblock \bibinfo{journal}{\emph{Annals of surgery}} \bibinfo{volume}{270}, \bibinfo{number}{2} (\bibinfo{year}{2019}), \bibinfo{pages}{384--389}.
\newblock


\bibitem[Shi et~al\mbox{.}(2024)]{shiFullsceneVolumetricVideo2024}
\bibfield{author}{\bibinfo{person}{Jianxin Shi}, \bibinfo{person}{Miao Zhang}, \bibinfo{person}{Linfeng Shen}, \bibinfo{person}{Jiangchuan Liu}, \bibinfo{person}{Yuan Zhang}, \bibinfo{person}{Lingjun Pu}, {and} \bibinfo{person}{Jingdong Xu}.} \bibinfo{year}{2024}\natexlab{}.
\newblock \showarticletitle{Towards {{Full-scene Volumetric Video Streaming}} via {{Spatially Layered Representation}} and {{NeRF Generation}}}. In \bibinfo{booktitle}{\emph{Proceedings of the 34th Edition of the {{Workshop}} on {{Network}} and {{Operating System Support}} for {{Digital Audio}} and {{Video}}}} \emph{(\bibinfo{series}{{{NOSSDAV}} '24})}. \bibinfo{pages}{22--28}.
\newblock


\bibitem[Shi et~al\mbox{.}(2025)]{shiLapisGSLayeredProgressive2025}
\bibfield{author}{\bibinfo{person}{Yuang Shi}, \bibinfo{person}{G{\'e}raldine Morin}, \bibinfo{person}{Simone Gasparini}, {and} \bibinfo{person}{Wei~Tsang Ooi}.} \bibinfo{year}{2025}\natexlab{}.
\newblock \showarticletitle{{{LapisGS}}: {{Layered Progressive 3D Gaussian Splatting}} for {{Adaptive Streaming}}}. In \bibinfo{booktitle}{\emph{International {{Conference}} on {{3D Vision}} 2025}}.
\newblock


\bibitem[Sun et~al\mbox{.}(2025)]{sunLTSDASHStreaming2025}
\bibfield{author}{\bibinfo{person}{Yuan-Chun Sun}, \bibinfo{person}{Yuang Shi}, \bibinfo{person}{Cheng-Tse Lee}, \bibinfo{person}{Mufeng Zhu}, \bibinfo{person}{Wei~Tsang Ooi}, \bibinfo{person}{Yao Liu}, \bibinfo{person}{Chun-Ying Huang}, {and} \bibinfo{person}{Cheng-Hsin Hsu}.} \bibinfo{year}{2025}\natexlab{}.
\newblock \showarticletitle{{{LTS}}: {{A DASH Streaming System}} for {{Dynamic Multi-Layer 3D Gaussian Splatting Scenes}}}. In \bibinfo{booktitle}{\emph{Proceedings of the 16th {{ACM Multimedia Systems Conference}}}} \emph{(\bibinfo{series}{{{MMSys}} '25})}. \bibinfo{pages}{136--147}.
\newblock


\bibitem[Takikawa et~al\mbox{.}(2022)]{takikawaVariableBitrateNeural2022}
\bibfield{author}{\bibinfo{person}{Towaki Takikawa}, \bibinfo{person}{Alex Evans}, \bibinfo{person}{Jonathan Tremblay}, \bibinfo{person}{Thomas M{\"u}ller}, \bibinfo{person}{Morgan McGuire}, \bibinfo{person}{Alec Jacobson}, {and} \bibinfo{person}{Sanja Fidler}.} \bibinfo{year}{2022}\natexlab{}.
\newblock \showarticletitle{Variable {{Bitrate Neural Fields}}}. In \bibinfo{booktitle}{\emph{Special {{Interest Group}} on {{Computer Graphics}} and {{Interactive Techniques Conference Proceedings}}}}. \bibinfo{pages}{1--9}.
\newblock


\bibitem[Tao et~al\mbox{.}(2018)]{tao2018digital}
\bibfield{author}{\bibinfo{person}{Fei Tao}, \bibinfo{person}{He Zhang}, \bibinfo{person}{Ang Liu}, {and} \bibinfo{person}{Andrew~YC Nee}.} \bibinfo{year}{2018}\natexlab{}.
\newblock \showarticletitle{Digital twin in industry: State-of-the-art}.
\newblock \bibinfo{journal}{\emph{IEEE Transactions on industrial informatics}} \bibinfo{volume}{15}, \bibinfo{number}{4} (\bibinfo{year}{2018}), \bibinfo{pages}{2405--2415}.
\newblock


\bibitem[Wang et~al\mbox{.}(2022)]{wangNeRFSRHighQuality2022}
\bibfield{author}{\bibinfo{person}{Chen Wang}, \bibinfo{person}{Xian Wu}, \bibinfo{person}{Yuan-Chen Guo}, \bibinfo{person}{Song-Hai Zhang}, \bibinfo{person}{Yu-Wing Tai}, {and} \bibinfo{person}{Shi-Min Hu}.} \bibinfo{year}{2022}\natexlab{}.
\newblock \showarticletitle{{{NeRF-SR}}: {{High Quality Neural Radiance Fields}} Using {{Supersampling}}}. In \bibinfo{booktitle}{\emph{Proceedings of the 30th {{ACM International Conference}} on {{Multimedia}}}} \emph{(\bibinfo{series}{{{MM}} '22})}. \bibinfo{pages}{6445--6454}.
\newblock


\bibitem[Wang et~al\mbox{.}(2024a)]{wangV^3ViewingVolumetric2024}
\bibfield{author}{\bibinfo{person}{Penghao Wang}, \bibinfo{person}{Zhirui Zhang}, \bibinfo{person}{Liao Wang}, \bibinfo{person}{Kaixin Yao}, \bibinfo{person}{Siyuan Xie}, \bibinfo{person}{Jingyi Yu}, \bibinfo{person}{Minye Wu}, {and} \bibinfo{person}{Lan Xu}.} \bibinfo{year}{2024}\natexlab{a}.
\newblock \showarticletitle{V{\textasciicircum}3: {{Viewing Volumetric Videos}} on {{Mobiles}} via {{Streamable 2D Dynamic Gaussians}}}. In \bibinfo{booktitle}{\emph{{{SIGGRAPH Asia}} 2024 {{Conference Papers}}}}.
\newblock


\bibitem[Wang et~al\mbox{.}(2024b)]{wangBandwidthEfficientMobileVolumetric2024}
\bibfield{author}{\bibinfo{person}{Yizong Wang}, \bibinfo{person}{Dong Zhao}, \bibinfo{person}{Huanhuan Zhang}, \bibinfo{person}{Teng Gao}, \bibinfo{person}{Zixuan Guo}, \bibinfo{person}{Chenghao Huang}, {and} \bibinfo{person}{Huadong Ma}.} \bibinfo{year}{2024}\natexlab{b}.
\newblock \showarticletitle{Bandwidth-{{Efficient Mobile Volumetric Video Streaming}} by {{Exploiting Inter-Frame Correlation}}}.
\newblock \bibinfo{journal}{\emph{IEEE Transactions on Mobile Computing}} \bibinfo{volume}{23}, \bibinfo{number}{10} (\bibinfo{year}{2024}), \bibinfo{pages}{9410--9423}.
\newblock


\bibitem[Wegen et~al\mbox{.}(2024)]{wegenSurveyNonphotorealisticRendering2024}
\bibfield{author}{\bibinfo{person}{Ole Wegen}, \bibinfo{person}{Willy Scheibel}, \bibinfo{person}{Matthias Trapp}, \bibinfo{person}{Rico Richter}, {and} \bibinfo{person}{Jurgen Dollner}.} \bibinfo{year}{2024}\natexlab{}.
\newblock \showarticletitle{A {{Survey}} on {{Non-photorealistic Rendering Approaches}} for {{Point Cloud Visualization}}}.
\newblock \bibinfo{journal}{\emph{IEEE Transactions on Visualization and Computer Graphics}} (\bibinfo{year}{2024}), \bibinfo{pages}{1--20}.
\newblock
\showISSN{1941-0506}


\bibitem[Wu et~al\mbox{.}(2024)]{wu4DGaussianSplatting2024}
\bibfield{author}{\bibinfo{person}{Guanjun Wu}, \bibinfo{person}{Taoran Yi}, \bibinfo{person}{Jiemin Fang}, \bibinfo{person}{Lingxi Xie}, \bibinfo{person}{Xiaopeng Zhang}, \bibinfo{person}{Wei Wei}, \bibinfo{person}{Wenyu Liu}, \bibinfo{person}{Qi Tian}, {and} \bibinfo{person}{Xinggang Wang}.} \bibinfo{year}{2024}\natexlab{}.
\newblock \showarticletitle{{{4D Gaussian Splatting}} for {{Real-Time Dynamic Scene Rendering}}}. In \bibinfo{booktitle}{\emph{Proceedings of the {{IEEE}}/{{CVF Conference}} on {{Computer Vision}} and {{Pattern Recognition}}}}. \bibinfo{pages}{20310--20320}.
\newblock


\bibitem[Wu et~al\mbox{.}(2023)]{wuZGamingZeroLatency3D2023}
\bibfield{author}{\bibinfo{person}{Jiangkai Wu}, \bibinfo{person}{Yu Guan}, \bibinfo{person}{Qi Mao}, \bibinfo{person}{Yong Cui}, \bibinfo{person}{Zongming Guo}, {and} \bibinfo{person}{Xinggong Zhang}.} \bibinfo{year}{2023}\natexlab{}.
\newblock \showarticletitle{{{ZGaming}}: {{Zero-Latency 3D Cloud Gaming}} by {{Image Prediction}}}. In \bibinfo{booktitle}{\emph{Proceedings of the {{ACM SIGCOMM}} 2023 {{Conference}}}} \emph{(\bibinfo{series}{{{ACM SIGCOMM}} '23})}. \bibinfo{pages}{710--723}.
\newblock


\bibitem[Xie et~al\mbox{.}(2025)]{xieSizeGSSizeawareCompression2025}
\bibfield{author}{\bibinfo{person}{Shuzhao Xie}, \bibinfo{person}{Jiahang Liu}, \bibinfo{person}{Weixiang Zhang}, \bibinfo{person}{Shijia Ge}, \bibinfo{person}{Sicheng Pan}, \bibinfo{person}{Chen Tang}, \bibinfo{person}{Yunpeng Bai}, \bibinfo{person}{Cong Zhang}, \bibinfo{person}{Xiaoyi Fan}, {and} \bibinfo{person}{Zhi Wang}.} \bibinfo{year}{2025}\natexlab{}.
\newblock \showarticletitle{{{SizeGS}}: {{Size-aware Compression}} of {{3D Gaussian Splatting}} via {{Mixed Integer Programming}}}. In \bibinfo{booktitle}{\emph{Proceedings of the 33rd {{ACM International Conference}} on {{Multimedia}}}} \emph{(\bibinfo{series}{{{MM}} '25})}. \bibinfo{publisher}{Association for Computing Machinery}, \bibinfo{address}{New York, NY, USA}, \bibinfo{pages}{8214--8223}.
\newblock
\showISBNx{979-8-4007-2035-2}
\href{https://doi.org/10.1145/3746027.3755370}{doi:\nolinkurl{10.1145/3746027.3755370}}


\bibitem[Xu et~al\mbox{.}(2024a)]{xuGrid4D4DDecomposed2024}
\bibfield{author}{\bibinfo{person}{Jiawei Xu}, \bibinfo{person}{Zexin Fan}, \bibinfo{person}{Jian Yang}, {and} \bibinfo{person}{Jij Xie}.} \bibinfo{year}{2024}\natexlab{a}.
\newblock \showarticletitle{{{Grid4D}}: {{4D}} Decomposed Hash Encoding for High-Fidelity Dynamic {{Gaussian}} Splatting}. In \bibinfo{booktitle}{\emph{Proceedings of the 38th {{International Conference}} on {{Neural Information Processing Systems}}}} \emph{(\bibinfo{series}{{{NIPS}} '24}, Vol.~\bibinfo{volume}{37})}. \bibinfo{pages}{123787--123811}.
\newblock
\showISBNx{979-8-3313-1438-5}


\bibitem[Xu et~al\mbox{.}(2024b)]{xuRepresentingLongVolumetric2024}
\bibfield{author}{\bibinfo{person}{Zhen Xu}, \bibinfo{person}{Yinghao Xu}, \bibinfo{person}{Zhiyuan Yu}, \bibinfo{person}{Sida Peng}, \bibinfo{person}{Jiaming Sun}, \bibinfo{person}{Hujun Bao}, {and} \bibinfo{person}{Xiaowei Zhou}.} \bibinfo{year}{2024}\natexlab{b}.
\newblock \showarticletitle{Representing {{Long Volumetric Video}} with {{Temporal Gaussian Hierarchy}}}.
\newblock \bibinfo{journal}{\emph{ACM Trans. Graph.}} \bibinfo{volume}{43}, \bibinfo{number}{6} (\bibinfo{year}{2024}), \bibinfo{pages}{171:1--171:18}.
\newblock


\bibitem[Yan et~al\mbox{.}(2024b)]{yan4DGaussianSplatting2024}
\bibfield{author}{\bibinfo{person}{Jinbo Yan}, \bibinfo{person}{Rui Peng}, \bibinfo{person}{Luyang Tang}, {and} \bibinfo{person}{Ronggang Wang}.} \bibinfo{year}{2024}\natexlab{b}.
\newblock \showarticletitle{{{4D Gaussian Splatting}} with {{Scale-aware Residual Field}} and {{Adaptive Optimization}} for {{Real-time Rendering}} of {{Temporally Complex Dynamic Scenes}}}. In \bibinfo{booktitle}{\emph{Proceedings of the 32nd {{ACM International Conference}} on {{Multimedia}}}} \emph{(\bibinfo{series}{{{MM}} '24})}. \bibinfo{pages}{7871--7880}.
\newblock


\bibitem[Yan et~al\mbox{.}(2024a)]{yanMultiScale3DGaussian2024}
\bibfield{author}{\bibinfo{person}{Zhiwen Yan}, \bibinfo{person}{Weng~Fei Low}, \bibinfo{person}{Yu Chen}, {and} \bibinfo{person}{Gim~Hee Lee}.} \bibinfo{year}{2024}\natexlab{a}.
\newblock \showarticletitle{Multi-{{Scale 3D Gaussian Splatting}} for {{Anti-Aliased Rendering}}}. In \bibinfo{booktitle}{\emph{Proceedings of the {{IEEE}}/{{CVF Conference}} on {{Computer Vision}} and {{Pattern Recognition}}}}. \bibinfo{pages}{20923--20931}.
\newblock


\bibitem[Yin et~al\mbox{.}(2025)]{yinTrackerSplatExploitingPoint2025}
\bibfield{author}{\bibinfo{person}{Daheng Yin}, \bibinfo{person}{Isaac Ding}, \bibinfo{person}{Yili Jin}, \bibinfo{person}{Jianxin Shi}, {and} \bibinfo{person}{Jiangchuan Liu}.} \bibinfo{year}{2025}\natexlab{}.
\newblock \showarticletitle{{{TrackerSplat}}: {{Exploiting Point Tracking}} for {{Fast}} and {{Robust Dynamic 3D Gaussians Reconstruction}}}. In \bibinfo{booktitle}{\emph{Proceedings of the {{SIGGRAPH Asia}} 2025 {{Conference Papers}}}} \emph{(\bibinfo{series}{{{SA Conference Papers}} '25})}. \bibinfo{pages}{1--11}.
\newblock


\bibitem[Yin et~al\mbox{.}(2024)]{yinFSVFGImmersiveFullScene2024}
\bibfield{author}{\bibinfo{person}{Daheng Yin}, \bibinfo{person}{Jianxin Shi}, \bibinfo{person}{Miao Zhang}, \bibinfo{person}{Zhaowu Huang}, \bibinfo{person}{Jiangchuan Liu}, {and} \bibinfo{person}{Fang Dong}.} \bibinfo{year}{2024}\natexlab{}.
\newblock \showarticletitle{{{FSVFG}}: {{Towards Immersive Full-Scene Volumetric Video Streaming}} with {{Adaptive Feature Grid}}}. In \bibinfo{booktitle}{\emph{Proceedings of the 32nd {{ACM International Conference}} on {{Multimedia}}}} \emph{(\bibinfo{series}{{{MM}} '24})}. \bibinfo{pages}{11089--11098}.
\newblock


\bibitem[Zhang et~al\mbox{.}(2022)]{zhangYuZuNeuralEnhancedVolumetric2022}
\bibfield{author}{\bibinfo{person}{Anlan Zhang}, \bibinfo{person}{Chendong Wang}, \bibinfo{person}{Bo Han}, {and} \bibinfo{person}{Feng Qian}.} \bibinfo{year}{2022}\natexlab{}.
\newblock \showarticletitle{{{YuZu}}: {{Neural-Enhanced Volumetric Video Streaming}}}. In \bibinfo{booktitle}{\emph{19th {{USENIX Symposium}} on {{Networked Systems Design}} and {{Implementation}} ({{NSDI}} 22)}}. \bibinfo{pages}{137--154}.
\newblock


\bibitem[Zhang et~al\mbox{.}(2024)]{zhangHabitusBoostingMobile2024}
\bibfield{author}{\bibinfo{person}{Anlan Zhang}, \bibinfo{person}{Chendong Wang}, \bibinfo{person}{Yuming Hu}, \bibinfo{person}{Ahmad Hassan}, \bibinfo{person}{Zejun Zhang}, \bibinfo{person}{Bo Han}, \bibinfo{person}{Feng Qian}, {and} \bibinfo{person}{Shichang Xu}.} \bibinfo{year}{2024}\natexlab{}.
\newblock \showarticletitle{Habitus: {{Boosting Mobile Immersive Content Delivery}} through {{Full-body Pose Tracking}} and {{Multipath Networking}}}. In \bibinfo{booktitle}{\emph{21st {{USENIX Symposium}} on {{Networked Systems Design}} and {{Implementation}} ({{NSDI}} 24)}}. \bibinfo{pages}{1677--1695}.
\newblock


\bibitem[Zhang et~al\mbox{.}(2019)]{zhangDeepExemplarBasedVideo2019}
\bibfield{author}{\bibinfo{person}{Bo Zhang}, \bibinfo{person}{Mingming He}, \bibinfo{person}{Jing Liao}, \bibinfo{person}{Pedro~V. Sander}, \bibinfo{person}{Lu Yuan}, \bibinfo{person}{Amine Bermak}, {and} \bibinfo{person}{Dong Chen}.} \bibinfo{year}{2019}\natexlab{}.
\newblock \showarticletitle{Deep {{Exemplar-Based Video Colorization}}}. In \bibinfo{booktitle}{\emph{Proceedings of the {{IEEE}}/{{CVF Conference}} on {{Computer Vision}} and {{Pattern Recognition}}}}. \bibinfo{pages}{8052--8061}.
\newblock


\bibitem[Zhang et~al\mbox{.}(2021)]{zhangEditableFreeviewpointVideo2021}
\bibfield{author}{\bibinfo{person}{Jiakai Zhang}, \bibinfo{person}{Xinhang Liu}, \bibinfo{person}{Xinyi Ye}, \bibinfo{person}{Fuqiang Zhao}, \bibinfo{person}{Yanshun Zhang}, \bibinfo{person}{Minye Wu}, \bibinfo{person}{Yingliang Zhang}, \bibinfo{person}{Lan Xu}, {and} \bibinfo{person}{Jingyi Yu}.} \bibinfo{year}{2021}\natexlab{}.
\newblock \showarticletitle{Editable Free-Viewpoint Video Using a Layered Neural Representation}.
\newblock \bibinfo{journal}{\emph{ACM Transactions on Graphics}} \bibinfo{volume}{40}, \bibinfo{number}{4} (\bibinfo{year}{2021}), \bibinfo{pages}{149:1--149:18}.
\newblock


\bibitem[Zhang et~al\mbox{.}(2018)]{zhang2018perceptual}
\bibfield{author}{\bibinfo{person}{Richard Zhang}, \bibinfo{person}{Phillip Isola}, \bibinfo{person}{Alexei~A Efros}, \bibinfo{person}{Eli Shechtman}, {and} \bibinfo{person}{Oliver Wang}.} \bibinfo{year}{2018}\natexlab{}.
\newblock \showarticletitle{The Unreasonable Effectiveness of Deep Features as a Perceptual Metric}. In \bibinfo{booktitle}{\emph{CVPR}}.
\newblock


\bibitem[Zhong et~al\mbox{.}(2024)]{zhongVQNeRFNeuralReflectance2024}
\bibfield{author}{\bibinfo{person}{Hongliang Zhong}, \bibinfo{person}{Jingbo Zhang}, {and} \bibinfo{person}{Jing Liao}.} \bibinfo{year}{2024}\natexlab{}.
\newblock \showarticletitle{{{VQ-NeRF}}: {{Neural Reflectance Decomposition}} and {{Editing With Vector Quantization}}}.
\newblock \bibinfo{journal}{\emph{IEEE Transactions on Visualization and Computer Graphics}} \bibinfo{volume}{30}, \bibinfo{number}{9} (\bibinfo{year}{2024}), \bibinfo{pages}{6247--6260}.
\newblock
\showISSN{1941-0506}


\bibitem[Zhou et~al\mbox{.}(2024)]{zhouGaussianSplattingNeural2024}
\bibfield{author}{\bibinfo{person}{Zhi Zhou}, \bibinfo{person}{Junke Zhu}, {and} \bibinfo{person}{Zhangjin Huang}.} \bibinfo{year}{2024}\natexlab{}.
\newblock \showarticletitle{Gaussian {{Splatting}} with {{Neural Basis Extension}}}. In \bibinfo{booktitle}{\emph{Proceedings of the 32nd {{ACM International Conference}} on {{Multimedia}}}} \emph{(\bibinfo{series}{{{MM}} '24})}. \bibinfo{pages}{6043--6052}.
\newblock


\bibitem[Zhu et~al\mbox{.}(2025)]{zhuSGSSStreaming6DoF2025}
\bibfield{author}{\bibinfo{person}{Mufeng Zhu}, \bibinfo{person}{Mingju Liu}, \bibinfo{person}{Cunxi Yu}, \bibinfo{person}{Cheng-Hsin Hsu}, {and} \bibinfo{person}{Yao Liu}.} \bibinfo{year}{2025}\natexlab{}.
\newblock \showarticletitle{{{SGSS}}: {{Streaming}} 6-{{DoF Navigation}} of {{Gaussian Splat Scenes}}}. In \bibinfo{booktitle}{\emph{Proceedings of the 16th {{ACM Multimedia Systems Conference}}}} \emph{(\bibinfo{series}{{{MMSys}} '25})}. \bibinfo{pages}{46--56}.
\newblock


\bibitem[Zimmer et~al\mbox{.}(2024)]{zimmer2024tumtraf}
\bibfield{author}{\bibinfo{person}{Walter Zimmer}, \bibinfo{person}{Gerhard~Arya Wardana}, \bibinfo{person}{Suren Sritharan}, \bibinfo{person}{Xingcheng Zhou}, \bibinfo{person}{Rui Song}, {and} \bibinfo{person}{Alois~C Knoll}.} \bibinfo{year}{2024}\natexlab{}.
\newblock \showarticletitle{Tumtraf v2x cooperative perception dataset}. In \bibinfo{booktitle}{\emph{Proceedings of the IEEE/CVF conference on computer vision and pattern recognition}}. \bibinfo{pages}{22668--22677}.
\newblock


\end{thebibliography}

\appendix

\begin{figure*}[t]
	\centering
	\includegraphics[width=0.948\linewidth]{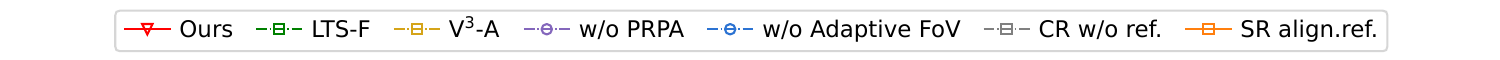}
	\\
	\includegraphics[width=0.0158\linewidth]{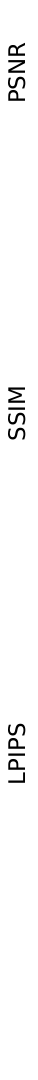}
	\begin{subfigure}[b]{0.158\linewidth}
		\centering
        \includegraphics[width=\linewidth]{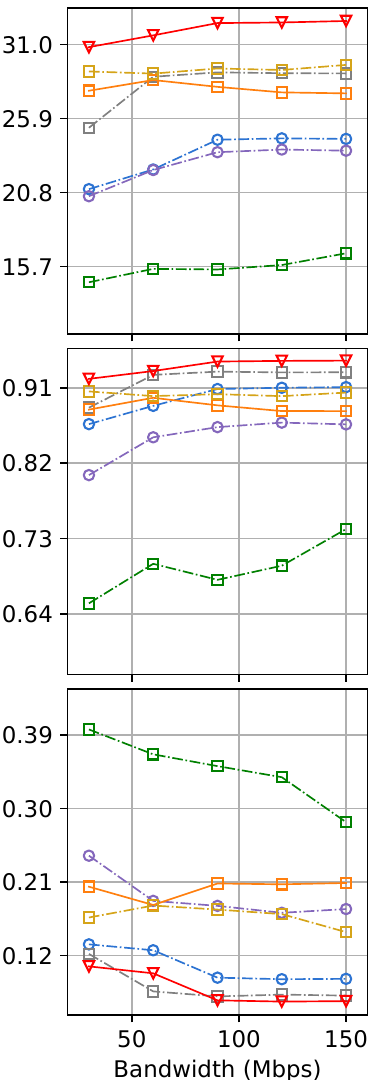}
		\caption{``cook spinach''}
	\end{subfigure}
	\begin{subfigure}[b]{0.158\linewidth}
		\centering
        \includegraphics[width=\linewidth]{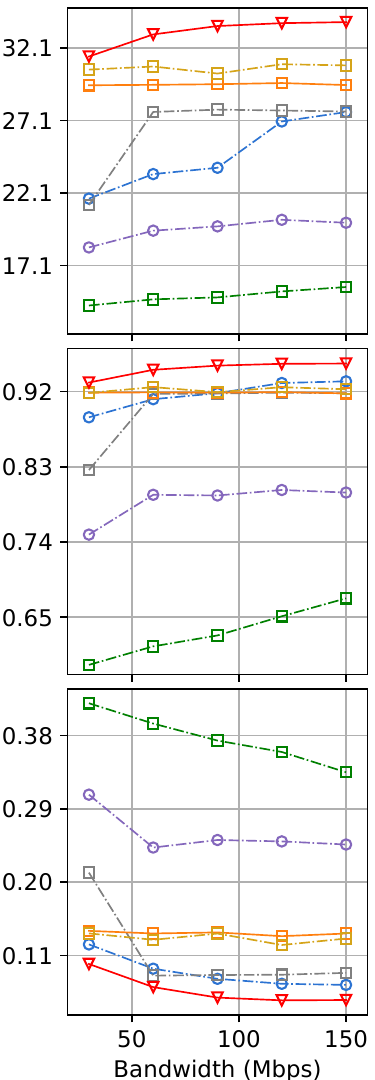}
		\caption{``cut roasted beef''}
	\end{subfigure}
	\begin{subfigure}[b]{0.158\linewidth}
		\centering
        \includegraphics[width=\linewidth]{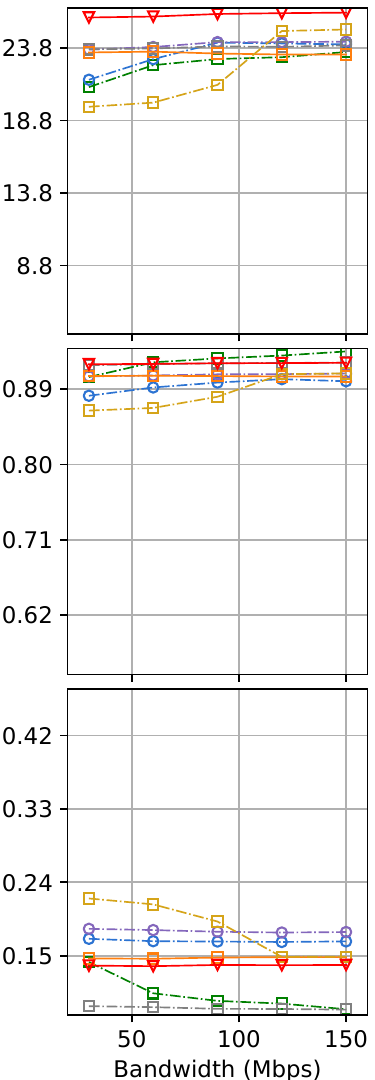}
		\caption{``walking''}\label{fig:walking}
	\end{subfigure}
	\begin{subfigure}[b]{0.158\linewidth}
		\centering
        \includegraphics[width=\linewidth]{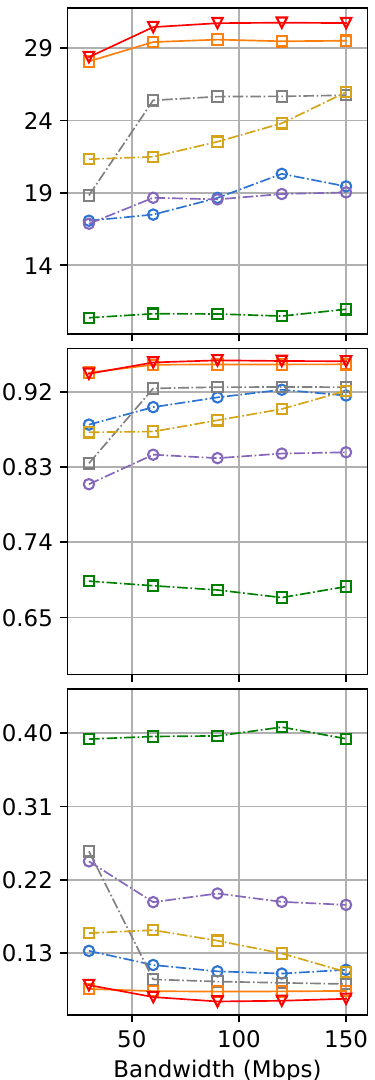}
		\caption{``discussion''}
	\end{subfigure}
	\begin{subfigure}[b]{0.158\linewidth}
		\centering
        \includegraphics[width=\linewidth]{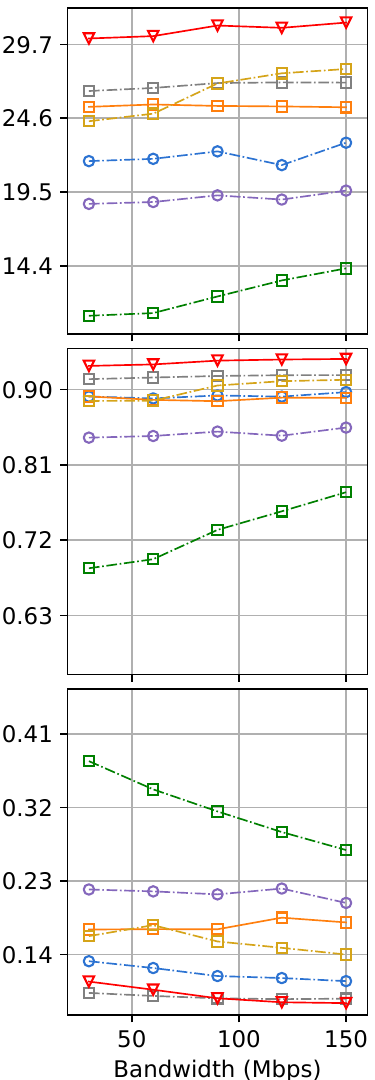}
		\caption{``stepin''}
	\end{subfigure}
	\begin{subfigure}[b]{0.158\linewidth}
		\centering
        \includegraphics[width=\linewidth]{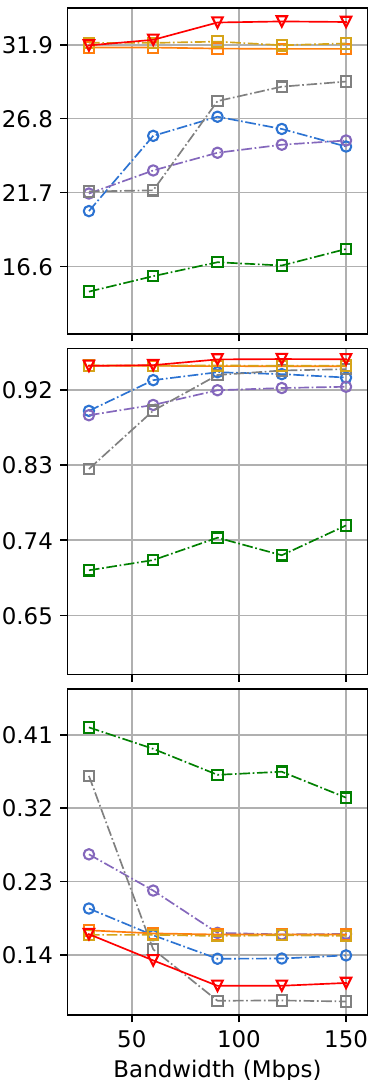}
		\caption{``basketball''}
	\end{subfigure}
	\caption{
		Comparison of visual quality under fixed bandwidth.
		The y-axis ranges vary across subplots while maintaining equal scale spans for each metric across videos.
		Results for all videos are included in the supplementary material.
	}\label{fig:FixedBandwidth}\end{figure*} 
\begin{figure*}[t]
	\centering
	\includegraphics[width=0.948\linewidth]{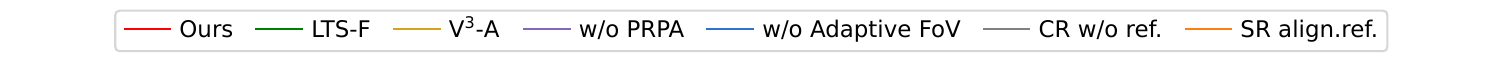}
	\\
	\includegraphics[width=0.0158\linewidth]{figures/FixedBandwidth/ylabel.pdf}
	\begin{subfigure}[b]{0.158\linewidth}
		\centering
        \includegraphics[width=\linewidth]{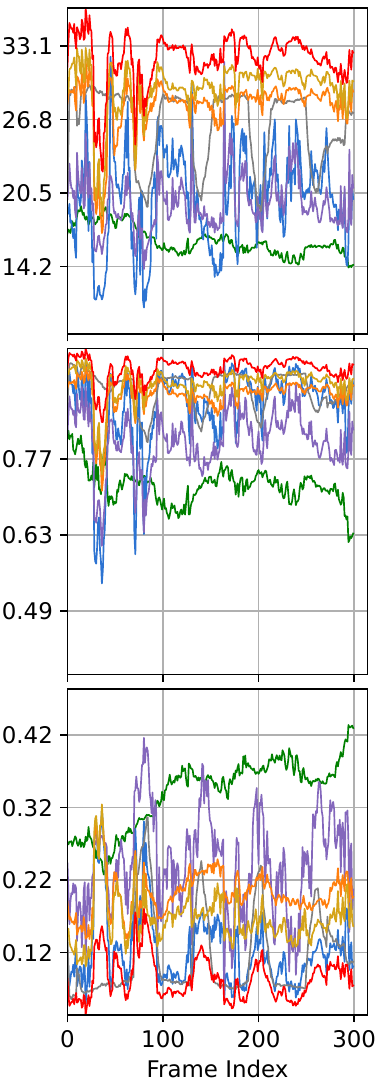}
		\caption{``cook spinach''}
	\end{subfigure}
	\begin{subfigure}[b]{0.158\linewidth}
		\centering
        \includegraphics[width=\linewidth]{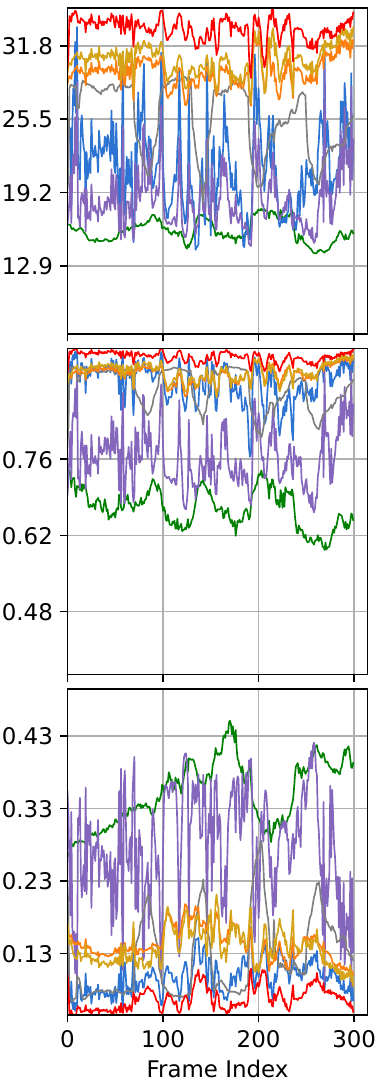}
		\caption{``cut roasted beef''}
	\end{subfigure}
	\begin{subfigure}[b]{0.158\linewidth}
		\centering
        \includegraphics[width=\linewidth]{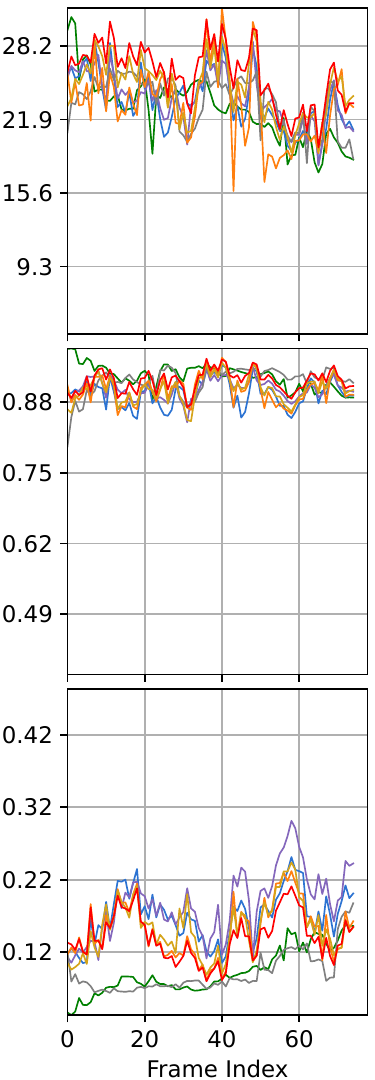}
		\caption{``walking''}
	\end{subfigure}
	\begin{subfigure}[b]{0.158\linewidth}
		\centering
        \includegraphics[width=\linewidth]{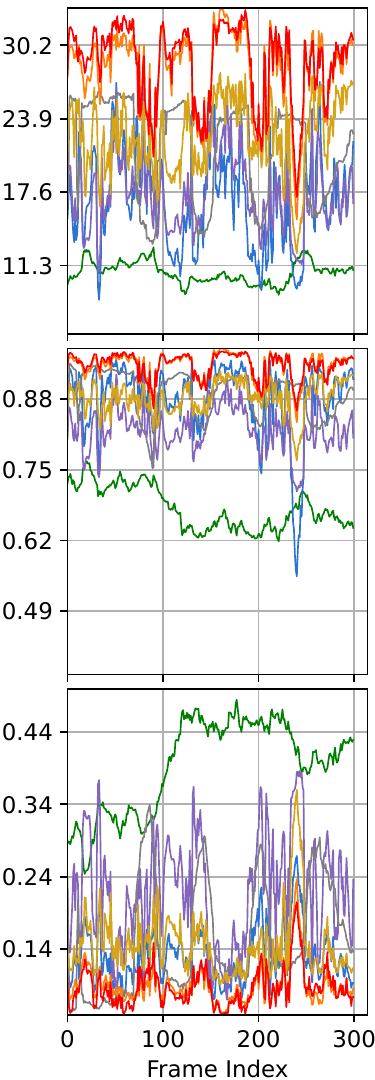}
		\caption{``discussion''}
	\end{subfigure}
	\begin{subfigure}[b]{0.158\linewidth}
		\centering
        \includegraphics[width=\linewidth]{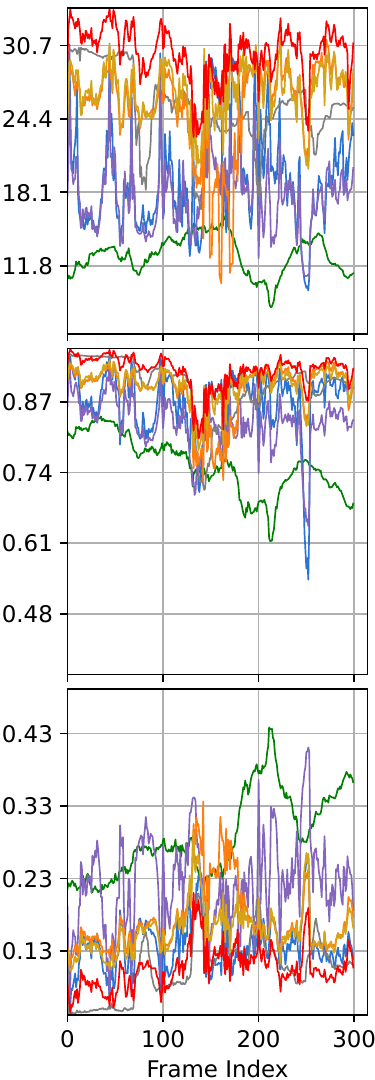}
		\caption{``stepin''}
	\end{subfigure}
	\begin{subfigure}[b]{0.158\linewidth}
		\centering
        \includegraphics[width=\linewidth]{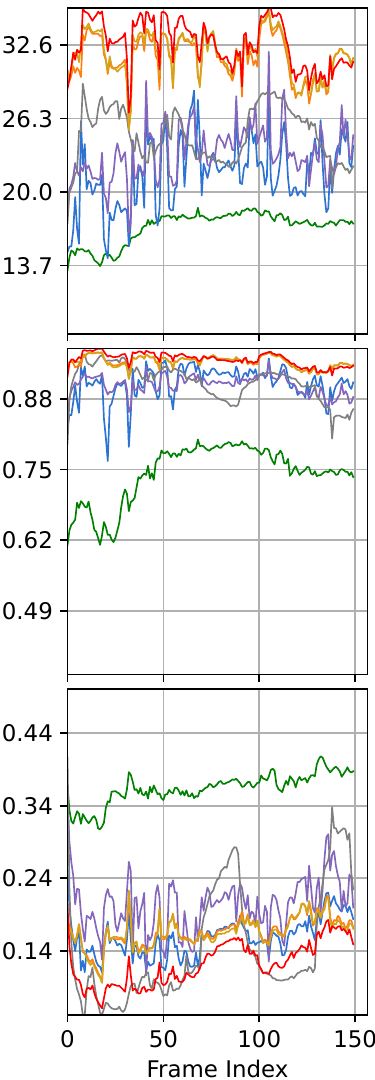}
		\caption{``basketball''}
	\end{subfigure}
	\caption{
		Comparison of visual quality under fluctuating bandwidth.
		The y-axis ranges vary across subplots while maintaining equal scale spans for each metric across videos.
		Results for all videos are included in the supplementary material.
	}\label{fig:DynamicBandwidth}\end{figure*} 
\begin{figure*}[t]
	\centering
	\begin{subfigure}[b]{0.245\linewidth}
		\centering
\scalebox{-1}[-1]{\includegraphics[width=\linewidth]{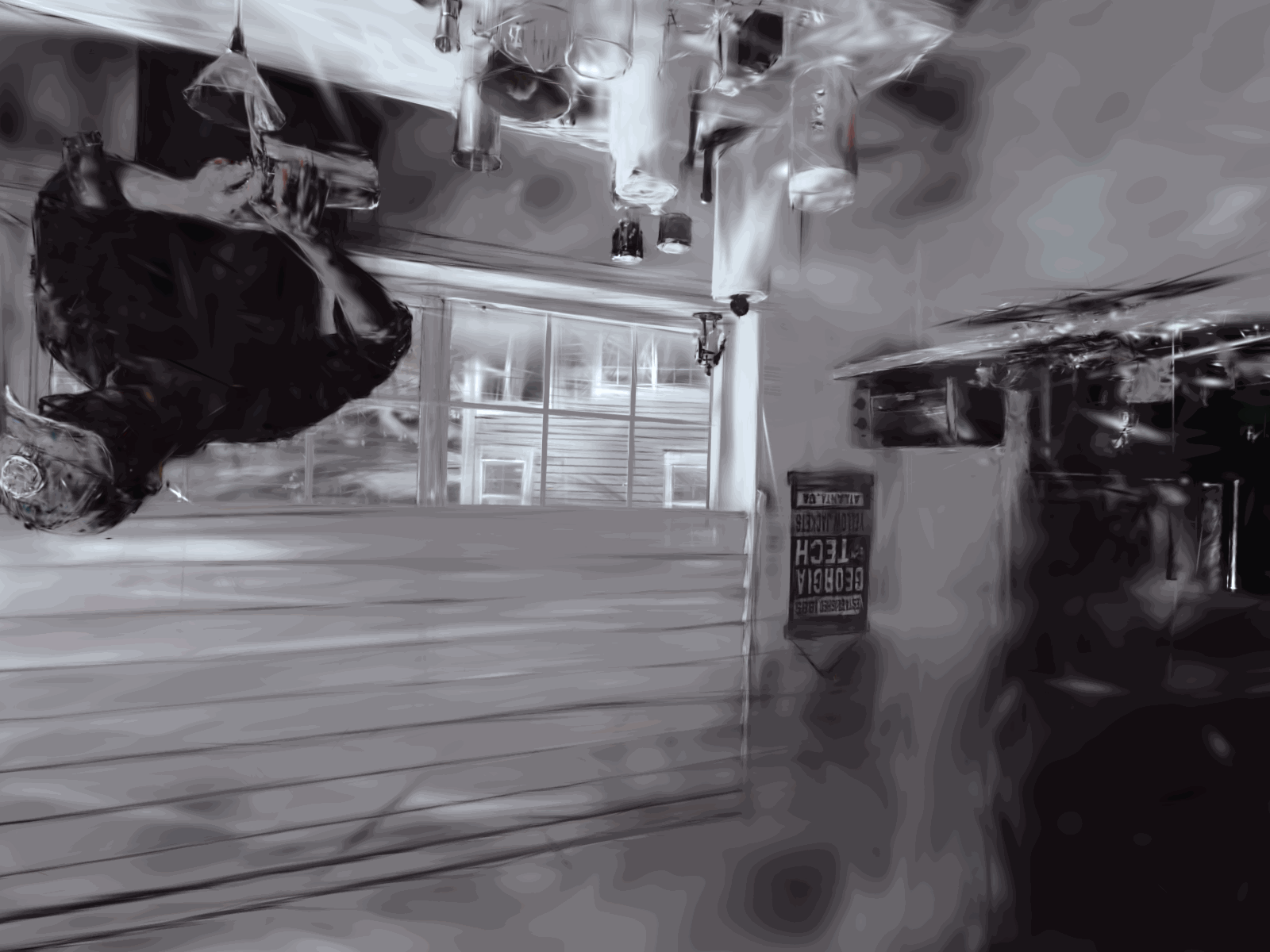}}
		\caption{Color-distorted (1600x1200)}\label{fig:visualdistorted}
	\end{subfigure}
	\begin{subfigure}[b]{0.245\linewidth}
		\centering
        \scalebox{-1}[-1]{\includegraphics[width=\linewidth]{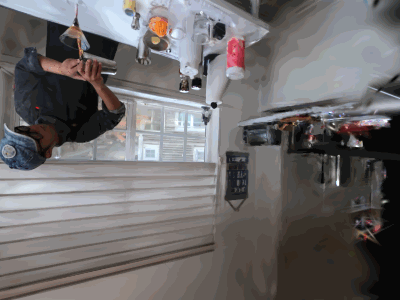}}
		\caption{Reference image (400x300)}
	\end{subfigure}
	\begin{subfigure}[b]{0.245\linewidth}
		\centering
        \scalebox{-1}[-1]{\includegraphics[width=\linewidth]{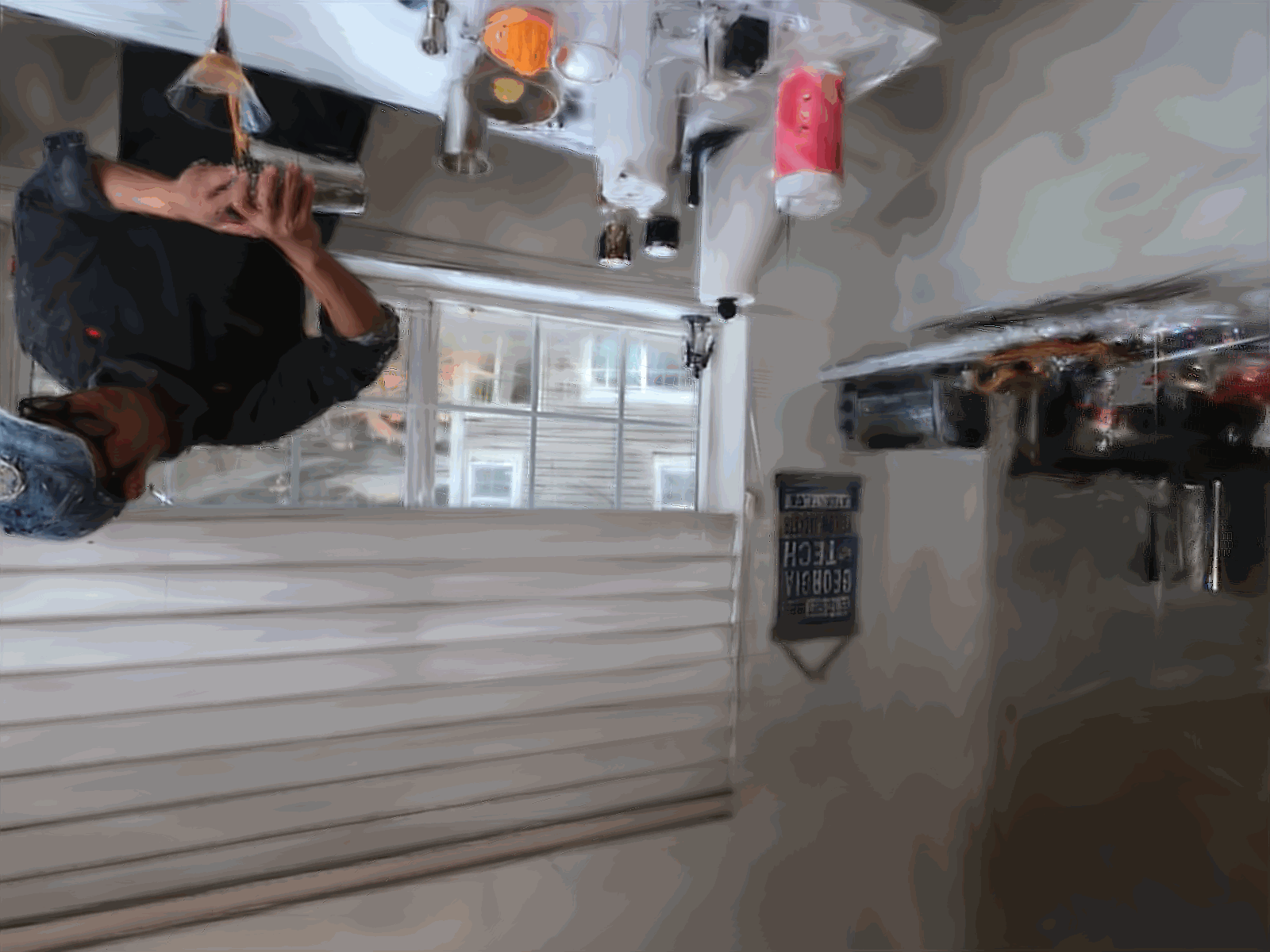}}
		\caption{Color-restored (1600x1200)}\label{fig:visualrestored}
	\end{subfigure}
	\begin{subfigure}[b]{0.245\linewidth}
		\centering
        \scalebox{-1}[-1]{\includegraphics[width=\linewidth]{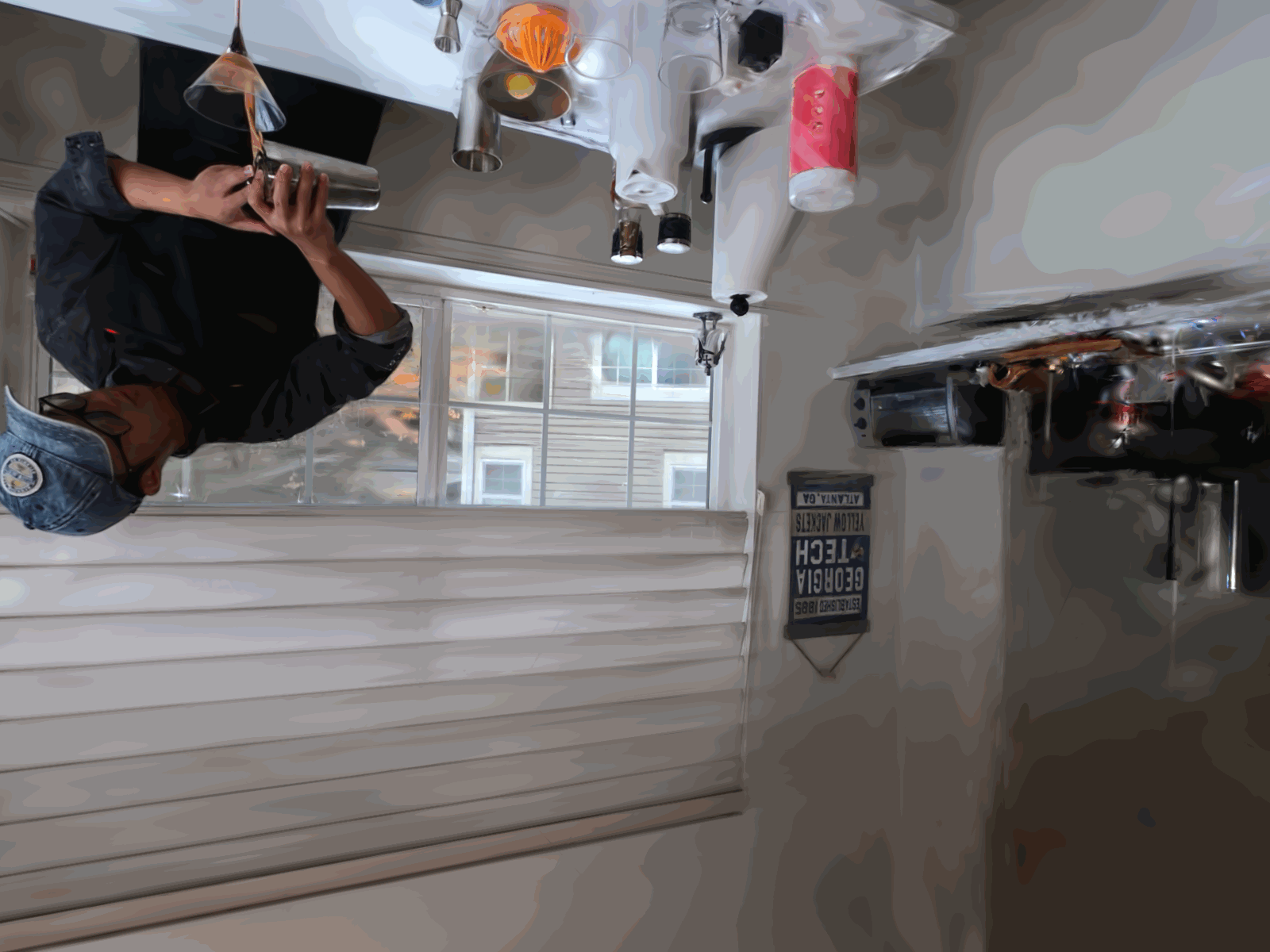}}
		\caption{Ground truth (1600x1200)}
	\end{subfigure}
	\caption{
		Visual results of the frame 64 in ``coffee martini'' under 30Mbps.
		Additional visual and video results are included in supplementary materials.
	}\label{fig:visual}
\end{figure*} 
\newcommand{\includesubfigureFixedBandwidth}[2]{\begin{subfigure}[b]{0.158\linewidth}
		\centering
        \includegraphics[width=\linewidth,trim={0 0 0 58mm},clip]{figures/FixedBandwidth/#1_no_lpips.pdf}
		\caption{``#2''}
	\end{subfigure}}
\newcommand{\includesubfigureDynamicBandwidth}[2]{\begin{subfigure}[b]{0.158\linewidth}
		\centering
        \includegraphics[width=\linewidth,trim={0 0 0 58mm},clip]{figures/DynamicBandwidth/#1_no_lpips.pdf}
		\caption{``#2''}
	\end{subfigure}}
\newcommand{\includeylabel}{
    \includegraphics[width=0.0158\linewidth,trim={0 61mm 0 0},clip]{figures/FixedBandwidth/ylabel.pdf}
}
\newcommand{\includelegend}{
    \includegraphics[width=0.948\linewidth]{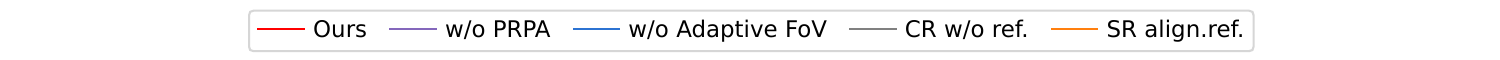}
}

\begin{figure*}[t]
	\centering
	\includelegend
	\\
	\includeylabel
	\includesubfigureFixedBandwidth{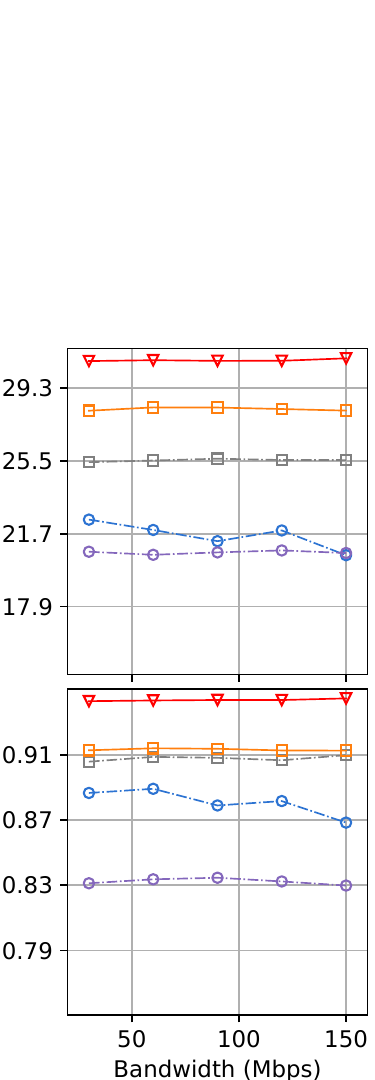}{cook spinach}
	\includesubfigureFixedBandwidth{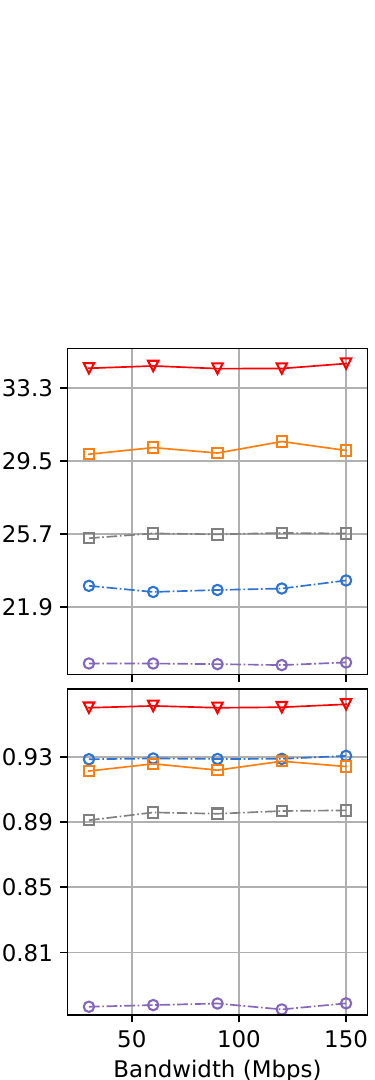}{cut roasted beef}
	\includesubfigureFixedBandwidth{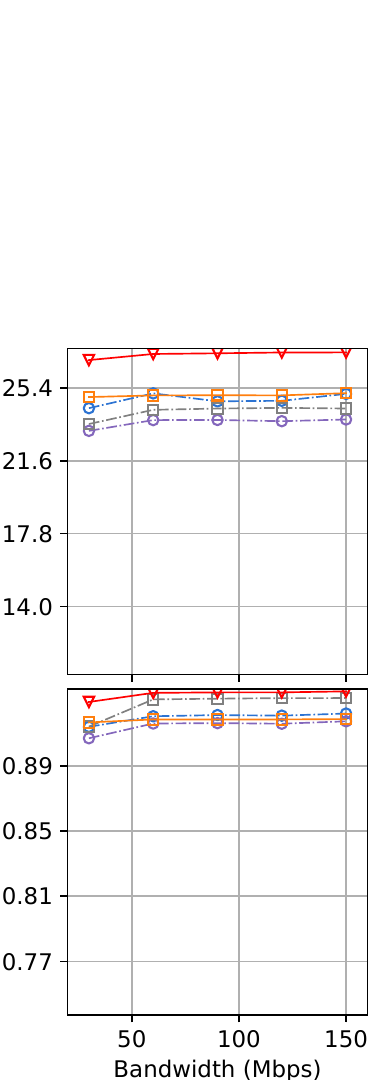}{walking}
	\includesubfigureFixedBandwidth{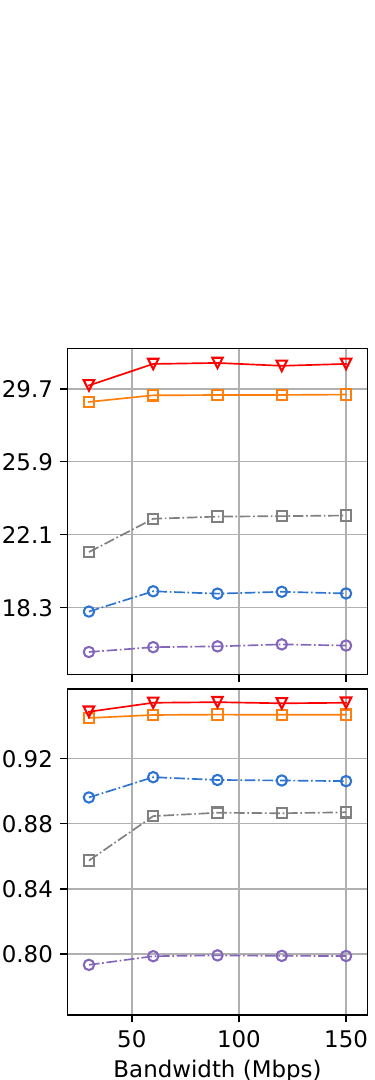}{discussion}
	\includesubfigureFixedBandwidth{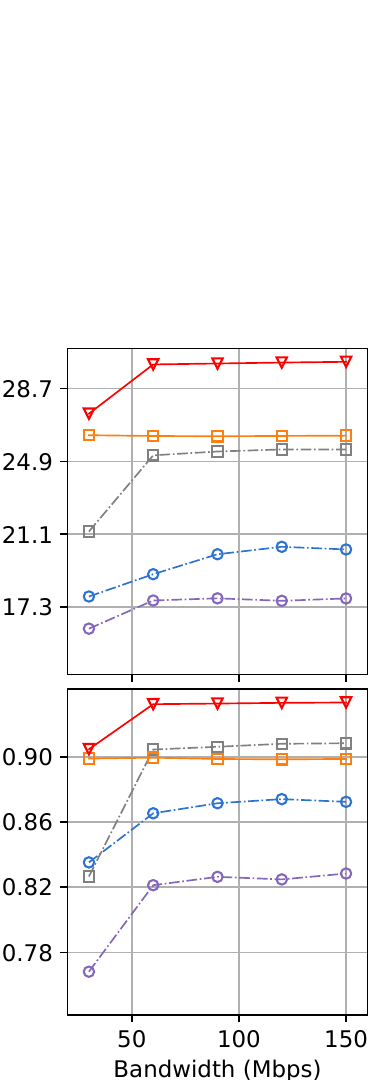}{stepin}
	\includesubfigureFixedBandwidth{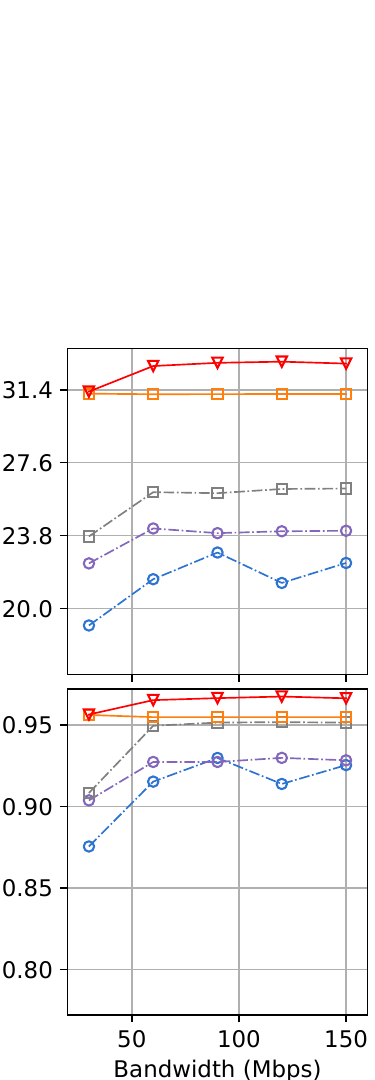}{basketball}
	\caption{
		Comparison of visual quality under fixed bandwidth on volumetric videos prepared by HiCoM~\cite{gaoHiCoMHierarchicalCoherent2024}.
		The y-axis ranges vary across subplots while maintaining equal scale spans for each metric across videos.
		Results for all videos (including volumetric videos prepared by Dynamic 3DGS~\cite{luiten2023dynamic}, 4DGS~\cite{wu4DGaussianSplatting2024} and HiCoM~\cite{gaoHiCoMHierarchicalCoherent2024}) under fixed bandwidth conditions are included in the supplementary material.
	}\label{fig:DynamicBandwidth-dynamic3dgs}\end{figure*}

\begin{figure*}[t]
	\centering
	\includelegend
	\\
	\includeylabel
	\includesubfigureDynamicBandwidth{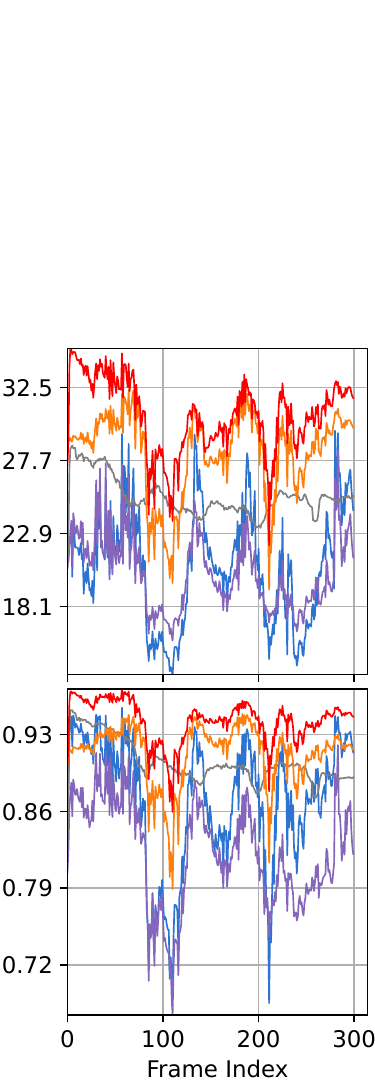}{cook spinach}
	\includesubfigureDynamicBandwidth{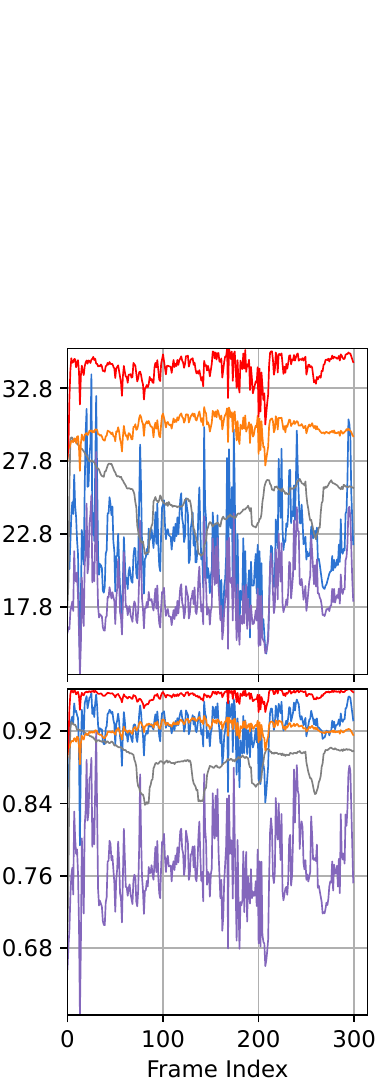}{cut roasted beef}
	\includesubfigureDynamicBandwidth{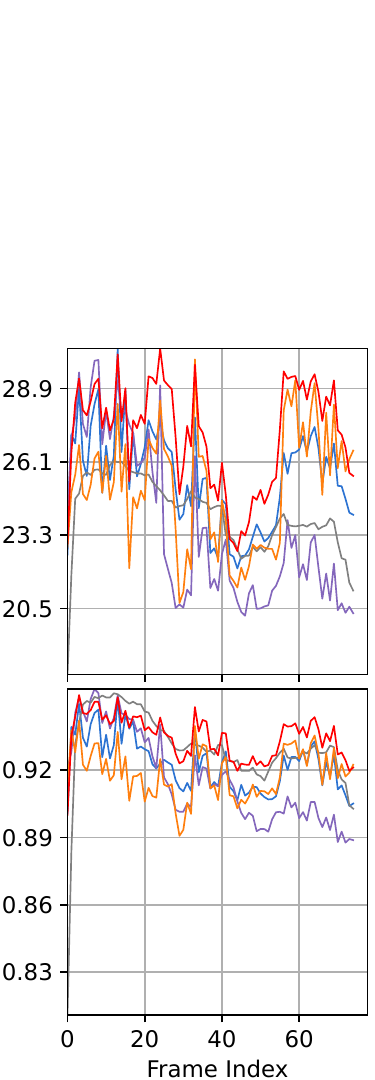}{walking}
	\includesubfigureDynamicBandwidth{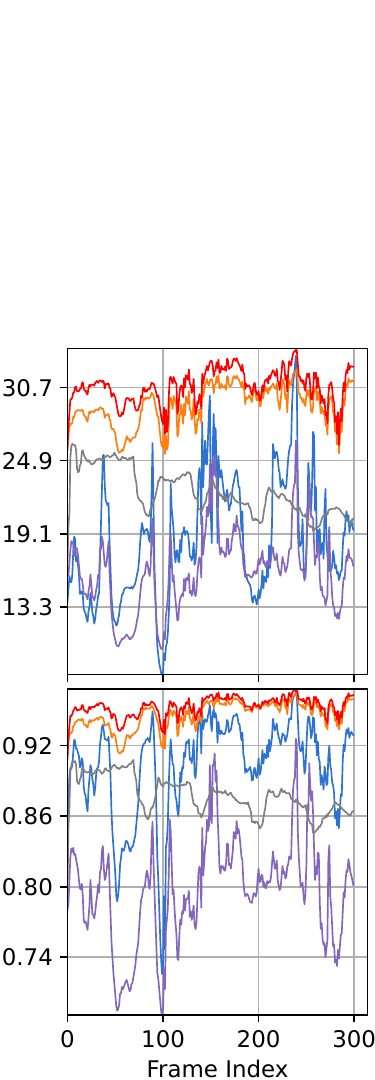}{discussion}
	\includesubfigureDynamicBandwidth{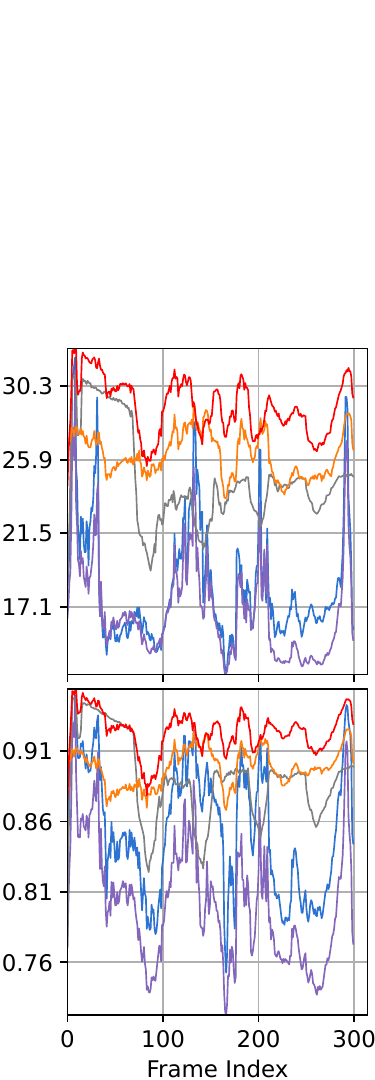}{stepin}
	\includesubfigureDynamicBandwidth{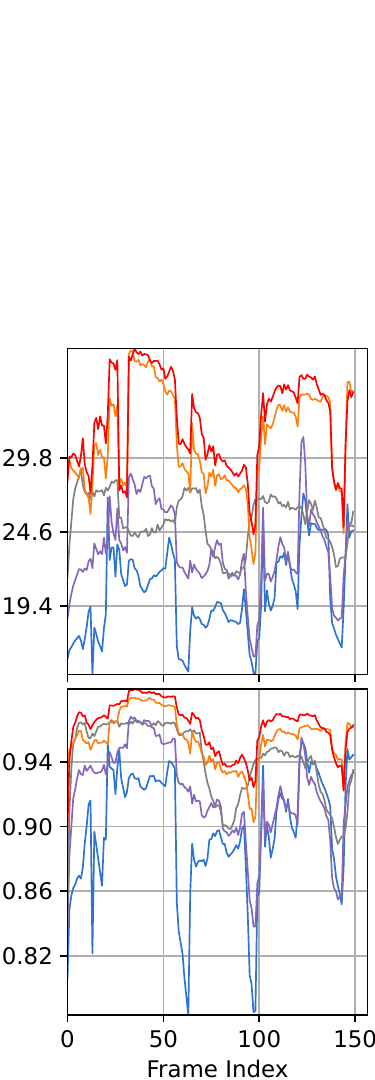}{basketball}
	\caption{
		Comparison of visual quality under fluctuating bandwidth on volumetric videos prepared by HiCoM~\cite{gaoHiCoMHierarchicalCoherent2024}.
		The y-axis ranges vary across subplots while maintaining equal scale spans for each metric across videos.
		Results for all videos (including volumetric videos prepared by Dynamic 3DGS~\cite{luiten2023dynamic}, 4DGS~\cite{wu4DGaussianSplatting2024} and HiCoM~\cite{gaoHiCoMHierarchicalCoherent2024}) under fluctuating bandwidth conditions are included in the supplementary material.
	}\label{fig:DynamicBandwidth-hicom}\end{figure*}  
\clearpage

\twocolumn[{
\sffamily\Huge Supplementary materials for ``CAGS: Color-Adaptive Volumetric Video Streaming with Dynamic 3D Gaussian Splatting''
\strut
}]

\section{Pseudocode of PRPA Algorithm}
\begin{algorithm}[H]
\caption{Post-Render Perspective Aligning (PRPA)}\label{alg:Feature Filtering}
\begin{algorithmic}[1]
\REQUIRE 
reference image $\bm A(i,j)$,
locally rendered depth map $\bm D(u,v)$,
camera intrinsic matrix $\bm K_l,\bm K_r$, rotation matrix $\bm R_l,\bm R_r$ and position $\bm t_l,\bm t_r$ of locally rendered image and reference image,
kernel size $k$ for average filtering,
depth threshold $T$ for occlusion detection.
\ENSURE aligned reference image $\bm A'(u,v)$ where the alignment error has been eroded.

\STATE $\bm P_c(u,v)\gets \bm K_l(\bm R_r\bm R_l^{-1}(\bm K_l^{-1}[u,v,1]^\top\bm D(u,v)-\bm t_l)+\bm t_r)$
\STATE $\bm D_r(u,v)\gets \bm P_c(u,v)_3$ \STATE $\bm P(u,v)\gets \bm P_c(u,v)_{1,2}/\bm P_c(u,v)_3$ \STATE $\bm A'(u,v)\gets \bm A(\bm P(u,v)_1,\bm P(u,v)_2)$ \COMMENT{Align with error}

\STATE $\bm C_r(i,j)\gets|\{(u,v)|\bm P(u,v)=(i,j)\}|$ \STATE $\bm C_l(u,v)\gets\bm C_r(\bm P(u,v)_1,\bm P(u,v)_2)$ \STATE $\bm D_r^m(i,j)\gets\min(\{\bm D_r(u,v)|\bm P(u,v)=(i,j)\})$ \STATE $\bm D_l^m(u,v)\gets\bm D_r^m(\bm P(u,v)_1,\bm P(u,v)_2)$ \STATE $\bm P_o\gets\{(u,v)|\bm C_l(u,v)>0 \land |\bm D(u,v)-\bm D^m_l(u,v)|\leq T\}$ \STATE $\bm P_d\gets\{(u,v)|\bm C_l(u,v)>0 \land |\bm D(u,v)-\bm D^m_l(u,v)|>T\}$ 

\WHILE{$|\bm P_d|>0$}
\STATE $\bm E_d\gets$\CALL{MorphologyDilate}{$\bm P_d$}$\cap\complement\bm P_d$ \FOR{$(u_E,v_E)\in \bm E_d\cap\complement\bm P_o$}
\STATE $\bm K\gets[u_E-k,u_E+k]\times[v_E-k,v_E+k]$
\STATE $a\gets\text{mean}(\{\bm A'(u,v)|(u,v)\in\bm K\cap\complement\bm P_o\cap\complement\bm P_d\})$
\FOR{$(u,v)\in\bm K\cap\bm P_d$}
\STATE $\bm A'(u,v)\gets a$ \COMMENT{Erode error along edge}
\STATE delete $(u,v)$ from $\bm P_d$
\ENDFOR
\ENDFOR
\ENDWHILE
\STATE \RETURN $\bm A'(u,v)$
\end{algorithmic}
\end{algorithm} 
\section{Detailed Prototype Implementation}
\begin{figure*}[htbp]
	\centering
    \includegraphics[width=0.9\linewidth]{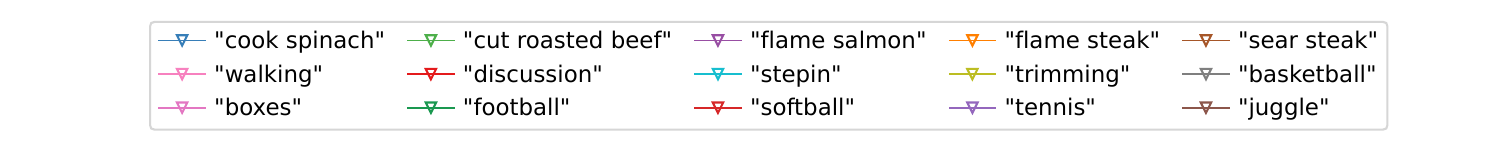}\\
    \includegraphics[width=0.387\linewidth]{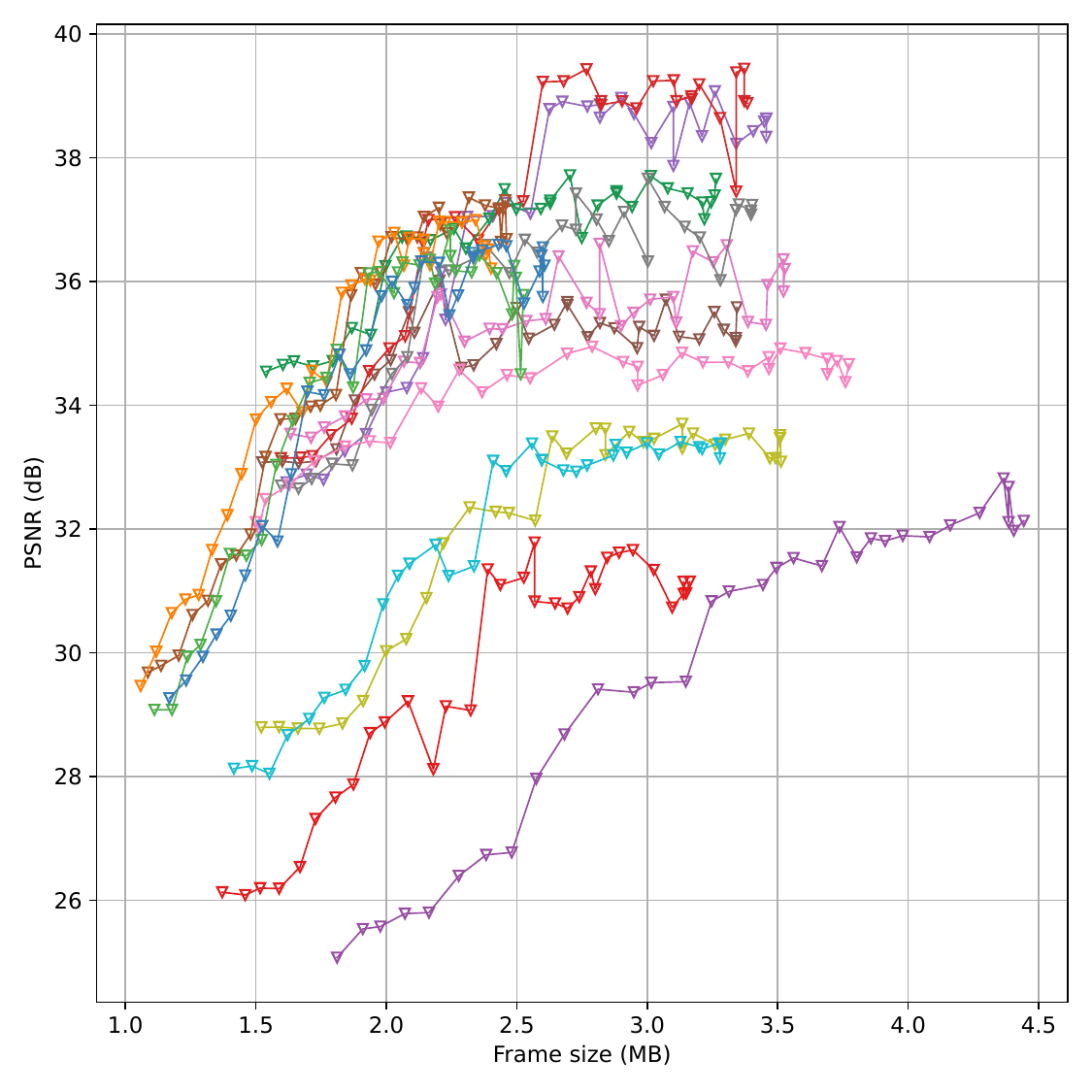}
    \includegraphics[width=0.387\linewidth]{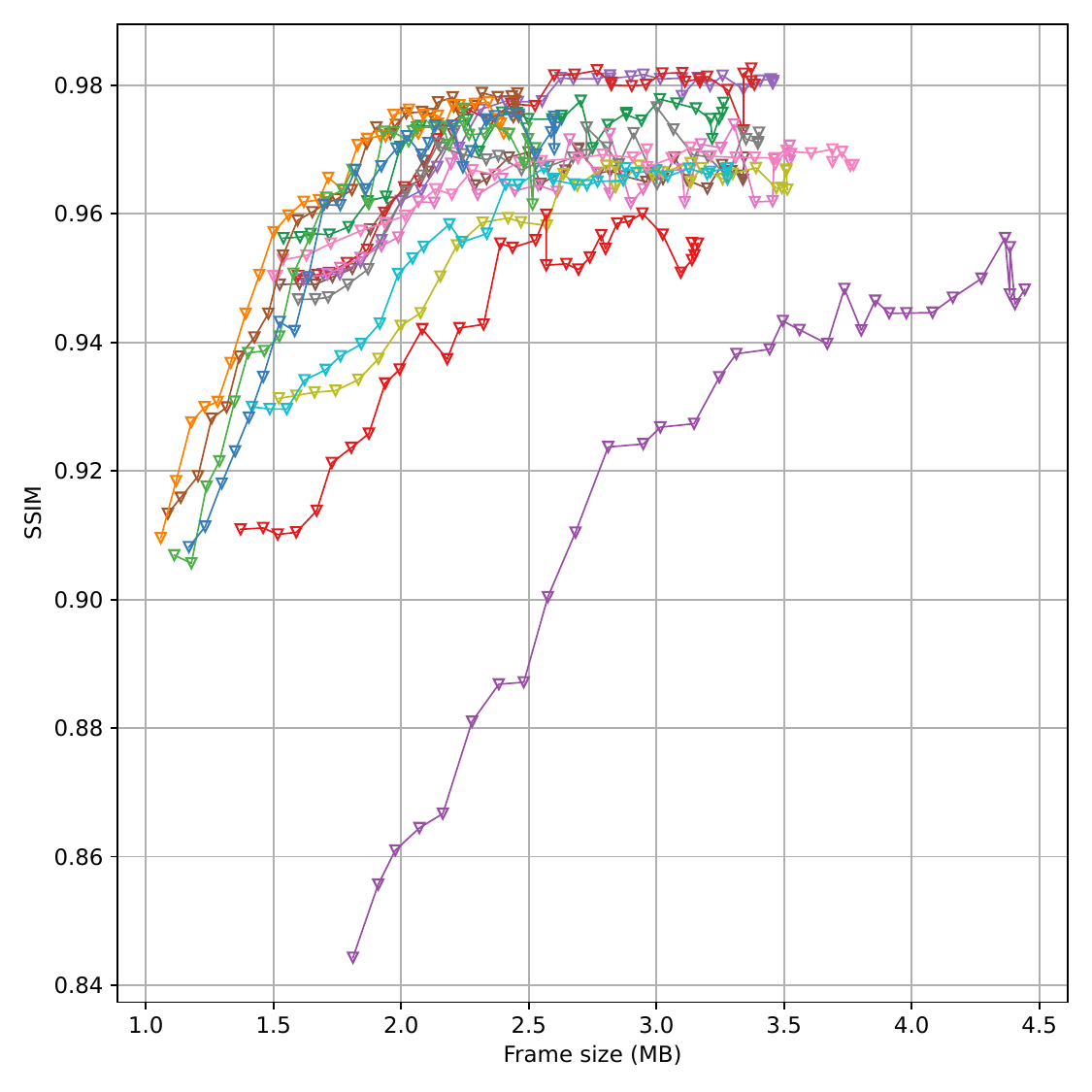}
	\caption{
		Visual quality versus frame size across LoDs constructed by interleaving SVQ layers, evaluated on the first frame of each video.
		Each point corresponds to one LoD.
		The x-axis shows the resulting frame size (excluding the codebook) after removing a given LoD and all higher LoDs, and the y-axis shows the resulting visual quality (PSNR/SSIM).
  }\label{fig:LayerQuality-LoDs}
\end{figure*}

\newcommand{\includepsnrssimpdf}[1]{\includegraphics[width=\linewidth,trim={0 0 0 58mm},clip]{#1}}
\begin{figure*}[htbp]
	\centering
\newcommand{\includelayerquality}[2]{
    \begin{subfigure}[b]{0.129\linewidth}
		\centering
        \includepsnrssimpdf{figures/#1.pdf}
		\caption{``#2''}
	\end{subfigure}
}
\newcommand{\includepsnrssimylabel}{
    \includegraphics[width=0.0129\linewidth,trim={0 61mm 0 0},clip]{figures/FixedBandwidth/ylabel.pdf}
}
    \includepsnrssimylabel
    \begin{subfigure}[b]{0.129\linewidth}
		\centering
        \includepsnrssimpdf{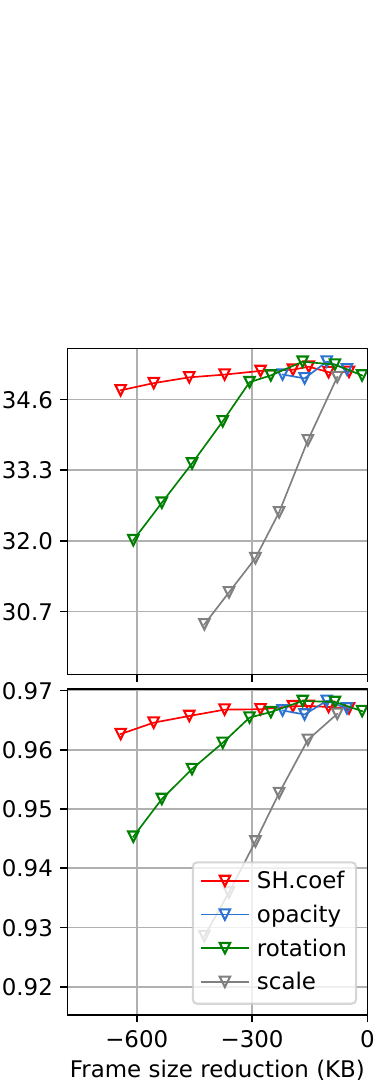}
		\caption{All Average}\label{fig:LayerQuality-all}
	\end{subfigure}
    \\\includepsnrssimylabel
\includelayerquality{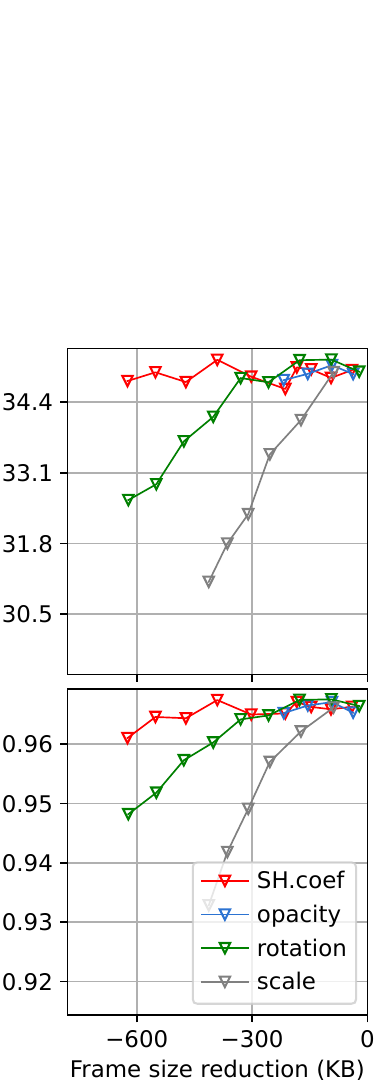}{cook spinach}
    \includelayerquality{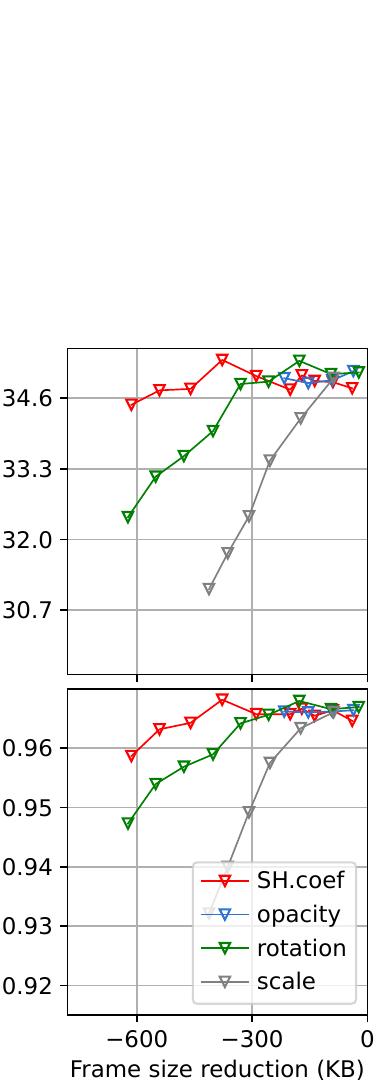}{cut roast beef}
    \includelayerquality{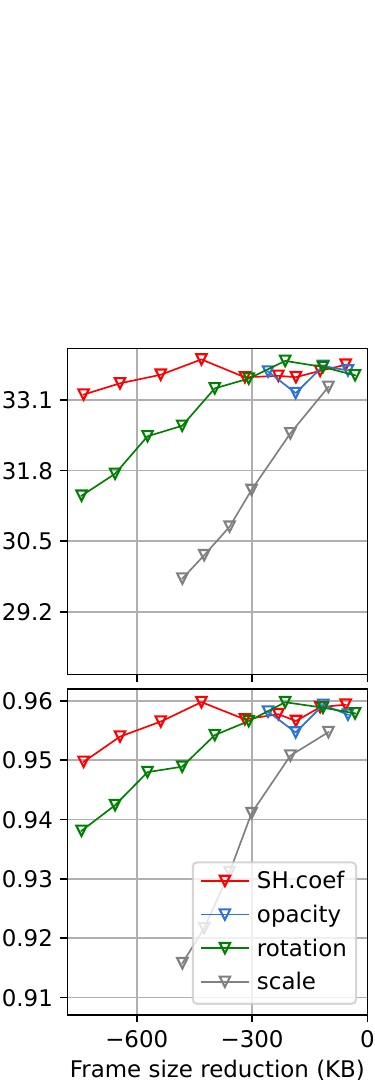}{flame salmon}
    \includelayerquality{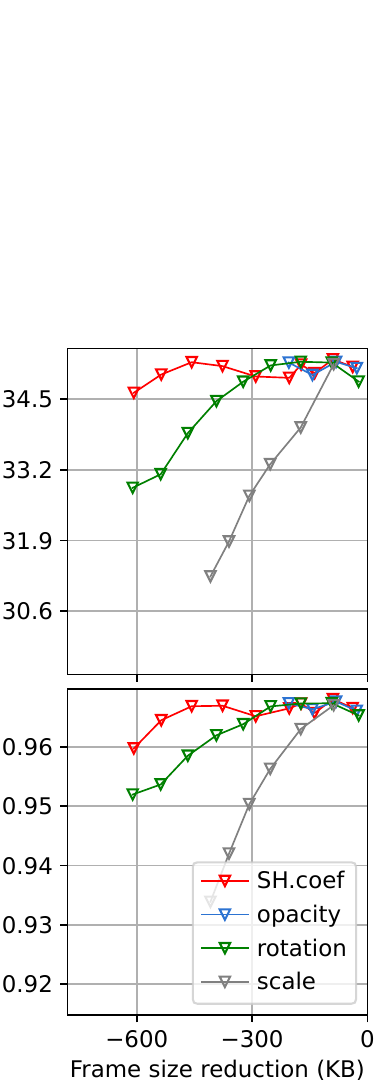}{flame steak}
    \includelayerquality{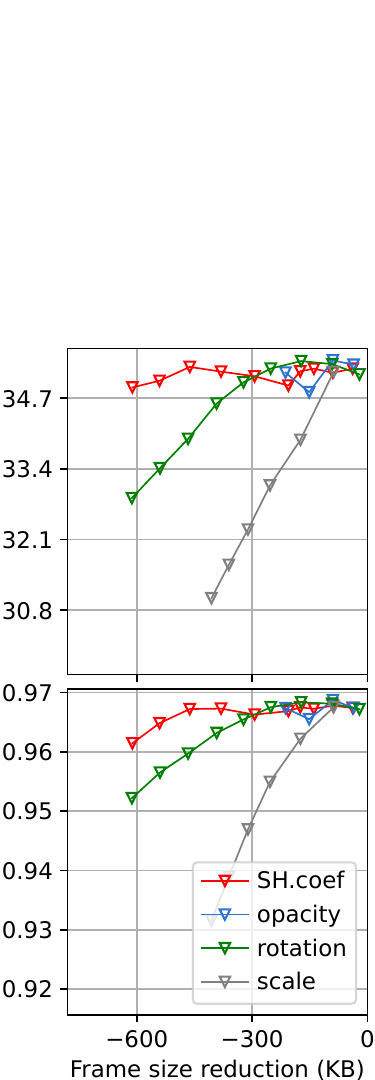}{sear steak}
    \\\includepsnrssimylabel
    \includelayerquality{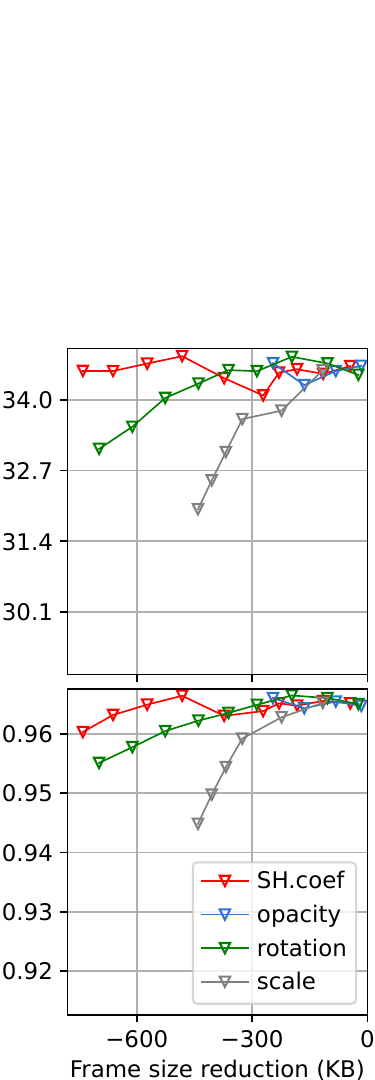}{walking}
\includelayerquality{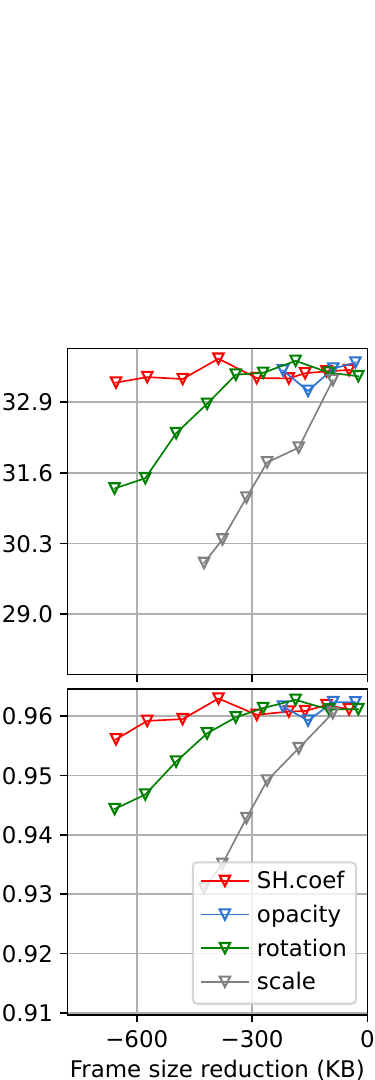}{discussion}
    \includelayerquality{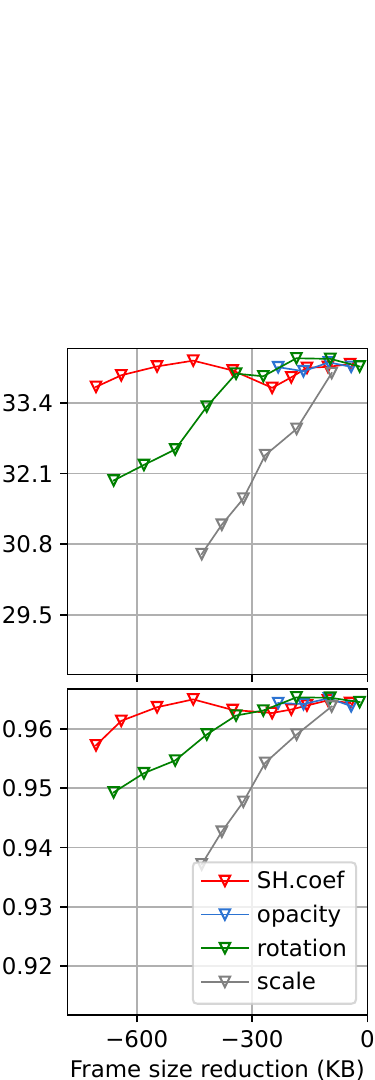}{stepin}
    \includelayerquality{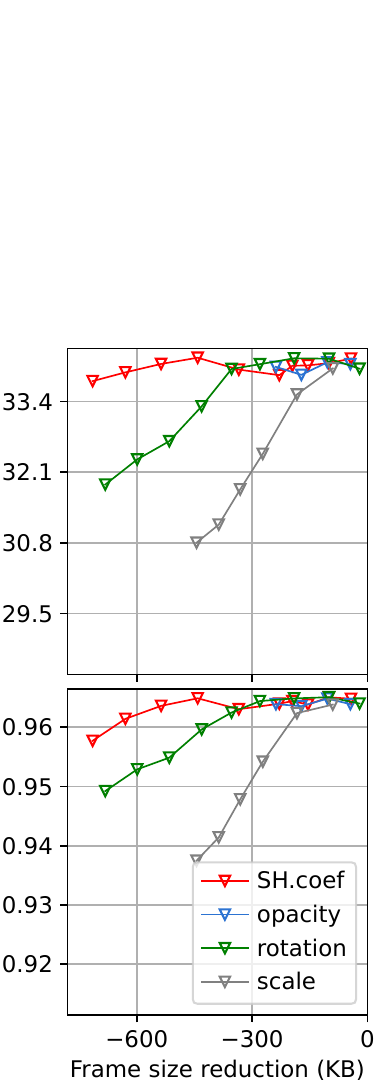}{trimming}
\includelayerquality{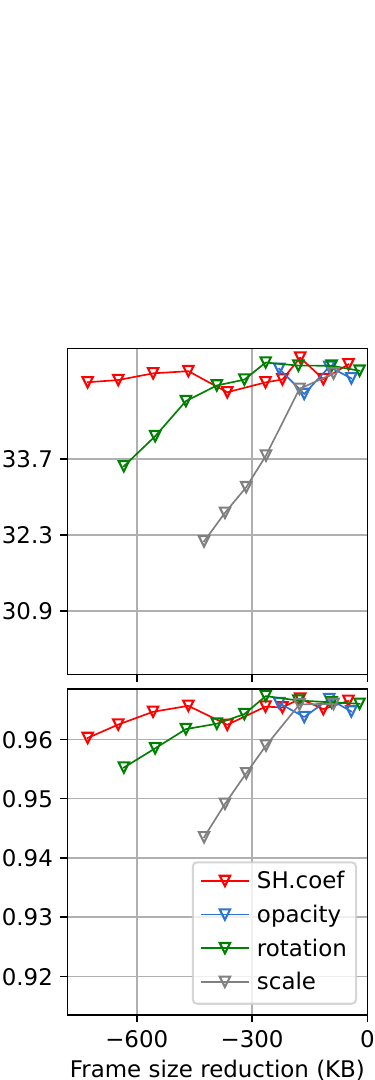}{basketball}
    \\\includepsnrssimylabel
    \includelayerquality{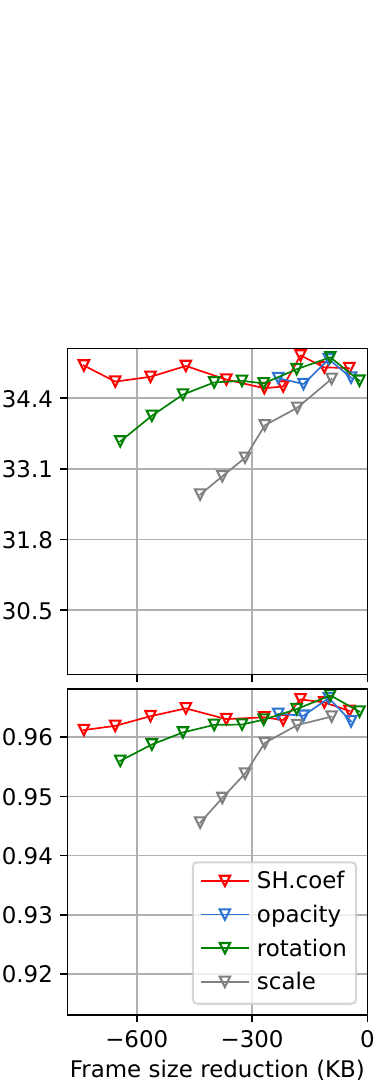}{boxes}
    \includelayerquality{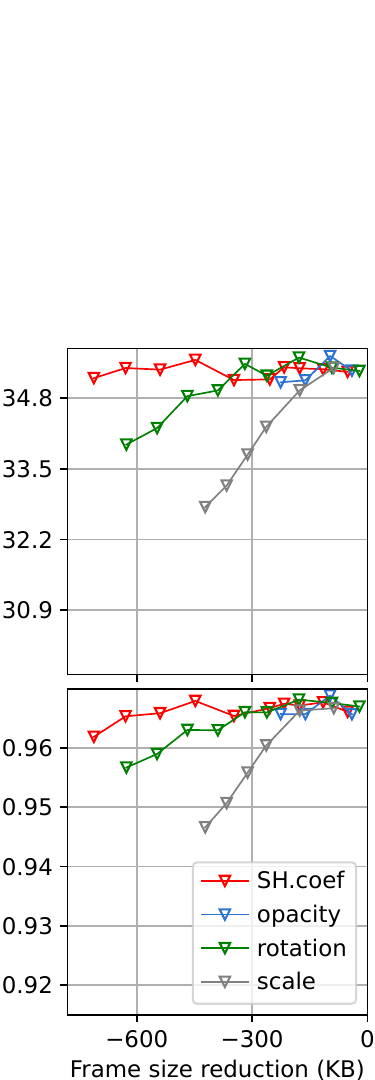}{football}
    \includelayerquality{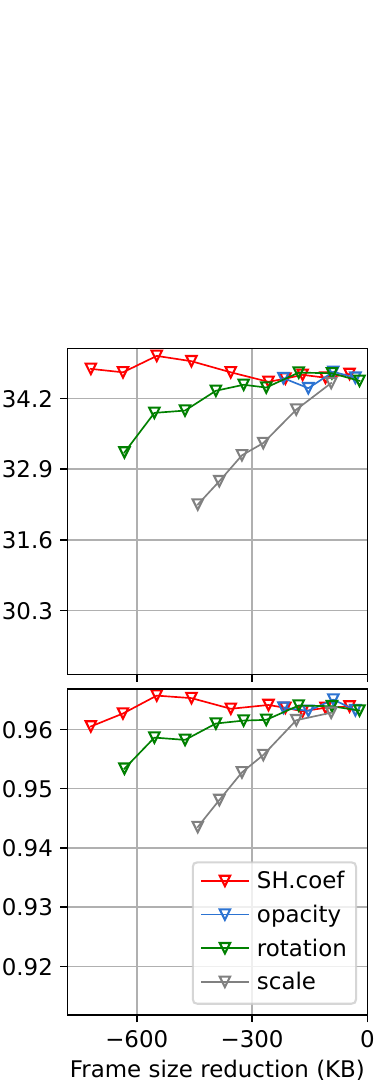}{juggle}
    \includelayerquality{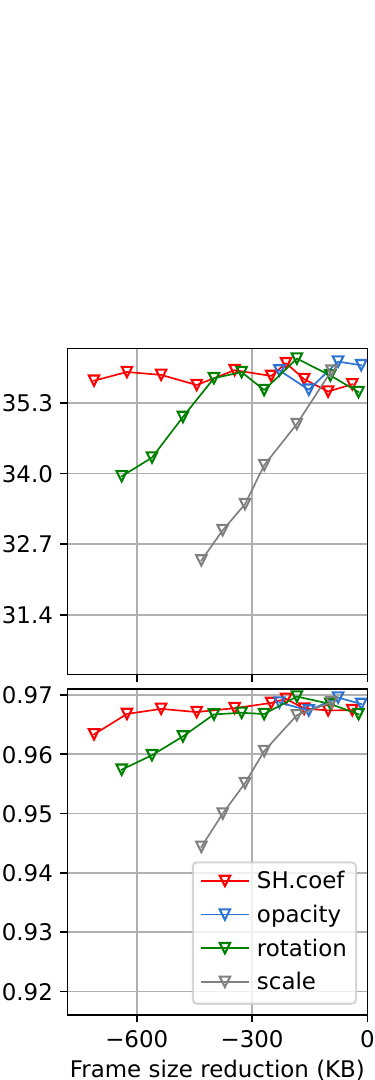}{softball}
    \includelayerquality{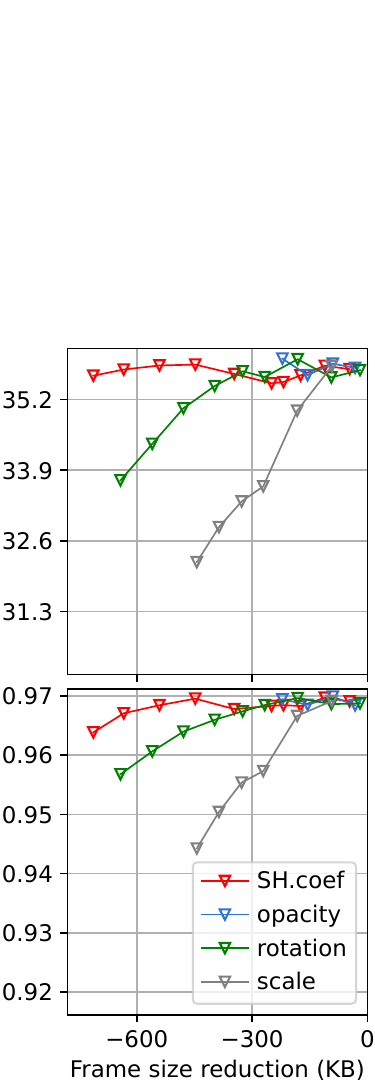}{tennis}
	\caption{
		Visual quality versus data size for individual SVQ enhancement layers.
		Each point corresponds to one SVQ layer.
		The x-axis shows data size reduction (excluding the codebook) after removing the given layer and all subsequent layers, while the y-axis reports the resulting rendering quality (PSNR/SSIM) for that video.
		SH.coef: coefficient for spherical harmonics.
  }\label{fig:LayerQuality}
\end{figure*} \label{sec:implement}
We implemented a prototype CAGS system to evaluate its effectiveness.
In this section, we describe engineering details and underlying rationales of each component to facilitate reproducibility.

\subsection{Preparing Volumetric Video}\label{sec:prepare}
Following TrackerSplat~\cite{yinTrackerSplatExploitingPoint2025}, we represent each dynamic scene as a set of Gaussians whose positions and attributes are updated across frames.
For each frame, we store only the Gaussians whose attributes change relative to the previous frame.
Concretely, the first frame is treated as a keyframe containing all Gaussians, and each subsequent frame stores only the updated Gaussians.
To further reduce the data footprint without sacrificing fidelity, we incorporate importance-based pruning from Reduced3DGS~\cite{papantonakisReducingMemoryFootprint2024} and GSNB~\cite{zhouGaussianSplattingNeural2024} into the keyframe training loop, removing low-contribution Gaussians while preserving rendering quality.

\subsection{SVQ}

We quantize each Gaussian attribute independently using SVQ.
We use Table~\ref{tab:sota} to determine suitable bit widths for cluster initialization.
As shown in Table~\ref{tab:sota}, SVQ achieves compression ratios comparable to recent scalable compression methods, while avoiding their heavy decoding cost.
Although a 76-bit initialization provides better quality than 66-bit, the improvement is not significant.
Given that 66-bit initialization already provides competitive performance, we adopt it as the default setting in our system.
In detail, the 66 bits are allocated as follows: scaling (10 bits), rotation (quaternion real: 8 bits, imaginary: 10 bits), opacity (8 bits), coefficients for level-0 SH (9 bits), and the remaining three SH levels (8, 7, and 6 bits).
Finally, the initial clusters of each attribute are merged into 16 clusters (4 bits) respectively (Sec.~\ref{sec:merging}).

\subsection{LoD Assignment}
Although each attribute can be compressed independently by SVQ, adaptive streaming requires a linear LoD structure, where levels are transmitted sequentially (LoD 0, 1, 2, ...) until reaching bandwidth limits~\cite{sunLTSDASHStreaming2025}.
This linearity avoids combinatorial decisions and supports real-time bitrate control by simply accumulating layer sizes.
Since lower LoDs are always delivered earlier, layers with larger visual impact must be prioritized to maximize quality under limited bandwidth.

We determine the layer ordering by evaluating the importance of enhancement layers for each attribute.
For each attribute, we progressively decode it with increasing numbers of enhancement layers while keeping other attributes at full quality.
For each decoded scene, we render the result, apply the restoration model in Sec.~\ref{sec:motivate-model}, and measure PSNR/SSIM.
Results in Figure~\ref{fig:LayerQuality} show that scaling has the greatest impact, followed by rotation, spherical harmonics, and opacity.
We also observe that higher layers provide diminishing returns, and the ranking is roughly stable across datasets.

Based on these measurements, we interleave attribute layers according to their PSNR gains in Figure~\ref{fig:LayerQuality-all} to construct a unified linear LoD hierarchy.
This design ensures early LoDs deliver the highest quality gains, while later LoDs provide progressively smaller refinements.

Figure~\ref{fig:LayerQuality-LoDs} plots quality versus frame size across LoDs under this hierarchy.
It shows a steep quality increase in the first few LoDs, then levels off, demonstrating that critical SVQ layers are positioned at lower LoDs, as required for effective adaptive streaming.

\subsection{Tiling}
Following LTS~\cite{sunLTSDASHStreaming2025}, we partition Gaussians into spatial tiles for viewport-adaptive streaming that transmits only visible tiles.
As our training integrates importance-based pruning, the resulting Gaussian distribution is notably sparser and less uniform than that in LTS.
As a result, directly applying the bounding-box tiling strategy in LTS would lead to many under-populated tiles, reducing compression efficiency.

To address this, we leverage the Morton-code sorting mechanism from TaoGS~\cite{jiangTopologyAwareOptimizationGaussian2025} to linearize Gaussians while preserving spatial locality.
For each frame, we sort Gaussians by Morton code and partition them into tiles with approximately equal size, so that each tile contains a similar number of Gaussians.
The tile size is set to 8,000 Gaussians for the first frame and 4,000 for subsequent frames.

\subsection{Compression}
After tiling, Gaussian positions and the lowest-LoD attributes are compressed using enhanced Draco~\cite{draco}, following LTS~\cite{sunLTSDASHStreaming2025}.
Higher LoDs and the SVQ codebook are compressed separately using Gzip.
In addition, we store unquantized frames (also compressed with Draco) on the server to support reference rendering.

\subsection{Network Adaptation Strategy}\label{sec:abr}
For evaluation, we implement a straightforward network adaptation strategy based on bandwidth limit to handle fluctuating bandwidth conditions:
1) On the streaming server, identify visible Gaussians based on rendered reference images.
2) Select tiles that cover all visible Gaussians from recent frames.
3) For tiles not yet transmitted, progressively increase their LoDs until their average LoD matches that of already-sent tiles or the bandwidth limit is reached.
4) For tiles already transmitted, further increase their LoDs in descending order according to the count of visible Gaussians, until the bandwidth limit is reached.

\subsection{Streaming}
Both the streaming server and client run on machines equipped with RTX 3080 GPUs.
Before streaming starts, we allocate a short initialization period to collect initial viewport traces and download necessary components, including the codebook (81.9 KB), the color restoration model (219 KB), and the lowest LoD of the first frame ($\sim$1.5 MB).
The streaming frame rate is 30 FPS, and the client renders at 60 FPS.
The client uses a fixed rendering FoV of 1.48 rad (horizontal) and 1.20 rad (vertical) at 1600$\times$1200, while reference images are rendered at 400$\times$300.
Both the 3D Tile Buffer and the Reference Buffer are set to 6 frames, i.e., the server predicts the viewport and renders reference images 180 ms (6 frames) ahead of the actual viewing time.
 
\newpage
\onecolumn
\section{Additional Evaluation Results}

\begin{table*}[htbp]
\caption{
    Performance and quality comparison of our SVQ method (with 66-bit and 76-bit initialization) against state-of-the-art \textbf{scalable} compression methods at the highest quality level.
Best and second-best results are highlighted in \textbf{bold} and \underline{underline}, respectively.
    Size is in MB and decoding time ("dec.t") in seconds.
}
\begin{center}
\tabcolsep=1mm
\begin{tabular}{l|cccrr|cccrr|cccrr}
\hline
Datasets & 
\multicolumn{5}{c|}{\textbf{Neural 3D Video dataset}} & 
\multicolumn{5}{c|}{\textbf{Meeting Room dataset}} &
\multicolumn{5}{c}{\textbf{Dynamic 3DGS dataset}} \\
Methods & 
psnr & ssim & lpips & size & dec.t & 
psnr & ssim & lpips & size & dec.t & 
psnr & ssim & lpips & size & dec.t \\
\hline
SPZ low & 23.3 & .842 & .112 & 3.33 & 0.076 & 21.8 & .822 & \underline{.070} & 2.54 & 0.062 & 21.1 & .761 & \underline{.130} & 2.26 & 0.052 \\
SPZ high & 23.1 & .842 & .110 & 4.53 & 0.075 & 21.8 & .822 & \textbf{.068} & 3.41 & 0.057 & 21.0 & .760 & \textbf{.112} & 3.32 & 0.057 \\
CompGS & \underline{24.9} & .898 & .193 & 22.20 & 6.213 & 25.0 & .880 & .221 & 5.58 & 0.799 & 19.1 & .721 & .353 & 4.27 & 2.772 \\
HAC & 22.2 & .889 & .203 & 25.69 & 24.407 & 25.2 & .873 & .216 & 8.69 & 9.135 & 20.5 & .758 & .338 & 2.84 & 1.571 \\
HAC++ & 21.5 & .885 & .202 & 17.98 & 51.708 & 25.2 & .868 & .222 & 6.78 & 14.706 & 20.9 & .780 & .291 & \textbf{2.00} & 6.169 \\
\hline
SVQ 66bit & 24.5 & \underline{.901} & \underline{.108} & \textbf{2.00} & \textbf{0.015} & \underline{26.0} & \underline{.887} & .093 & \textbf{1.53} & \textbf{0.018} & \textbf{25.3} & \textbf{.883} & .153 & \underline{2.25} & \textbf{0.028} \\
SVQ 76bit & \textbf{25.1} & \textbf{.907} & \textbf{.100} & \underline{2.14} & \underline{0.015} & \textbf{26.4} & \textbf{.894} & .079 & \underline{1.69} & \underline{0.019} & \underline{23.3} & \underline{.874} & .200 & 2.42 & \underline{0.029} \\
\hline
\end{tabular}
\label{tab:sota}
\end{center}
\end{table*} 
\begin{table*}[htbp]
\caption{
    Performance and quality comparison of our SVQ method against recent compression methods.
Best and second-best results are shown in \textbf{bold} and \underline{underline}, respectively.
    The "scal." column indicates whether a method is scalable.
    Size is in MB and decoding time ("dec.t") in seconds.
    Our method achieves comparable rate-distortion performance with significantly lower decoding latency, which is uniquely suited for real-time streaming.
}
\begin{center}
\tabcolsep=1mm
\begin{tabular}{lc|cccrr|cccrr|cccrr}
\hline
\multicolumn{2}{c|}{\textbf{Datasets}} & 
\multicolumn{5}{c|}{\textbf{Neural 3D Video dataset}} & 
\multicolumn{5}{c|}{\textbf{Meeting Room dataset}} &
\multicolumn{5}{c}{\textbf{Dynamic 3DGS dataset}} \\
\hline
Methods & scal. & 
psnr & ssim & lpips & size & dec.t & 
psnr & ssim & lpips & size & dec.t & 
psnr & ssim & lpips & size & dec.t \\
\hline
GIFStream & $\times$ & \underline{23.6} & \underline{.860} & \underline{.130} & \textbf{0.58} & \underline{0.050} & \underline{24.3} & \underline{.879} & .093 & \textbf{0.48} & \underline{0.042} & 13.6 & .615 & .375 & \textbf{0.13} & \textbf{0.015} \\
QUEEN & $\times$ & \textbf{29.5} & \textbf{.933} & .181 & 17.36 & 0.997 & \textbf{31.4} & \textbf{.954} & \underline{.153} & 7.70 & 0.437 & \underline{18.3} & \textbf{.788} & \underline{.315} & 3.41 & 0.197 \\
\hline
SVQ & $\checkmark$ & 23.2 & .829 & \textbf{.108} & \underline{0.75} & \textbf{0.015} & 21.8 & .812 & \textbf{.093} & \underline{0.96} & \textbf{0.018} & \textbf{21.6} & \underline{.766} & \textbf{.153} & \underline{1.08} & \underline{0.028} \\
\hline
\end{tabular}
\label{tab:sota}
\end{center}
\end{table*} 
\begin{table*}[htbp]
\caption{
        Viewport prediction and Adaptive FoV precision of the across datasets, measured by Position MAE (translation error), Angular MAE (degrees), and FoV Coverage (\%).
}
\begin{center}
\tabcolsep=1.8mm
\begin{tabular}{l|c|c|c}
\hline
\textbf{Dataset} & \textbf{Position MAE} & \textbf{Angular MAE} & \textbf{FoV Coverage (\%)} \\
\hline
Neural 3D Video dataset & 0.1115 & 1.54 & 94.72 \\
ST-NeRF dataset & 0.0804 & 1.58 & 94.66 \\
Meeting Room dataset & 0.1841 & 1.90 & 94.20 \\
Dynamic 3DGS dataset & 0.0677 & 2.38 & 93.77 \\
\hline
\end{tabular}
\label{tab:view}
\end{center}
\end{table*}
 
\clearpage

\providecommand{\includesubfigure}{}
\providecommand{\includeylabel}{}
\providecommand{\includelegend}{}
\renewcommand{\includesubfigure}[2]{
    \begin{subfigure}[b]{0.12\linewidth}
		\centering
        \includegraphics[width=\linewidth]{figures/#1.pdf}
		\caption{``#2''}
	\end{subfigure}
}
\renewcommand{\includeylabel}{
    \includegraphics[width=0.012\linewidth]{figures/FixedBandwidth/ylabel.pdf}
}
\renewcommand{\includelegend}[1]{
    \includegraphics[width=0.9\linewidth]{figures/#1/legend.pdf}
}

\begin{figure*}[htbp]
	\centering
	\includelegend{FixedBandwidth}
	\\\includeylabel
\includesubfigure{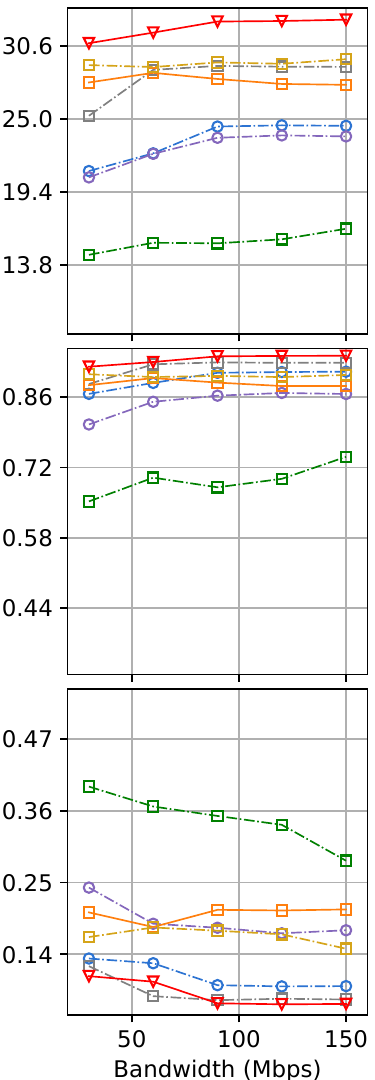}{cook spinach}
    \includesubfigure{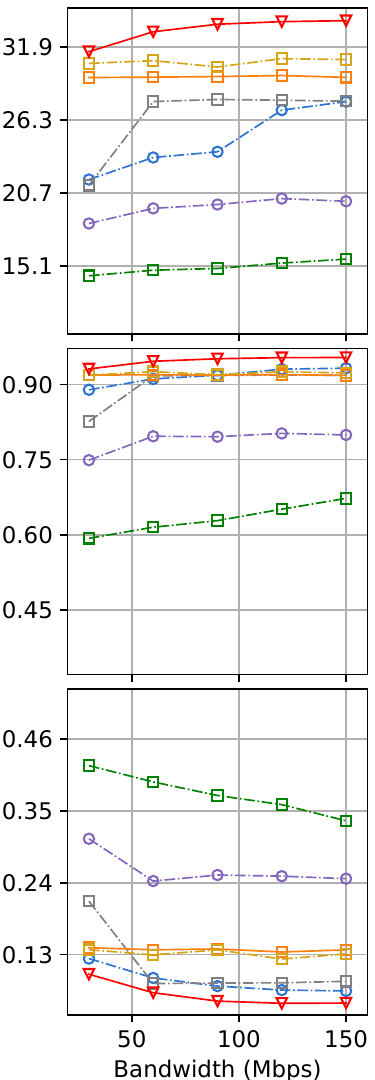}{cut roast beef}
    \includesubfigure{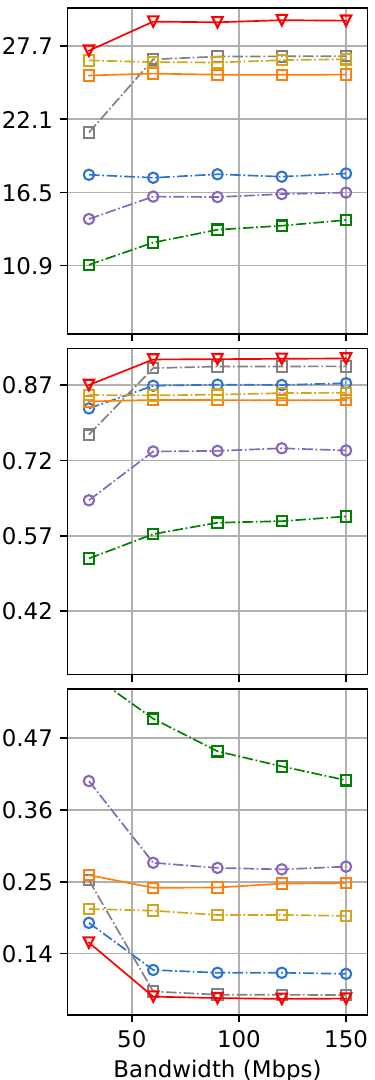}{flame salmon}
    \includesubfigure{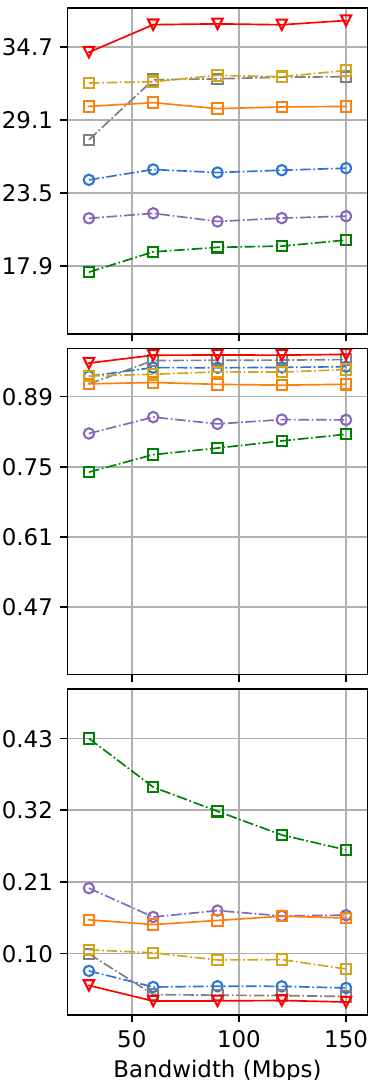}{flame steak}
    \includesubfigure{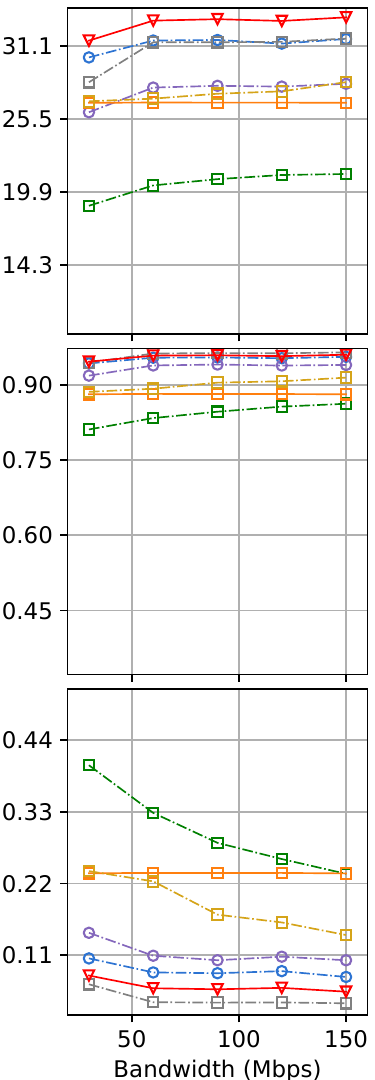}{sear steak}
	\\\includeylabel
    \includesubfigure{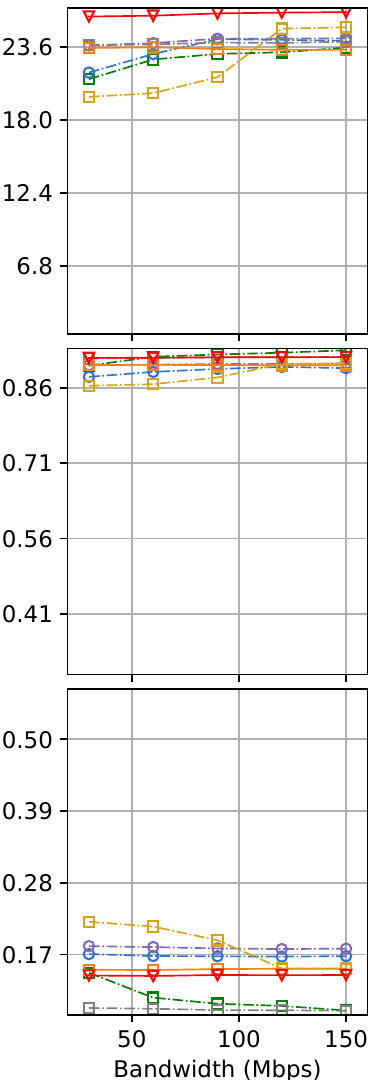}{walking}
\includesubfigure{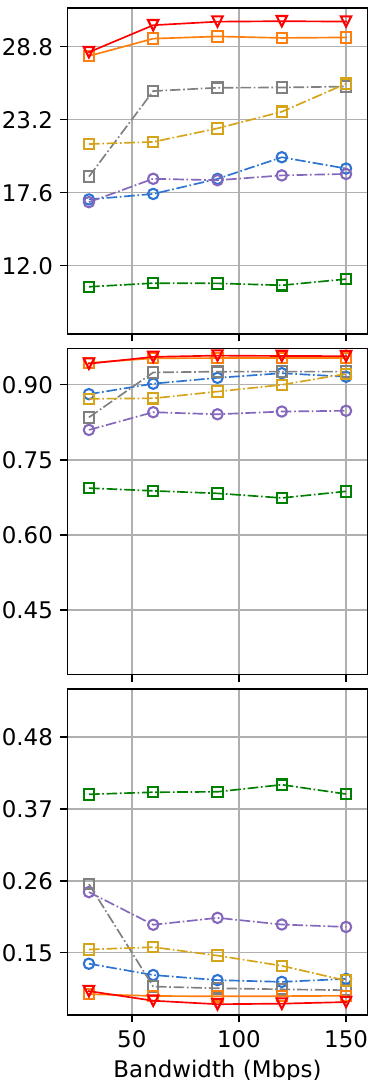}{discussion}
    \includesubfigure{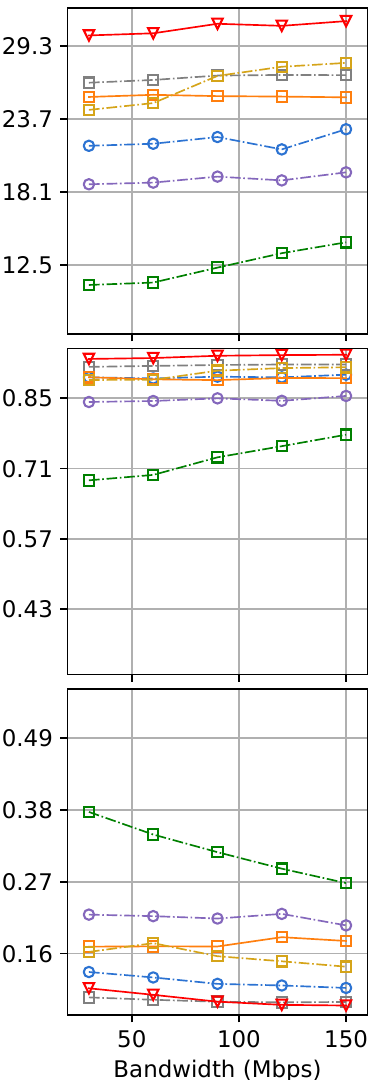}{stepin}
    \includesubfigure{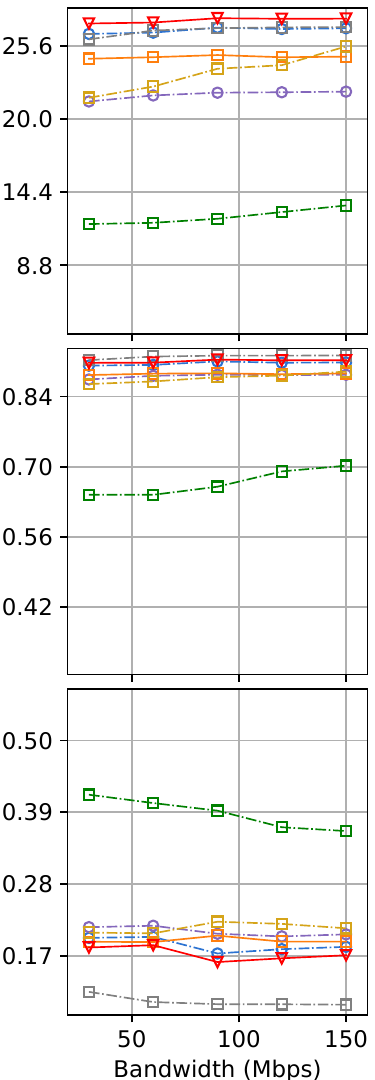}{trimming}
\includesubfigure{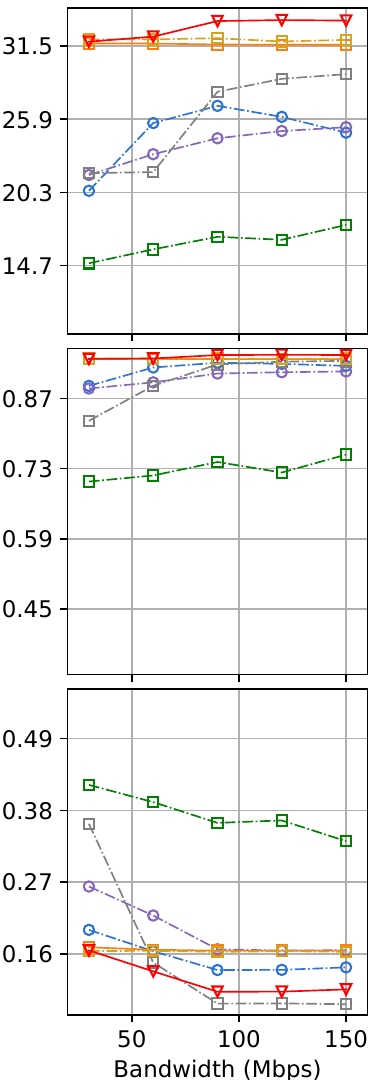}{basketball}
	\\\includeylabel
    \includesubfigure{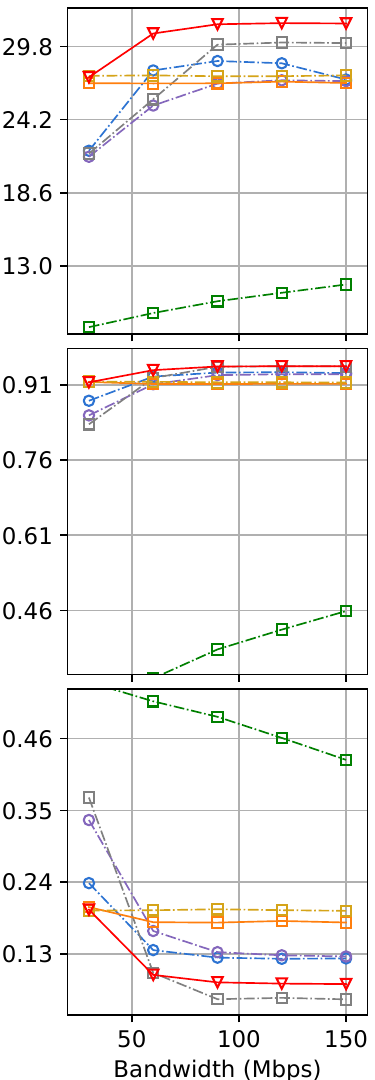}{boxes}
    \includesubfigure{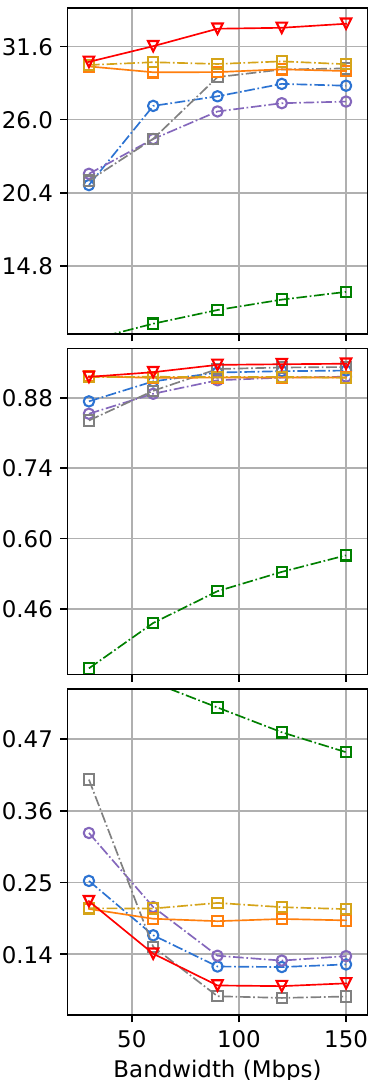}{football}
    \includesubfigure{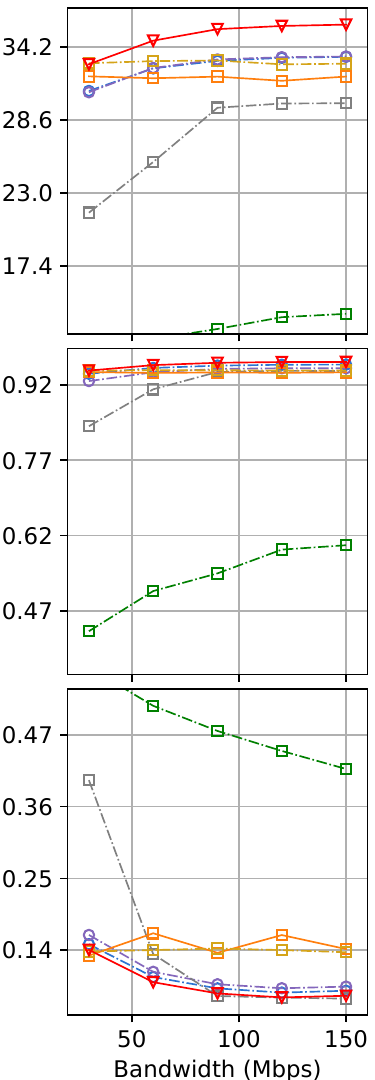}{juggle}
    \includesubfigure{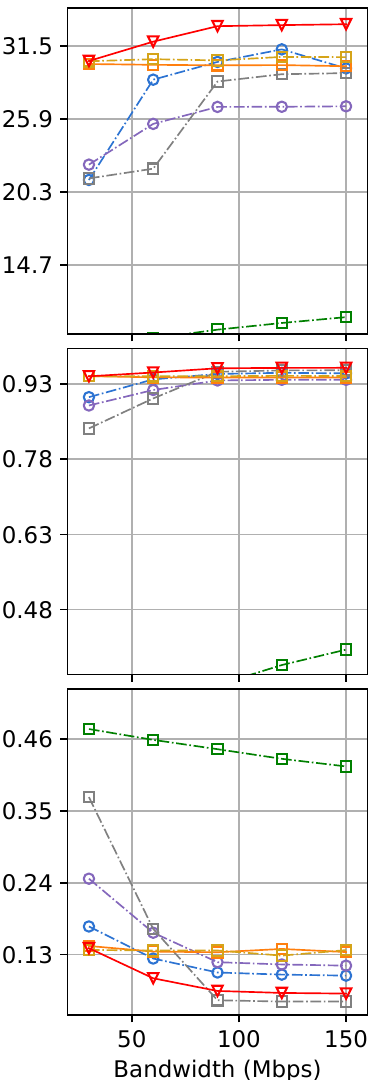}{softball}
    \includesubfigure{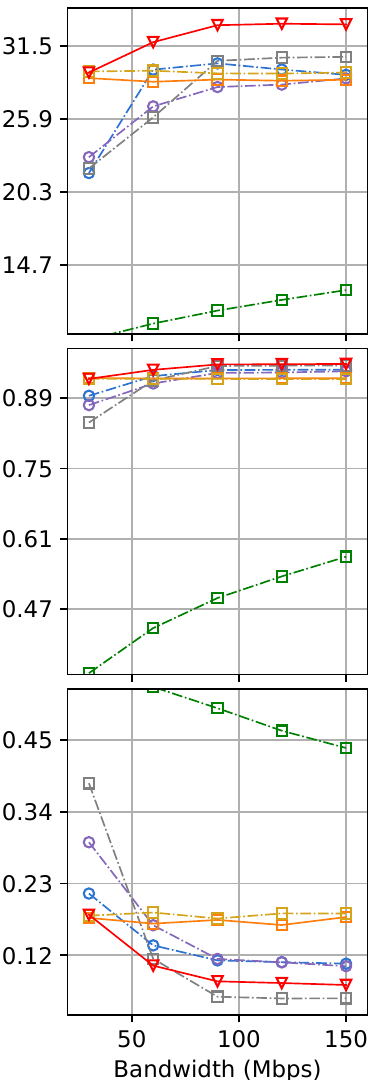}{tennis}
	\caption{
    Comparison of visual quality under fixed bandwidth.
		The y-axis ranges vary across subplots while maintaining equal scale spans for each metric across videos.
  }
\end{figure*}

\begin{figure*}[htbp]
	\centering
	\includelegend{DynamicBandwidth}
	\\\includeylabel
\includesubfigure{DynamicBandwidth/cook_spinach}{cook spinach}
    \includesubfigure{DynamicBandwidth/cut_roasted_beef}{cut roast beef}
    \includesubfigure{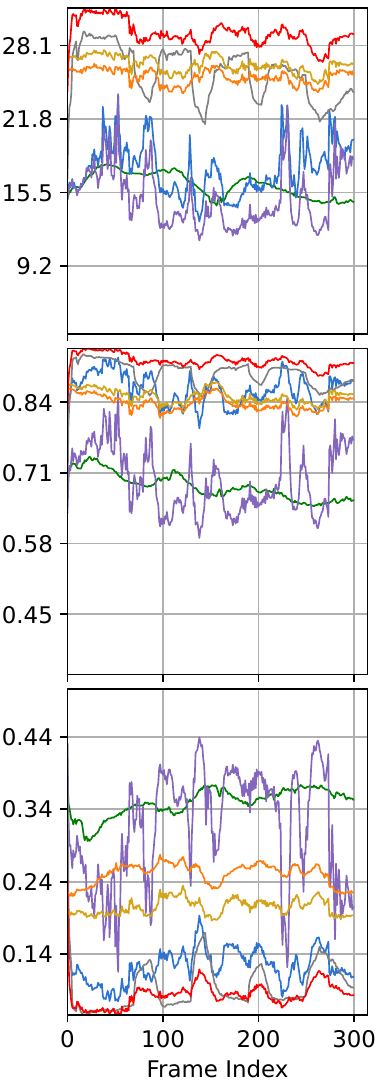}{flame salmon}
    \includesubfigure{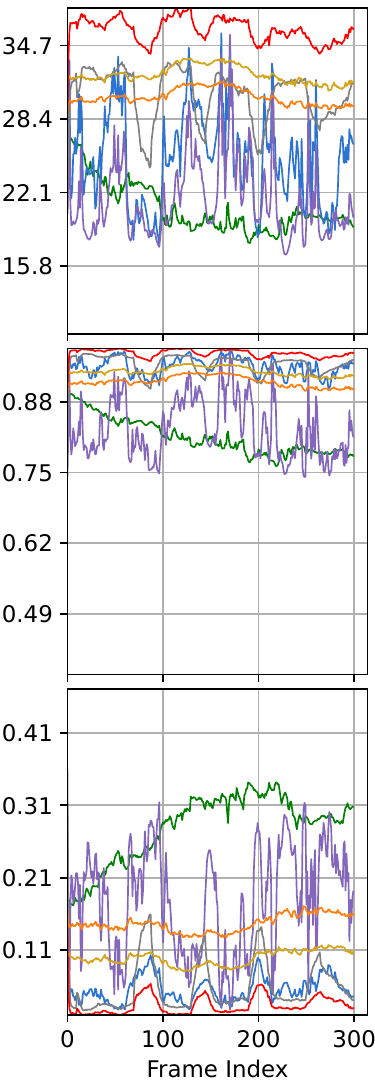}{flame steak}
    \includesubfigure{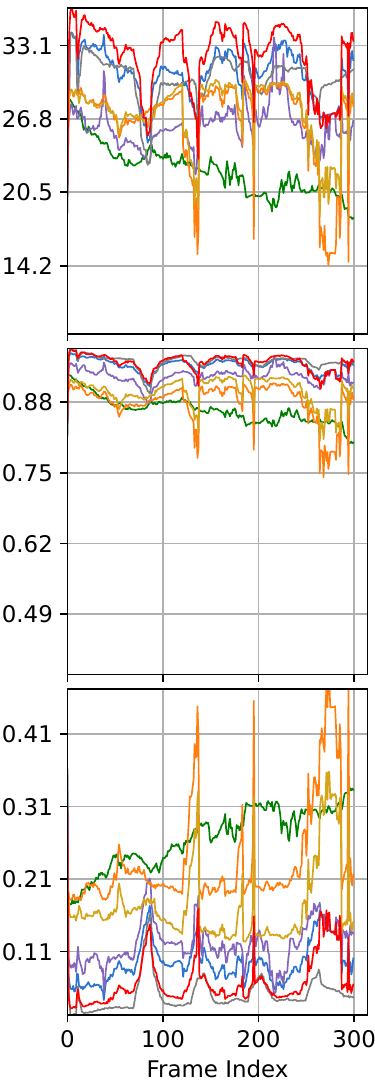}{sear steak}
	\\\includeylabel
    \includesubfigure{DynamicBandwidth/walking}{walking}
\includesubfigure{DynamicBandwidth/discussion}{discussion}
    \includesubfigure{DynamicBandwidth/stepin}{stepin}
    \includesubfigure{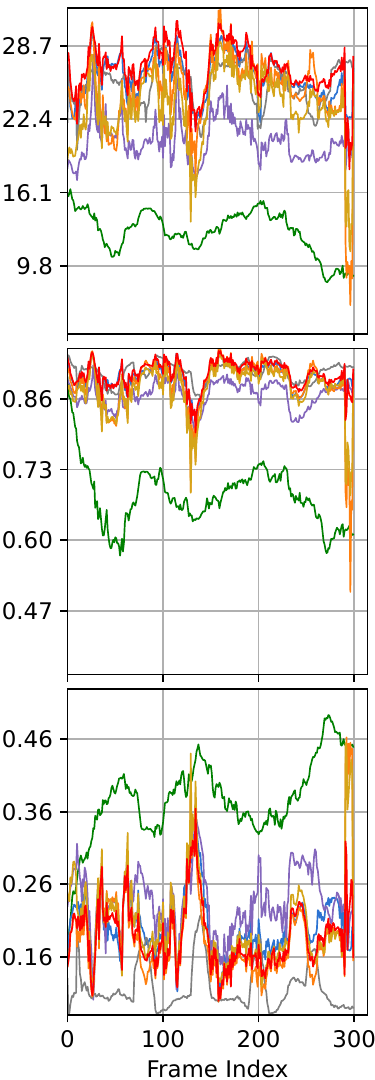}{trimming}
\includesubfigure{DynamicBandwidth/basketball}{basketball}
	\\\includeylabel
    \includesubfigure{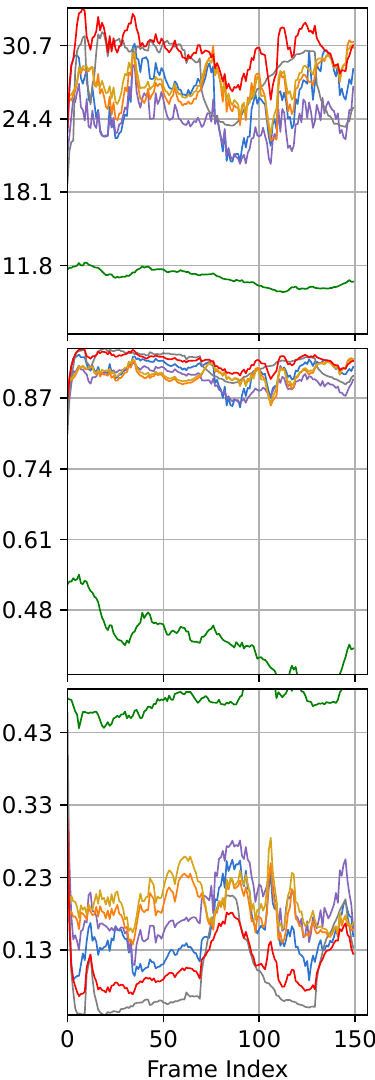}{boxes}
    \includesubfigure{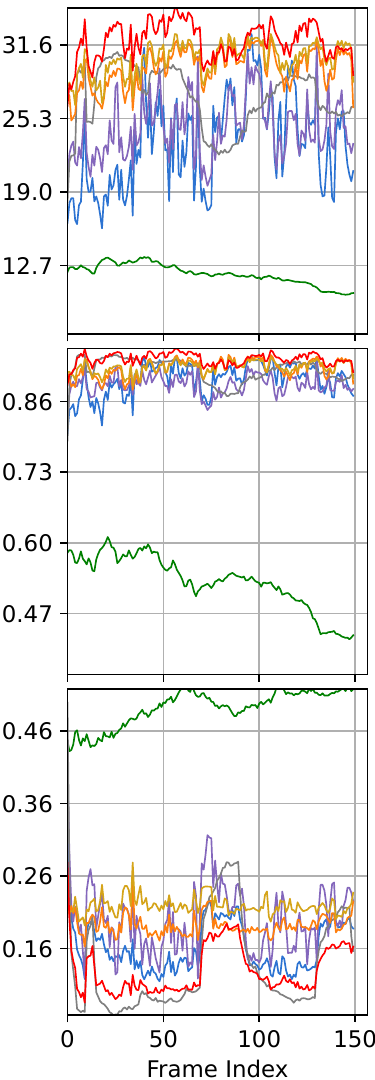}{football}
    \includesubfigure{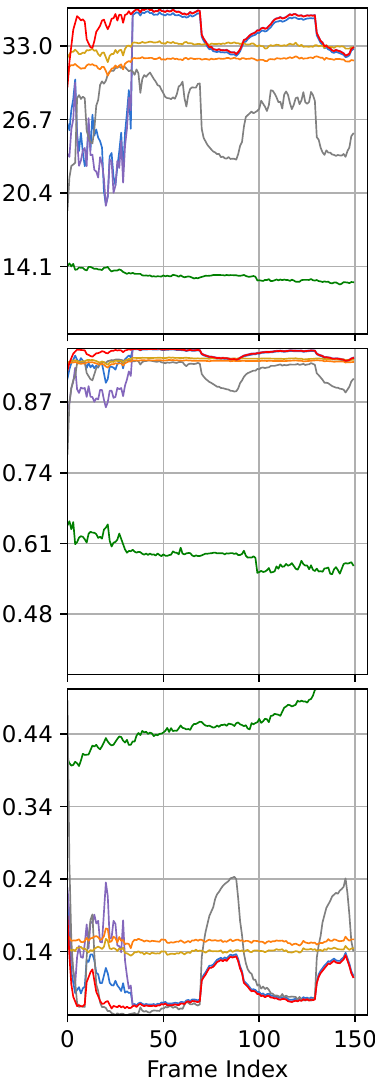}{juggle}
    \includesubfigure{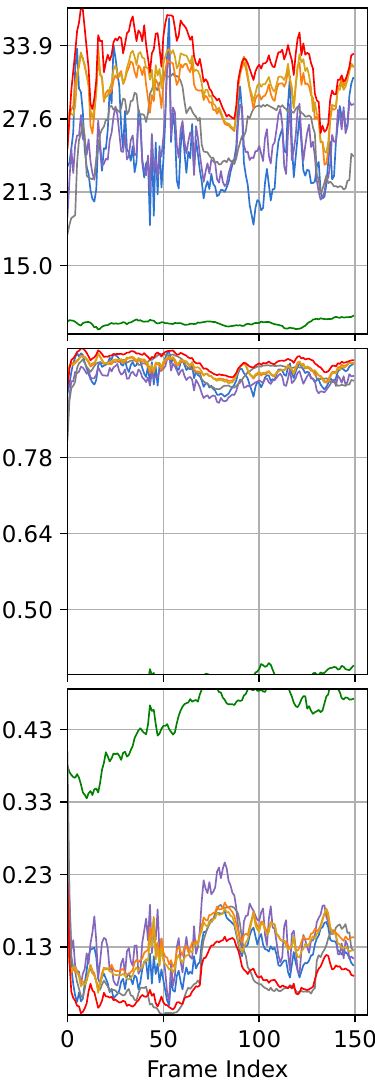}{softball}
    \includesubfigure{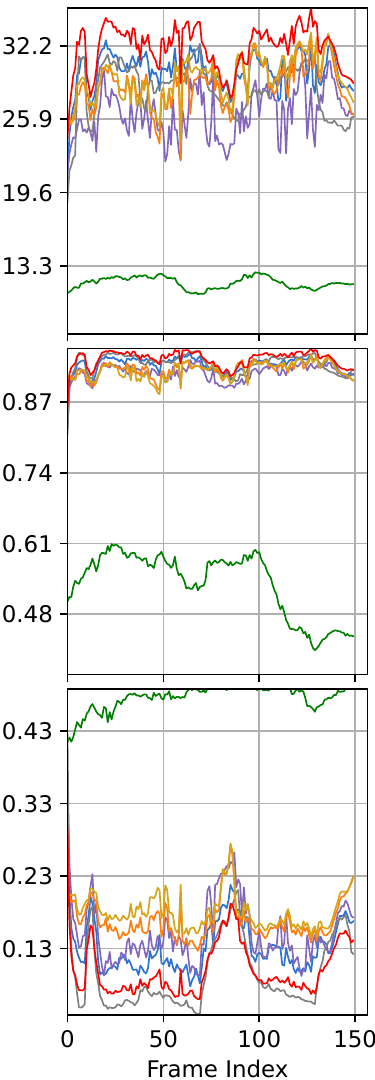}{tennis}
	\caption{
    Comparison of visual quality under fluctuating bandwidth.
		The y-axis ranges vary across subplots while maintaining equal scale spans for each metric across videos.
  }\end{figure*}

\begin{figure*}[htbp]
	\centering
	\includelegend{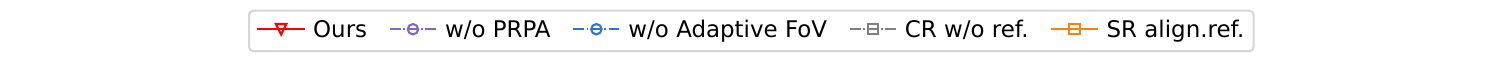}
	\\\includeylabel
\includesubfigure{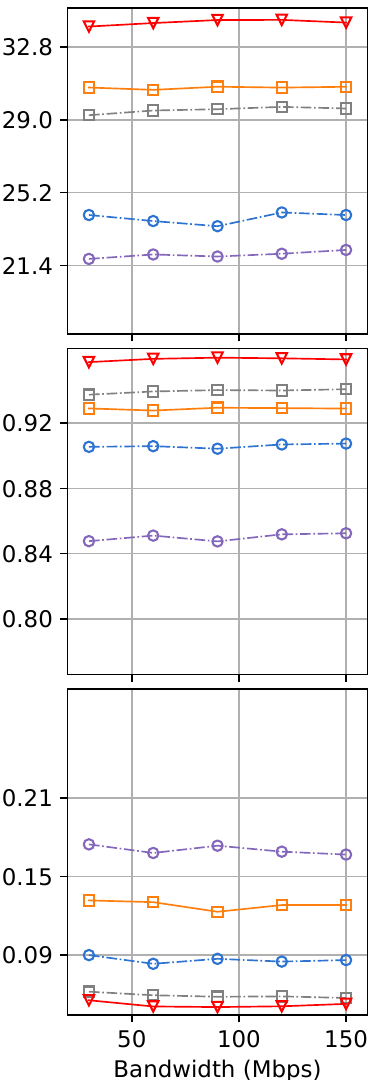}{cook spinach}
    \includesubfigure{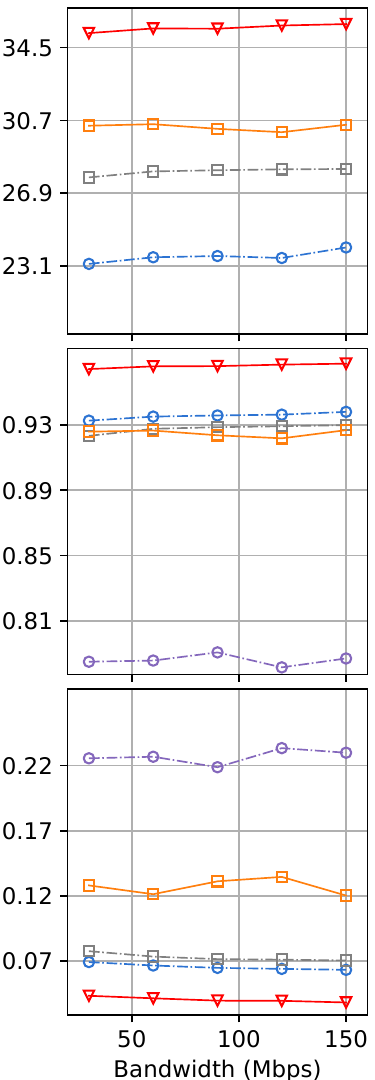}{cut roast beef}
    \includesubfigure{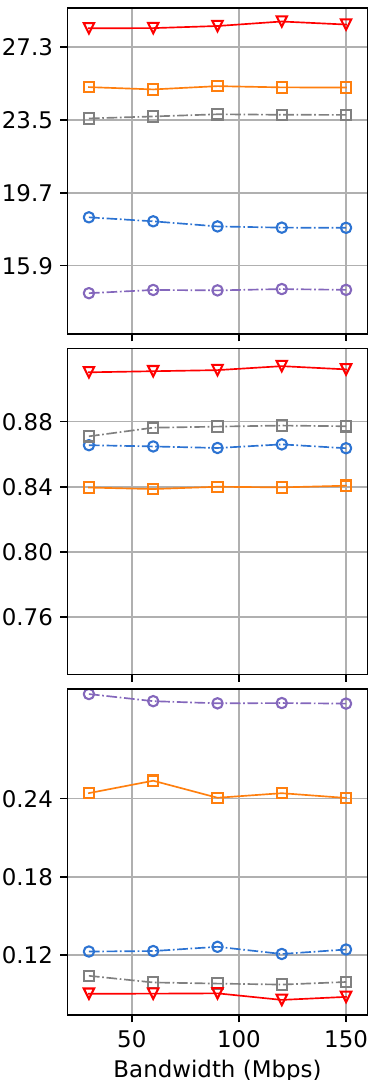}{flame salmon}
    \includesubfigure{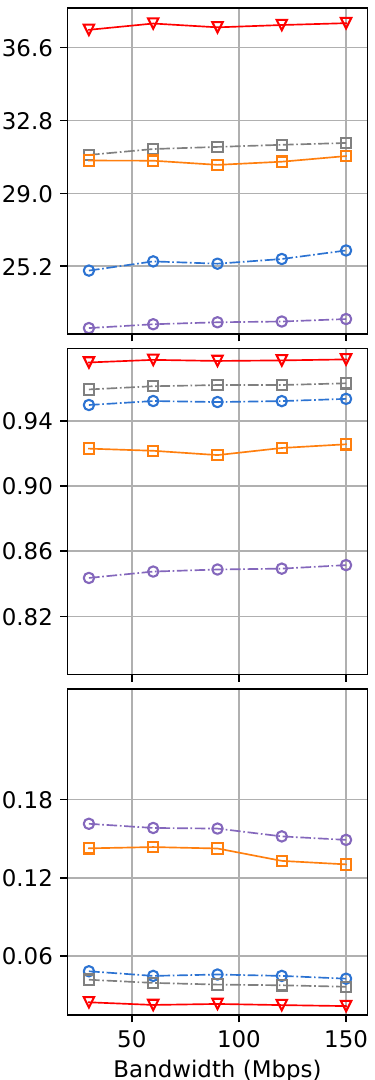}{flame steak}
    \includesubfigure{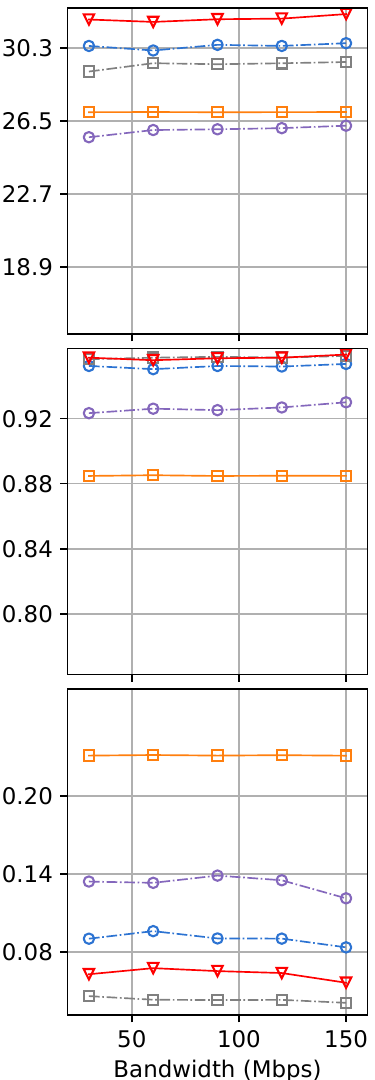}{sear steak}
	\\\includeylabel
    \includesubfigure{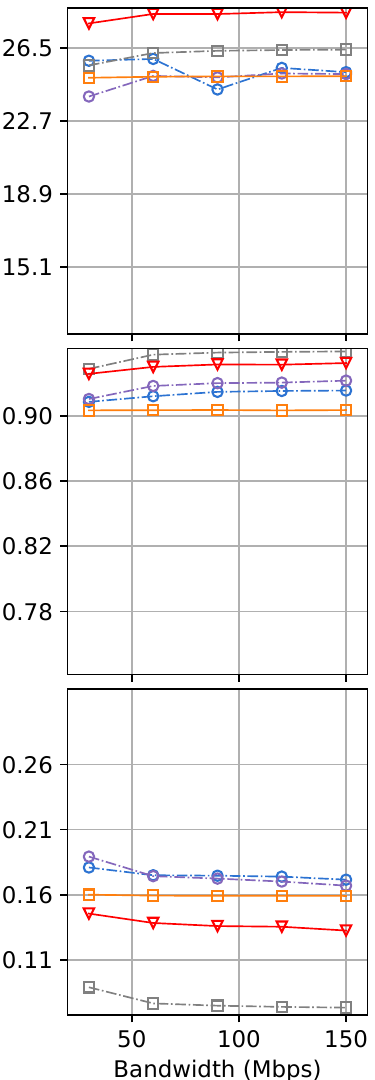}{walking}
\includesubfigure{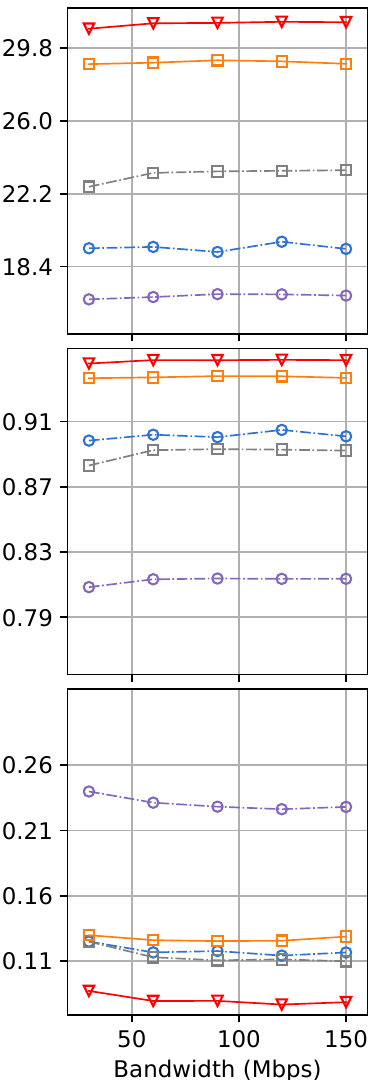}{discussion}
    \includesubfigure{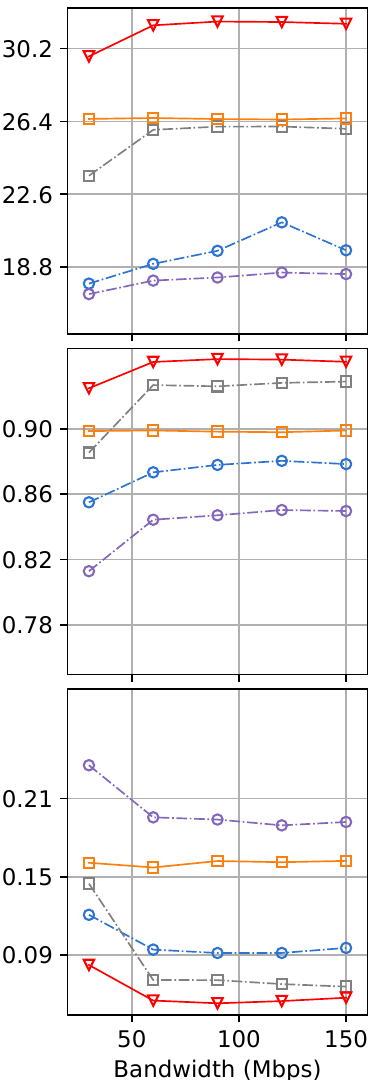}{stepin}
    \includesubfigure{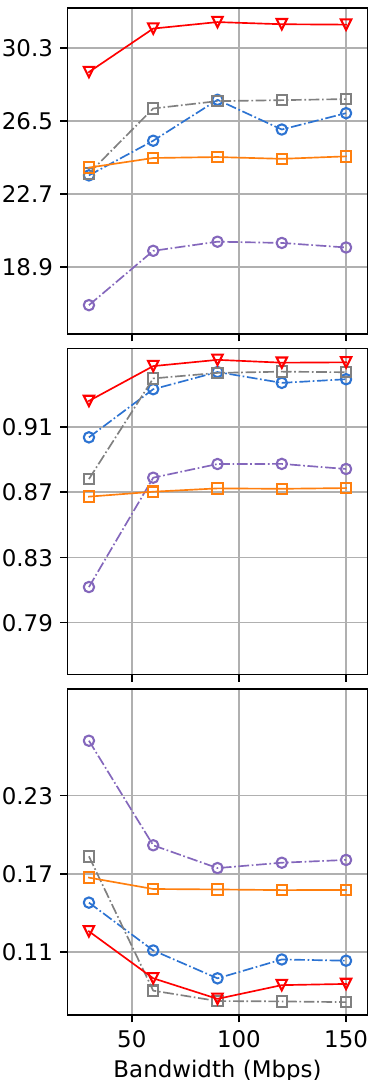}{trimming}
\includesubfigure{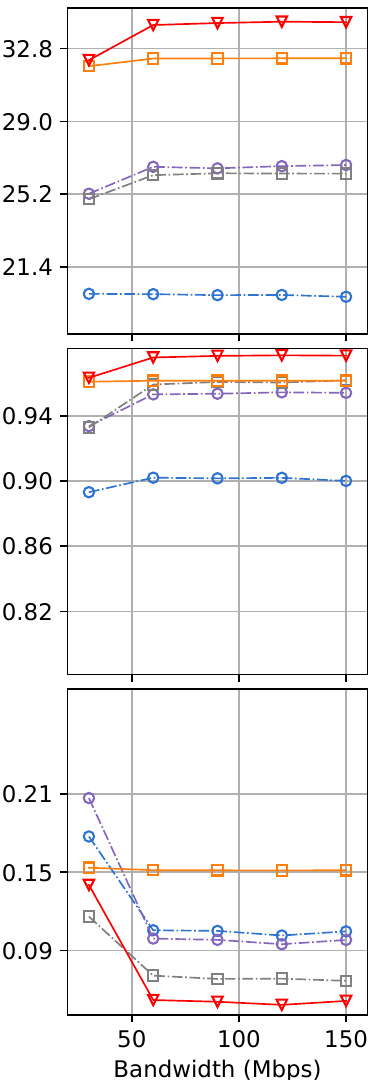}{basketball}
	\\\includeylabel
    \includesubfigure{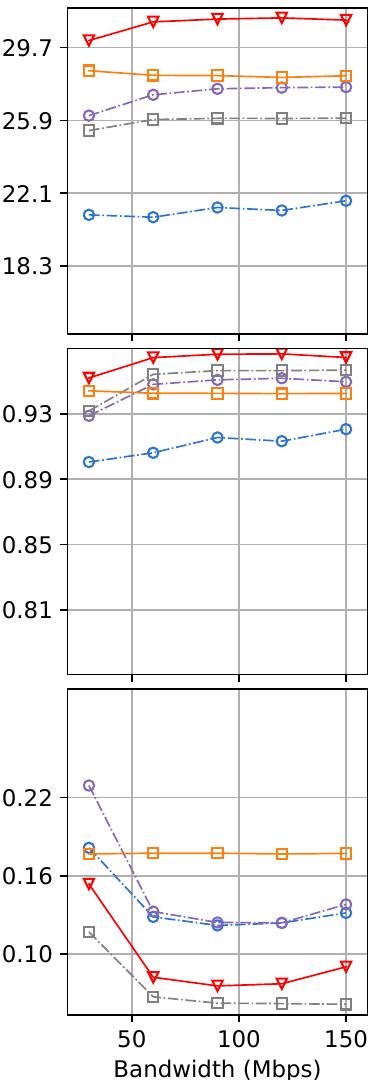}{boxes}
    \includesubfigure{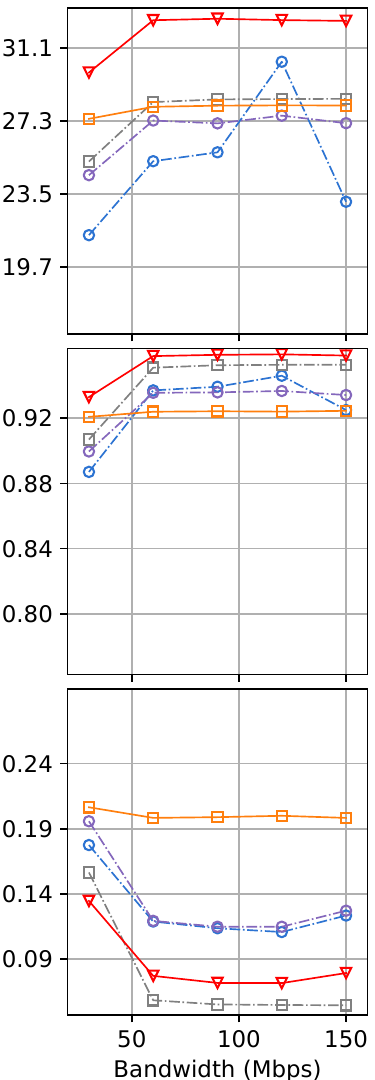}{football}
    \includesubfigure{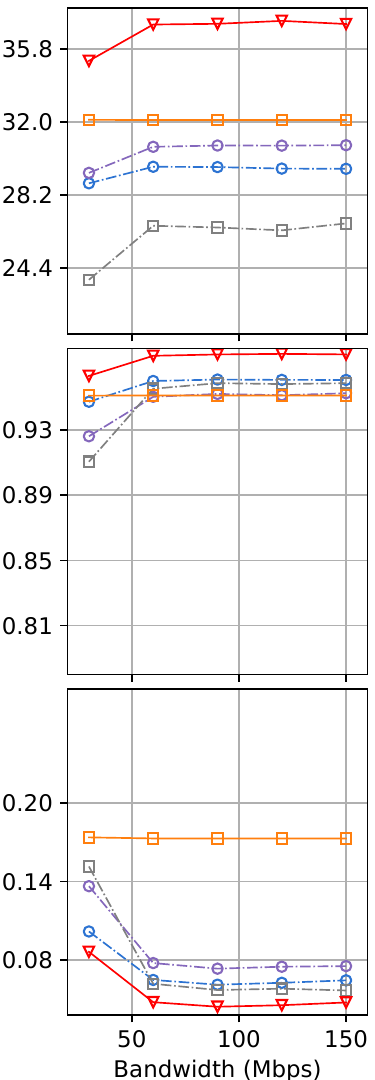}{juggle}
    \includesubfigure{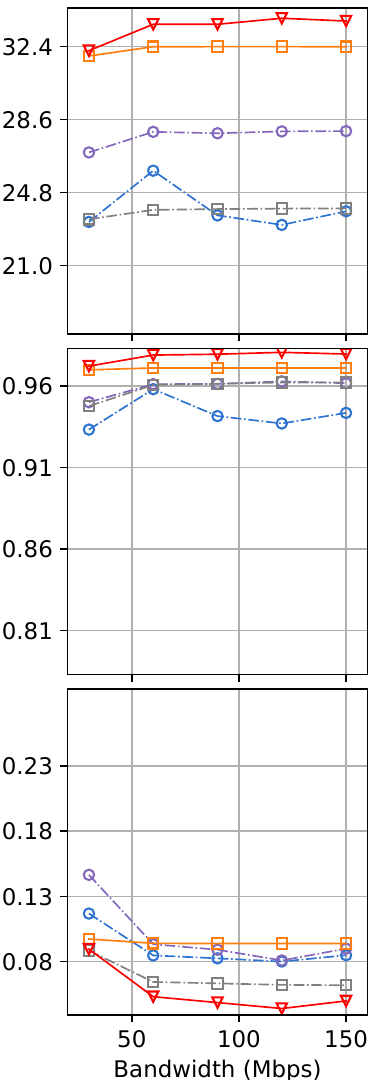}{softball}
    \includesubfigure{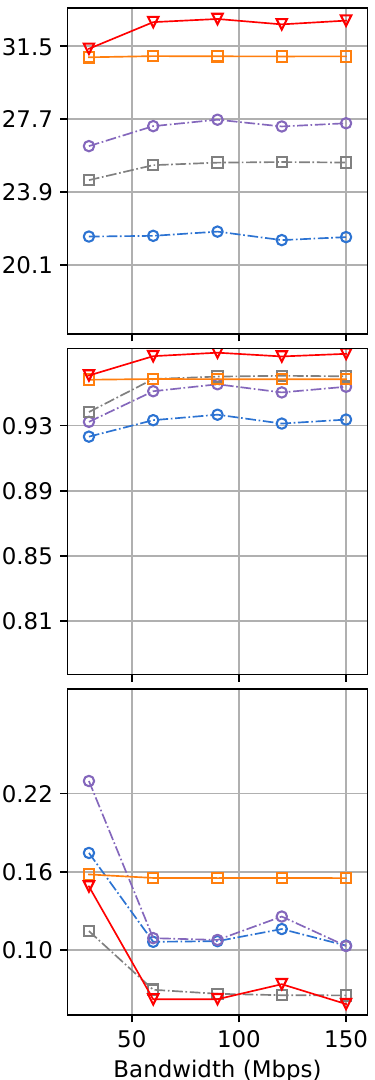}{tennis}
	\caption{
	Comparison of visual quality under fixed bandwidth on volumetric videos prepared by Dynamic 3DGS~\cite{luiten2023dynamic}.
		The y-axis ranges vary across subplots while maintaining equal scale spans for each metric across videos.
  }\end{figure*}

\begin{figure*}[htbp]
	\centering
	\includelegend{DynamicBandwidth/train/regularized}
	\\\includeylabel
\includesubfigure{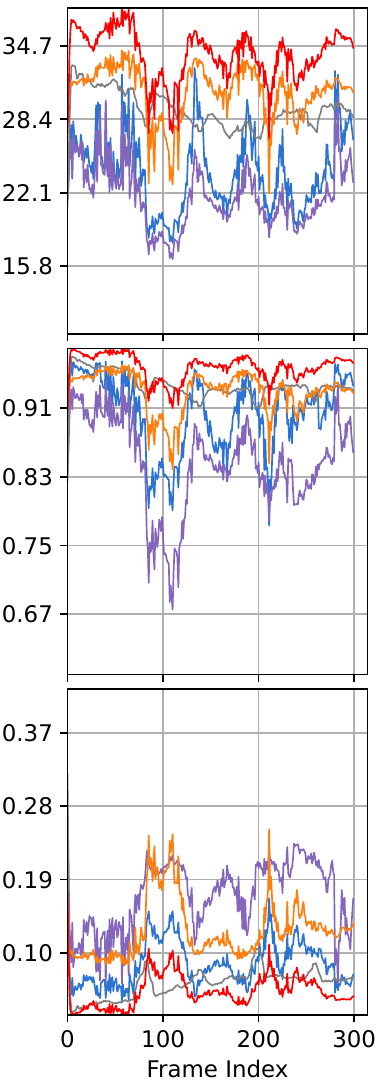}{cook spinach}
    \includesubfigure{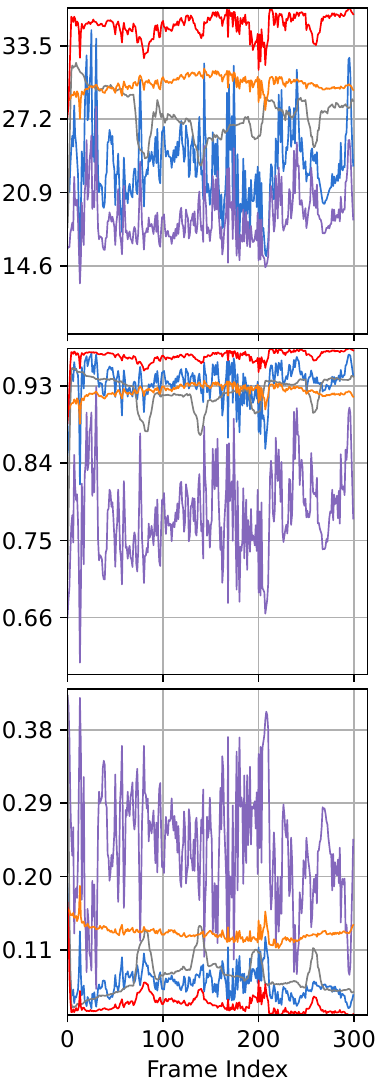}{cut roast beef}
    \includesubfigure{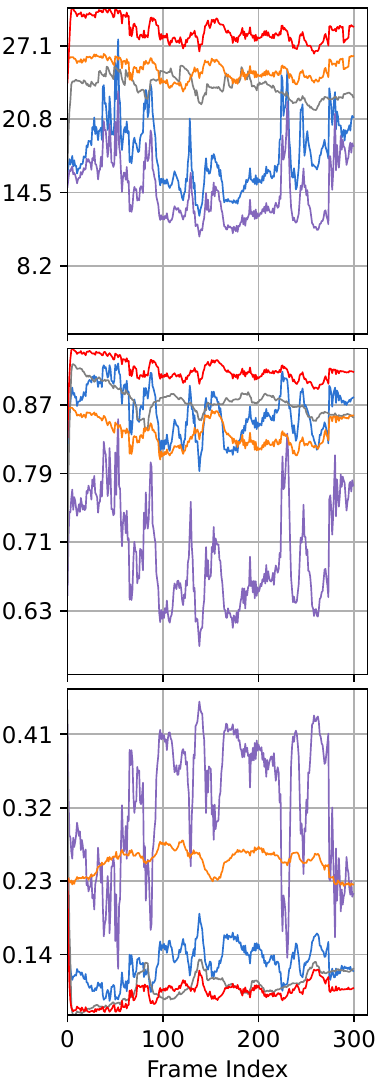}{flame salmon}
    \includesubfigure{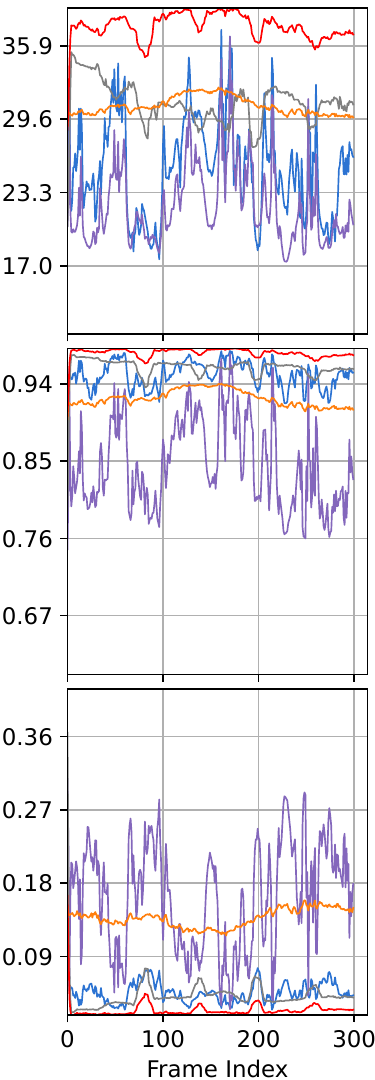}{flame steak}
    \includesubfigure{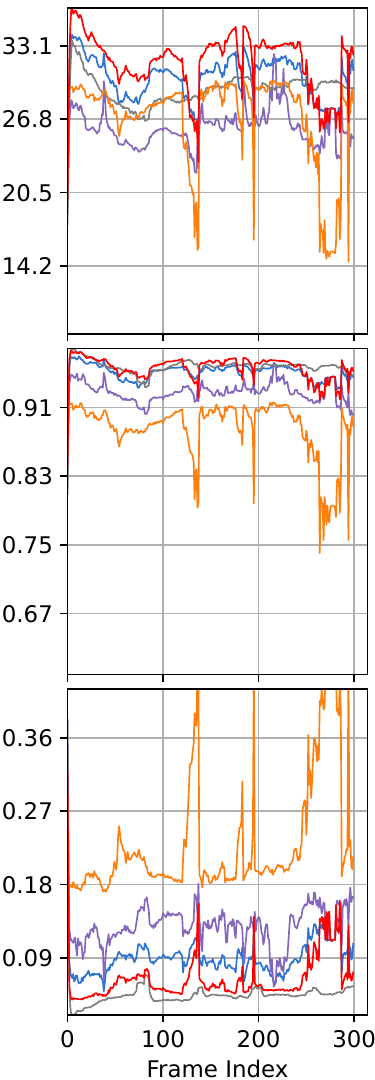}{sear steak}
	\\\includeylabel
    \includesubfigure{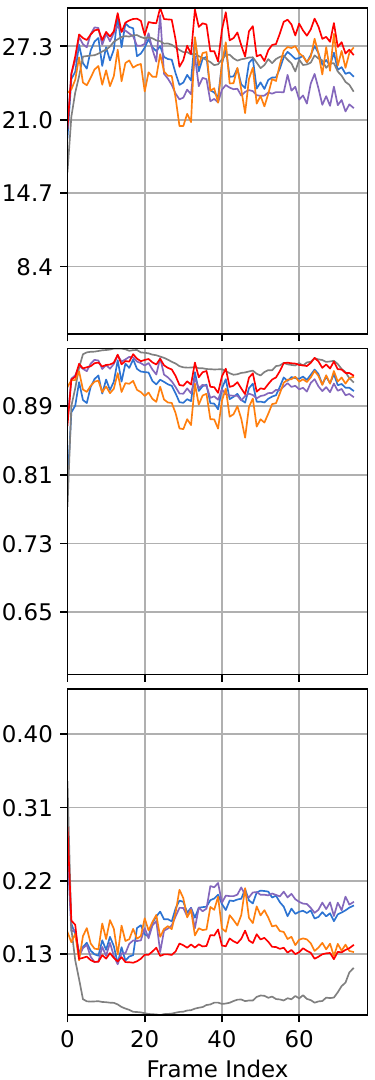}{walking}
\includesubfigure{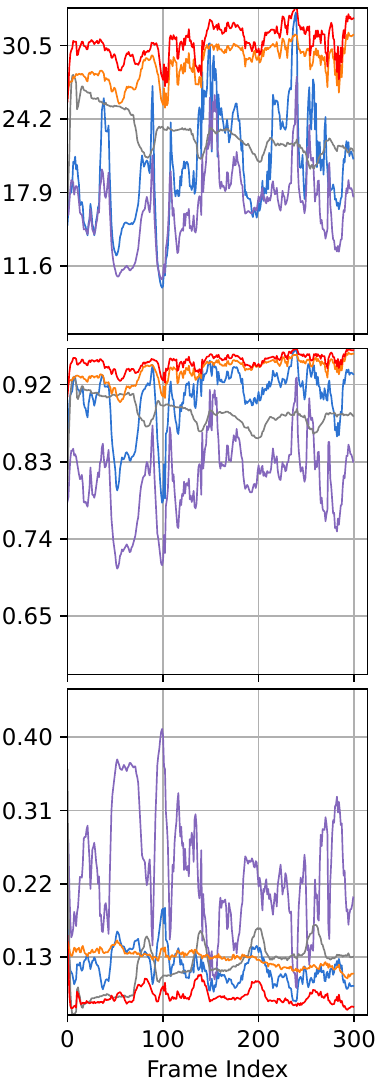}{discussion}
    \includesubfigure{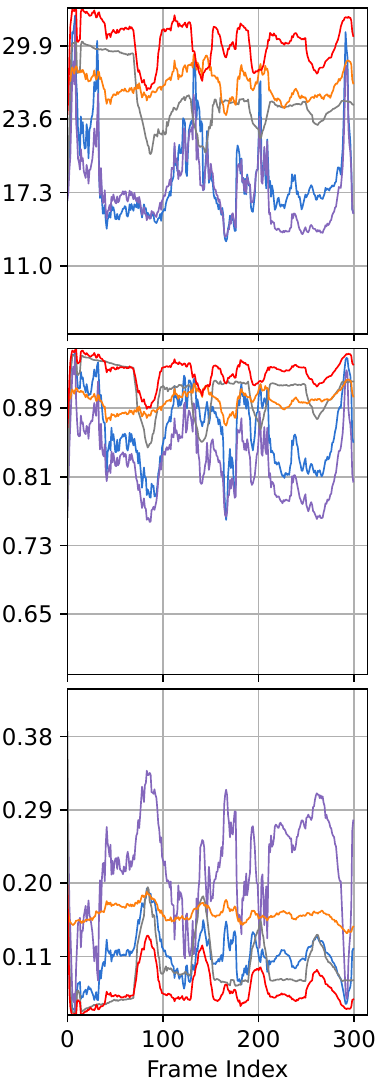}{stepin}
    \includesubfigure{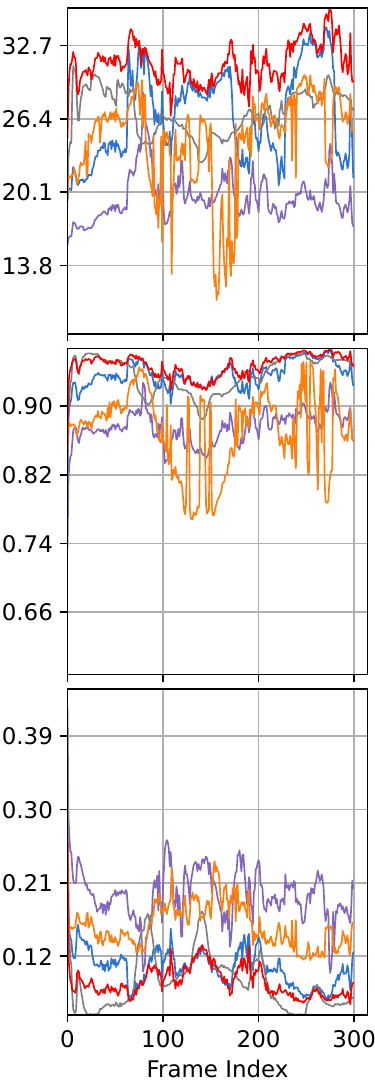}{trimming}
\includesubfigure{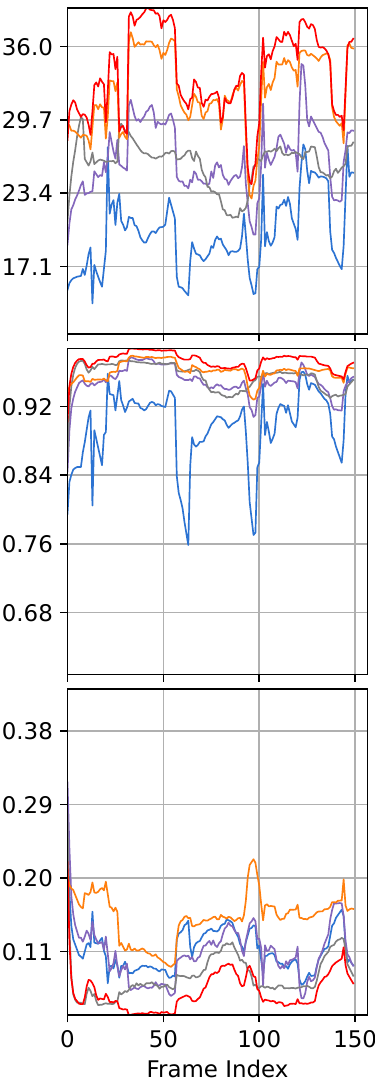}{basketball}
	\\\includeylabel
    \includesubfigure{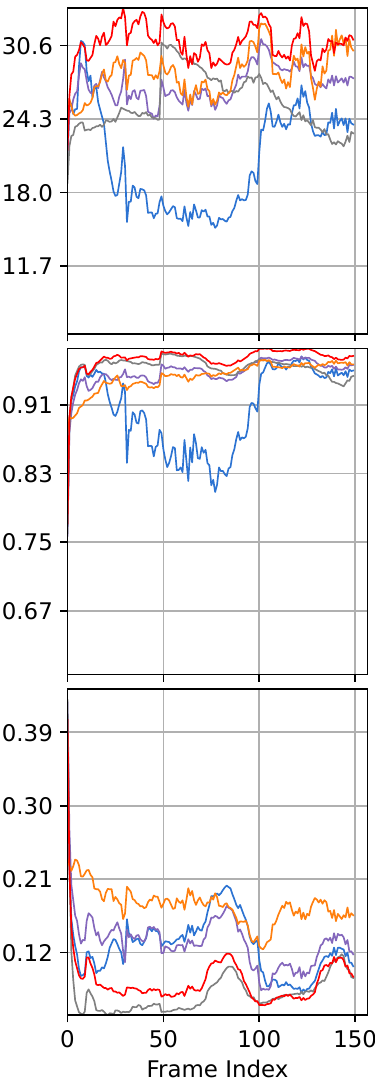}{boxes}
    \includesubfigure{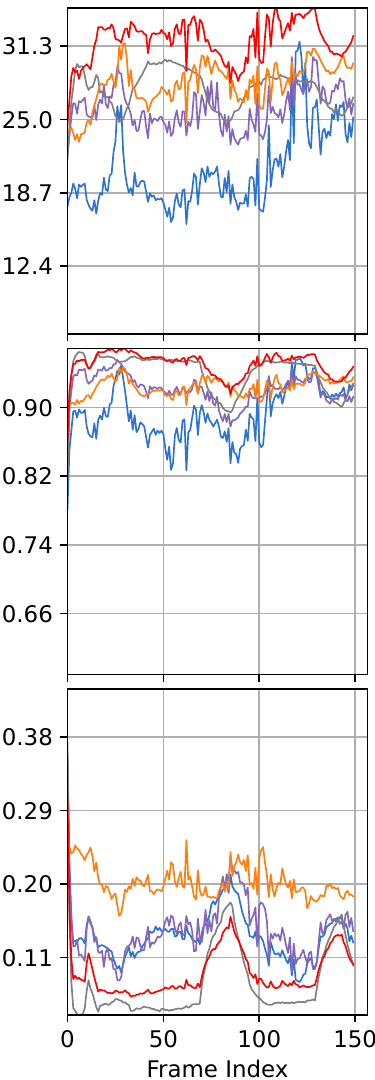}{football}
    \includesubfigure{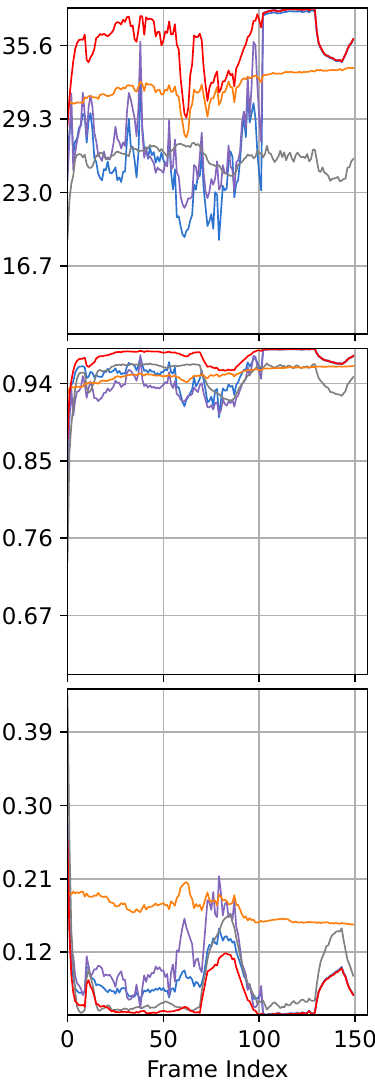}{juggle}
    \includesubfigure{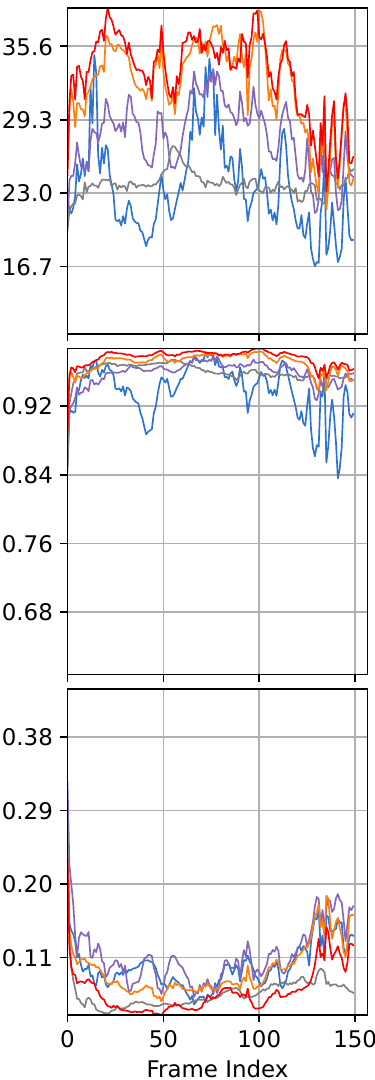}{softball}
    \includesubfigure{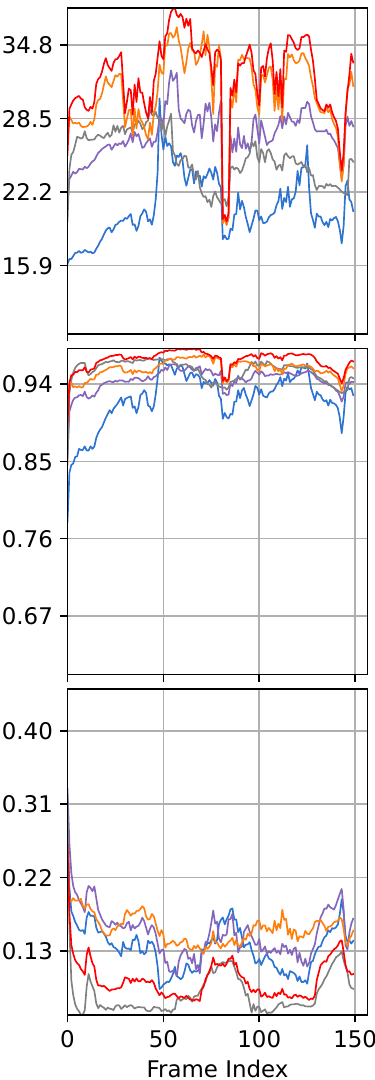}{tennis}
	\caption{
		Comparison of visual quality under fluctuating bandwidth on volumetric videos prepared by Dynamic 3DGS~\cite{luiten2023dynamic}.
		The y-axis ranges vary across subplots while maintaining equal scale spans for each metric across videos.
  }\end{figure*}

\begin{figure*}[htbp]
	\centering
	\includelegend{FixedBandwidth/train/regularized}
	\\\includeylabel
\includesubfigure{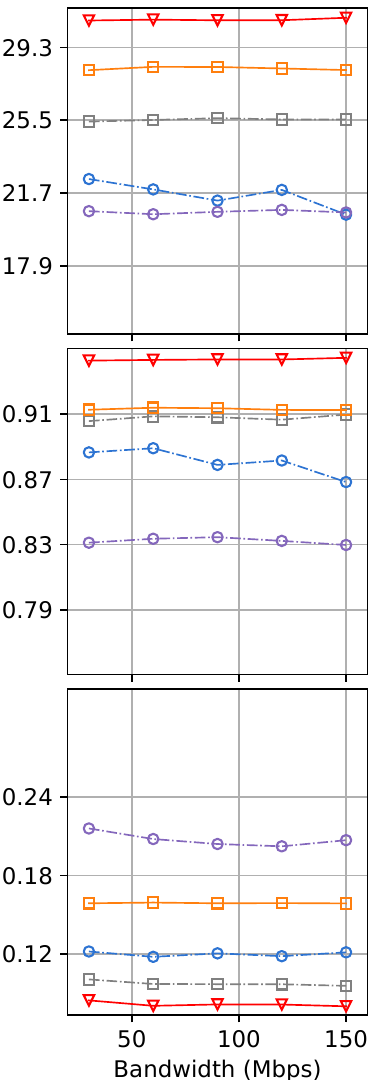}{cook spinach}
    \includesubfigure{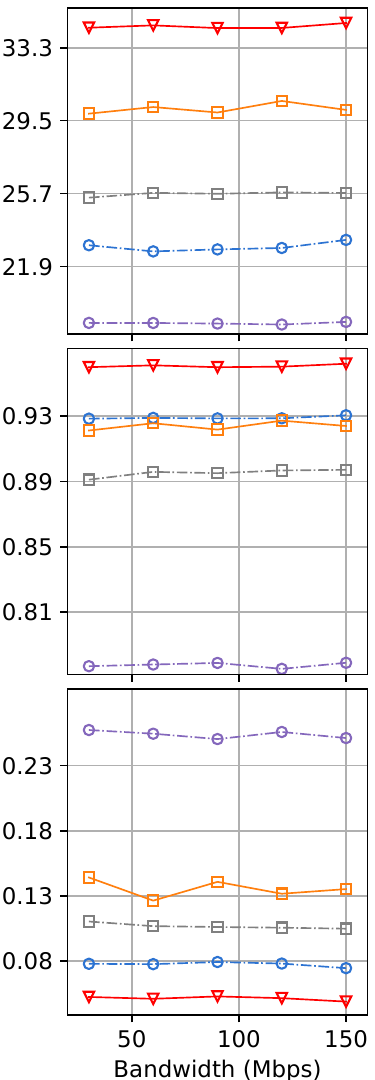}{cut roast beef}
    \includesubfigure{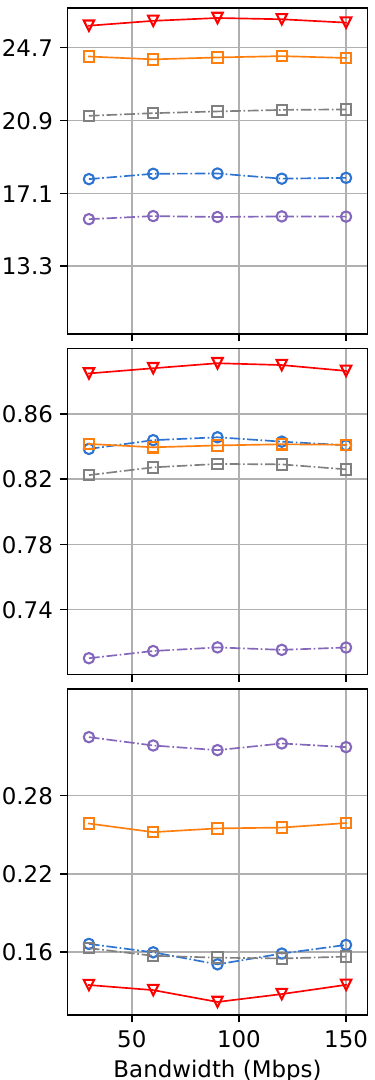}{flame salmon}
    \includesubfigure{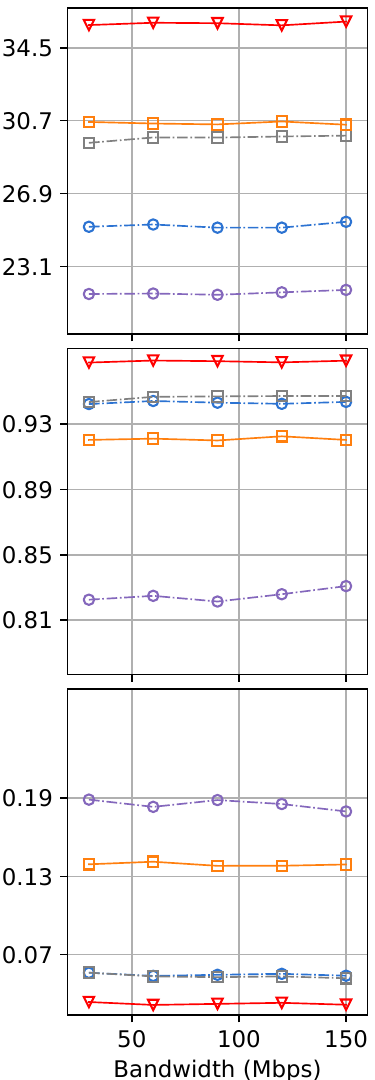}{flame steak}
    \includesubfigure{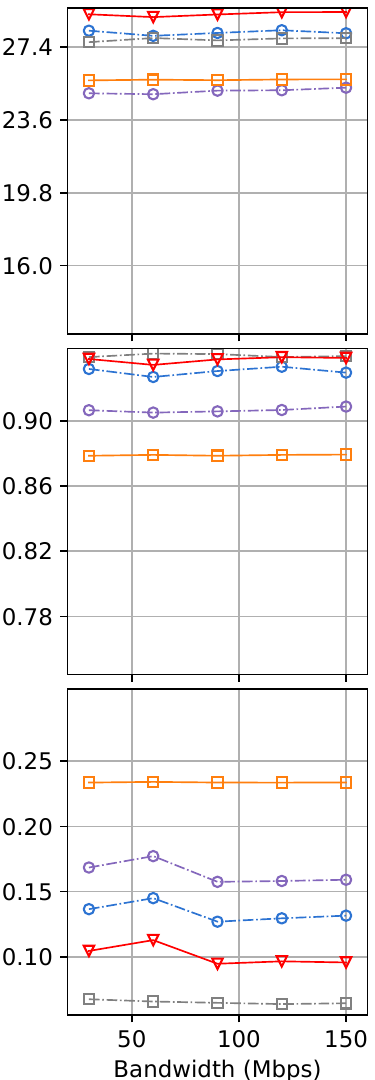}{sear steak}
	\\\includeylabel
    \includesubfigure{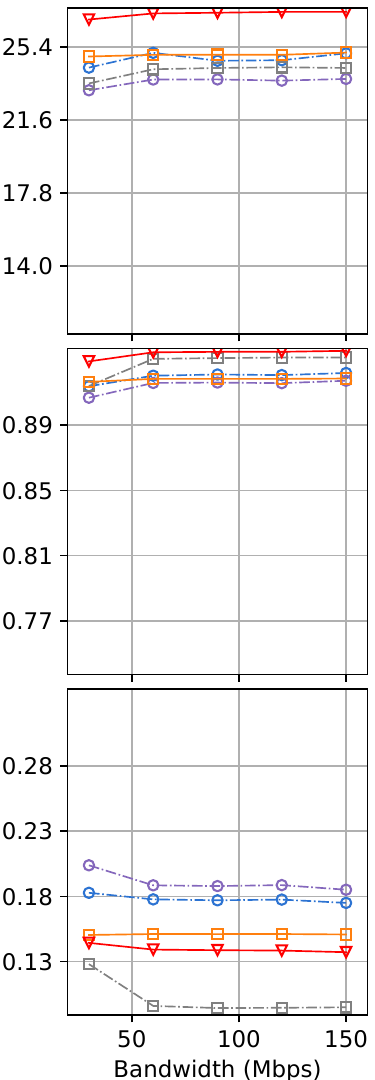}{walking}
\includesubfigure{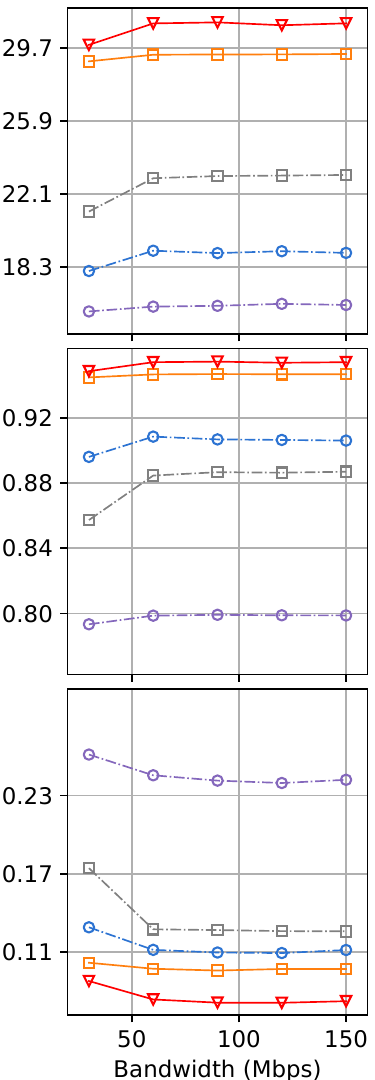}{discussion}
    \includesubfigure{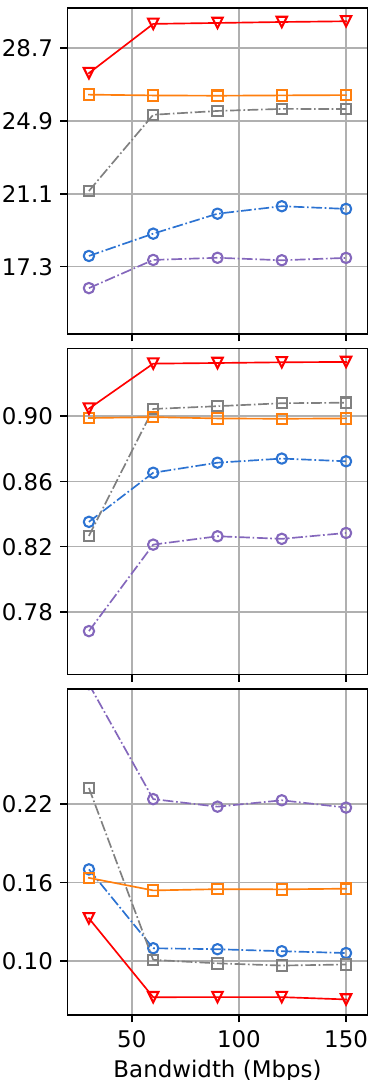}{stepin}
    \includesubfigure{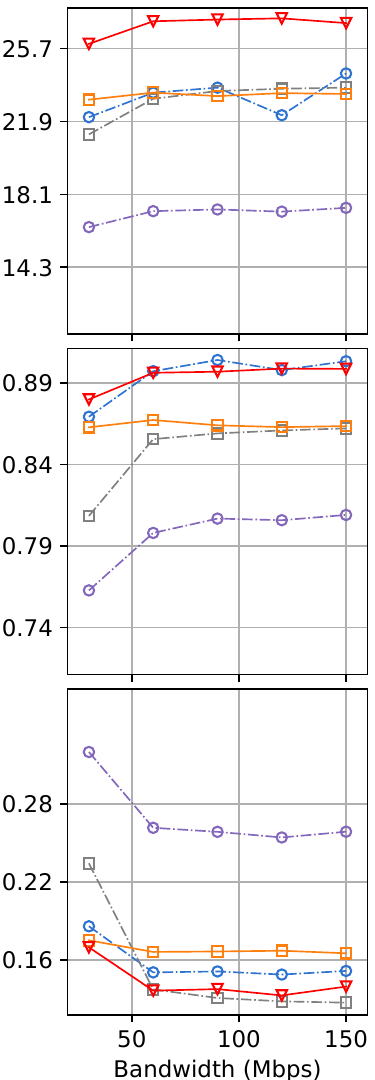}{trimming}
\includesubfigure{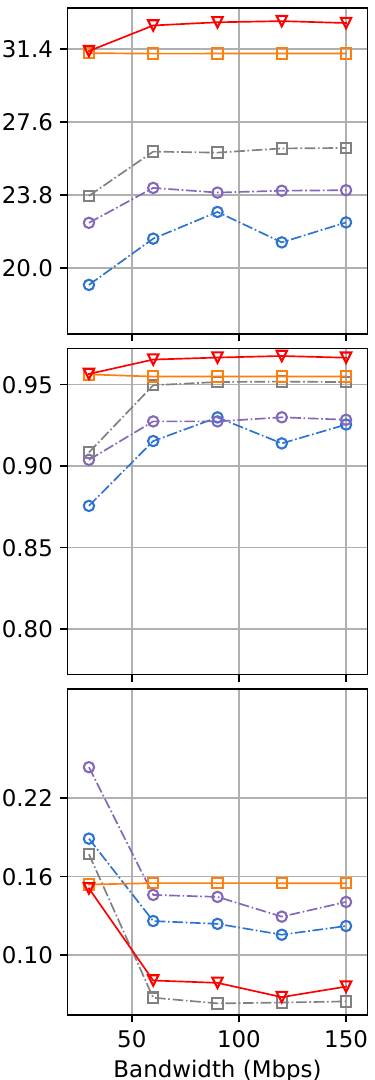}{basketball}
	\\\includeylabel
    \includesubfigure{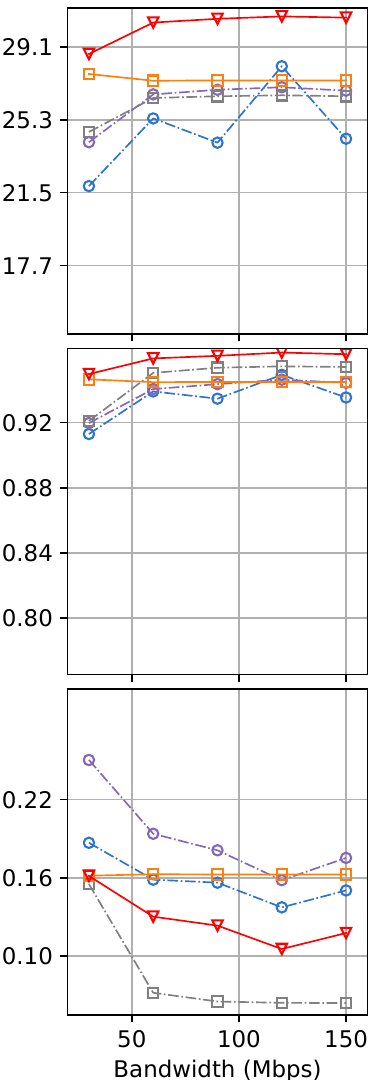}{boxes}
    \includesubfigure{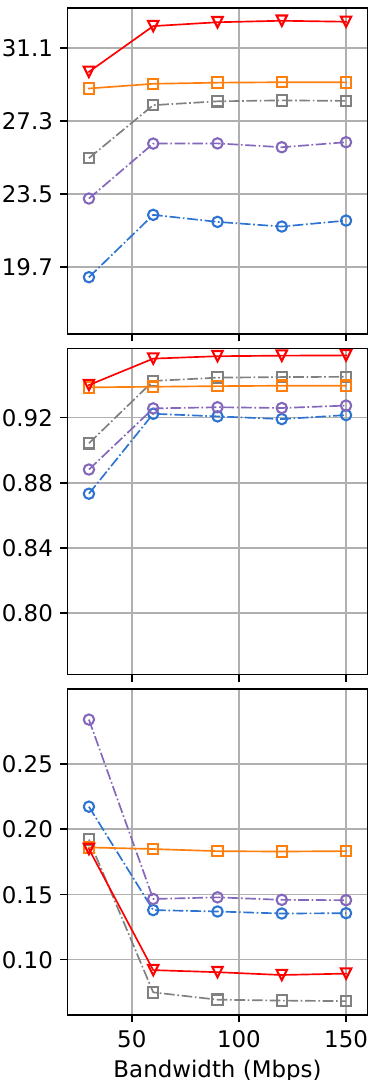}{football}
    \includesubfigure{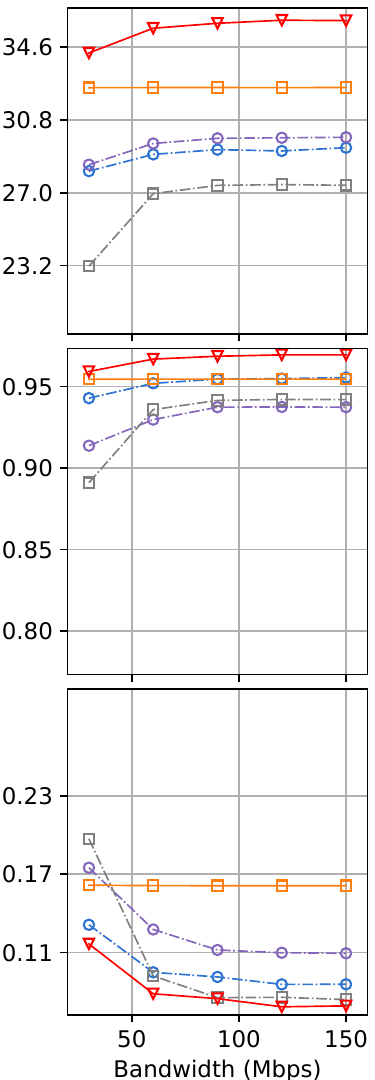}{juggle}
    \includesubfigure{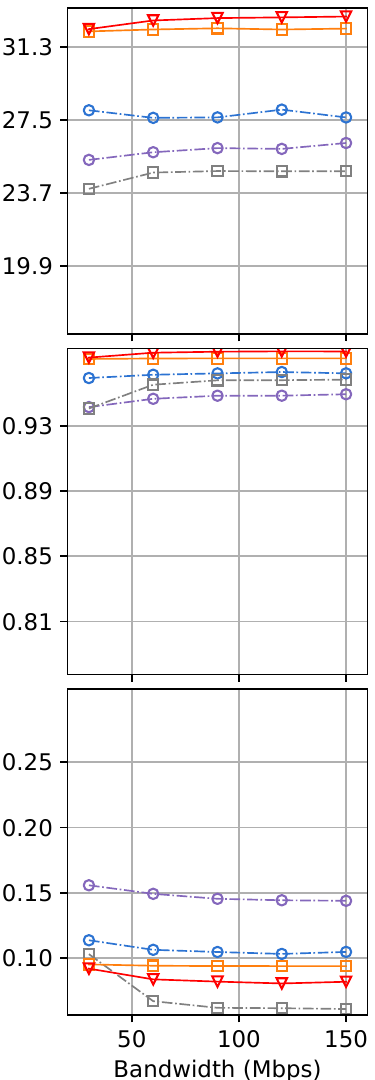}{softball}
    \includesubfigure{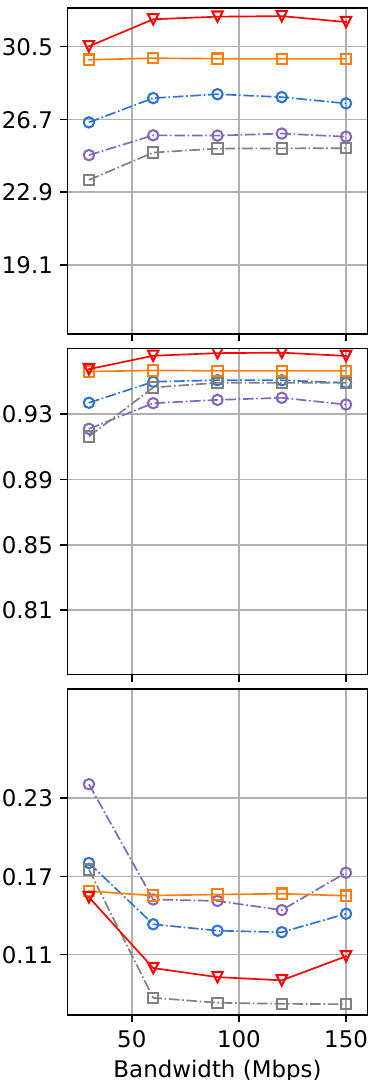}{tennis}
	\caption{
	Comparison of visual quality under fixed bandwidth on volumetric videos prepared by HiCoM~\cite{gaoHiCoMHierarchicalCoherent2024}.
		The y-axis ranges vary across subplots while maintaining equal scale spans for each metric across videos.
  }\end{figure*}

\begin{figure*}[htbp]
	\centering
	\includelegend{DynamicBandwidth/train/regularized}
	\\\includeylabel
\includesubfigure{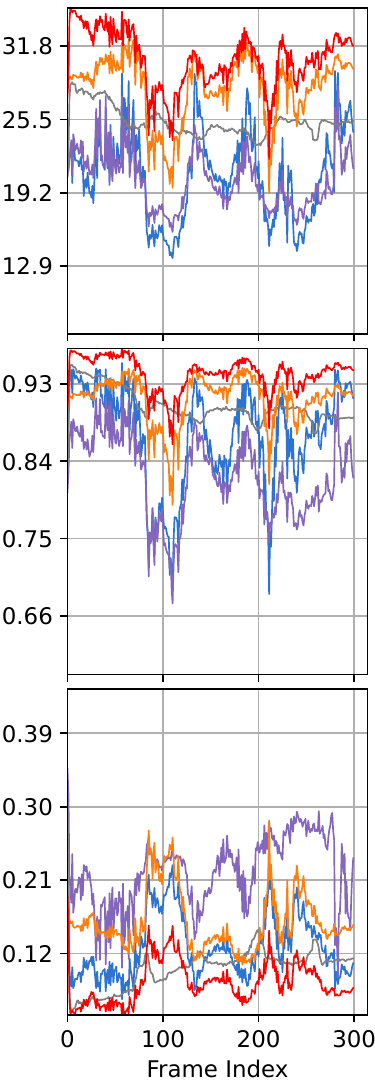}{cook spinach}
    \includesubfigure{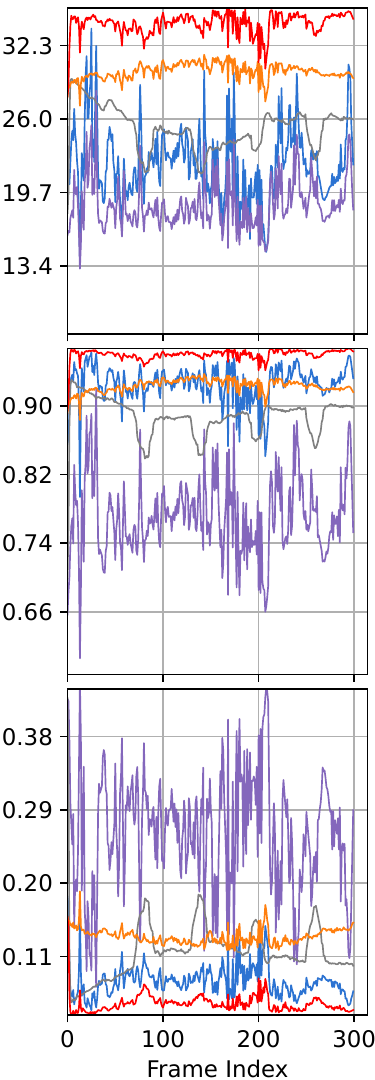}{cut roast beef}
    \includesubfigure{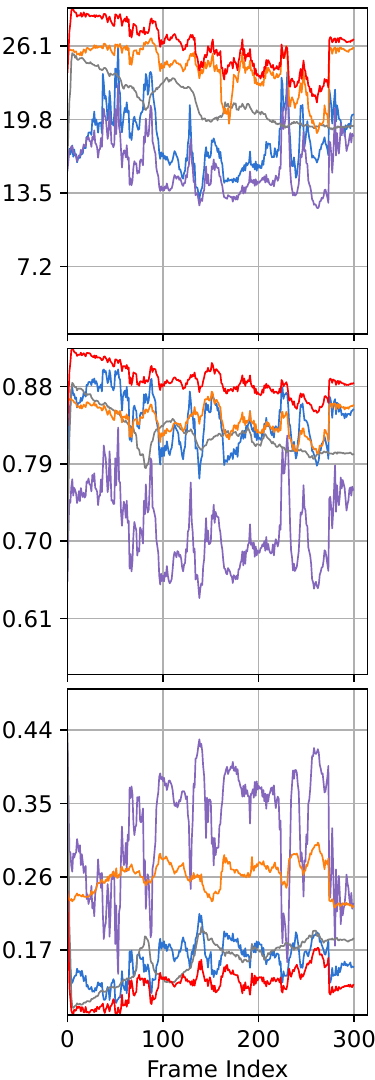}{flame salmon}
    \includesubfigure{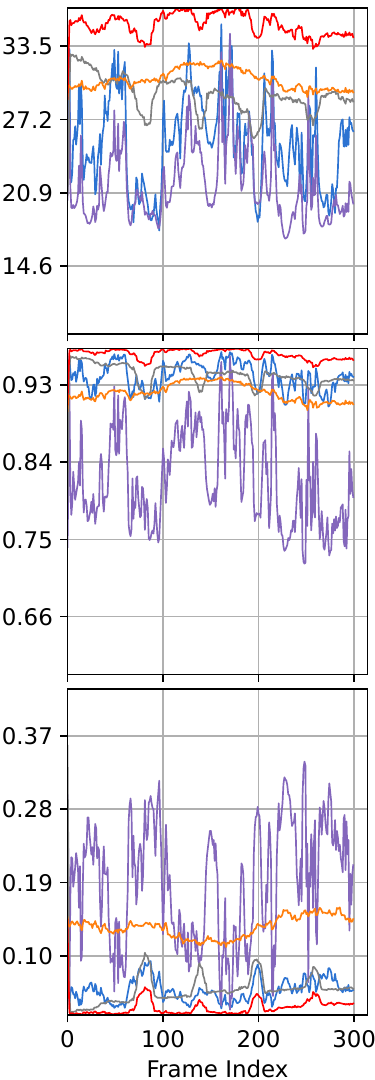}{flame steak}
    \includesubfigure{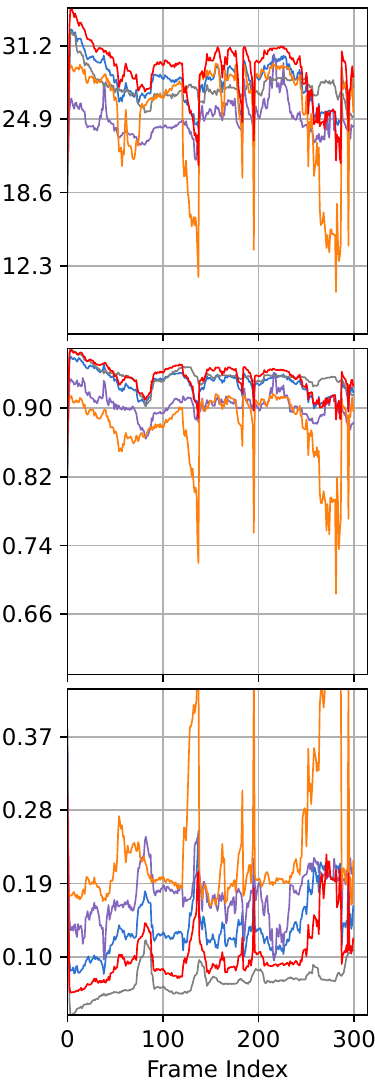}{sear steak}
	\\\includeylabel
    \includesubfigure{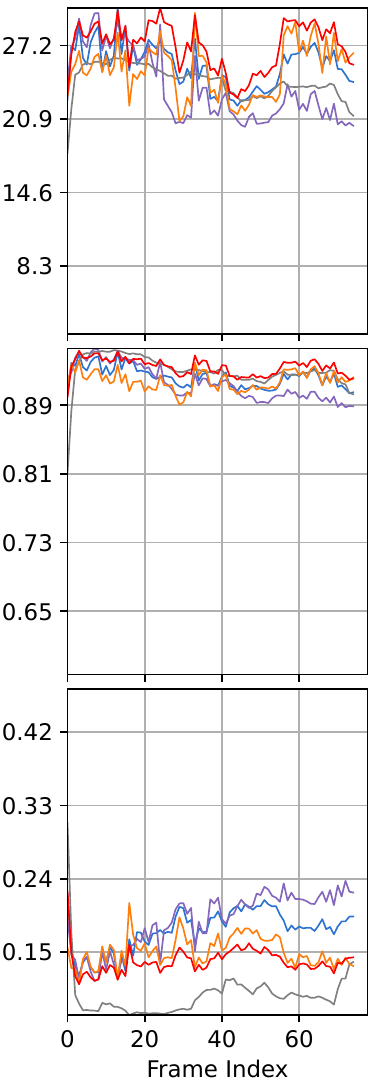}{walking}
\includesubfigure{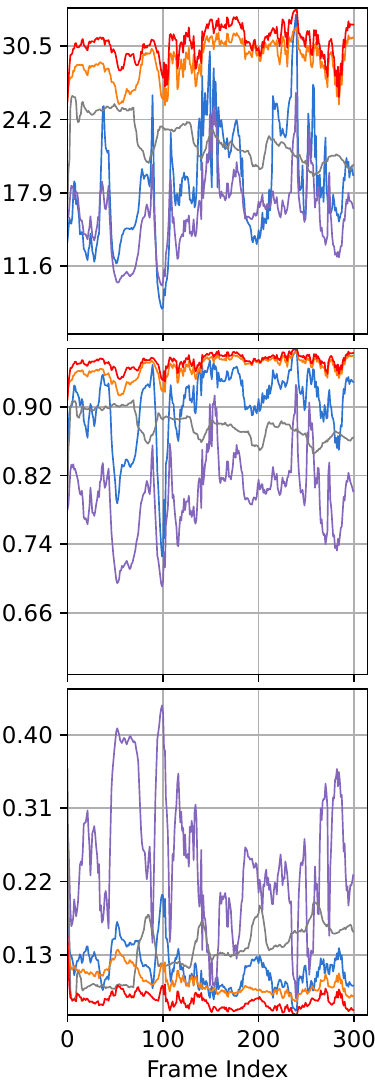}{discussion}
    \includesubfigure{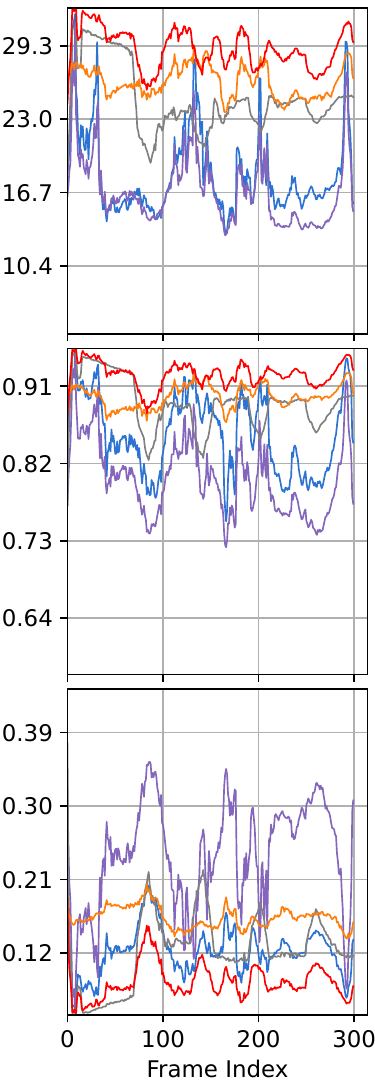}{stepin}
    \includesubfigure{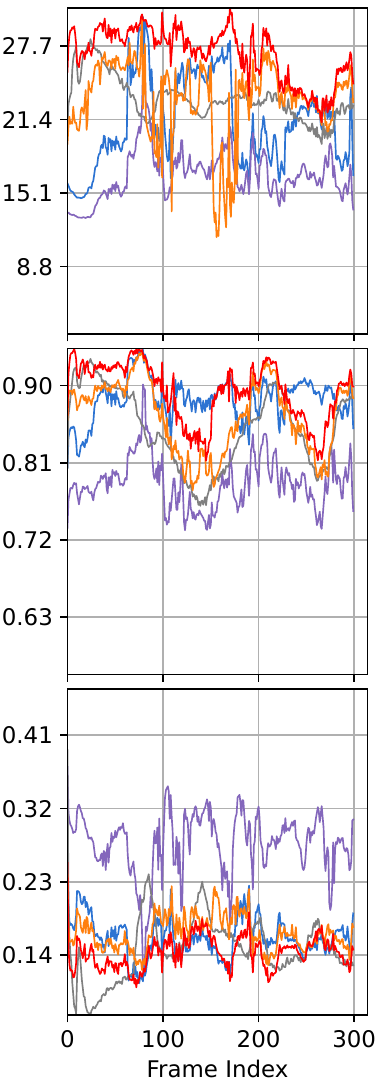}{trimming}
\includesubfigure{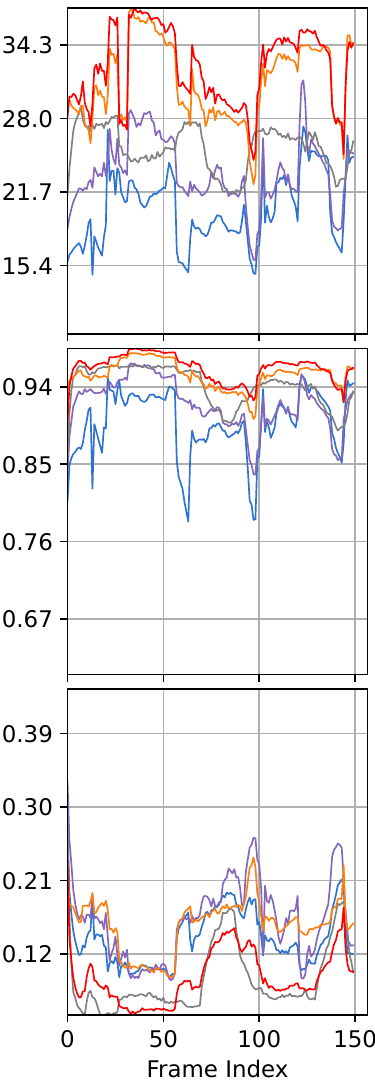}{basketball}
	\\\includeylabel
    \includesubfigure{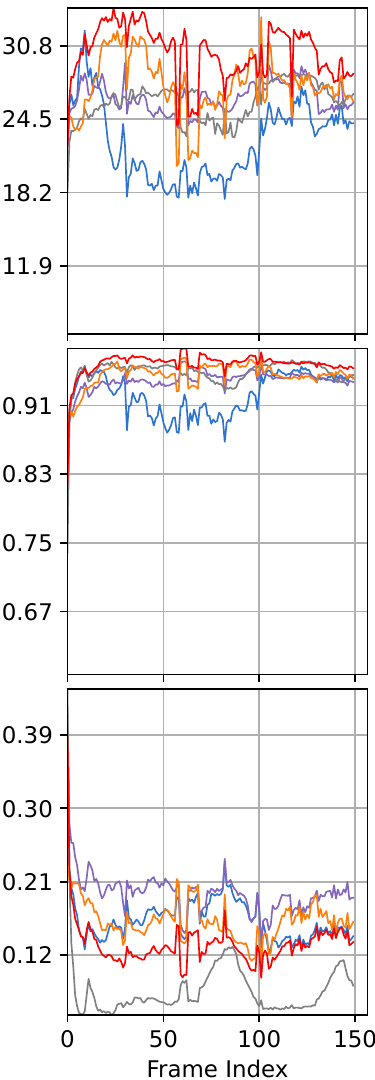}{boxes}
    \includesubfigure{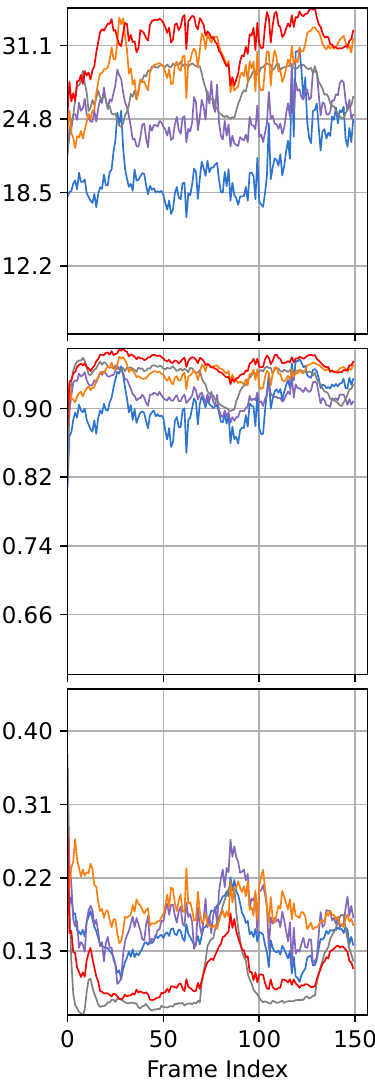}{football}
    \includesubfigure{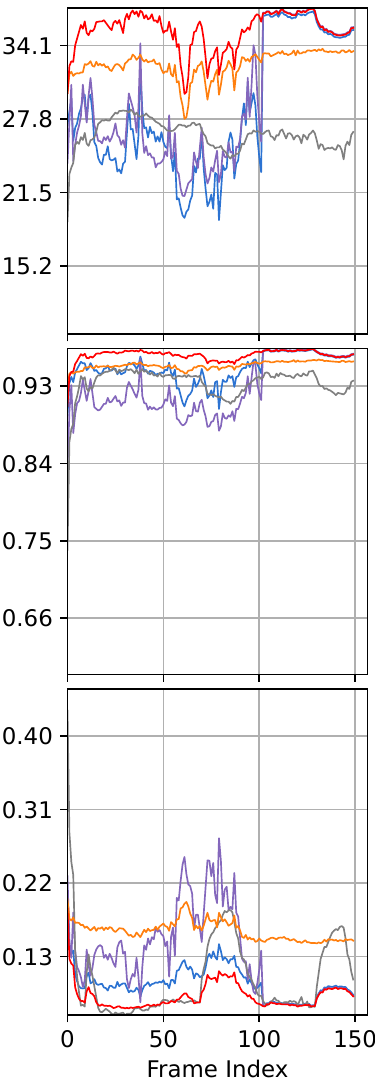}{juggle}
    \includesubfigure{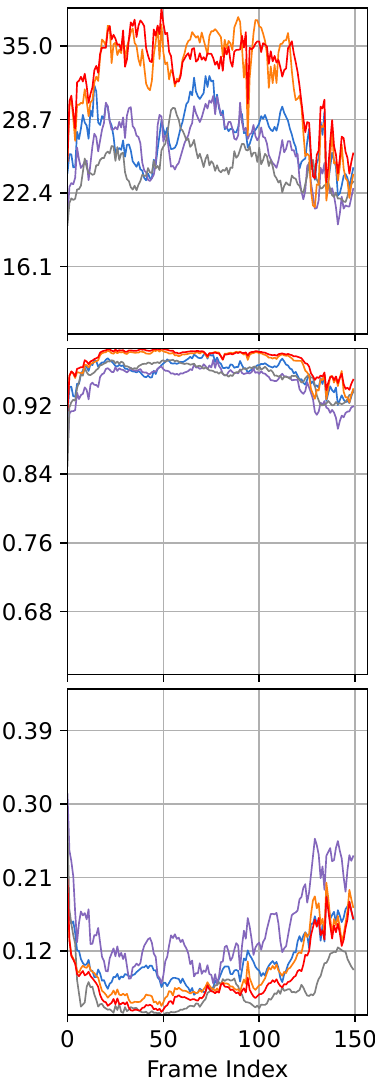}{softball}
    \includesubfigure{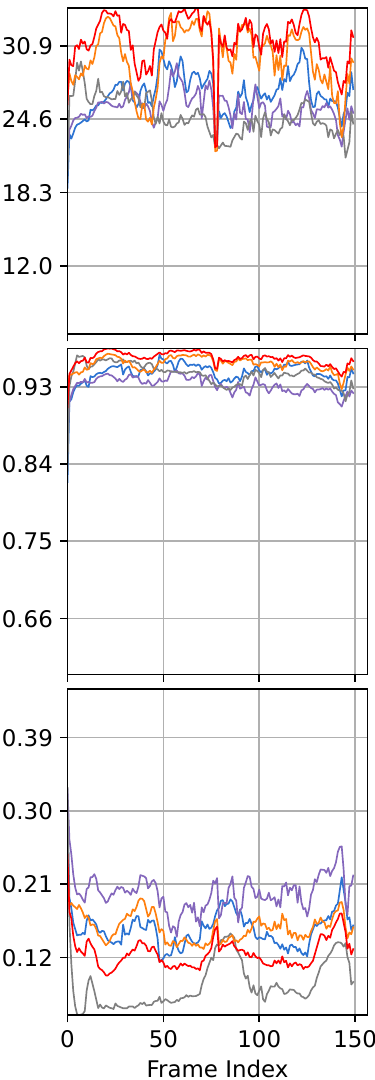}{tennis}
	\caption{
		Comparison of visual quality under fluctuating bandwidth on volumetric videos prepared by HiCoM~\cite{gaoHiCoMHierarchicalCoherent2024}.
		The y-axis ranges vary across subplots while maintaining equal scale spans for each metric across videos.
  }\end{figure*}

\begin{figure*}[htbp]
	\centering
	\includelegend{FixedBandwidth/train/regularized}
	\\\includeylabel
\includesubfigure{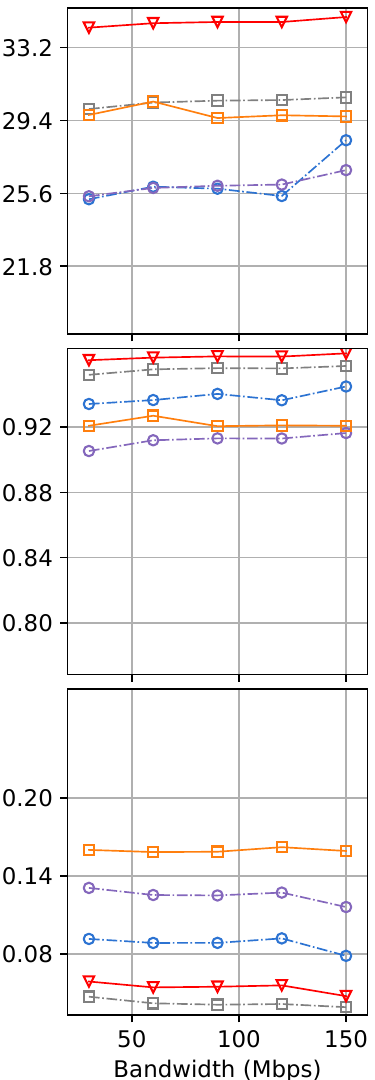}{cook spinach}
    \includesubfigure{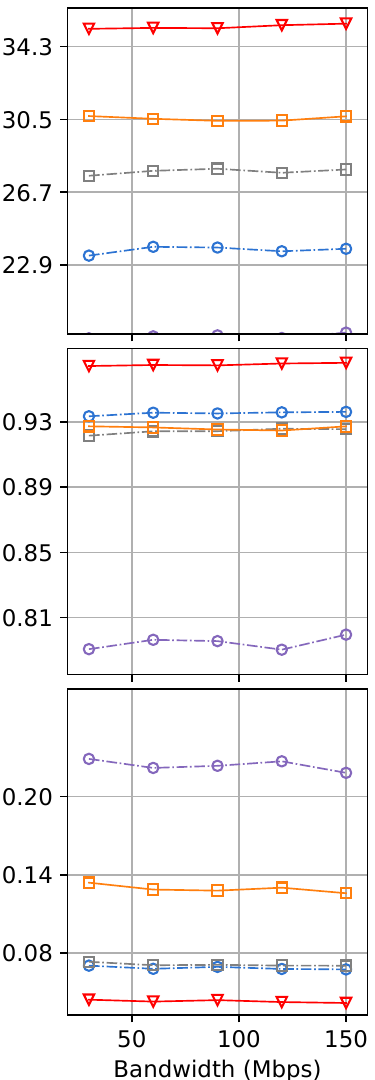}{cut roast beef}
    \includesubfigure{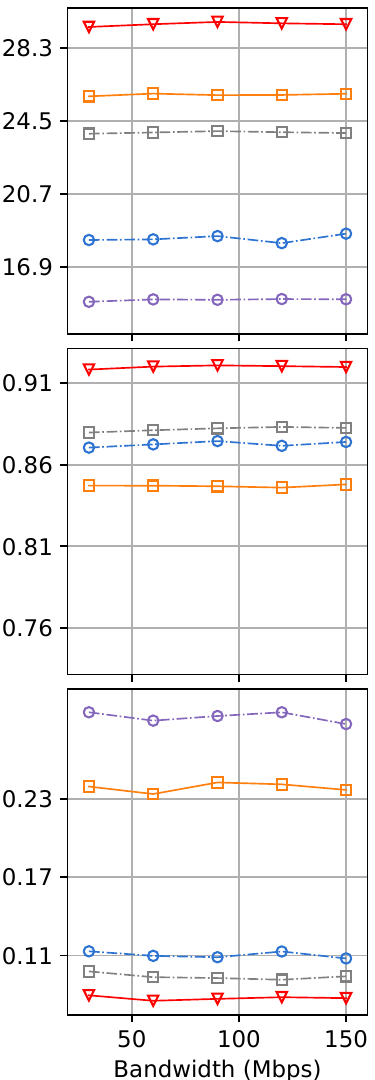}{flame salmon}
    \includesubfigure{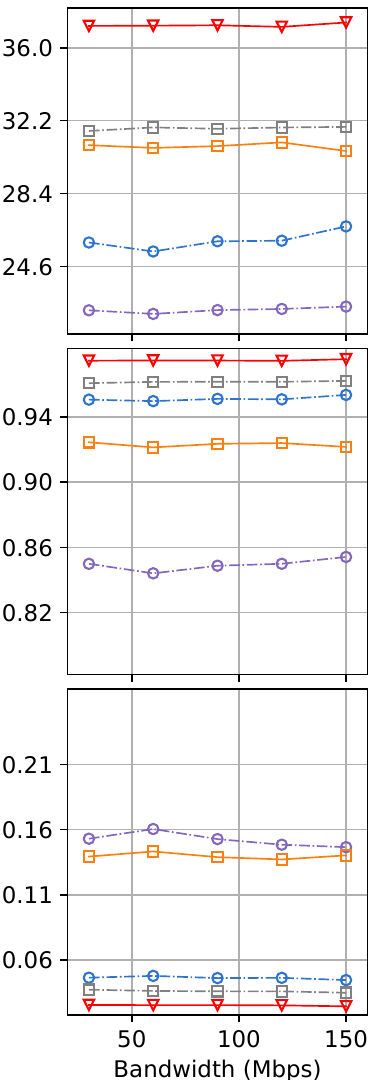}{flame steak}
    \includesubfigure{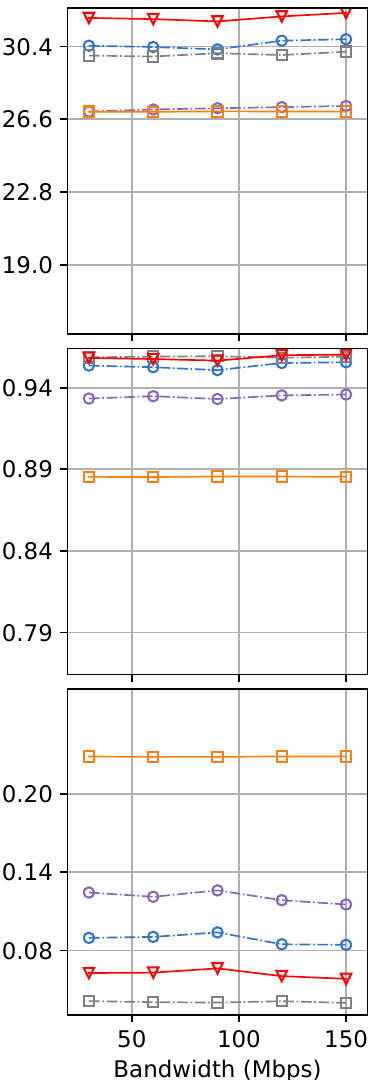}{sear steak}
	\\\includeylabel
    \includesubfigure{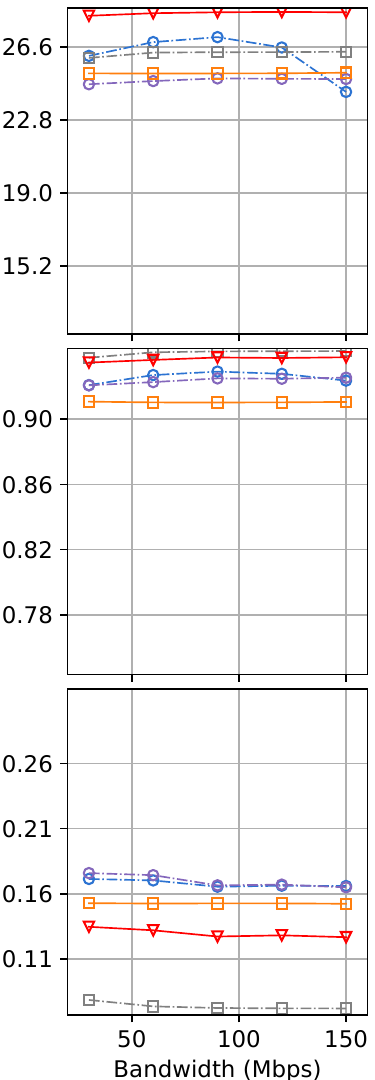}{walking}
\includesubfigure{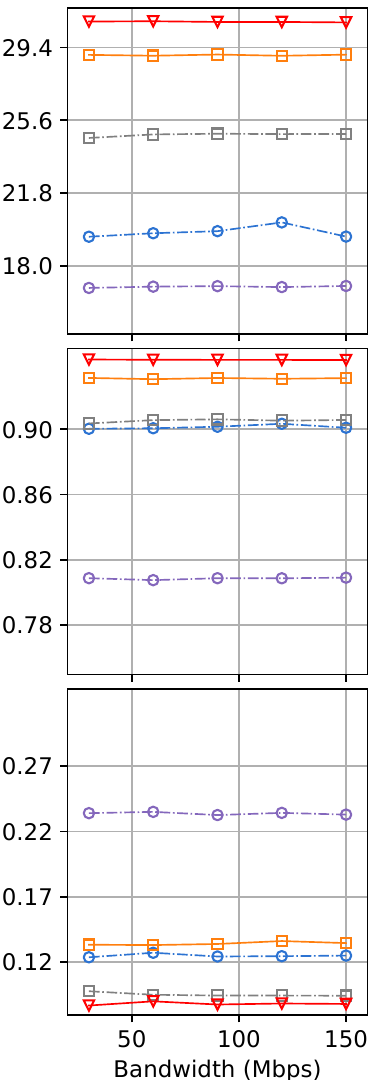}{discussion}
    \includesubfigure{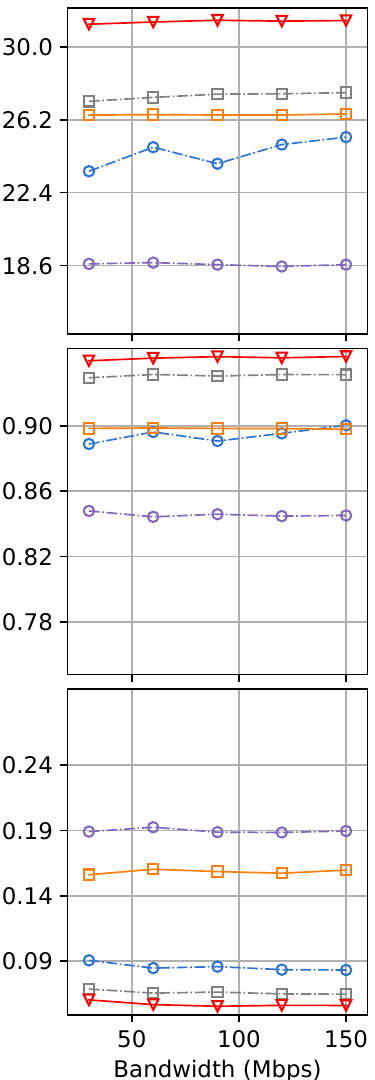}{stepin}
    \includesubfigure{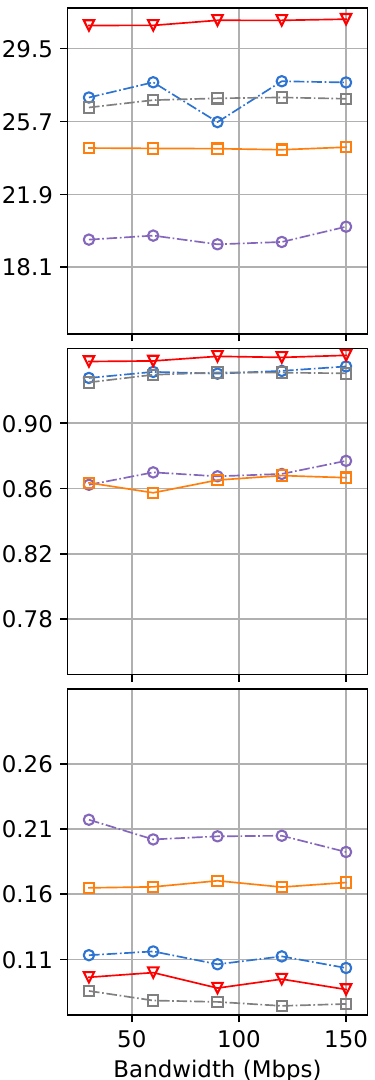}{trimming}
\includesubfigure{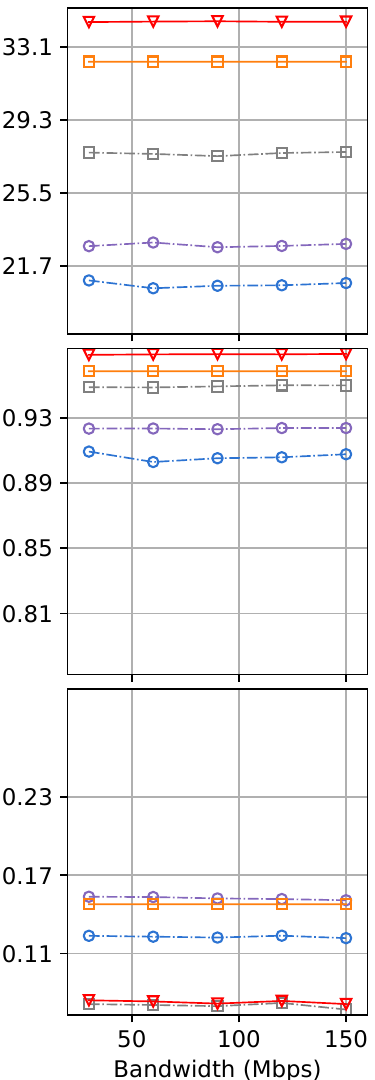}{basketball}
	\\\includeylabel
    \includesubfigure{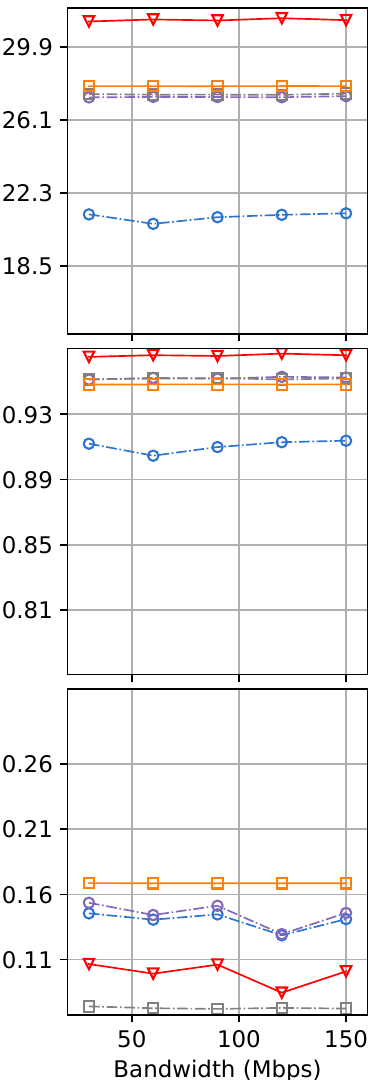}{boxes}
    \includesubfigure{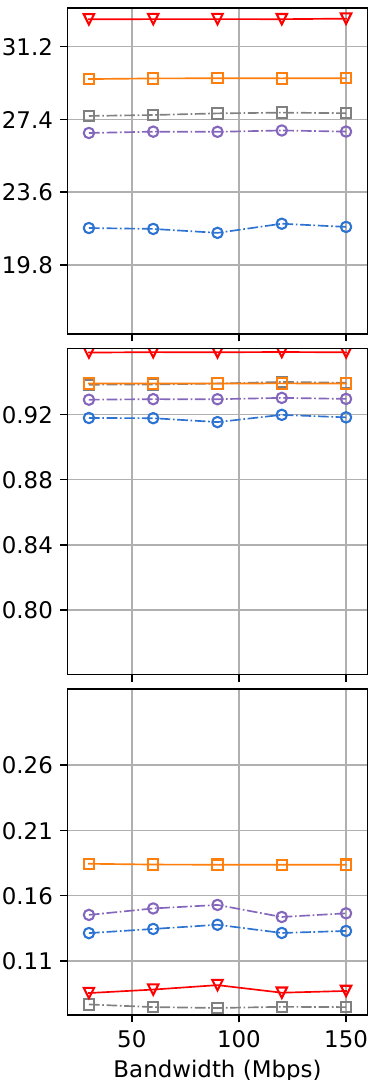}{football}
    \includesubfigure{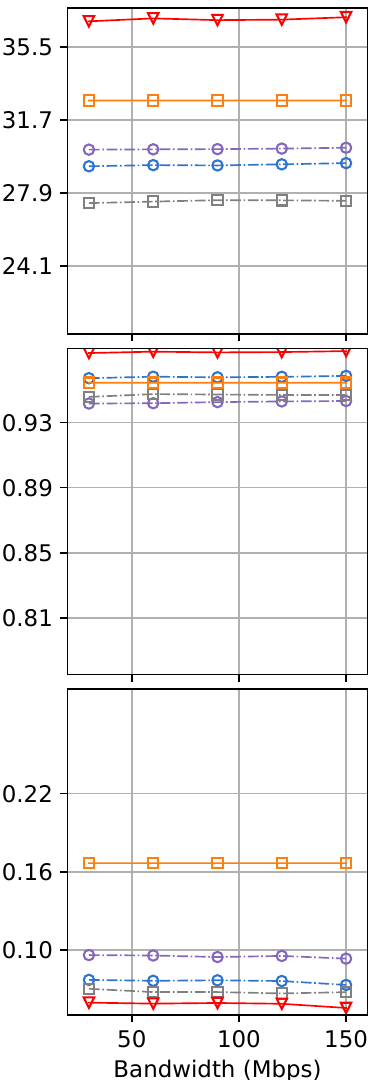}{juggle}
    \includesubfigure{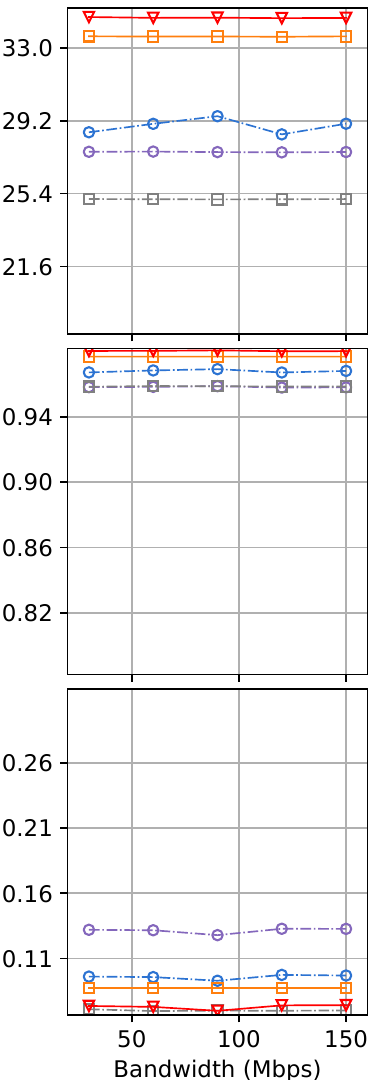}{softball}
    \includesubfigure{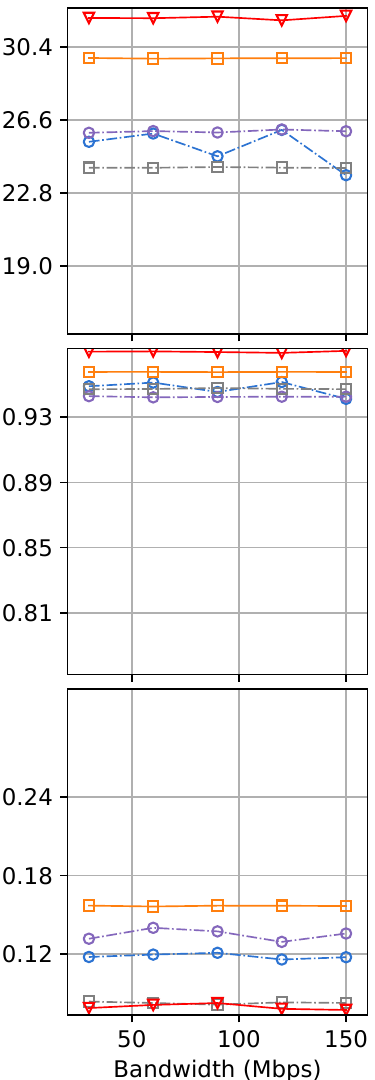}{tennis}
	\caption{
	Comparison of visual quality under fixed bandwidth on volumetric videos prepared by 4DGS~\cite{wu4DGaussianSplatting2024}.
		The y-axis ranges vary across subplots while maintaining equal scale spans for each metric across videos.
  }\end{figure*}

\begin{figure*}[htbp]
	\centering
	\includelegend{DynamicBandwidth/train/regularized}
	\\\includeylabel
\includesubfigure{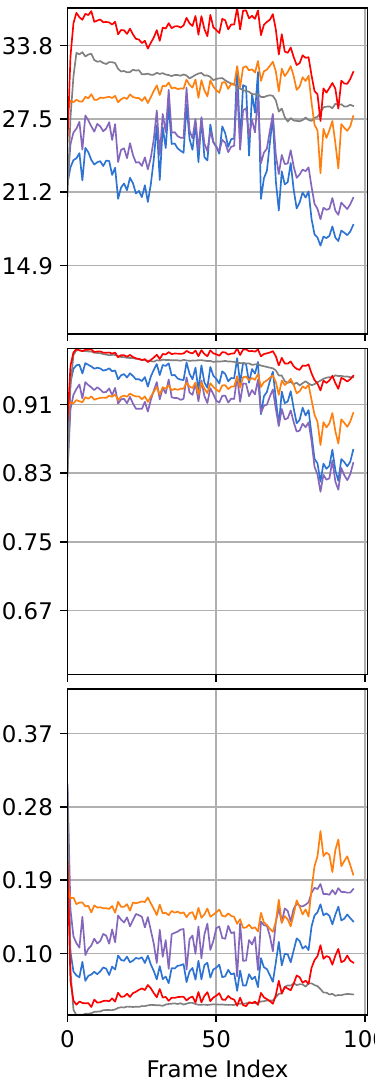}{cook spinach}
    \includesubfigure{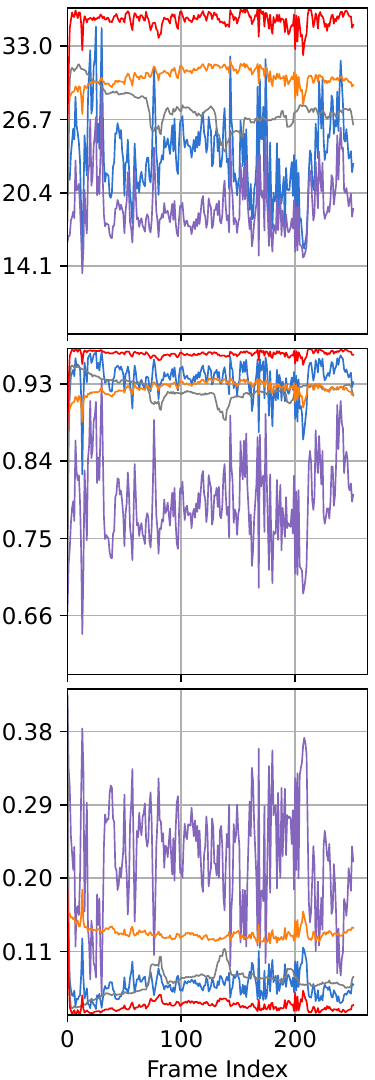}{cut roast beef}
    \includesubfigure{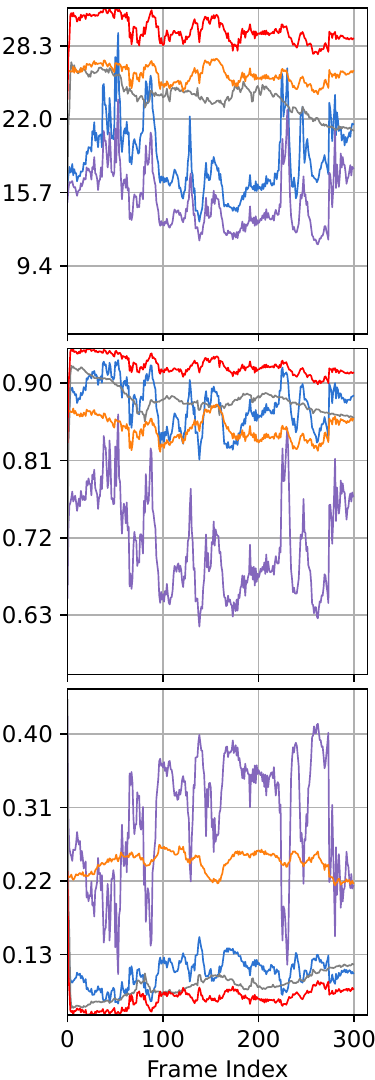}{flame salmon}
    \includesubfigure{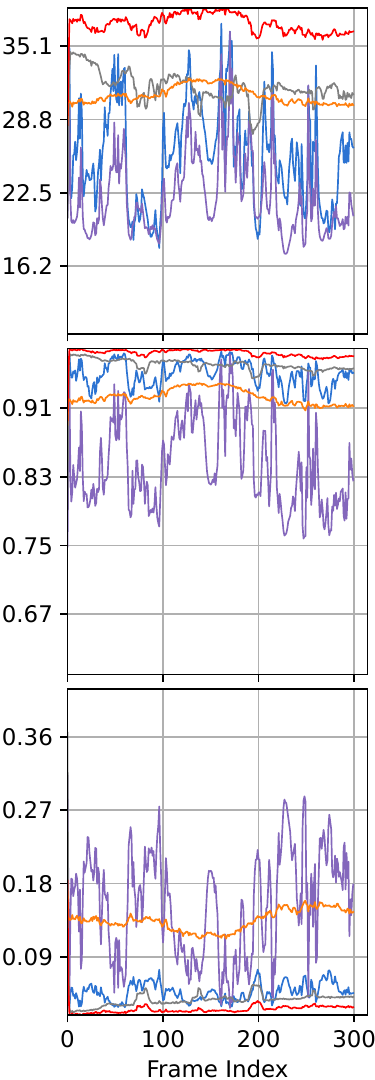}{flame steak}
    \includesubfigure{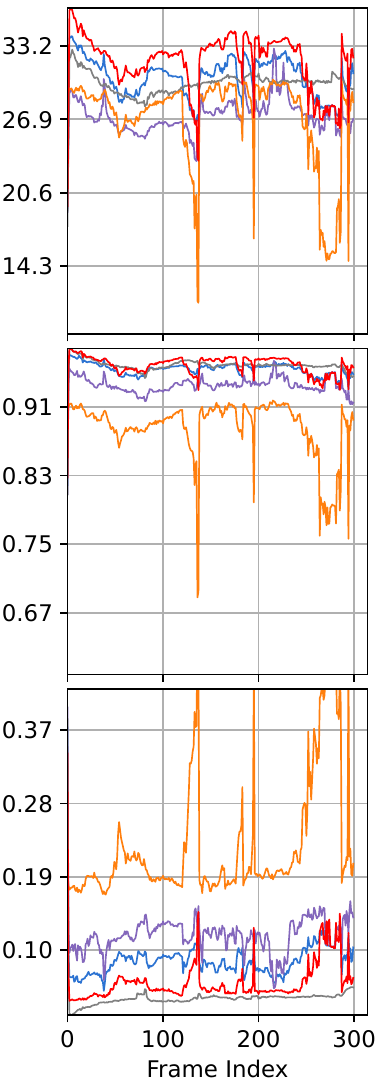}{sear steak}
	\\\includeylabel
    \includesubfigure{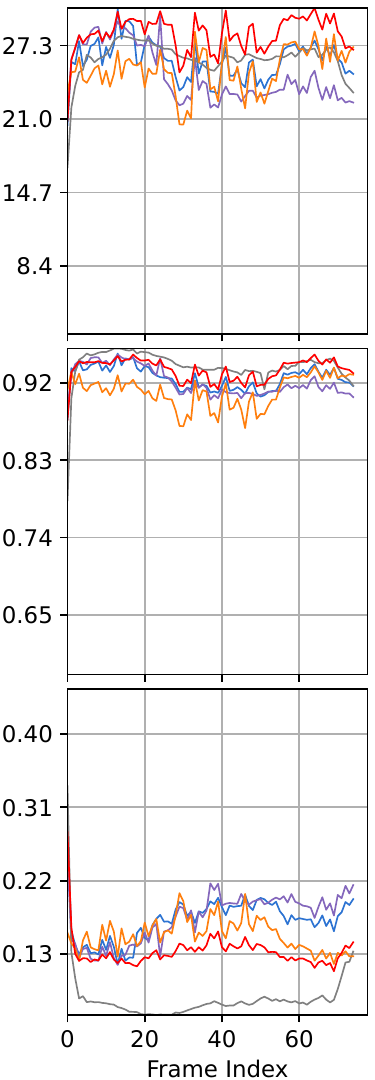}{walking}
\includesubfigure{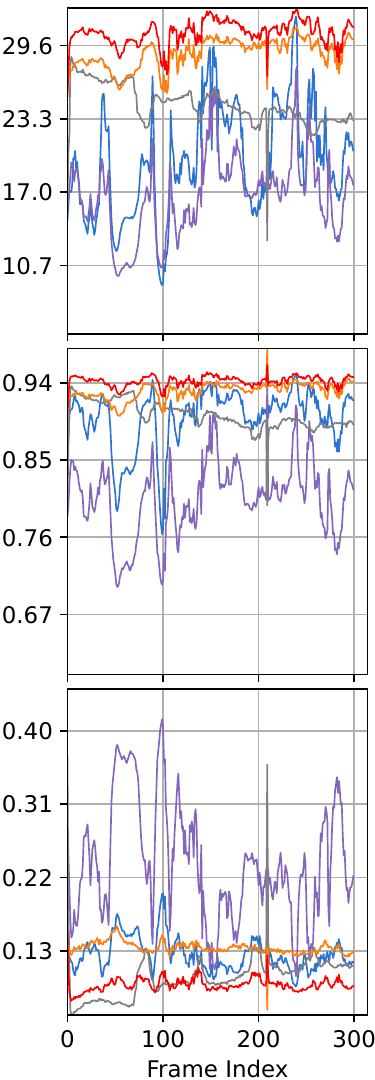}{discussion}
    \includesubfigure{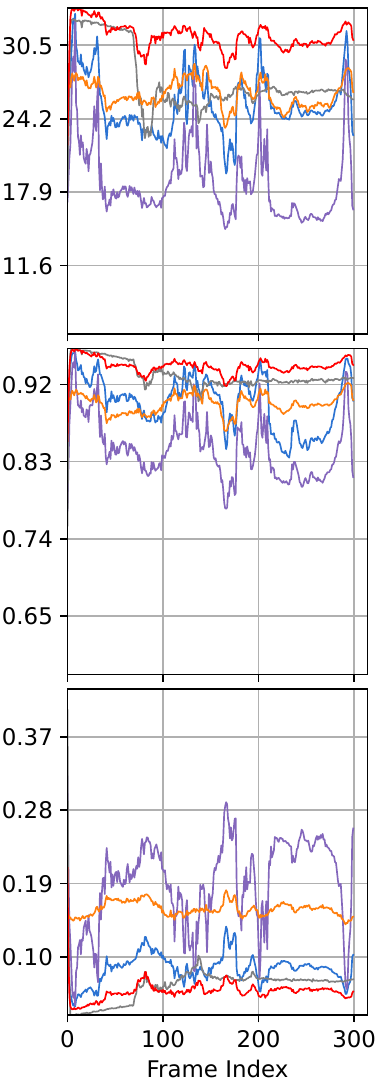}{stepin}
    \includesubfigure{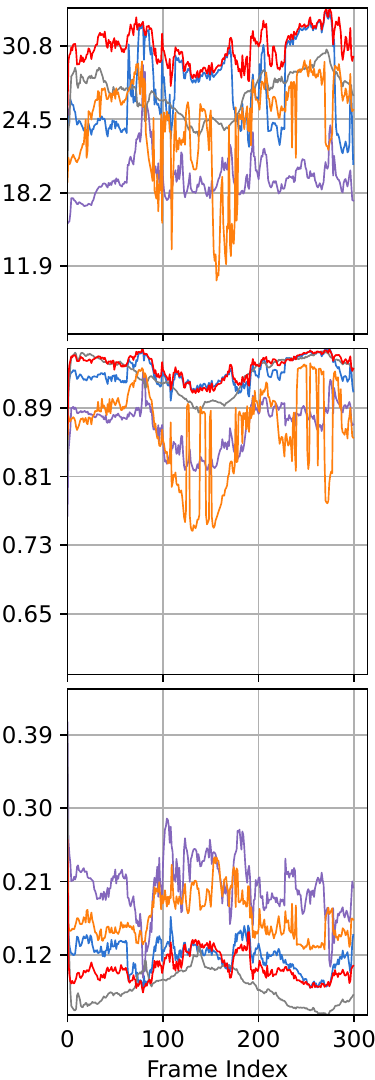}{trimming}
\includesubfigure{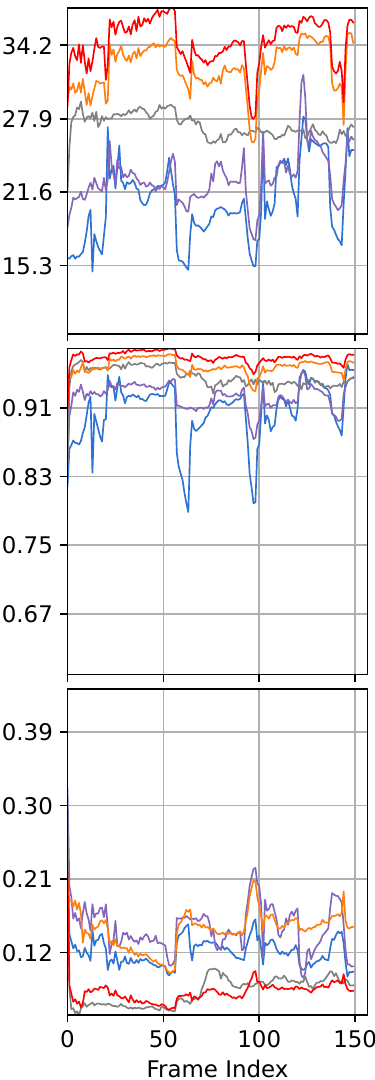}{basketball}
	\\\includeylabel
    \includesubfigure{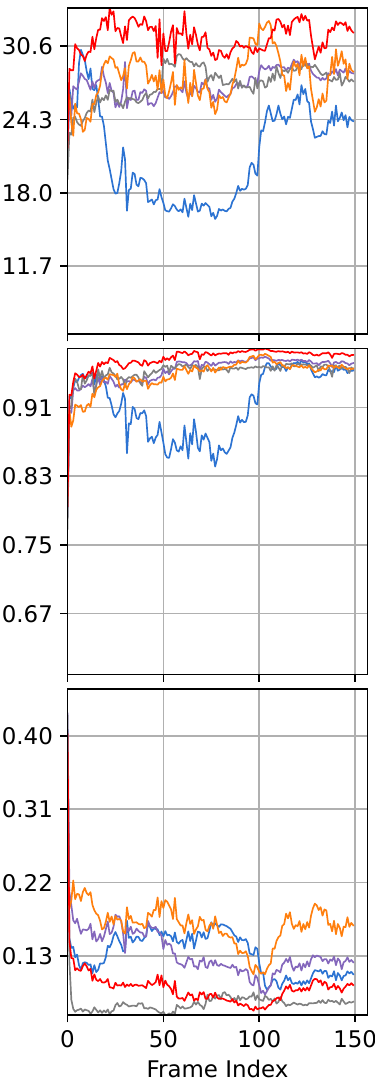}{boxes}
    \includesubfigure{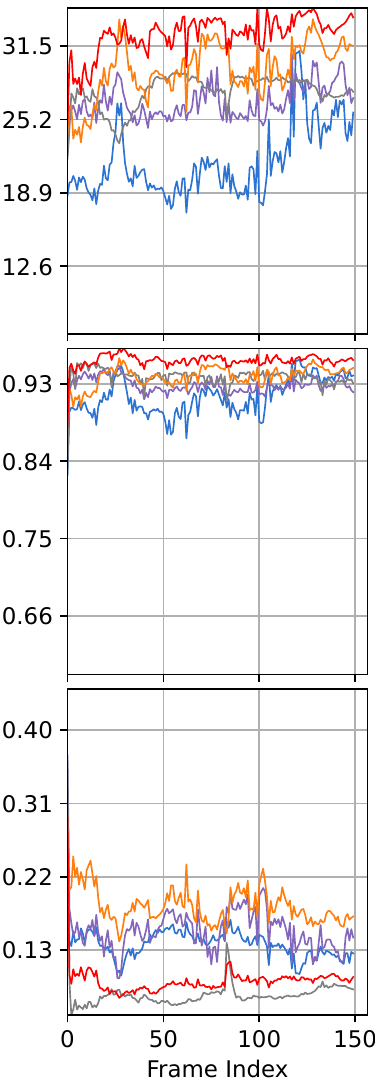}{football}
    \includesubfigure{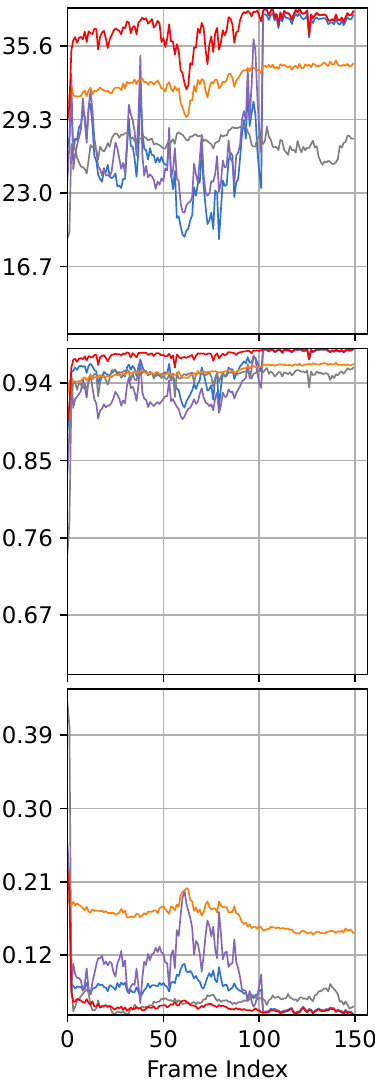}{juggle}
    \includesubfigure{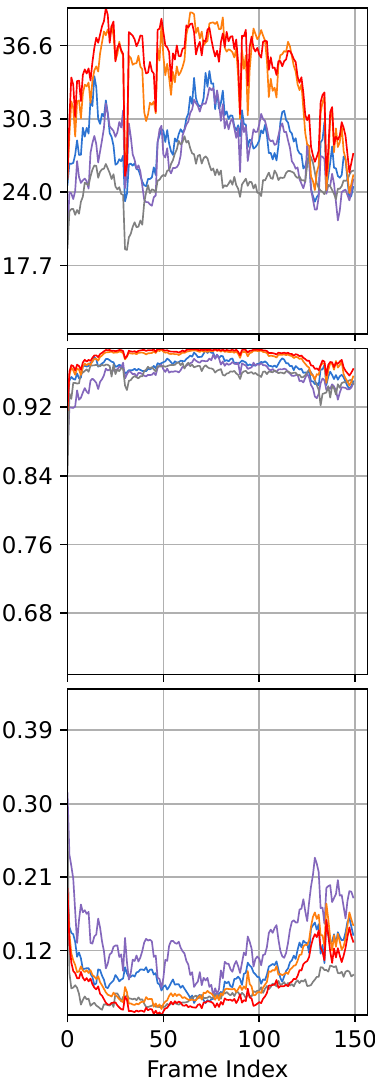}{softball}
    \includesubfigure{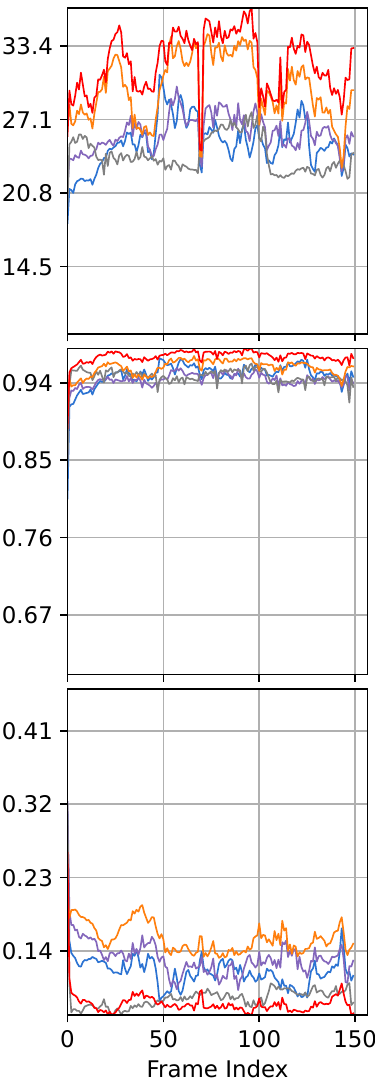}{tennis}
	\caption{
		Comparison of visual quality under fluctuating bandwidth on volumetric videos prepared by 4DGS~\cite{wu4DGaussianSplatting2024}.
		The y-axis ranges vary across subplots while maintaining equal scale spans for each metric across videos.
  }\end{figure*} 
\begin{figure*}[t]
    \centering
    \captionsetup[subfigure]{justification=centering}
    \begin{subfigure}[t]{0.138\linewidth}
        \centering
        \includegraphics[width=\linewidth]{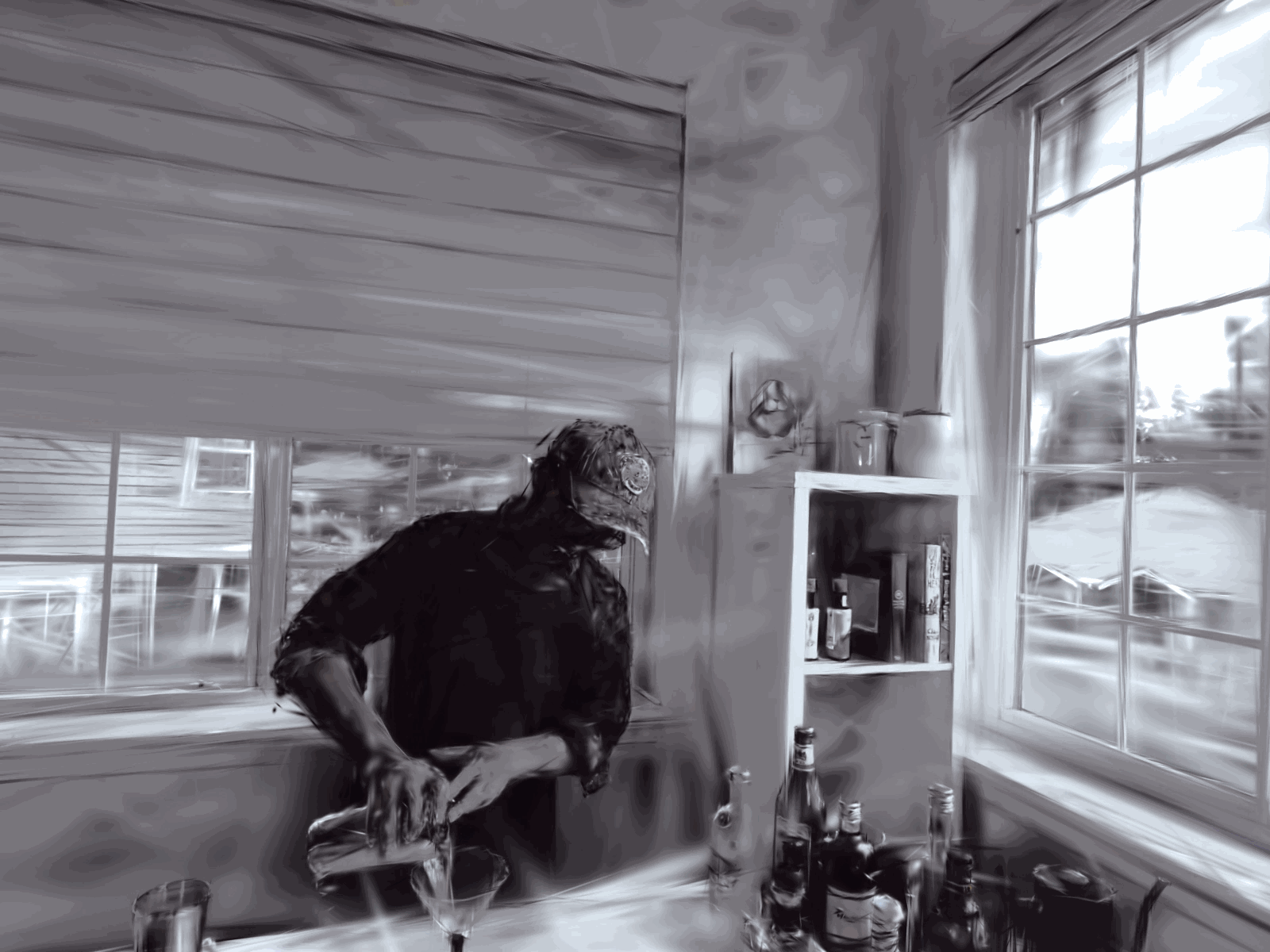}\\[-0.15em]
        \includegraphics[width=\linewidth]{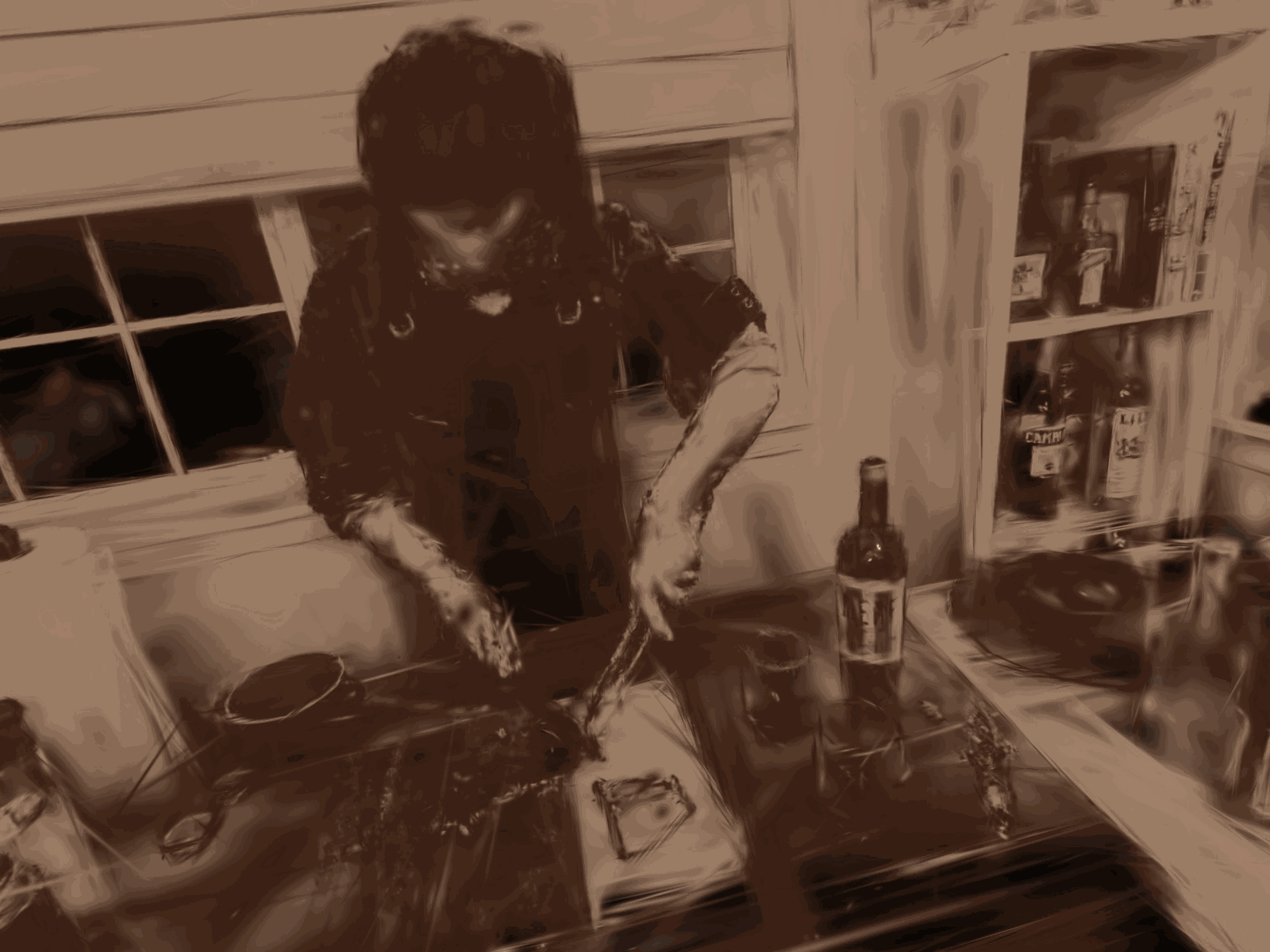}\\[-0.15em]
        \includegraphics[width=\linewidth]{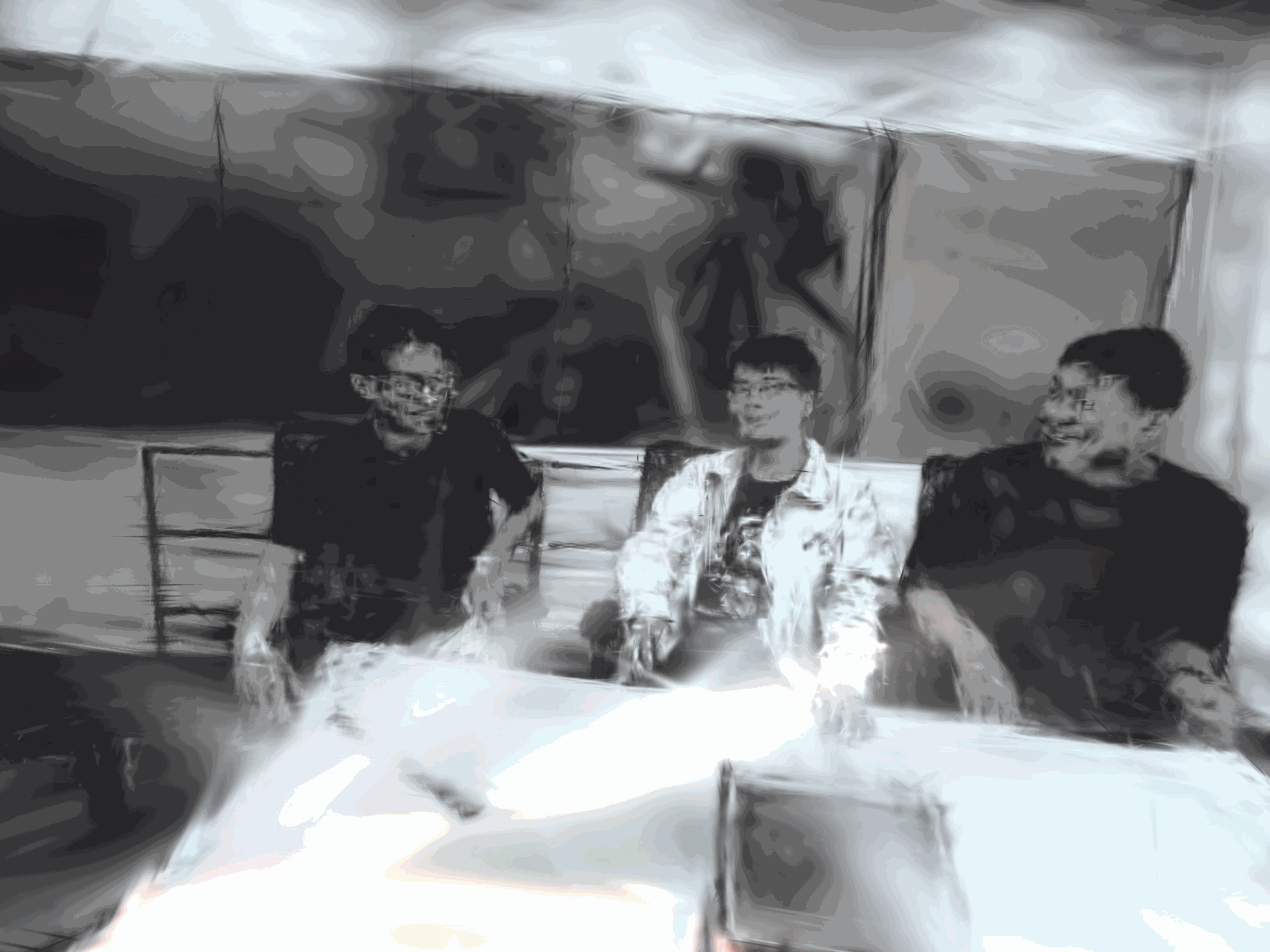}\\[-0.15em]
        \includegraphics[width=\linewidth]{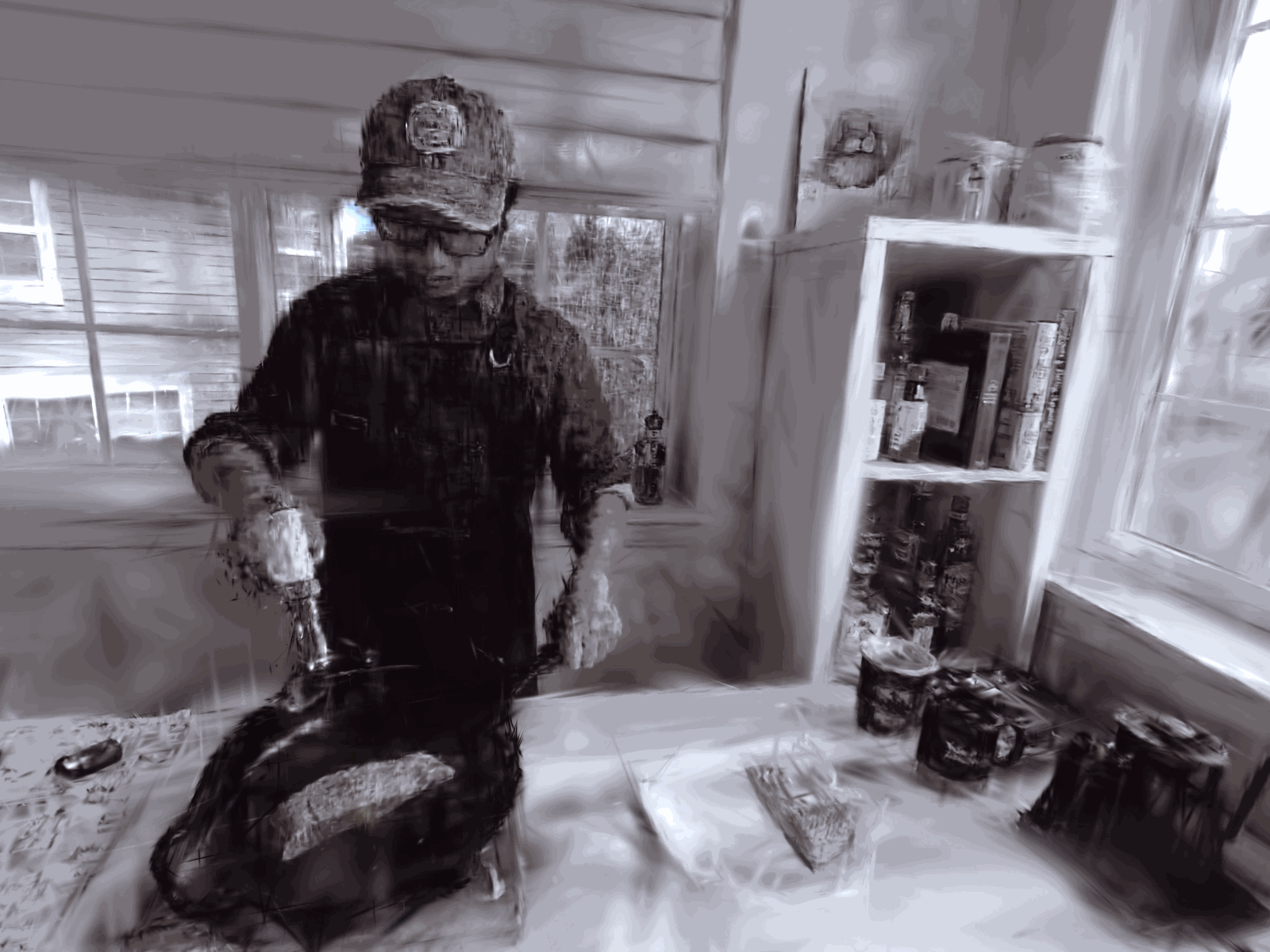}\\[-0.15em]
        \includegraphics[width=\linewidth]{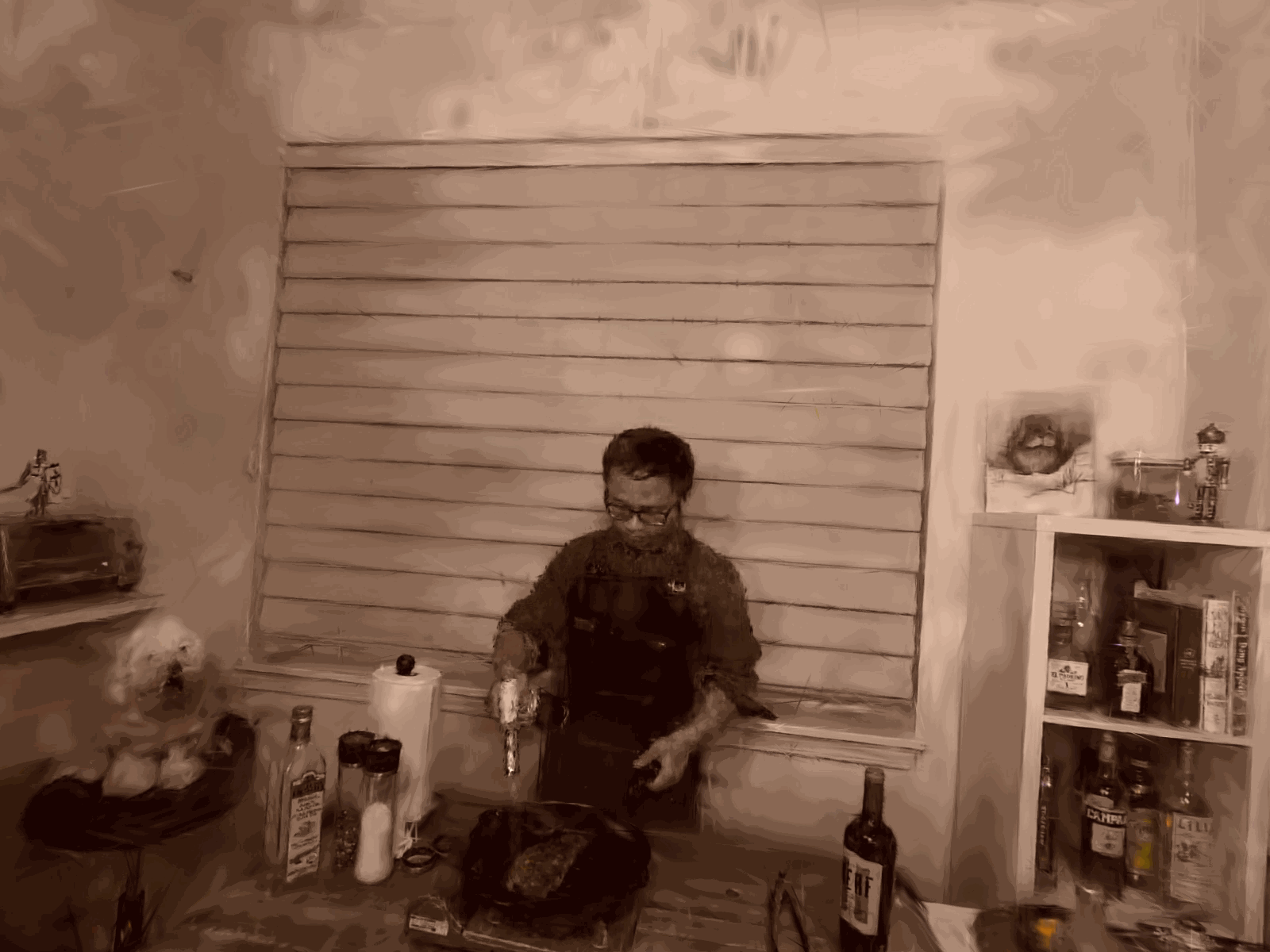}\\[-0.15em]
        \includegraphics[width=\linewidth]{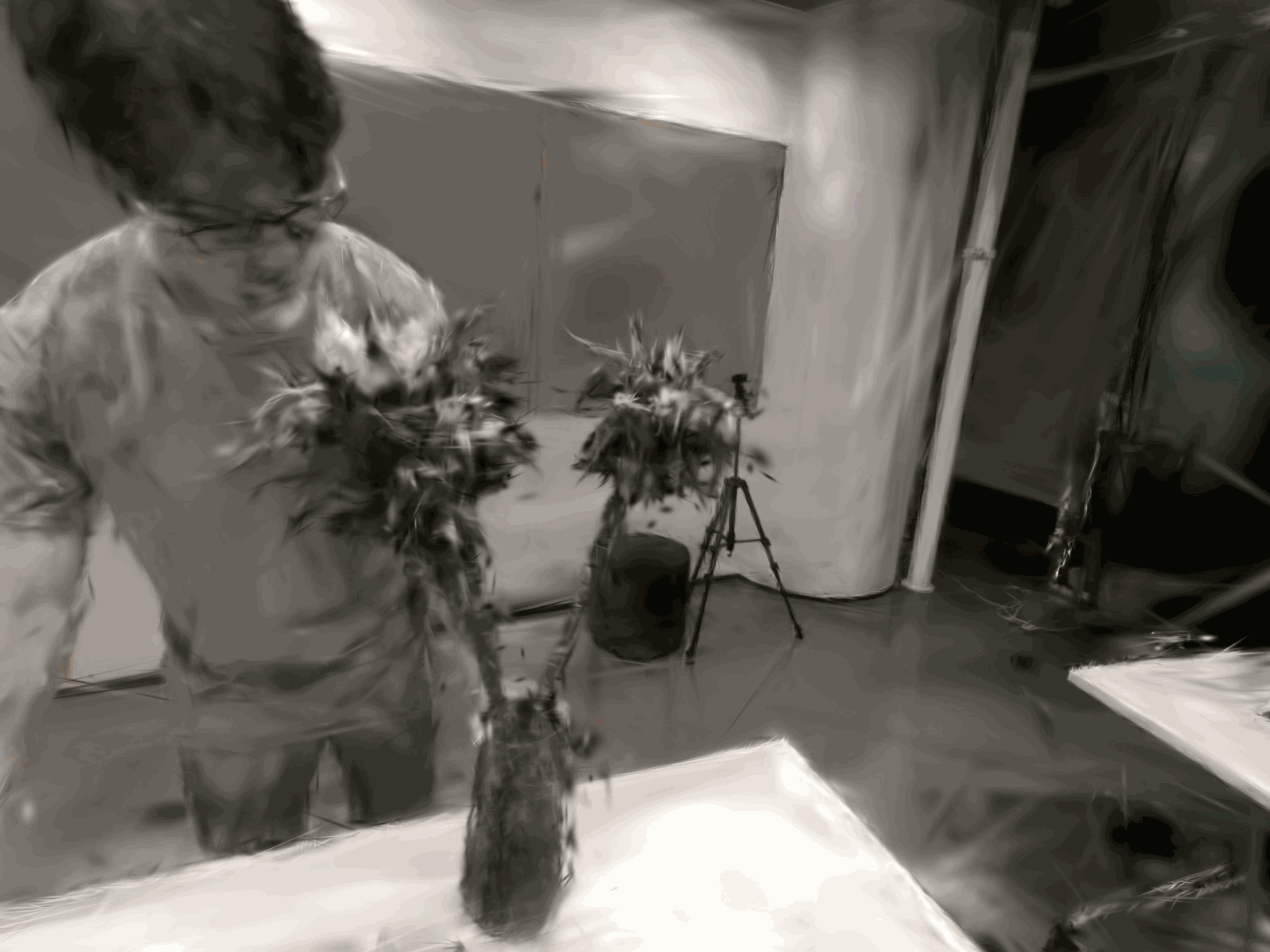}\\[-0.15em]
        \includegraphics[width=\linewidth]{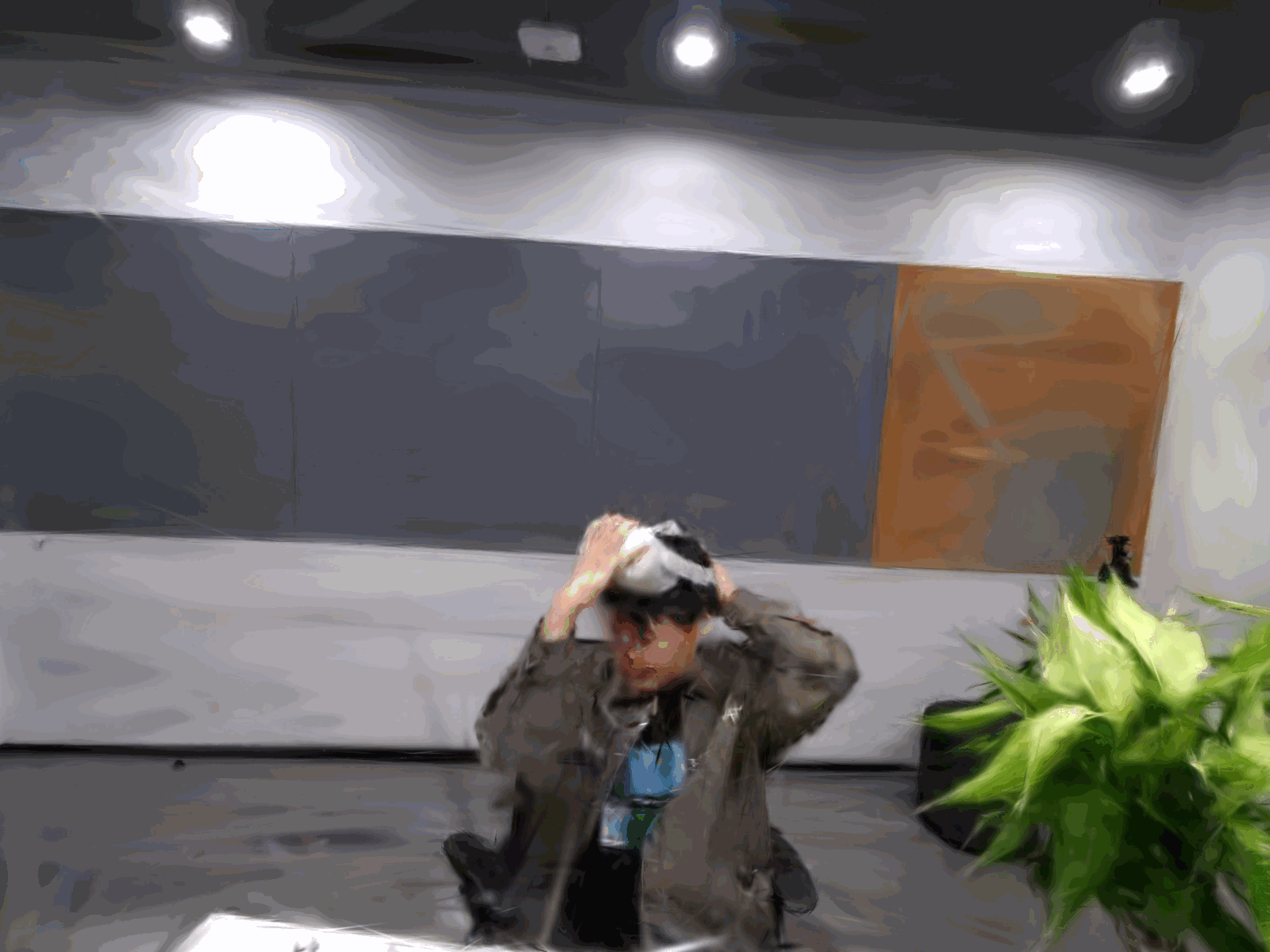}\\[-0.15em]
        \includegraphics[width=\linewidth]{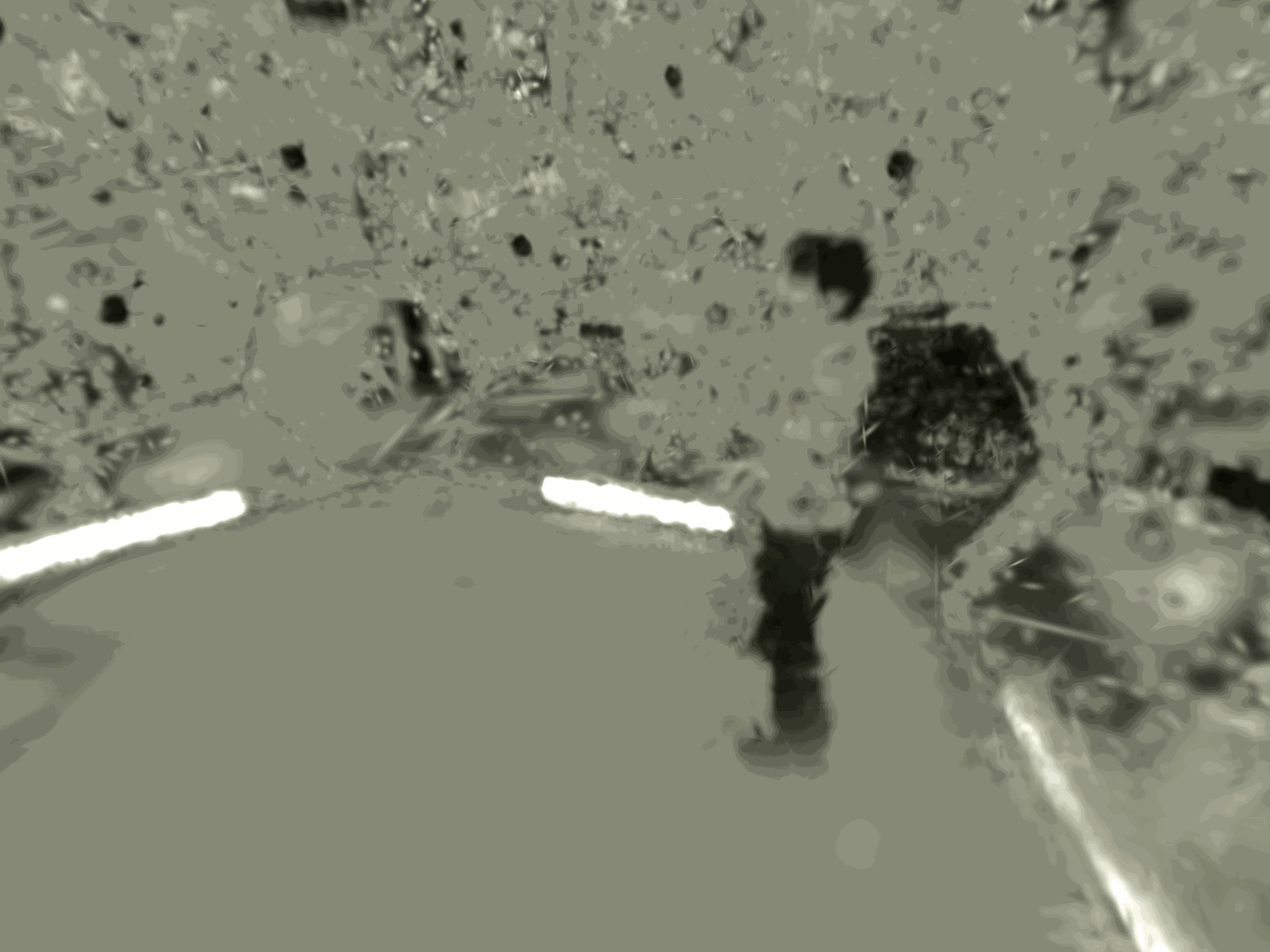}\\[-0.15em]
        \includegraphics[width=\linewidth]{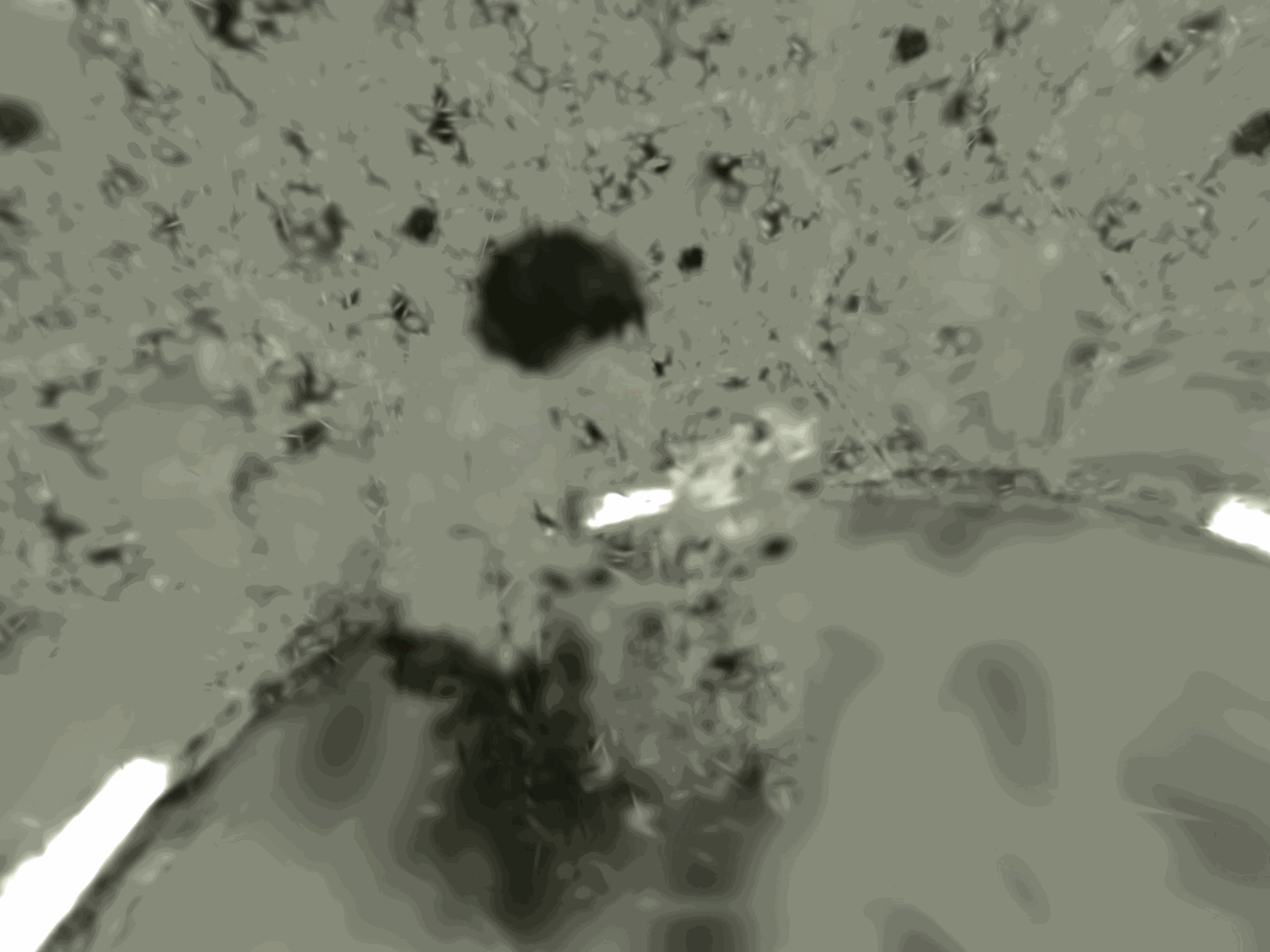}\\[-0.15em]
        \includegraphics[width=\linewidth]{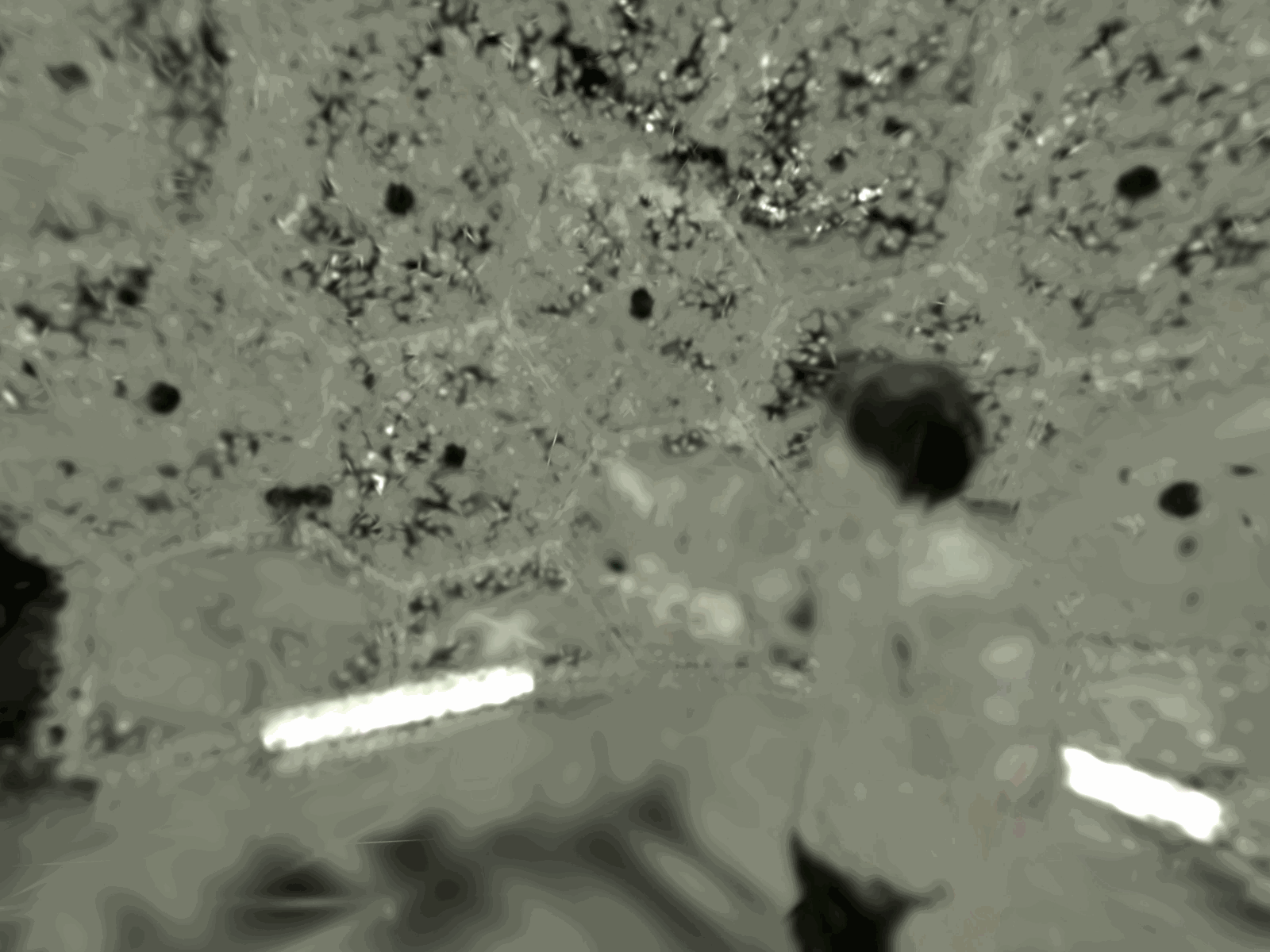}
        \caption{Color-distorted}
    \end{subfigure}\begin{subfigure}[t]{0.138\linewidth}
        \centering
        \includegraphics[width=\linewidth]{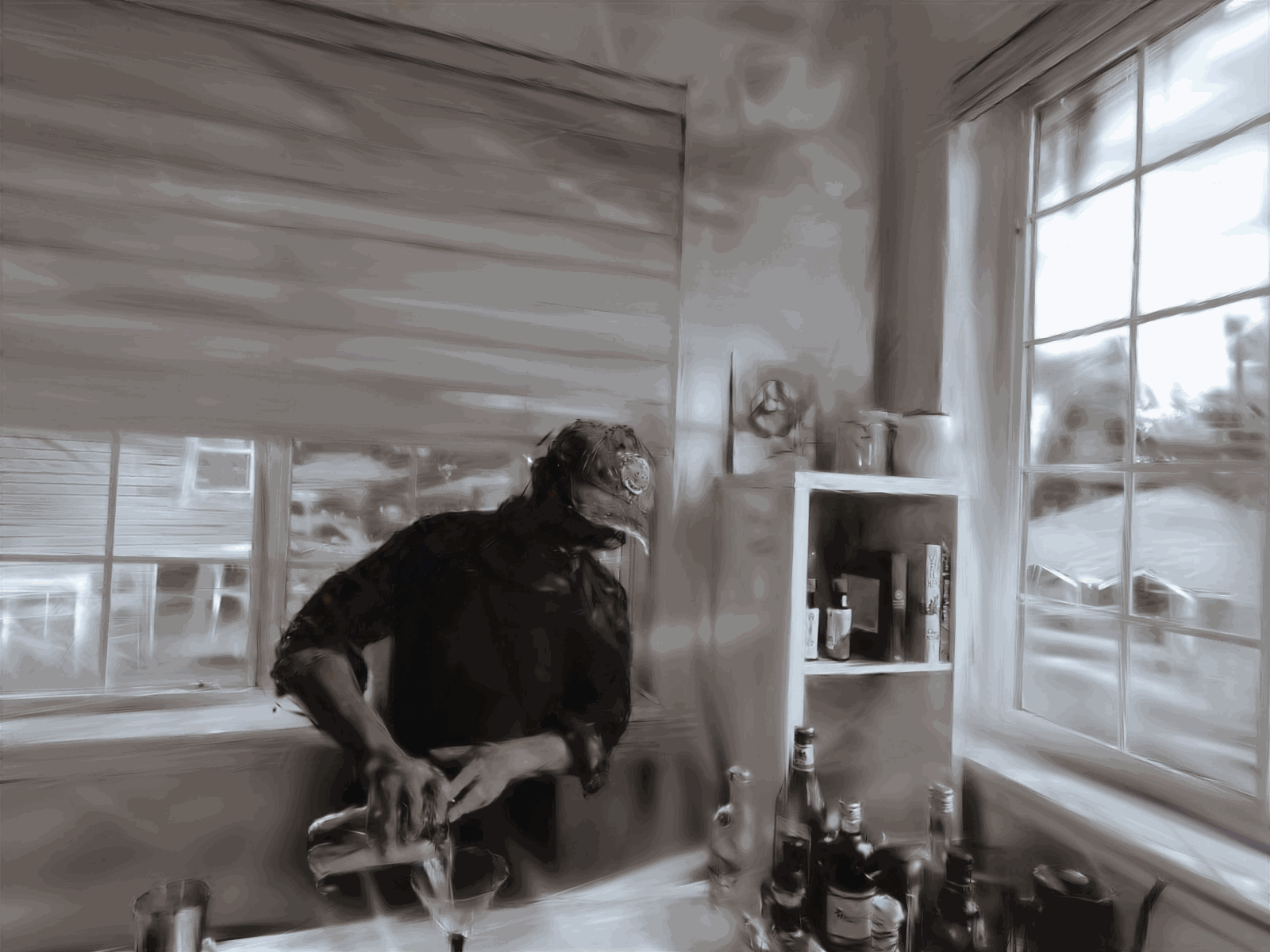}\\[-0.15em]
        \includegraphics[width=\linewidth]{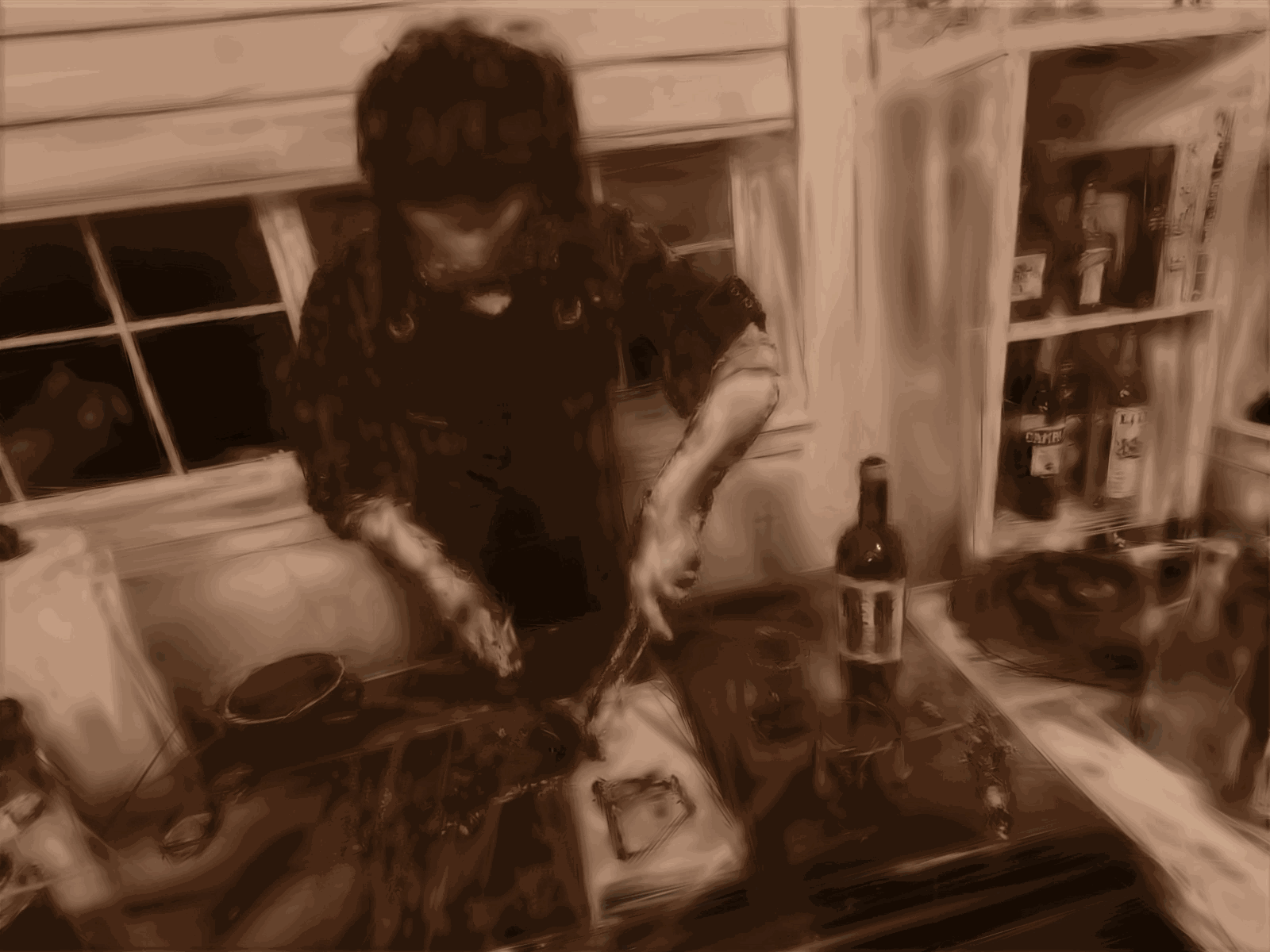}\\[-0.15em]
        \includegraphics[width=\linewidth]{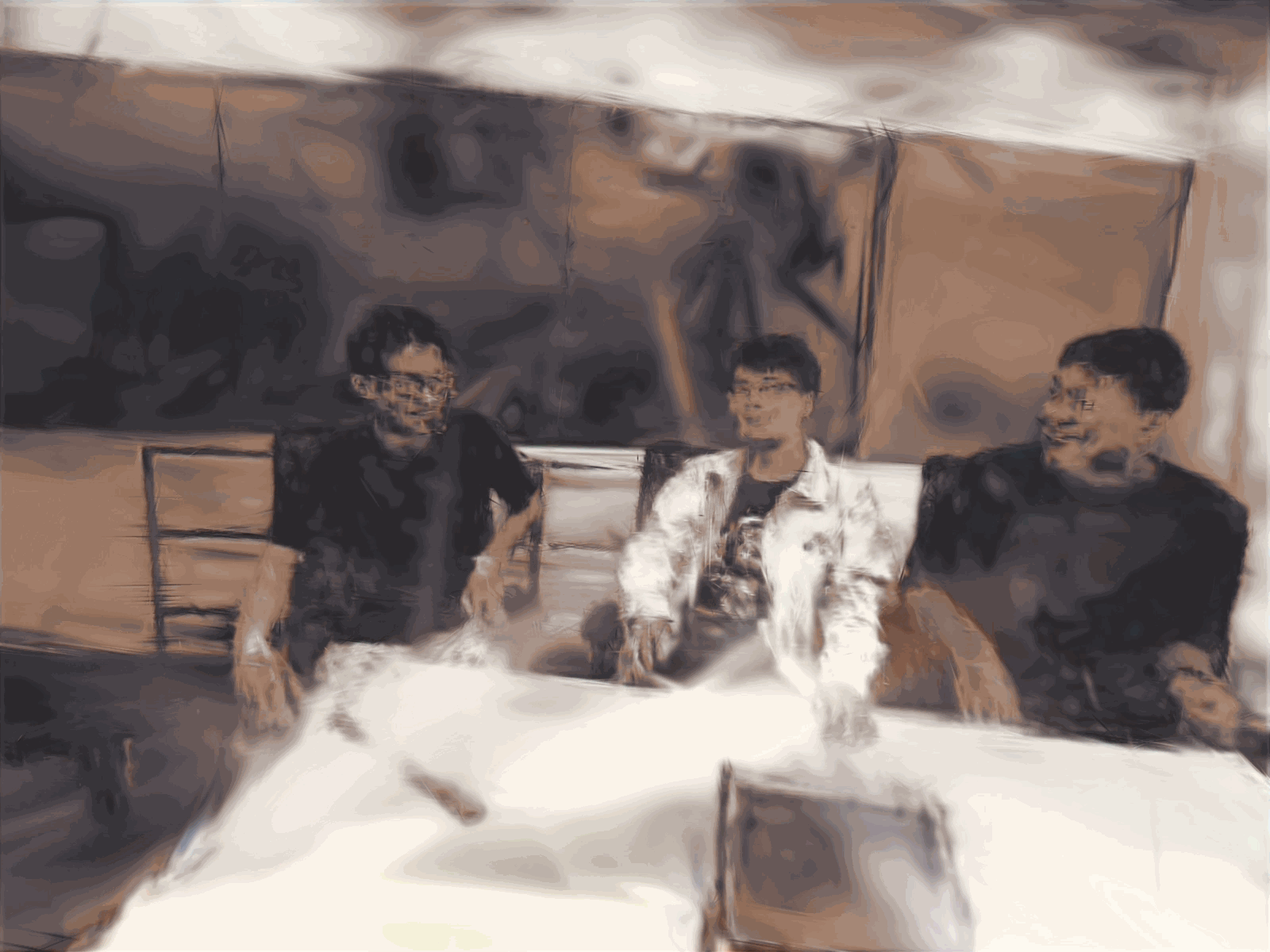}\\[-0.15em]
        \includegraphics[width=\linewidth]{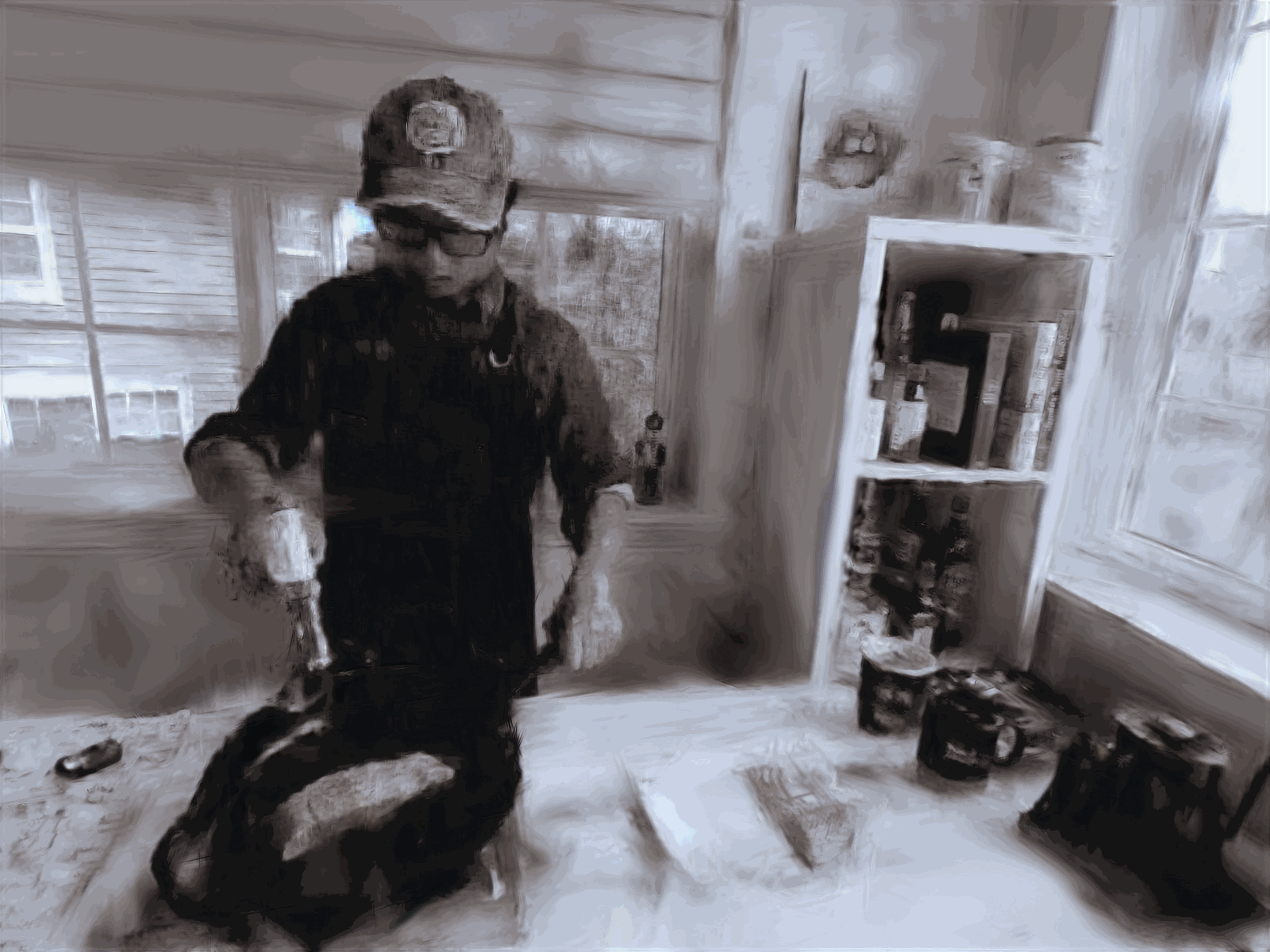}\\[-0.15em]
        \includegraphics[width=\linewidth]{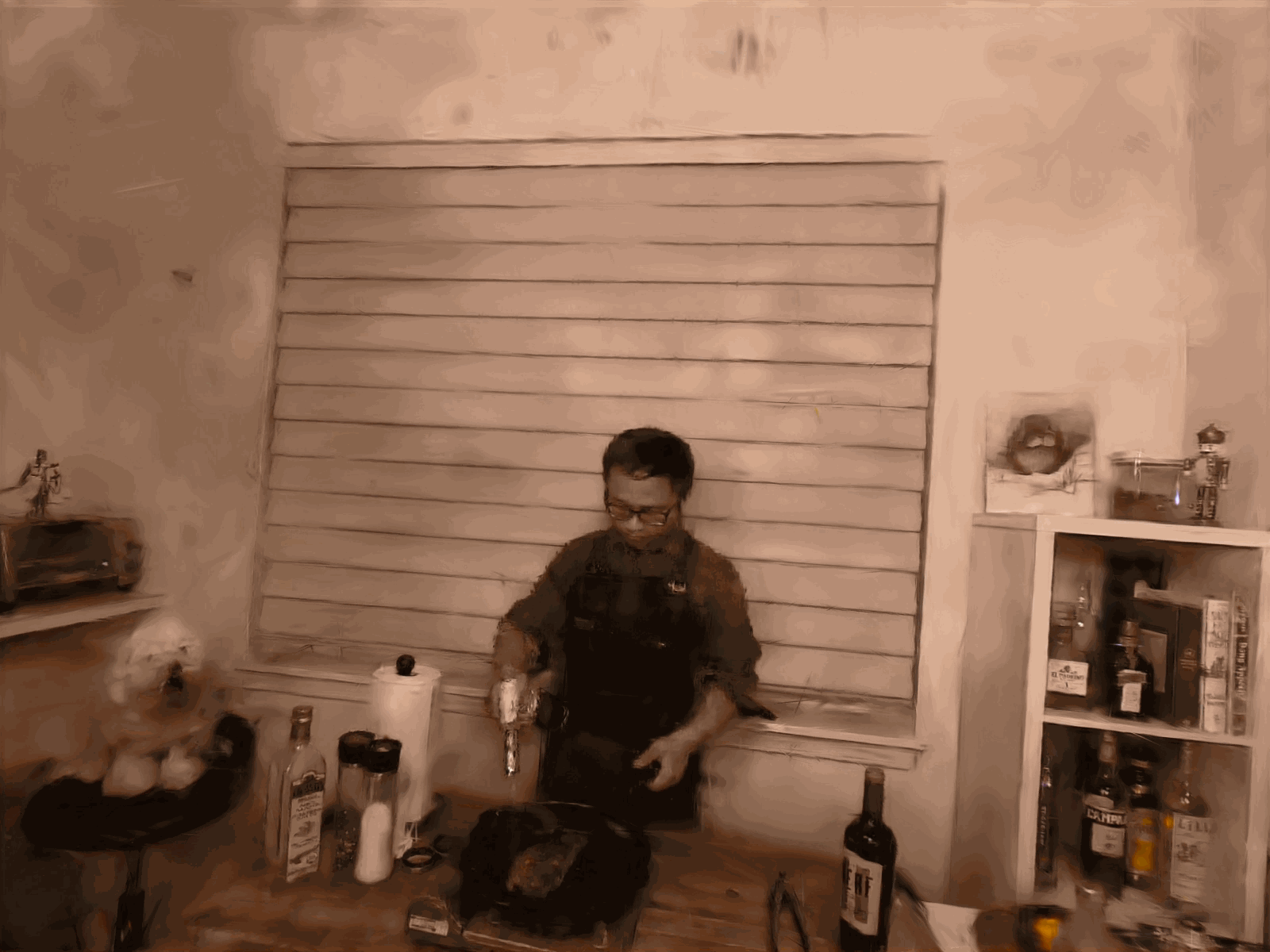}\\[-0.15em]
        \includegraphics[width=\linewidth]{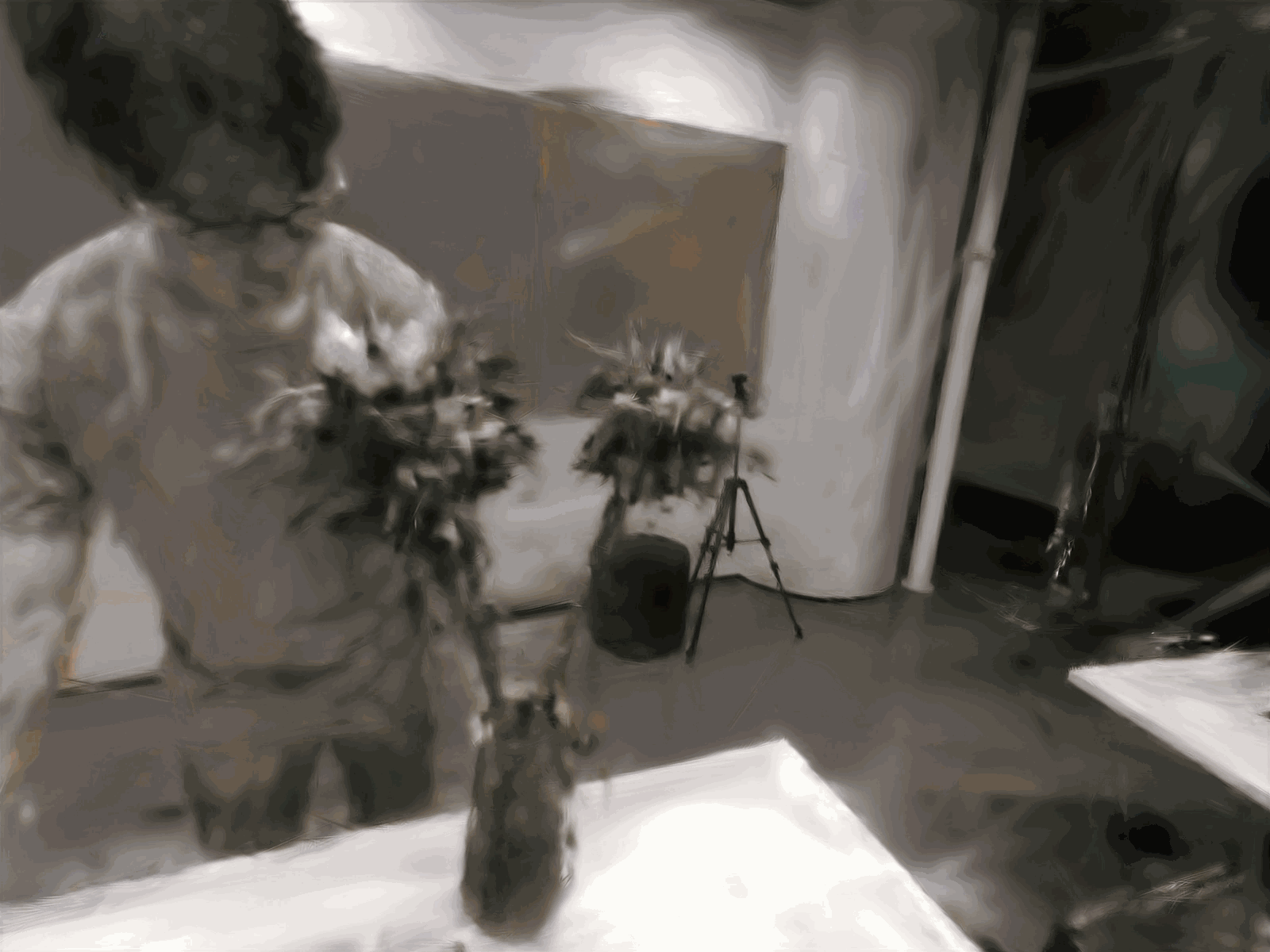}\\[-0.15em]
        \includegraphics[width=\linewidth]{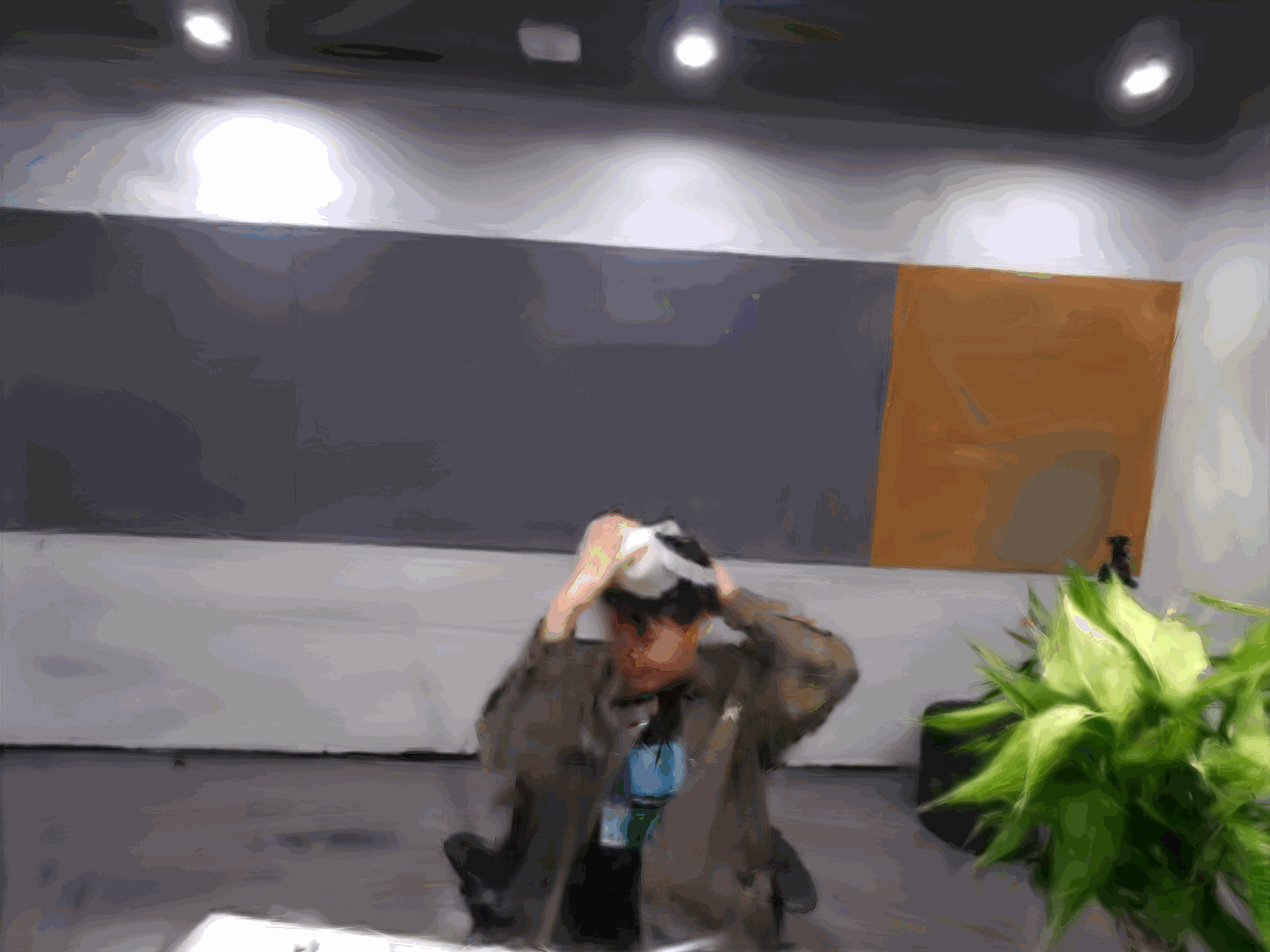}\\[-0.15em]
        \includegraphics[width=\linewidth]{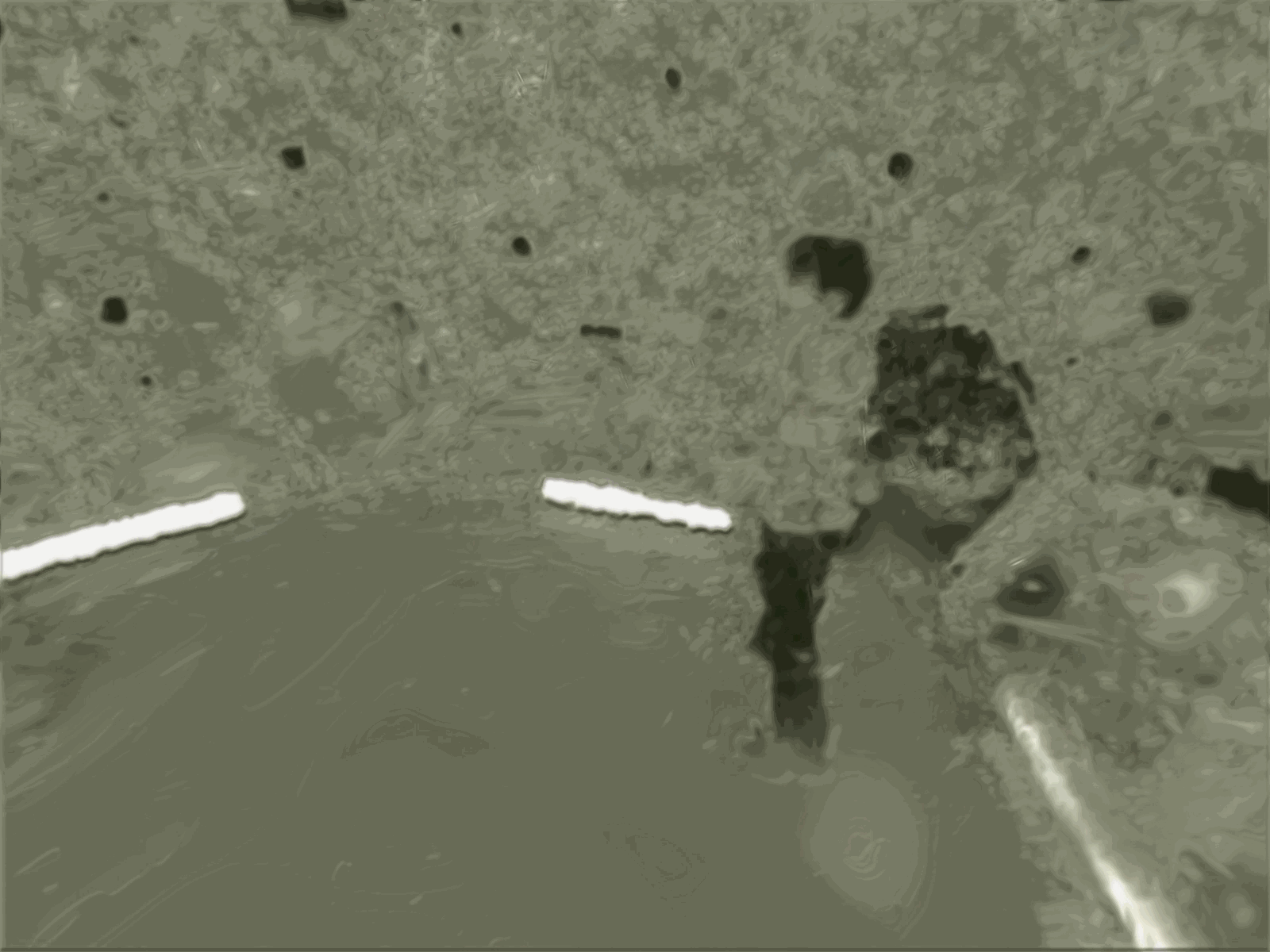}\\[-0.15em]
        \includegraphics[width=\linewidth]{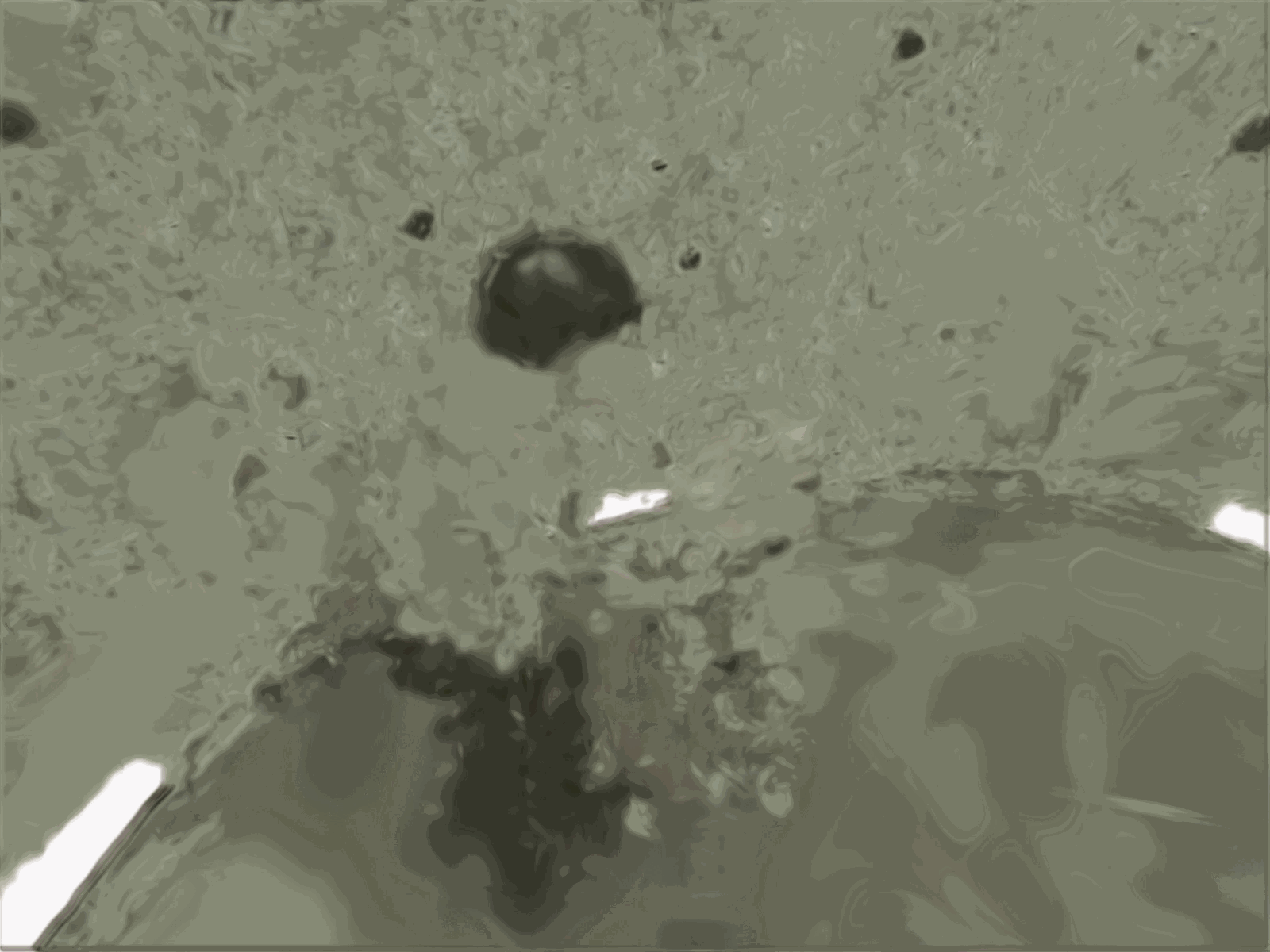}\\[-0.15em]
        \includegraphics[width=\linewidth]{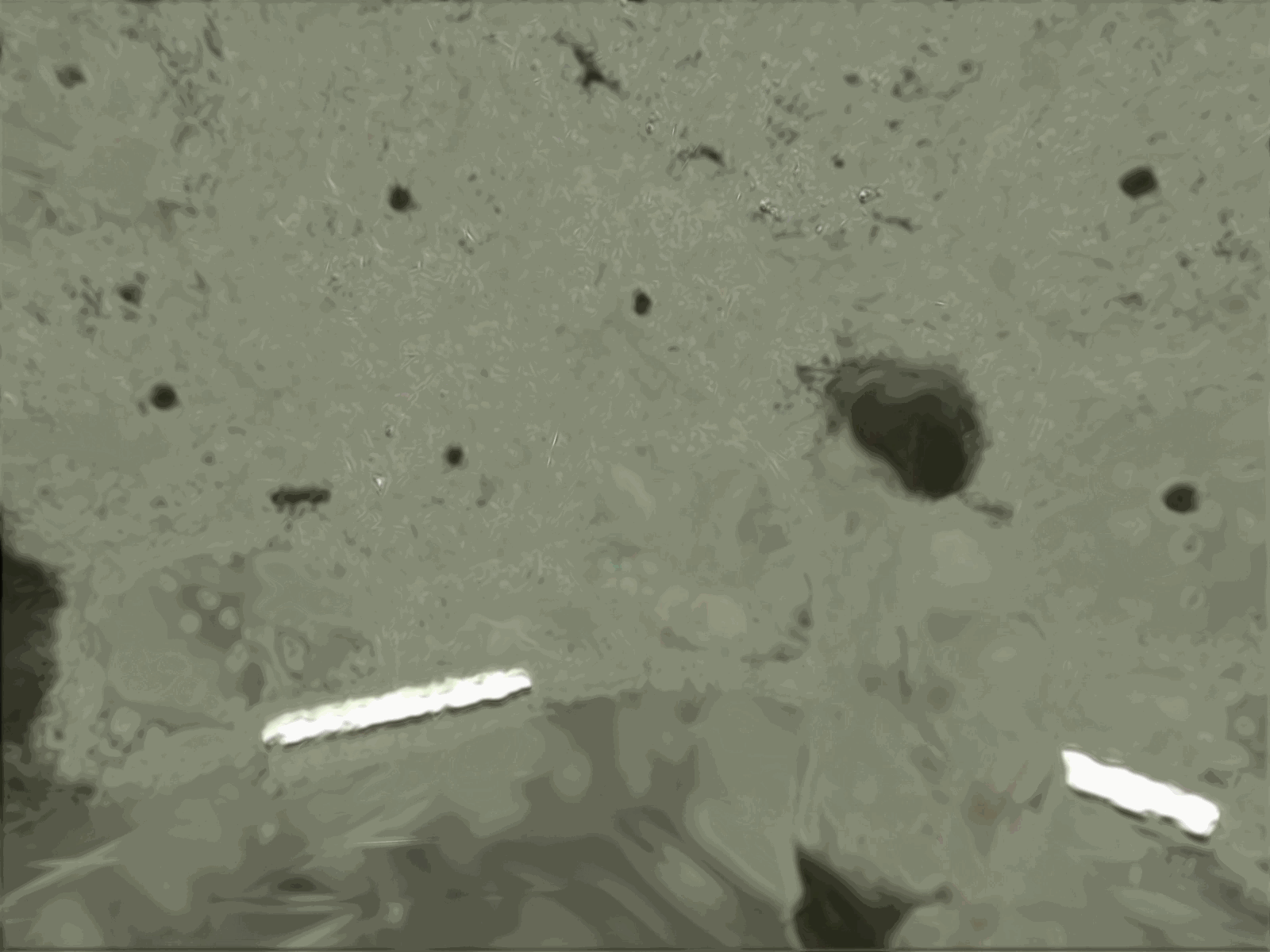}
        \caption{CR w/o ref.}
    \end{subfigure}\begin{subfigure}[t]{0.138\linewidth}
        \centering
        \includegraphics[width=\linewidth]{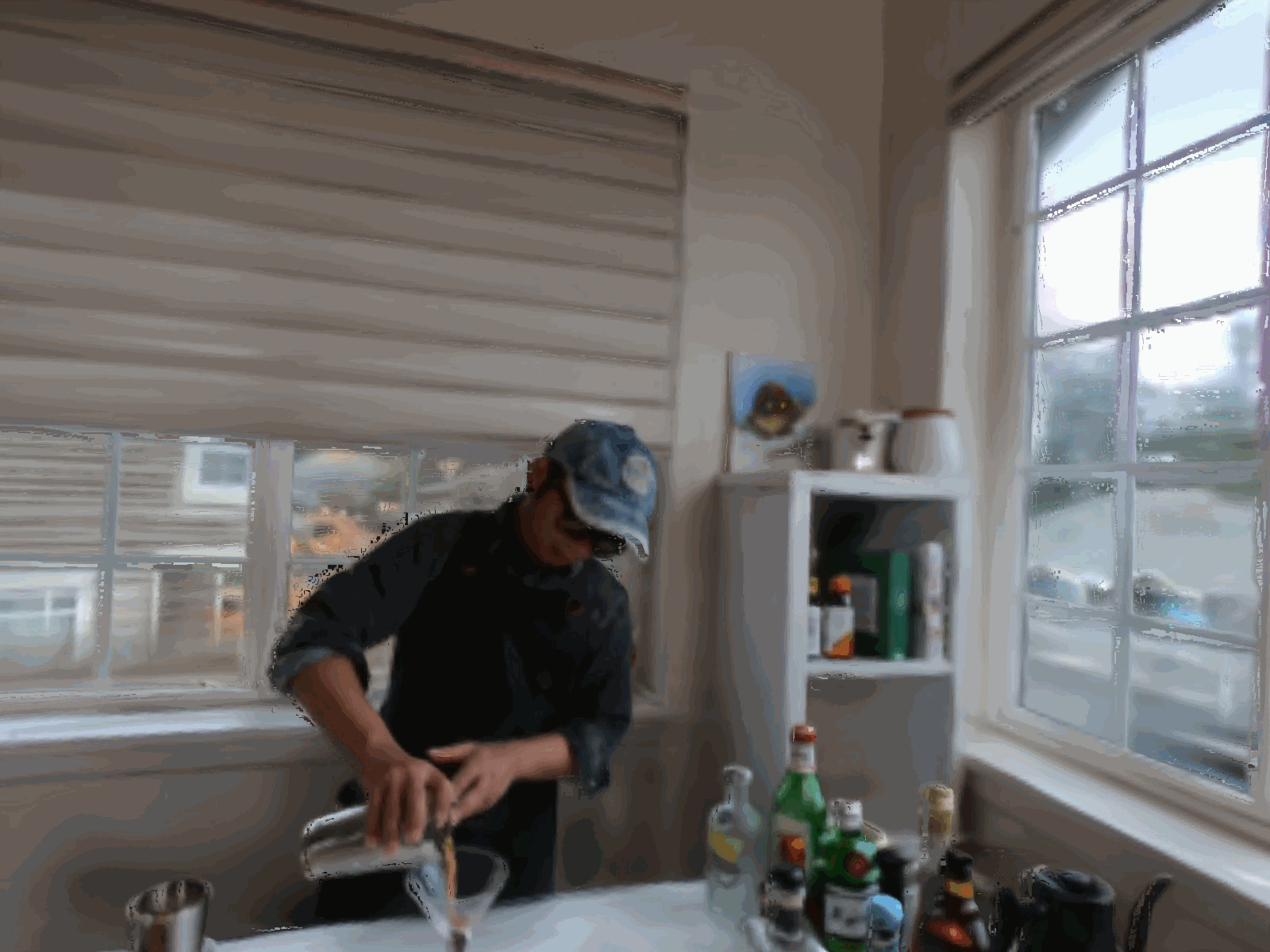}\\[-0.15em]
        \includegraphics[width=\linewidth]{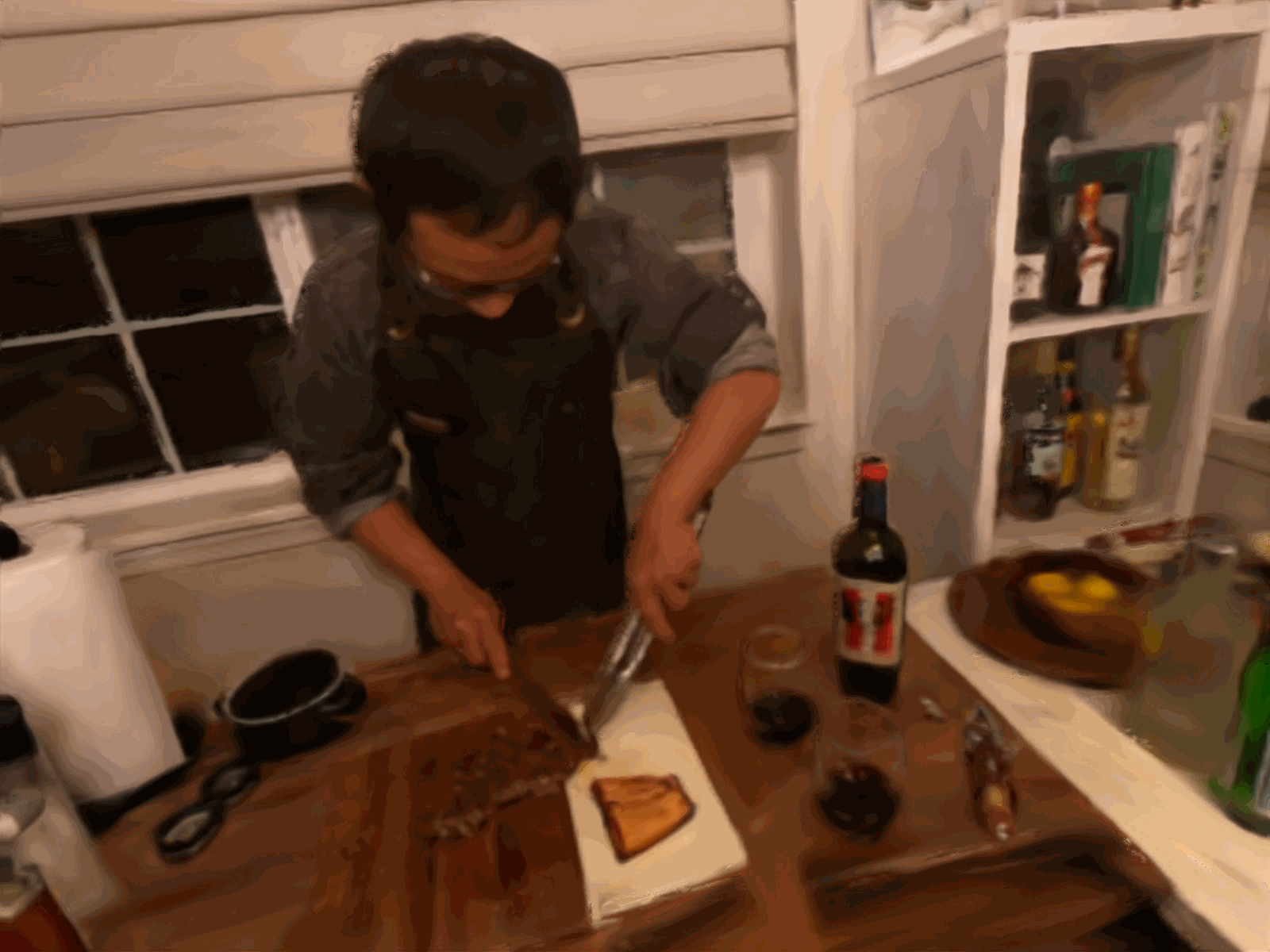}\\[-0.15em]
        \includegraphics[width=\linewidth]{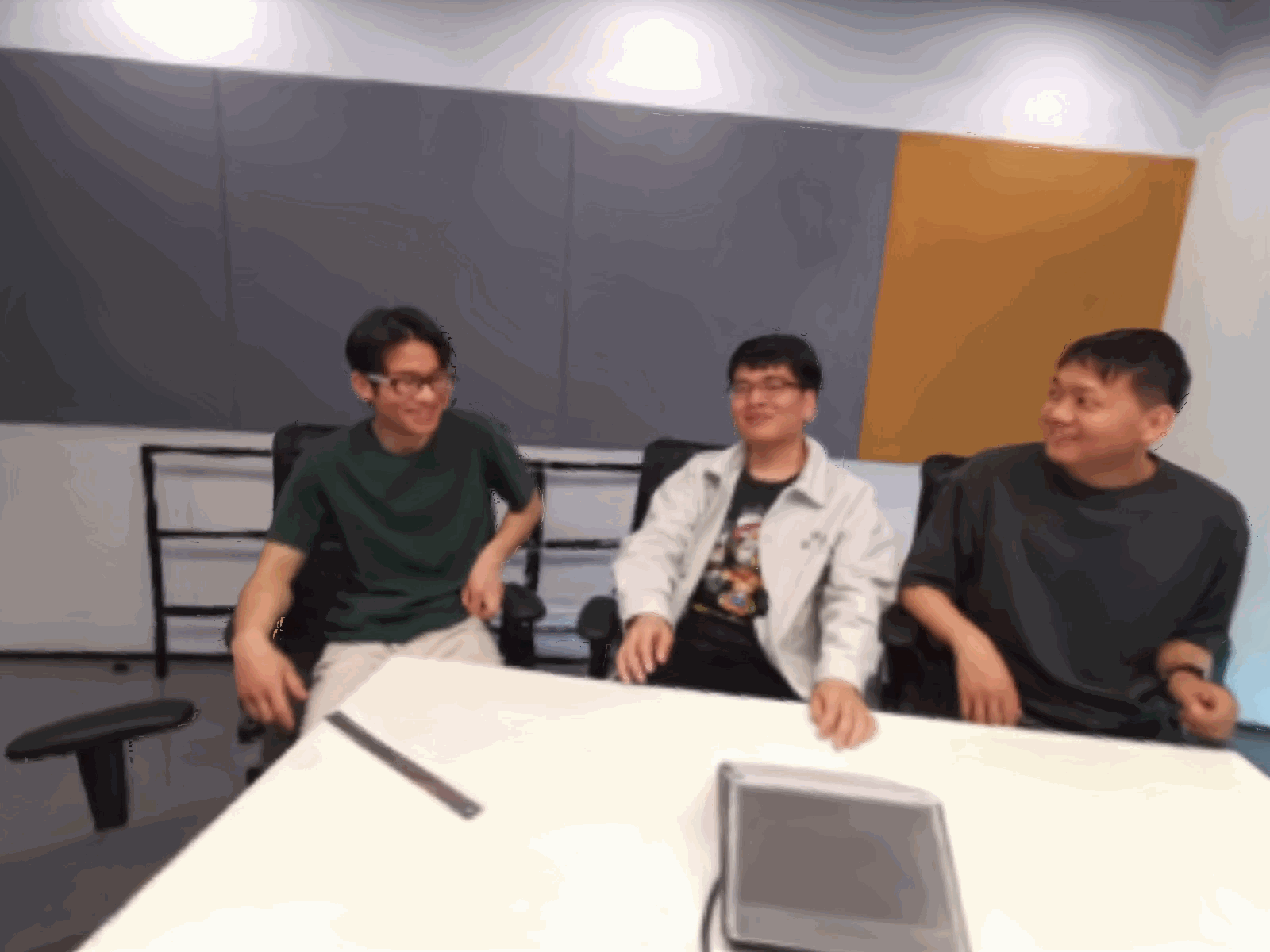}\\[-0.15em]
        \includegraphics[width=\linewidth]{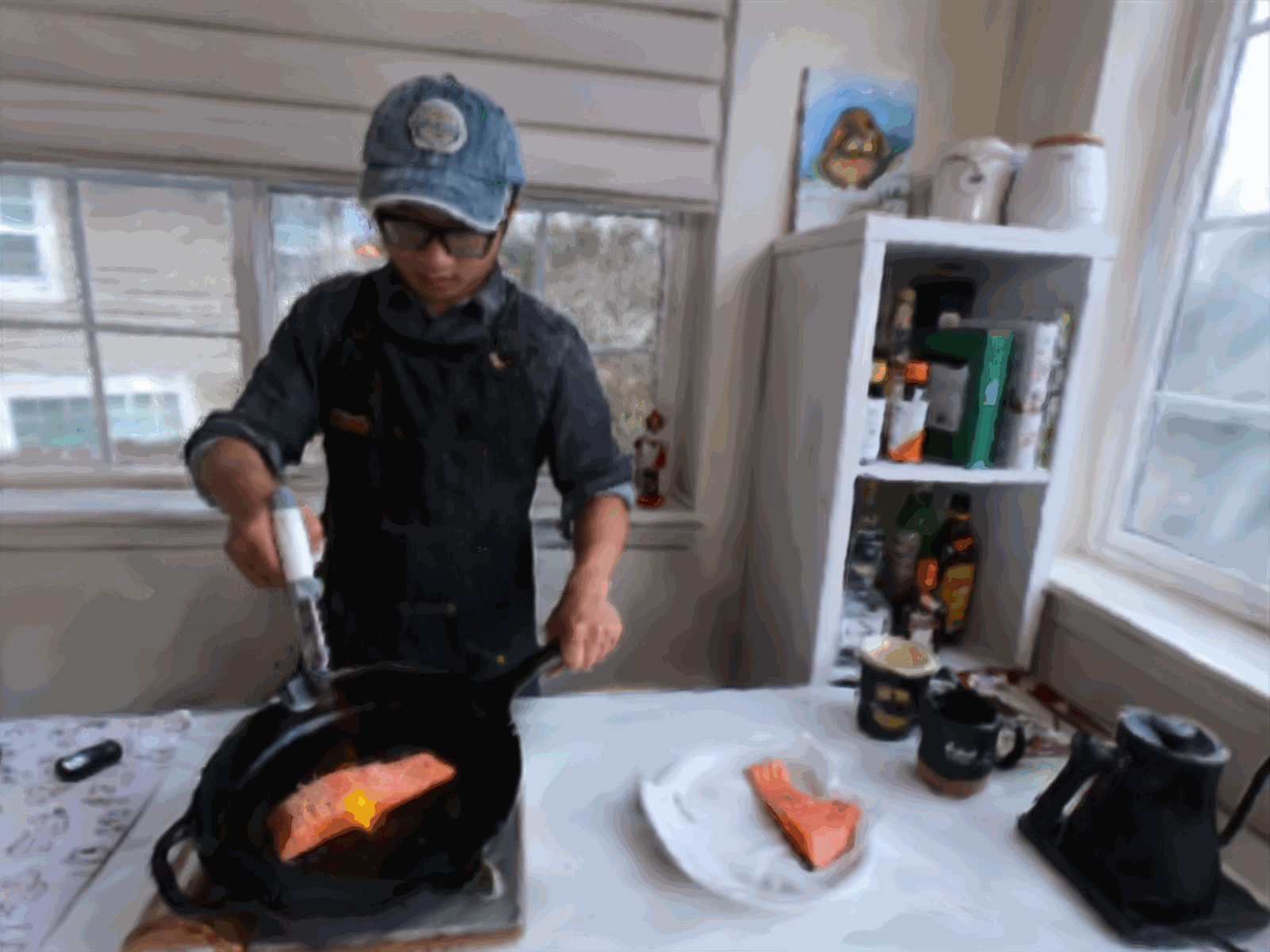}\\[-0.15em]
        \includegraphics[width=\linewidth]{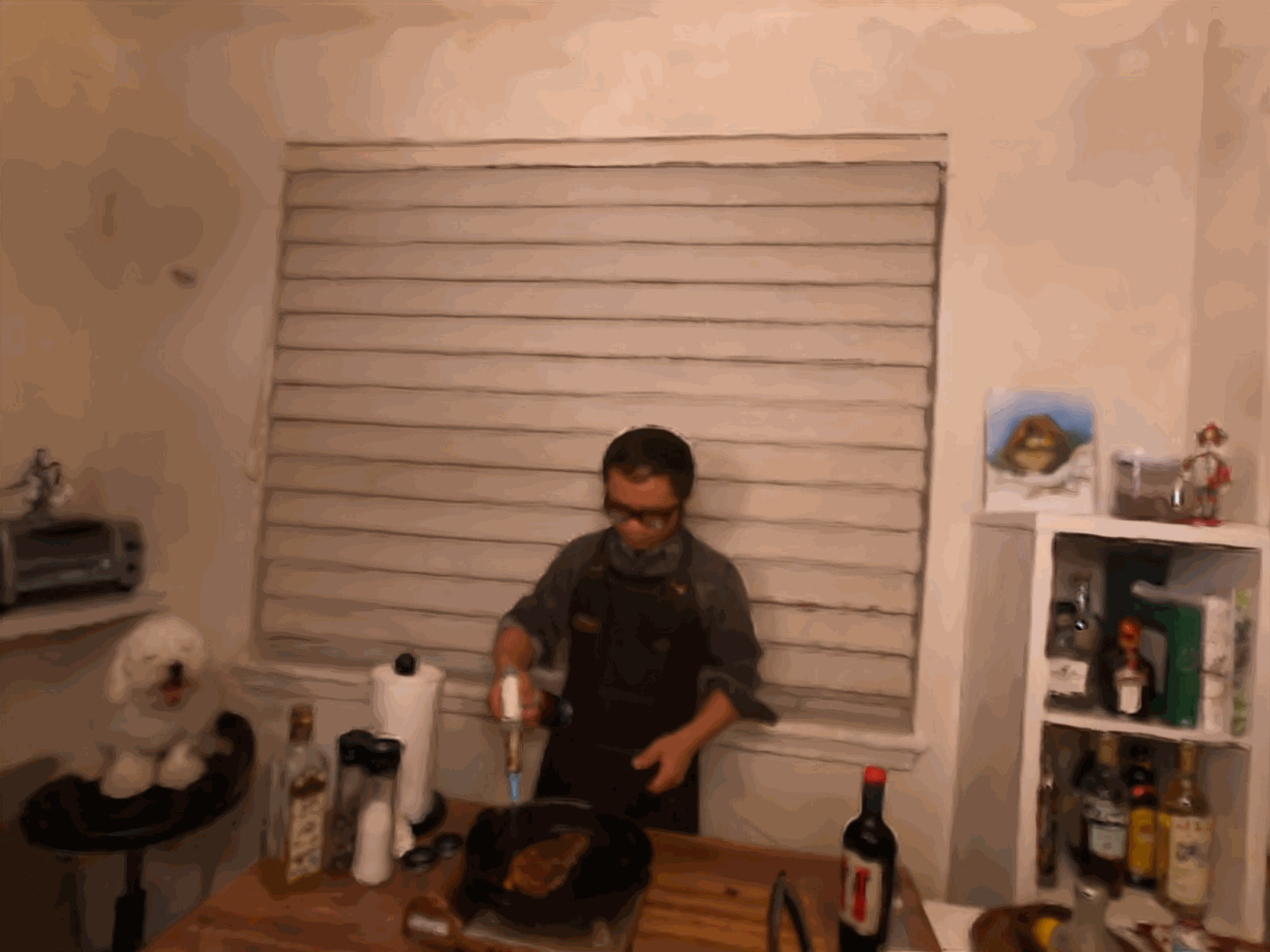}\\[-0.15em]
        \includegraphics[width=\linewidth]{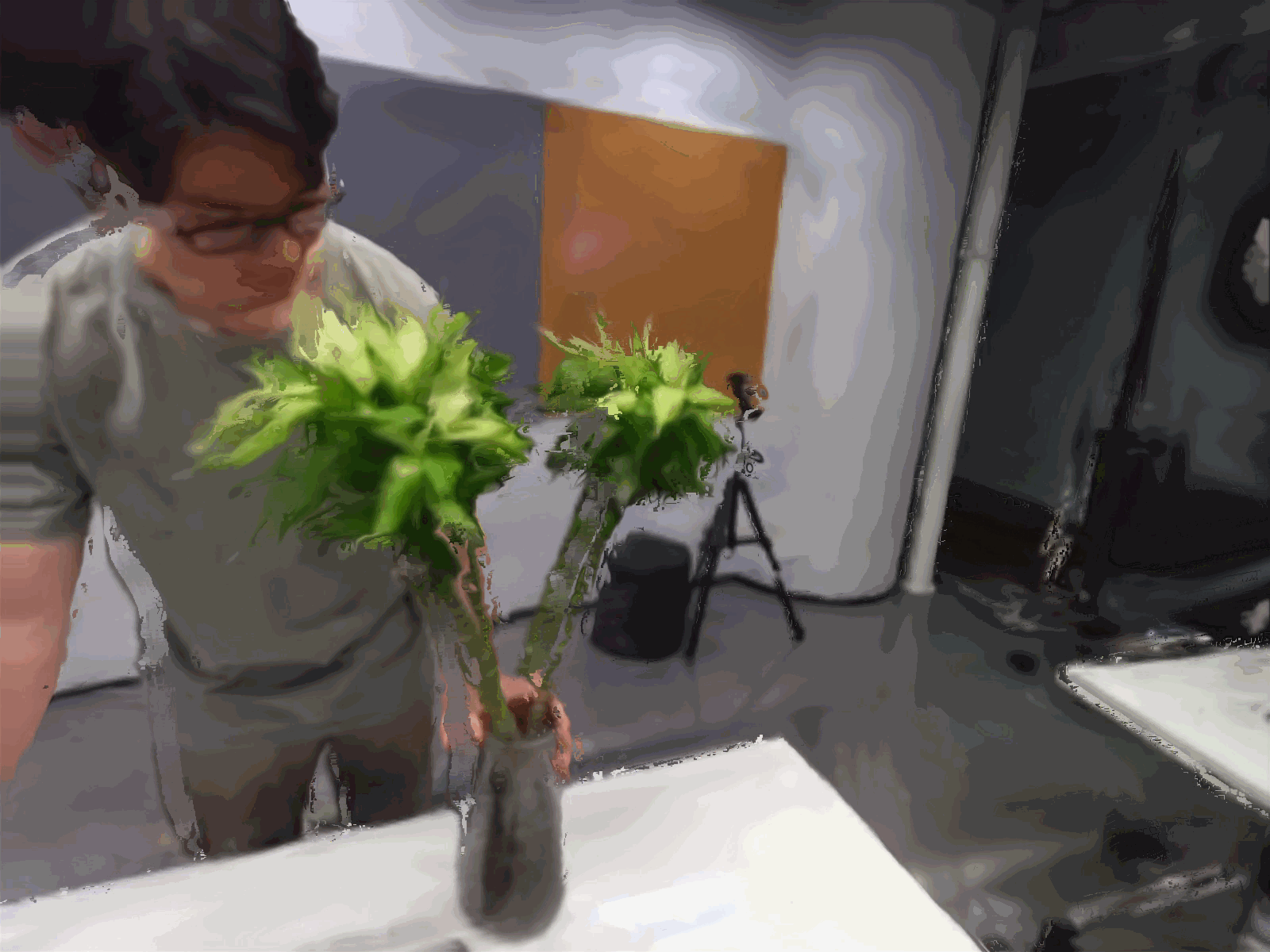}\\[-0.15em]
        \includegraphics[width=\linewidth]{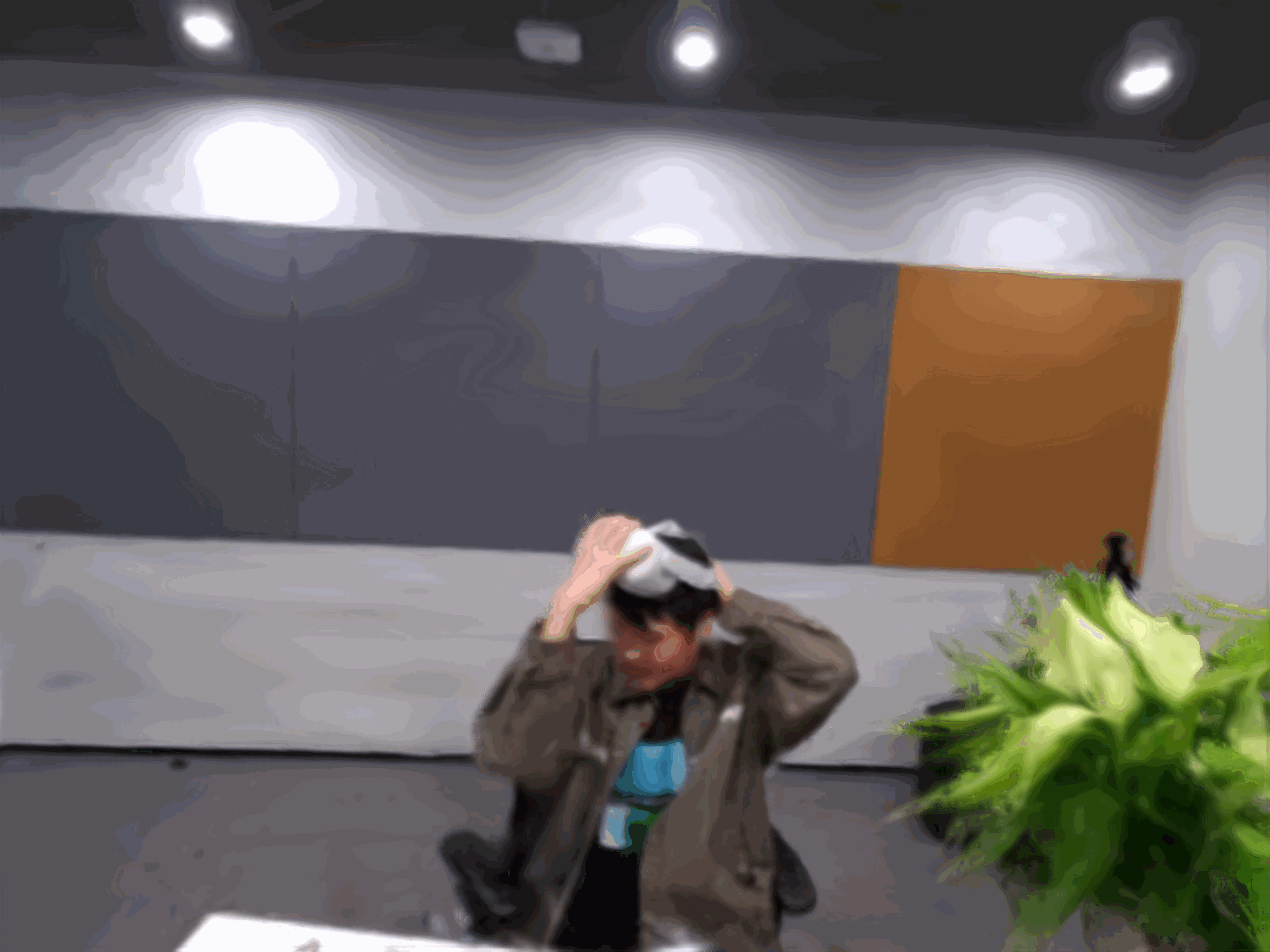}\\[-0.15em]
        \includegraphics[width=\linewidth]{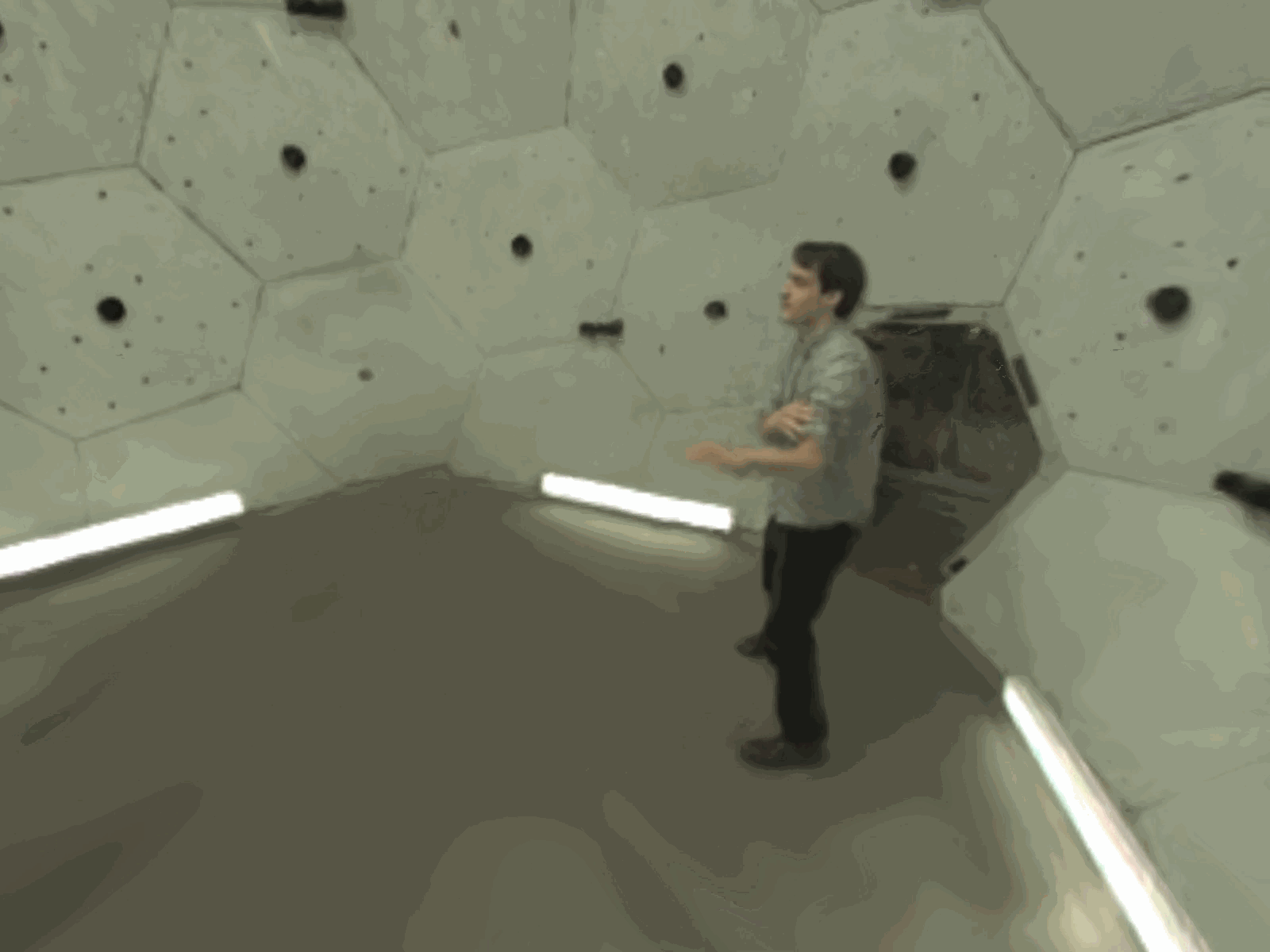}\\[-0.15em]
        \includegraphics[width=\linewidth]{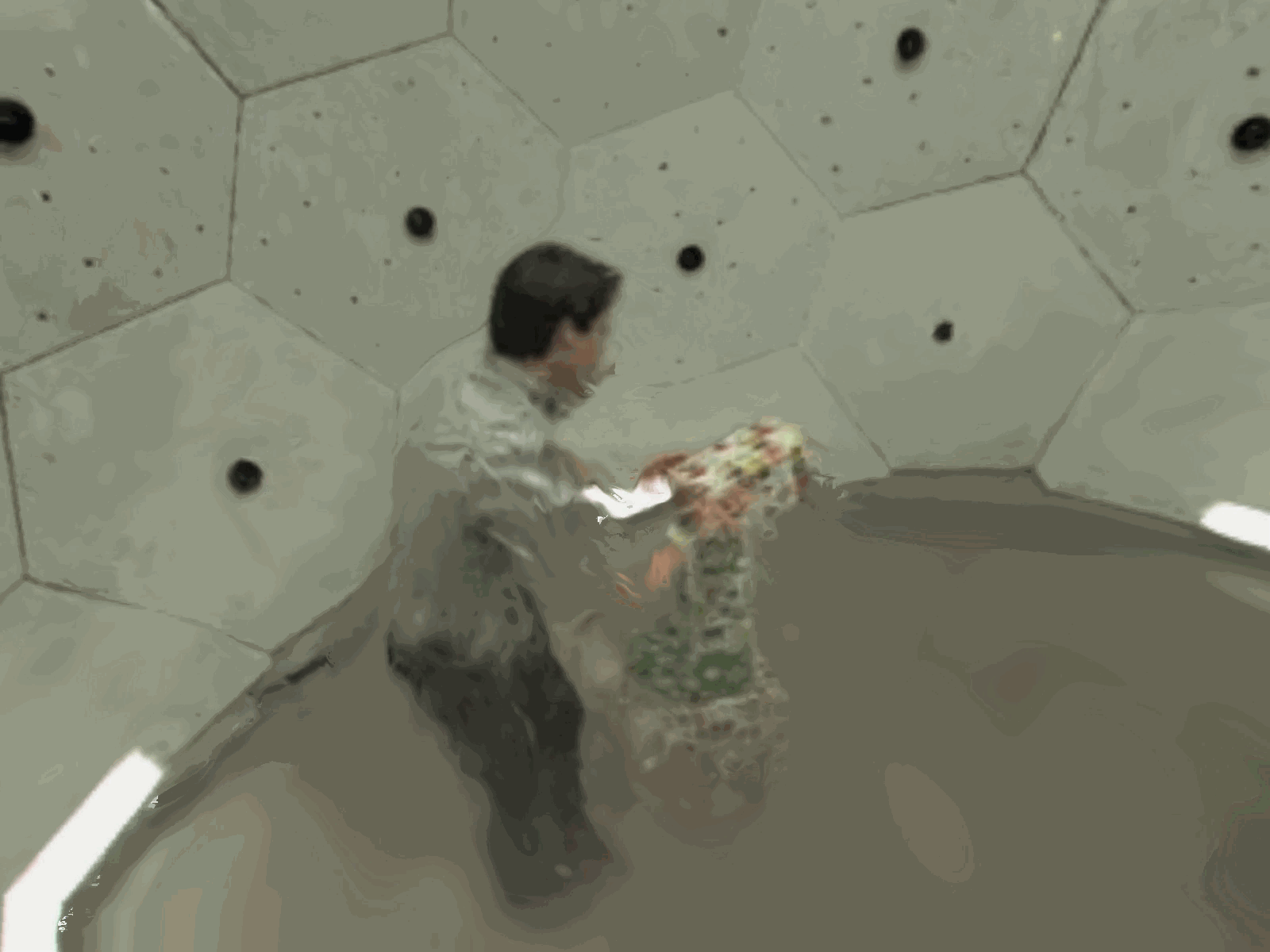}\\[-0.15em]
        \includegraphics[width=\linewidth]{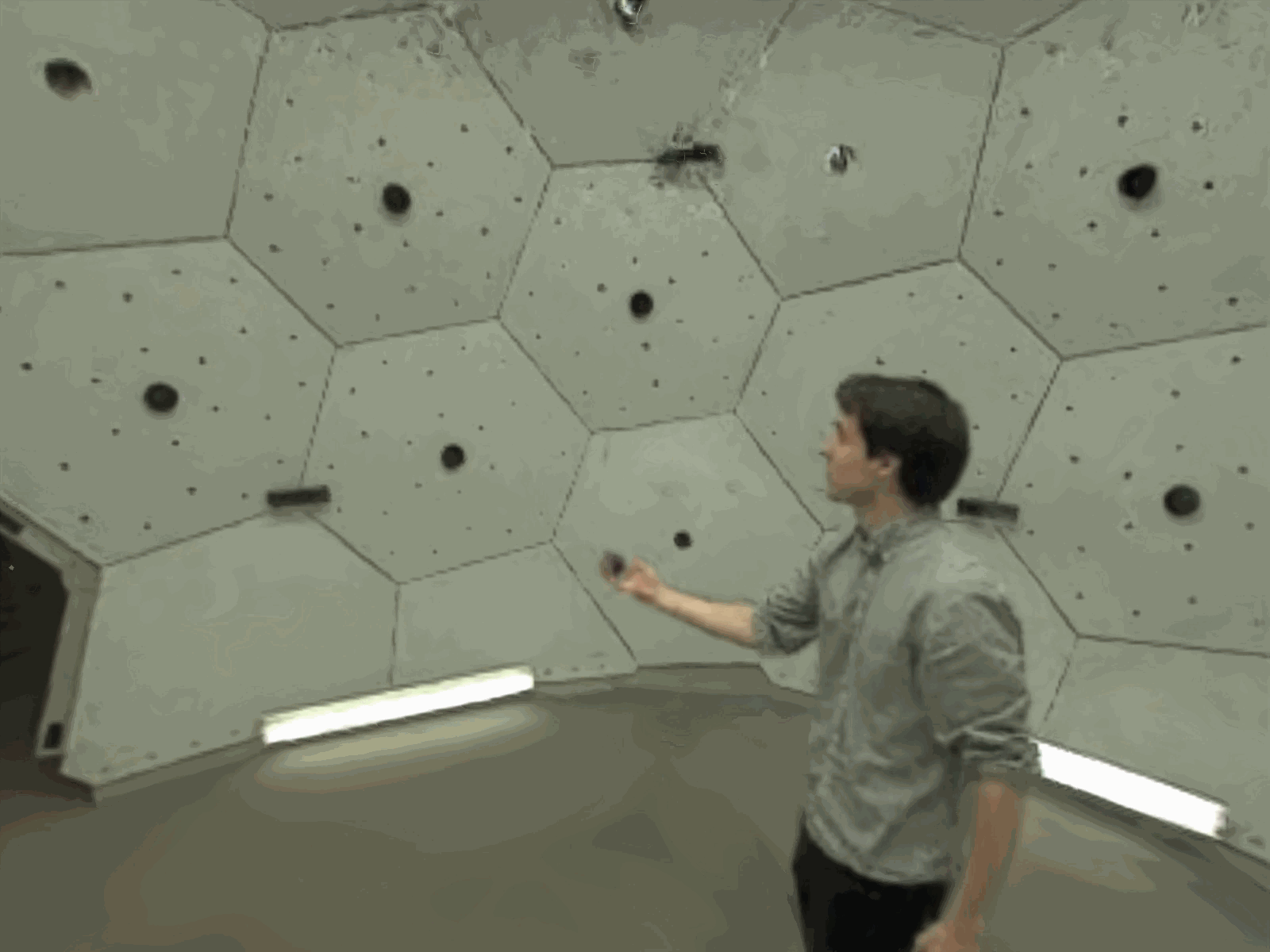}
        \caption{SR align ref.}
    \end{subfigure}\begin{subfigure}[t]{0.138\linewidth}
        \centering
        \includegraphics[width=\linewidth]{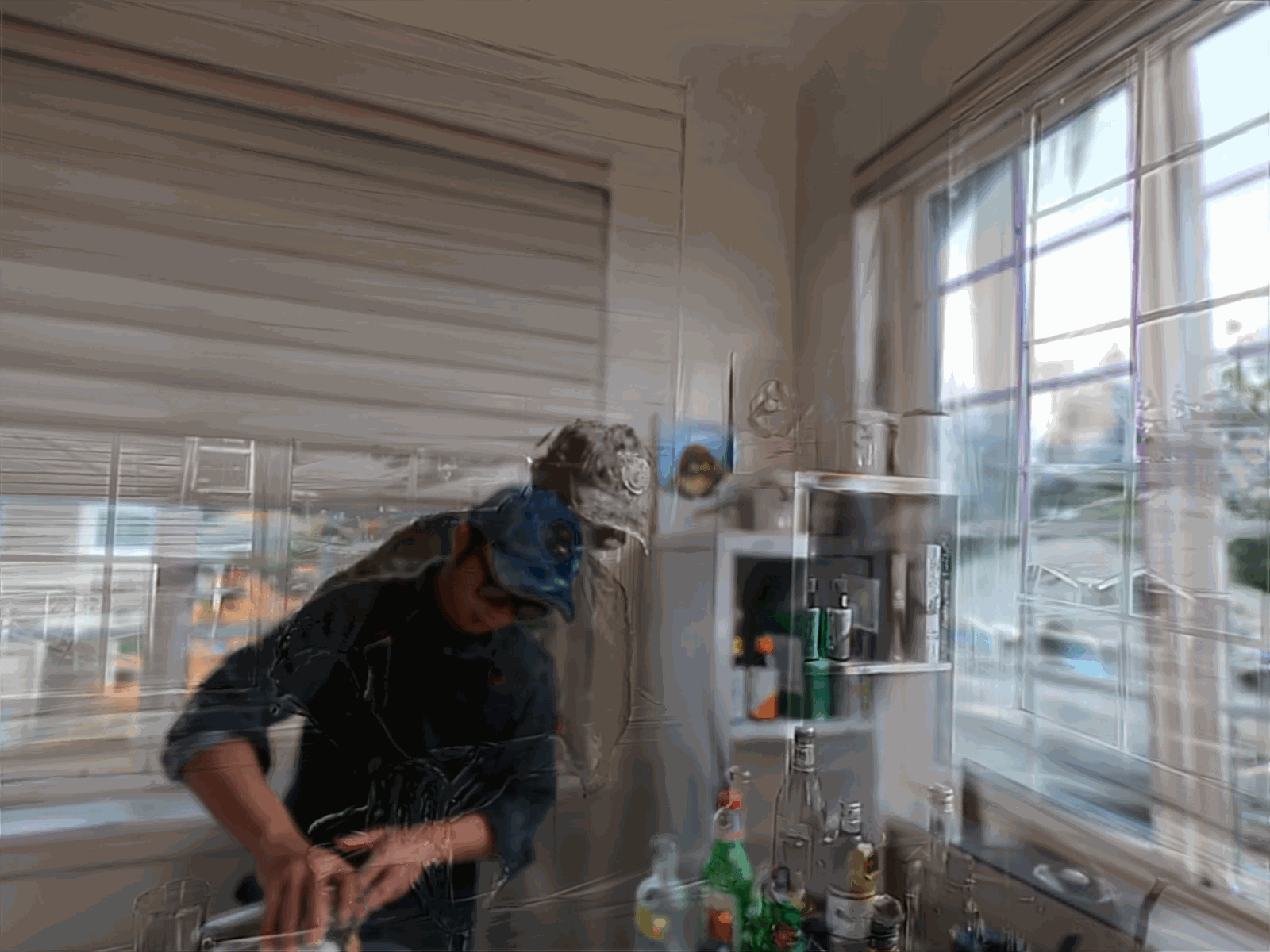}\\[-0.15em]
        \includegraphics[width=\linewidth]{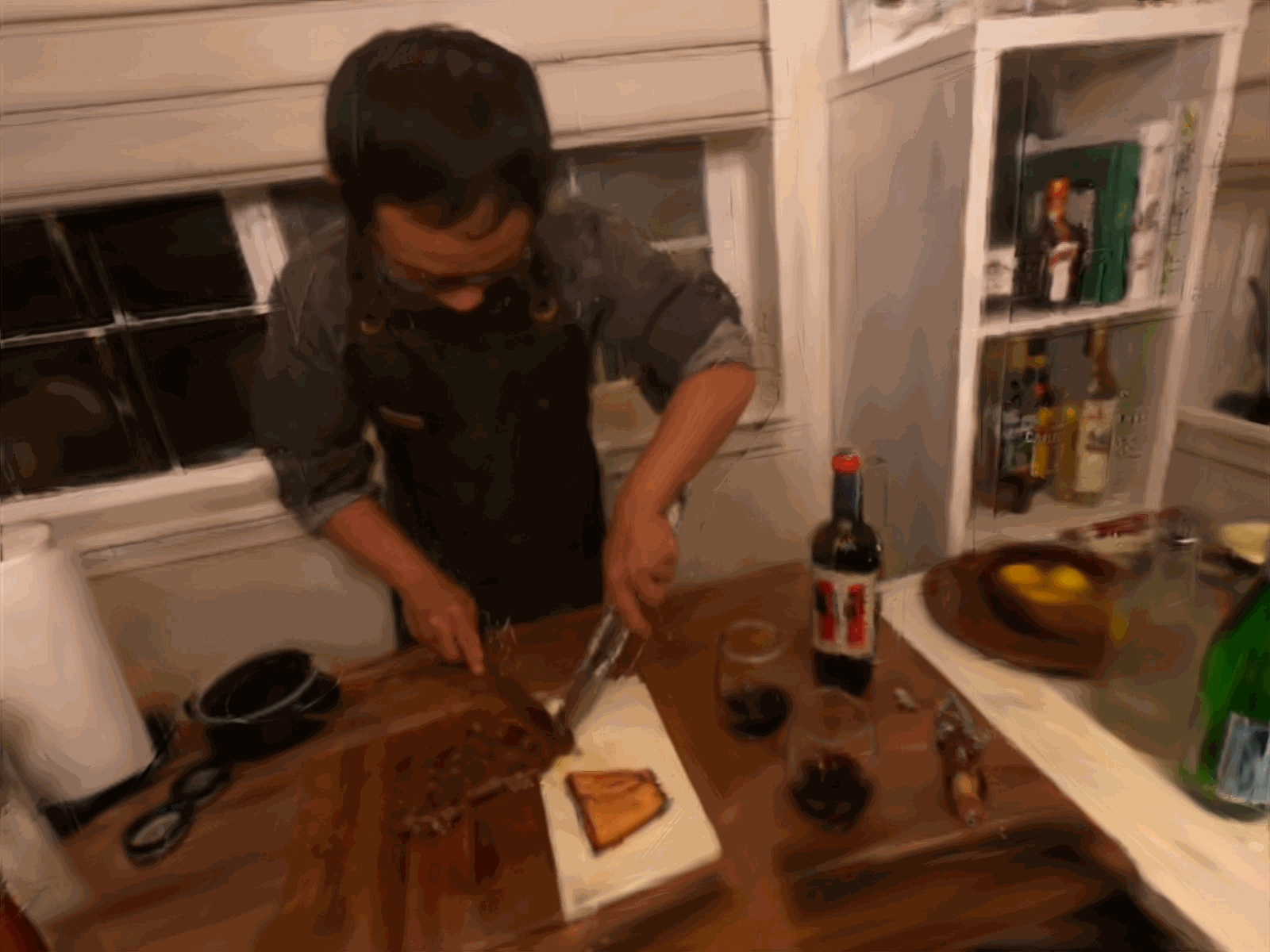}\\[-0.15em]
        \includegraphics[width=\linewidth]{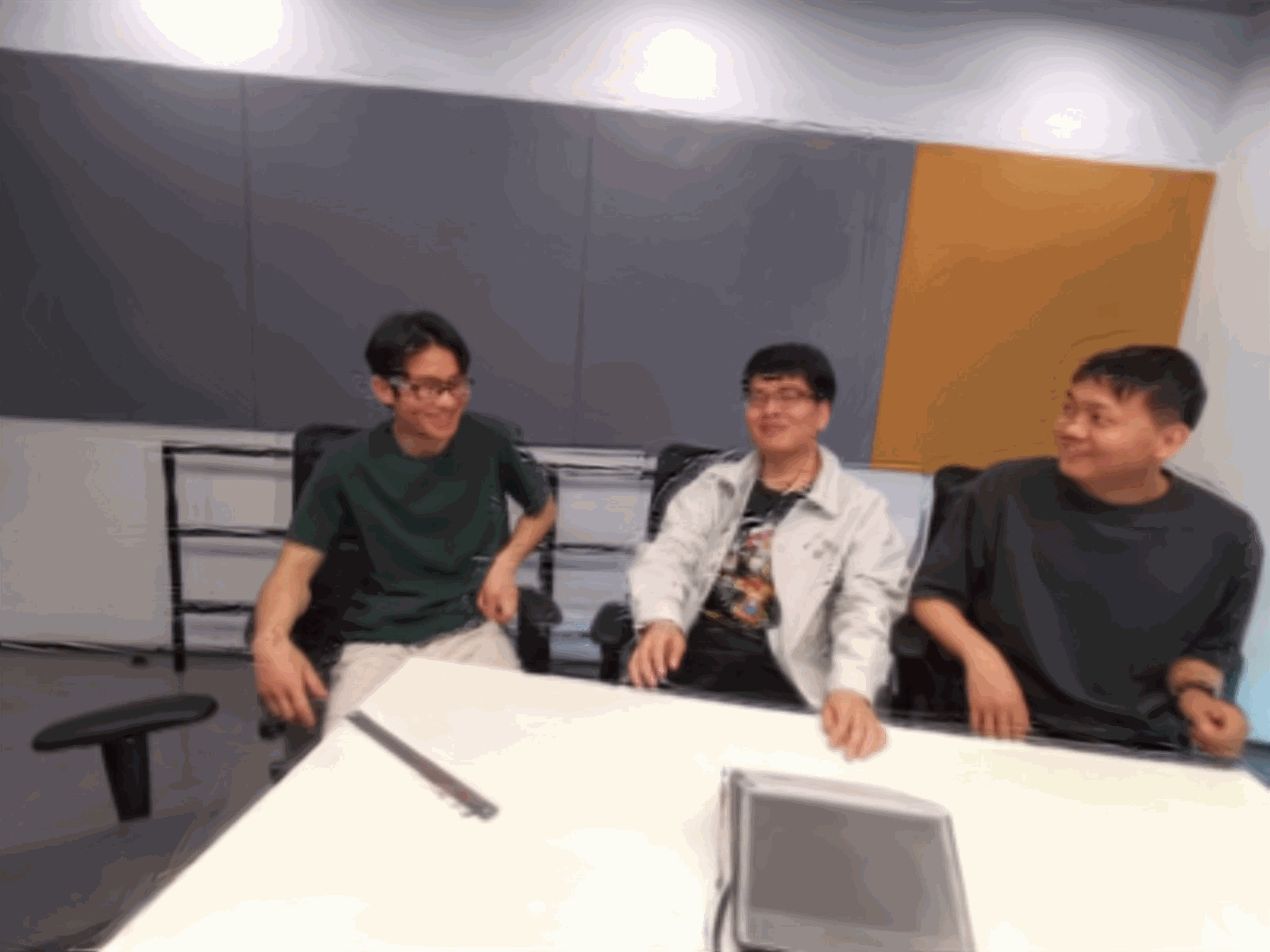}\\[-0.15em]
        \includegraphics[width=\linewidth]{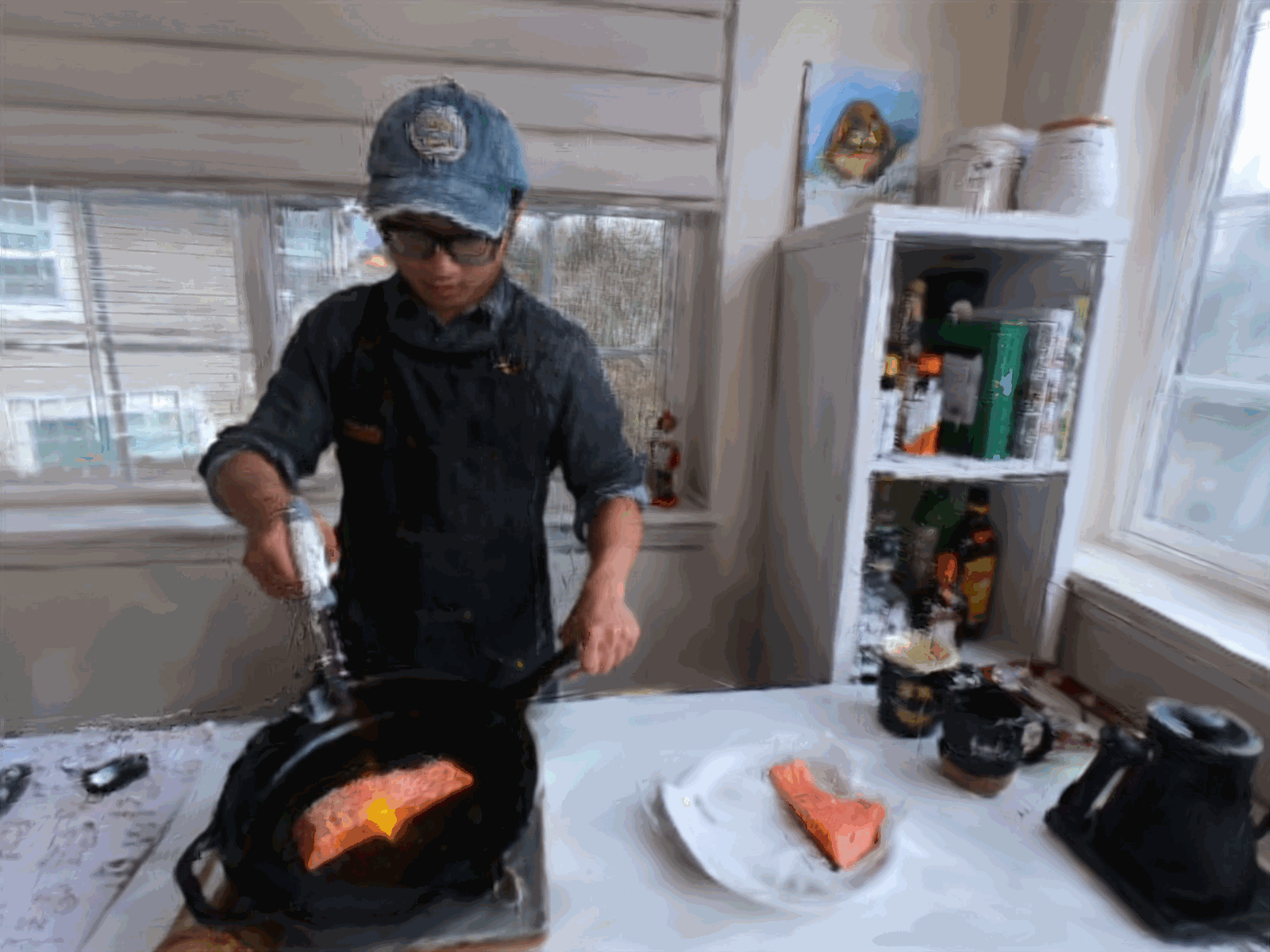}\\[-0.15em]
        \includegraphics[width=\linewidth]{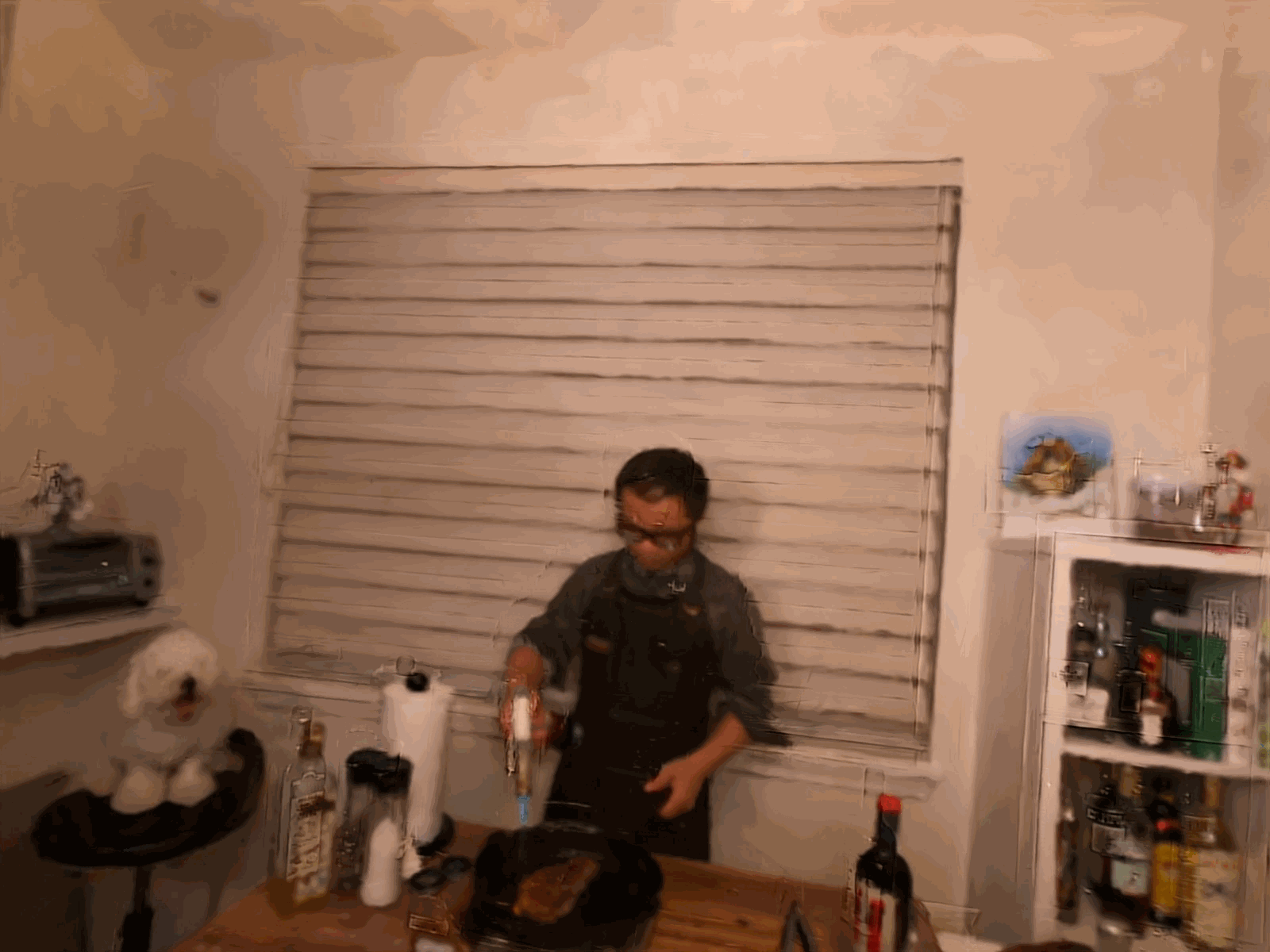}\\[-0.15em]
        \includegraphics[width=\linewidth]{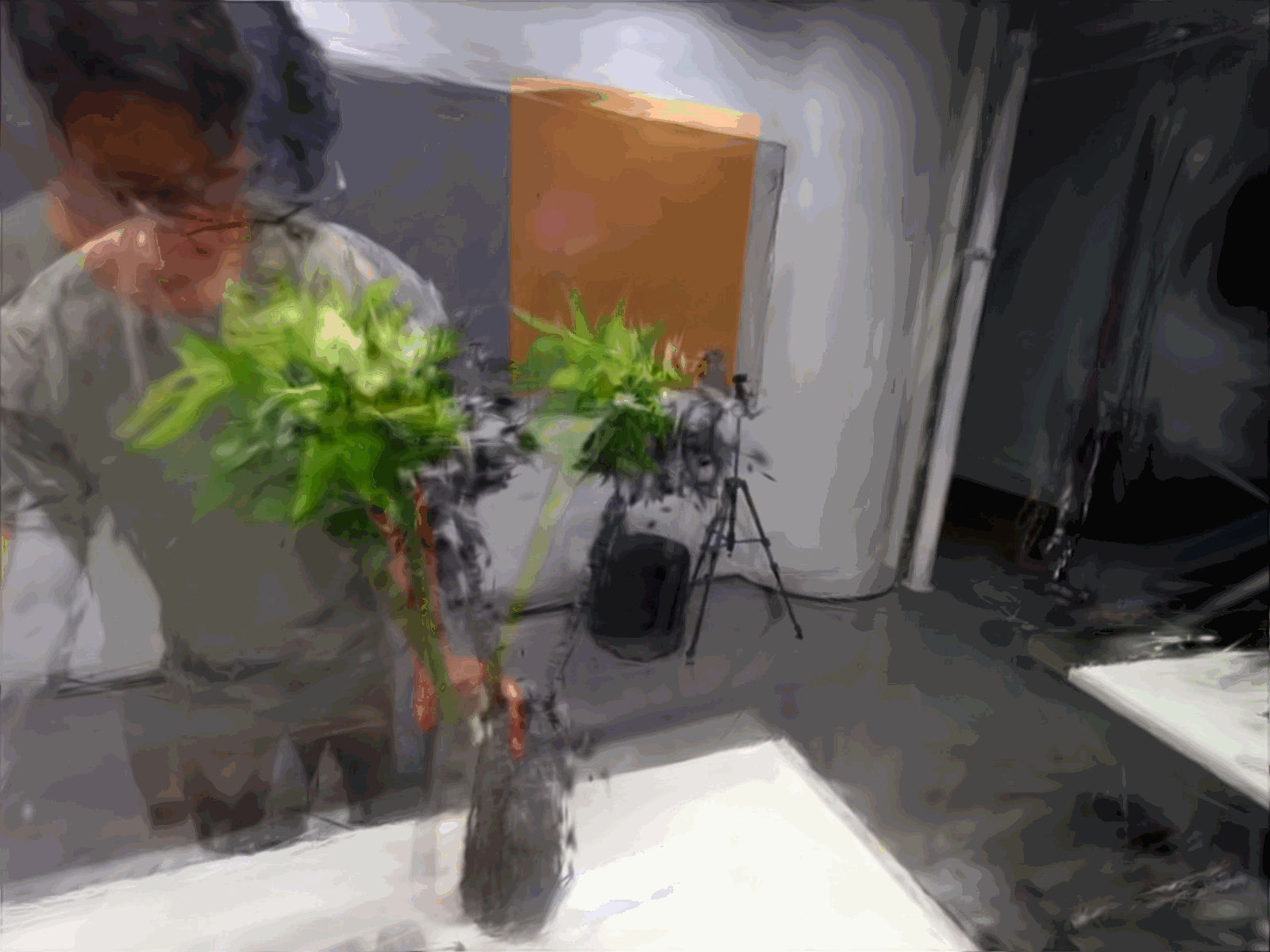}\\[-0.15em]
        \includegraphics[width=\linewidth]{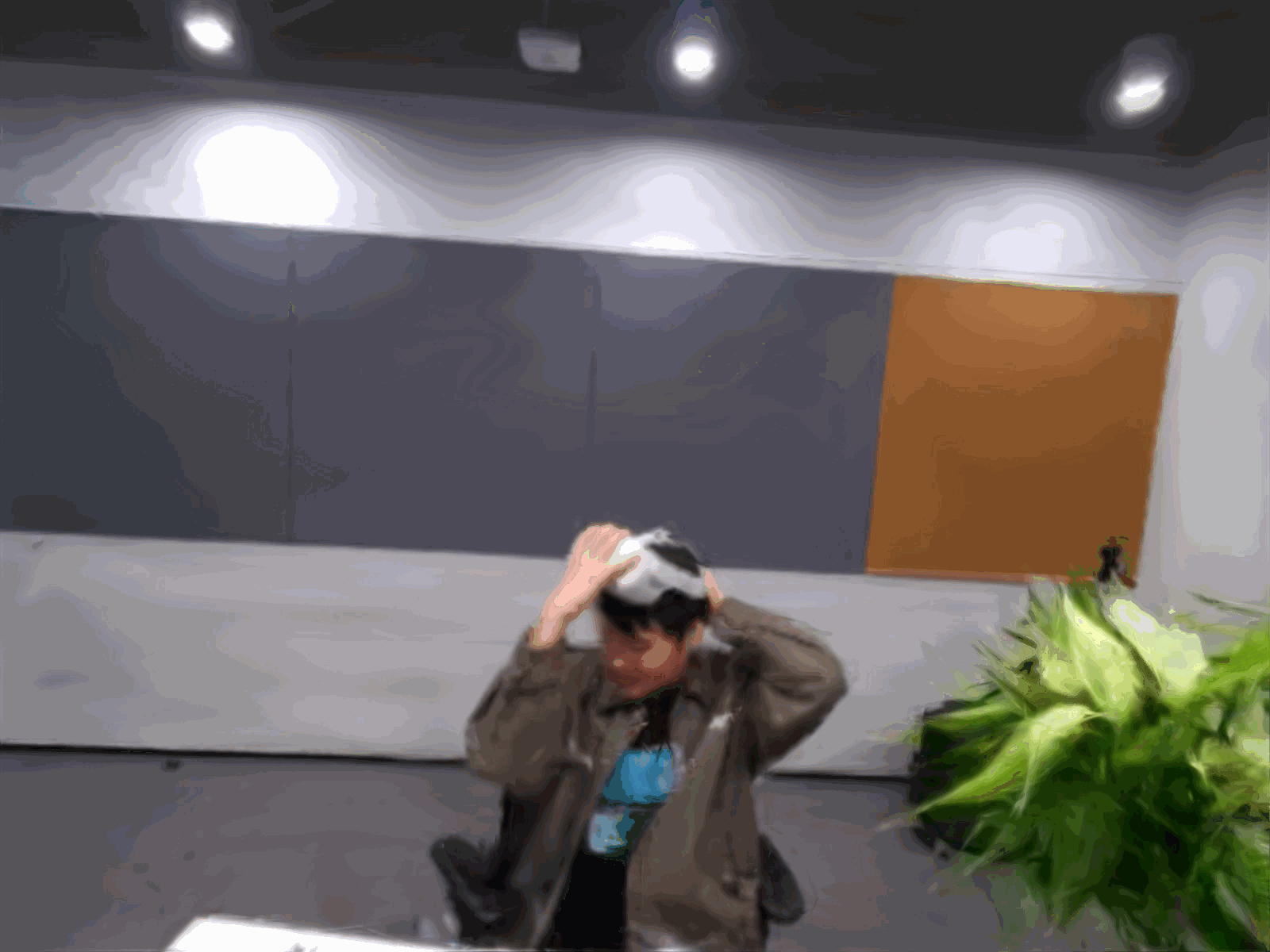}\\[-0.15em]
        \includegraphics[width=\linewidth]{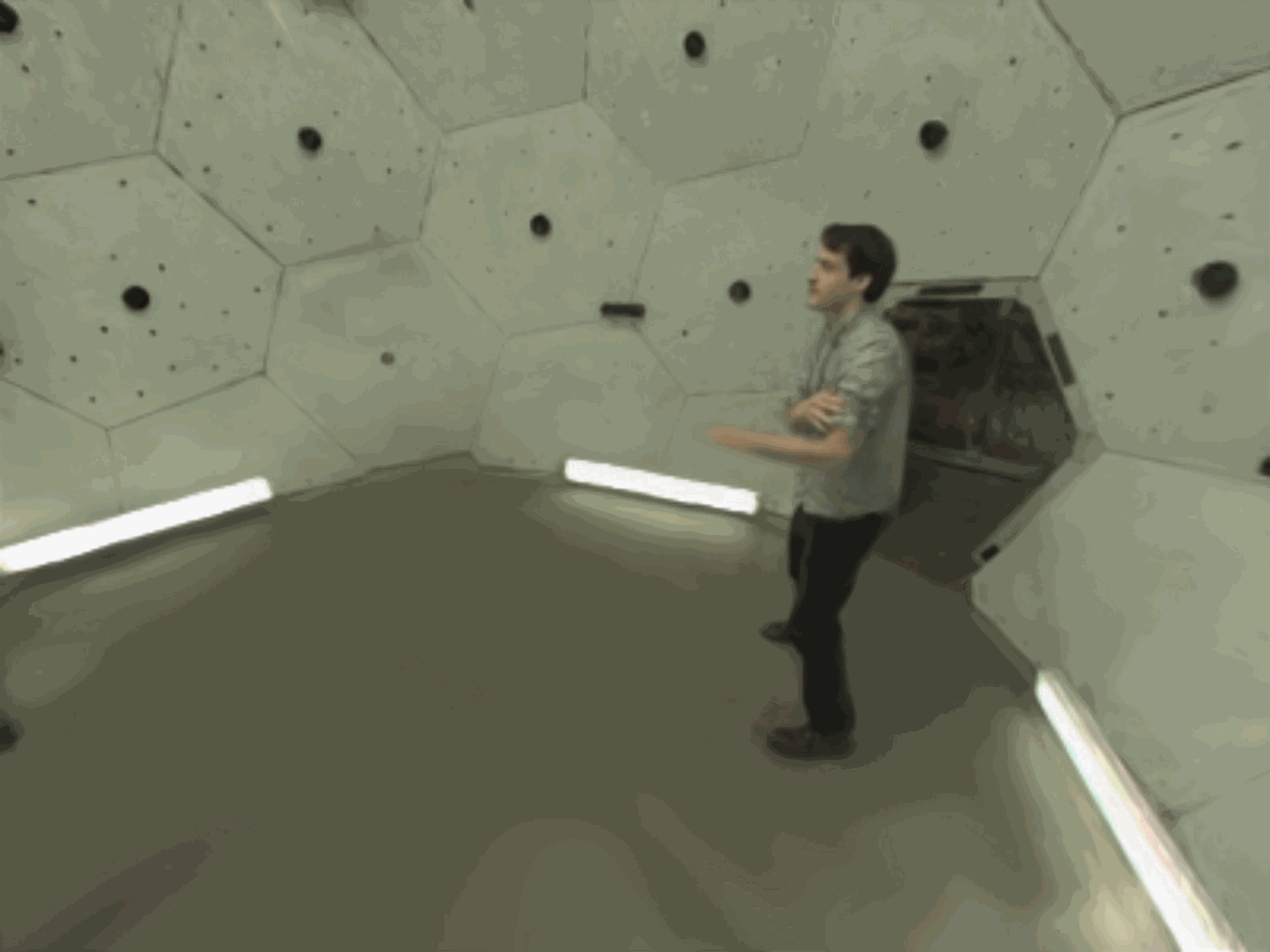}\\[-0.15em]
        \includegraphics[width=\linewidth]{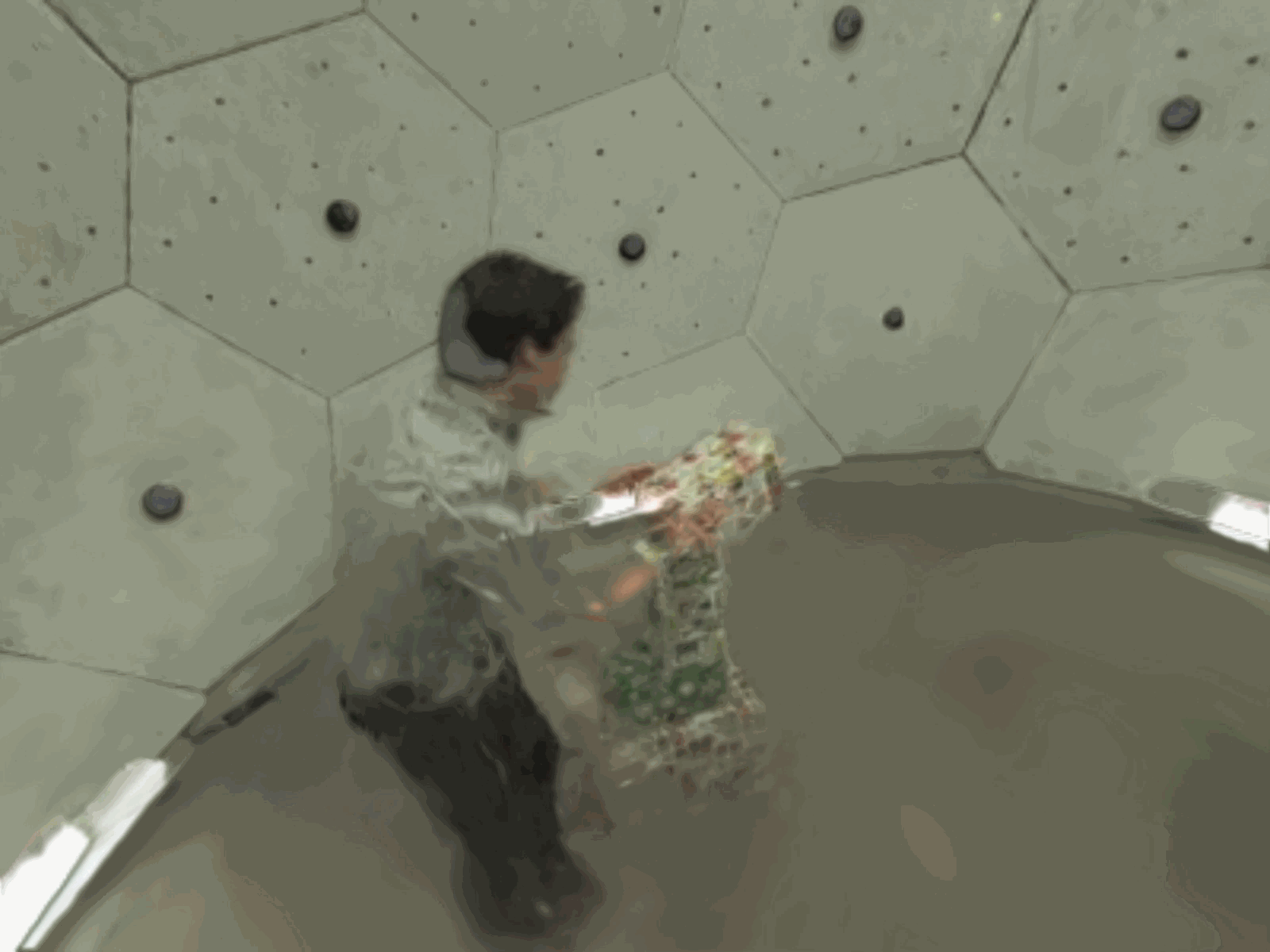}\\[-0.15em]
        \includegraphics[width=\linewidth]{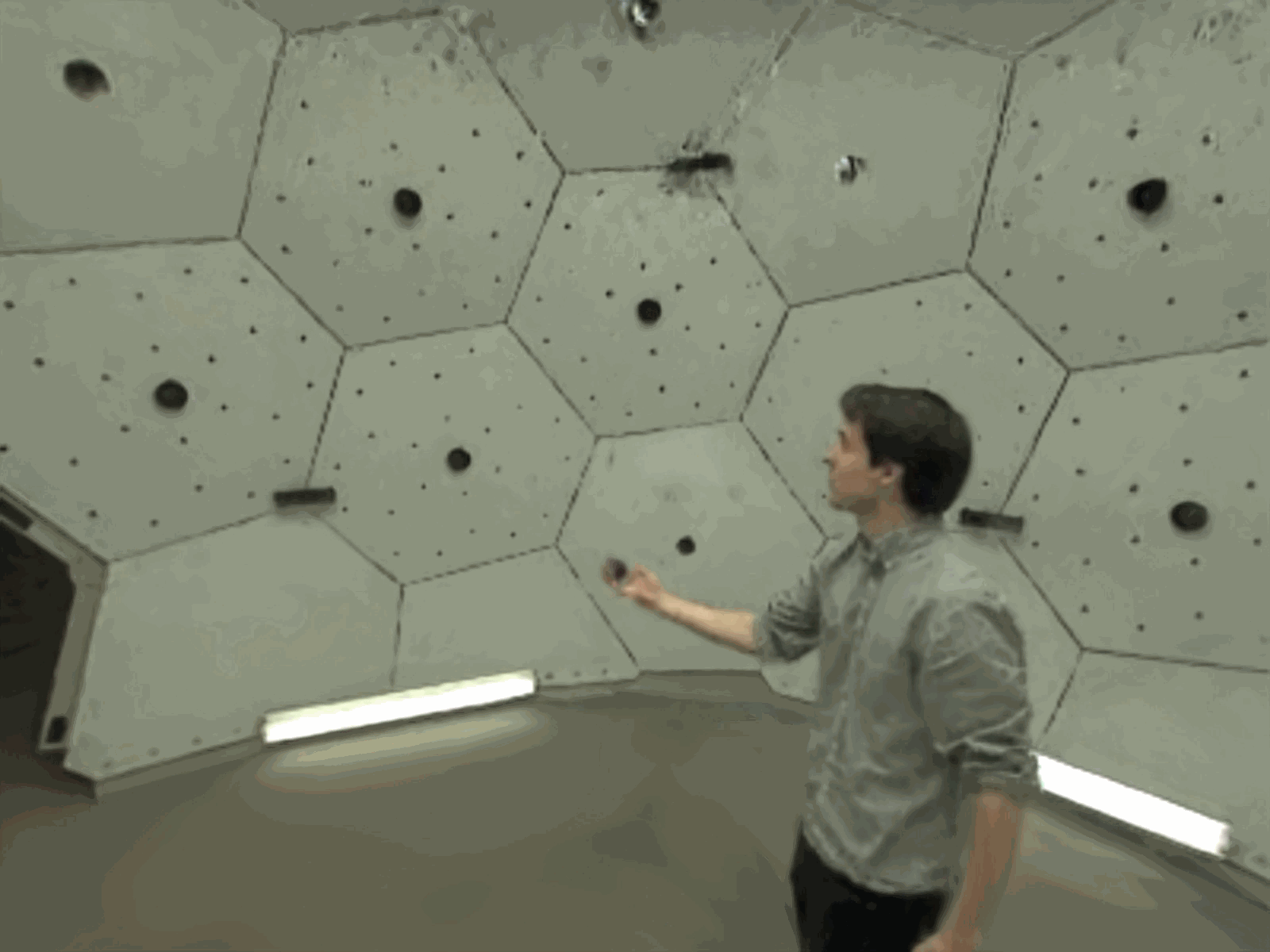}
        \caption{w/o PRPA}
    \end{subfigure}\begin{subfigure}[t]{0.138\linewidth}
        \centering
        \includegraphics[width=\linewidth]{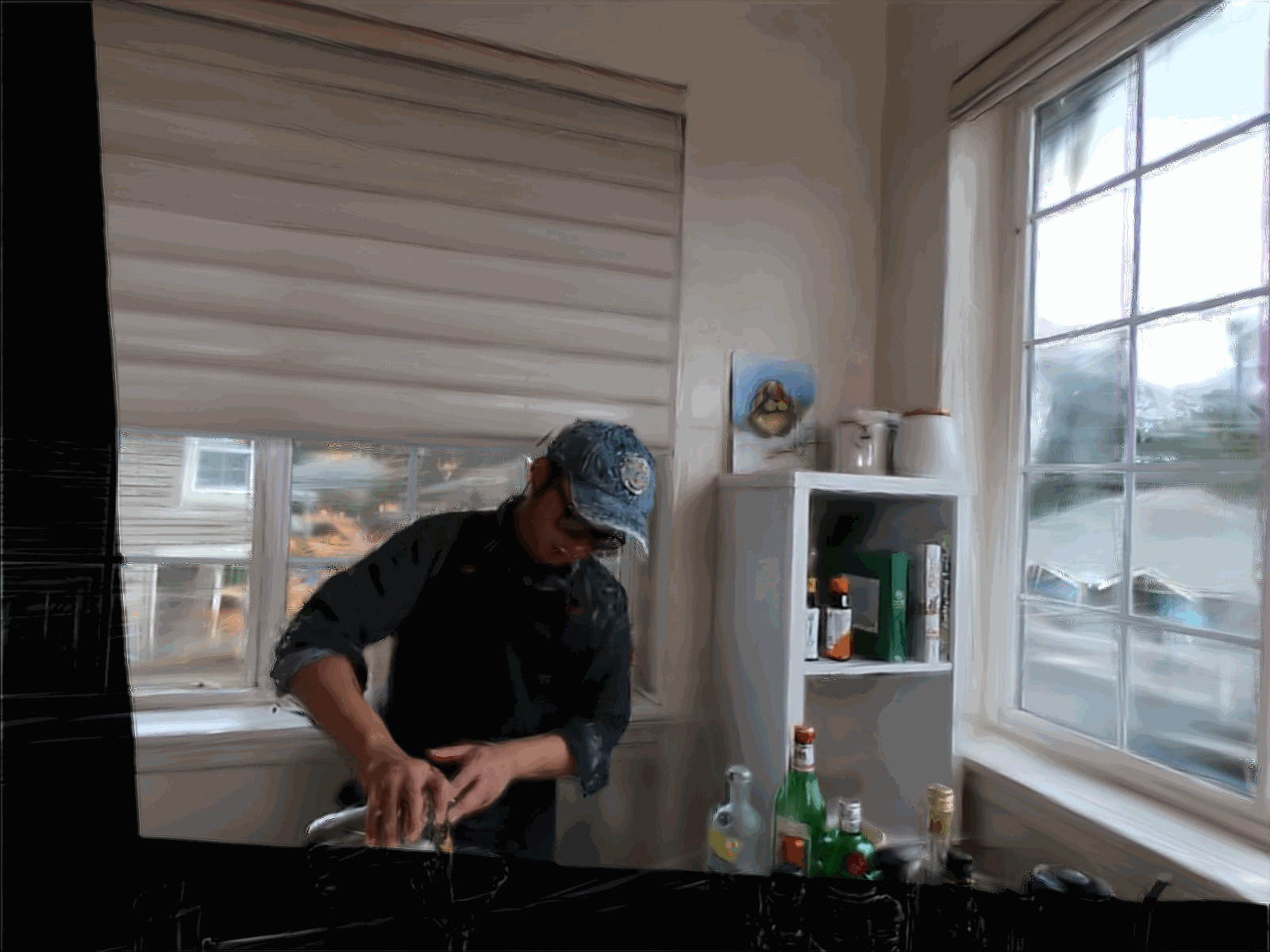}\\[-0.15em]
        \includegraphics[width=\linewidth]{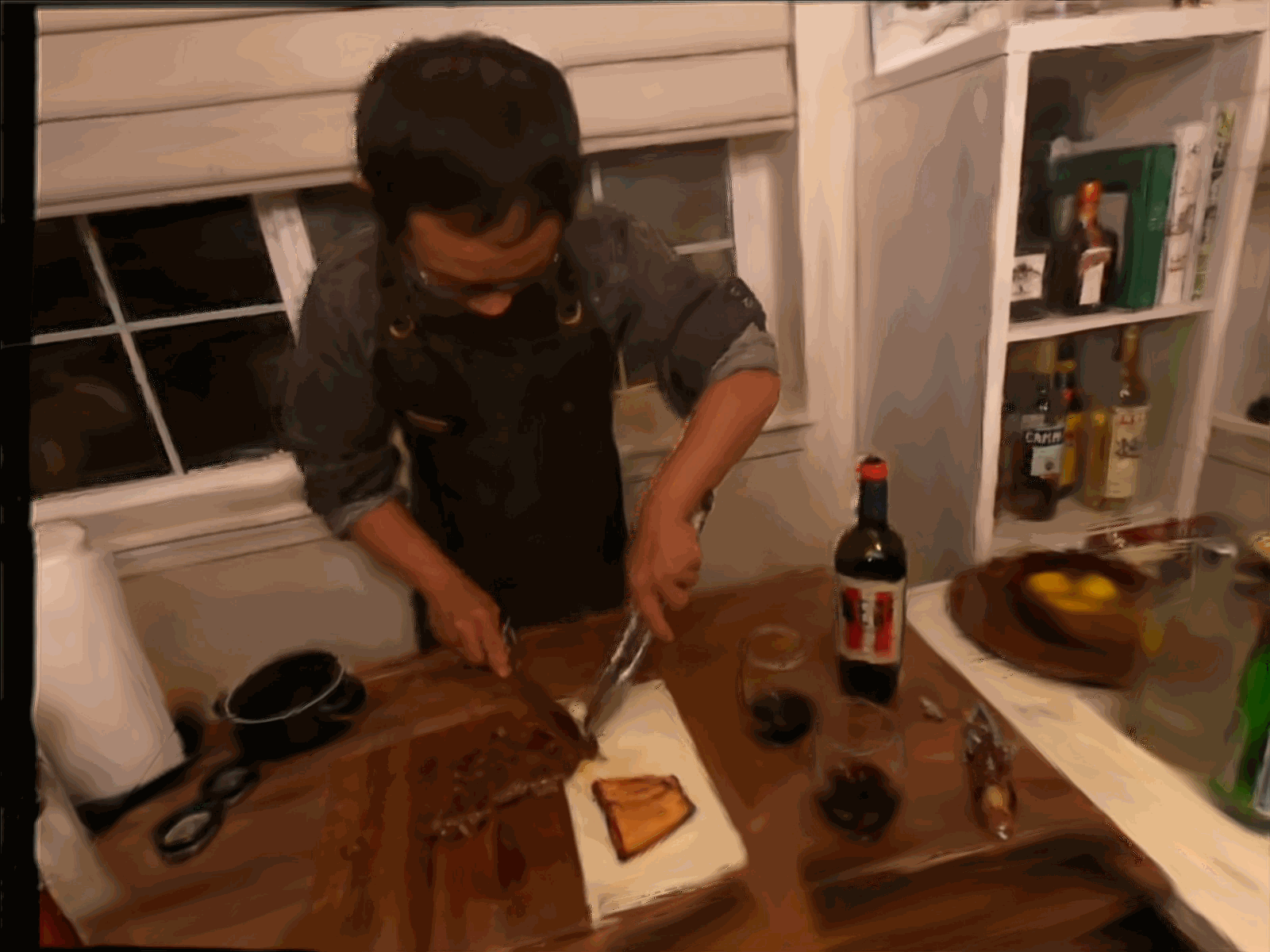}\\[-0.15em]
        \includegraphics[width=\linewidth]{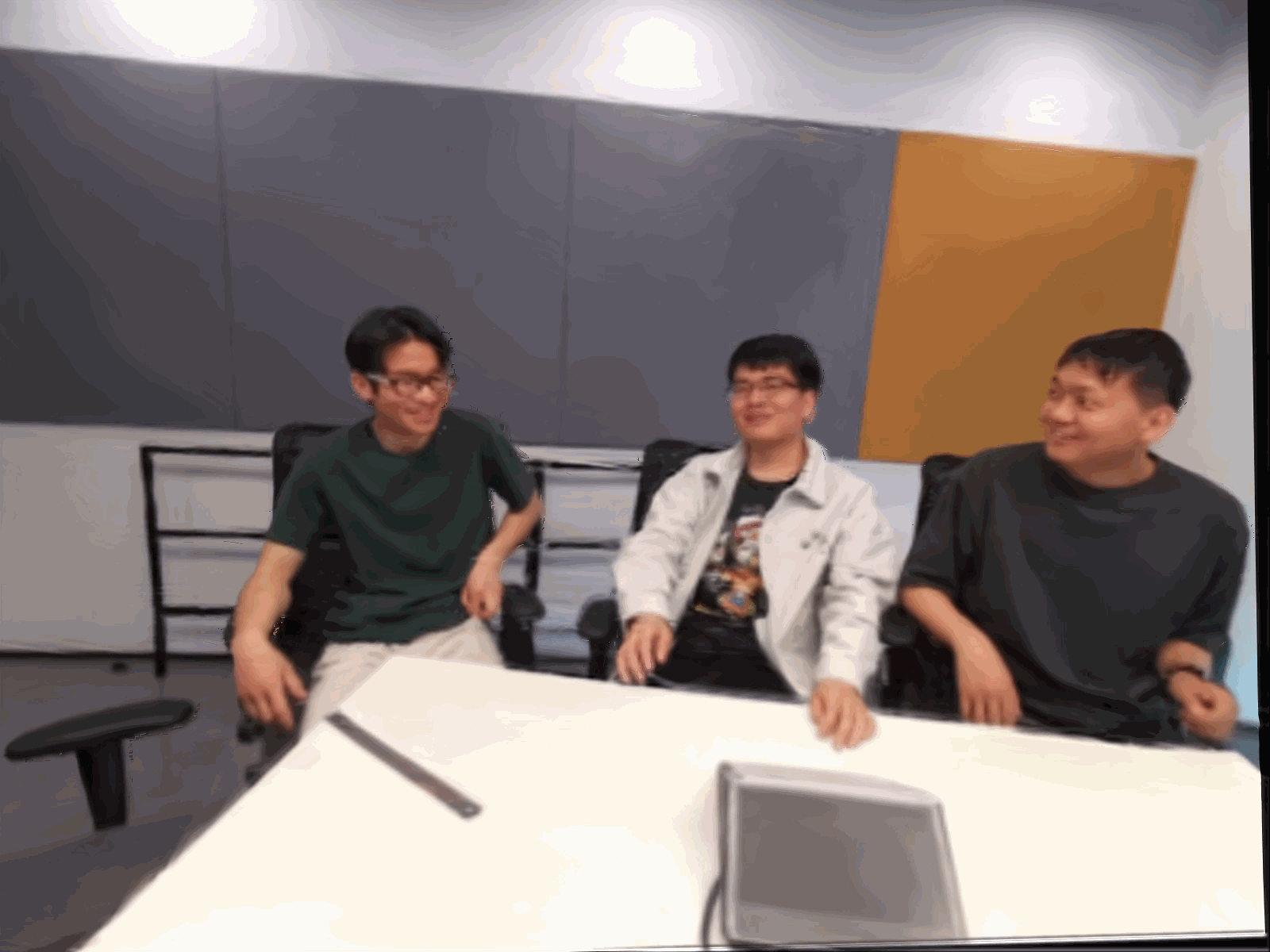}\\[-0.15em]
        \includegraphics[width=\linewidth]{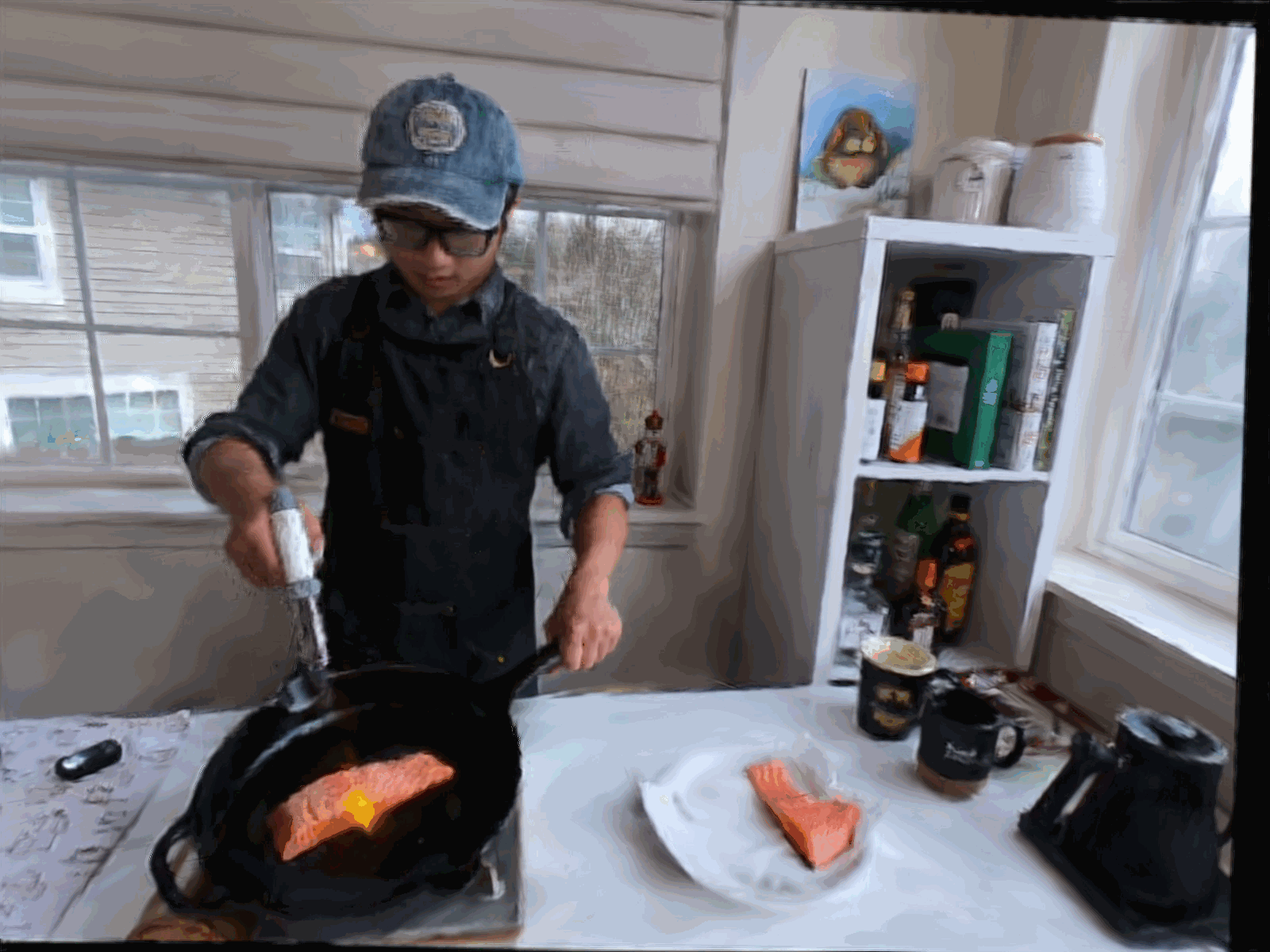}\\[-0.15em]
        \includegraphics[width=\linewidth]{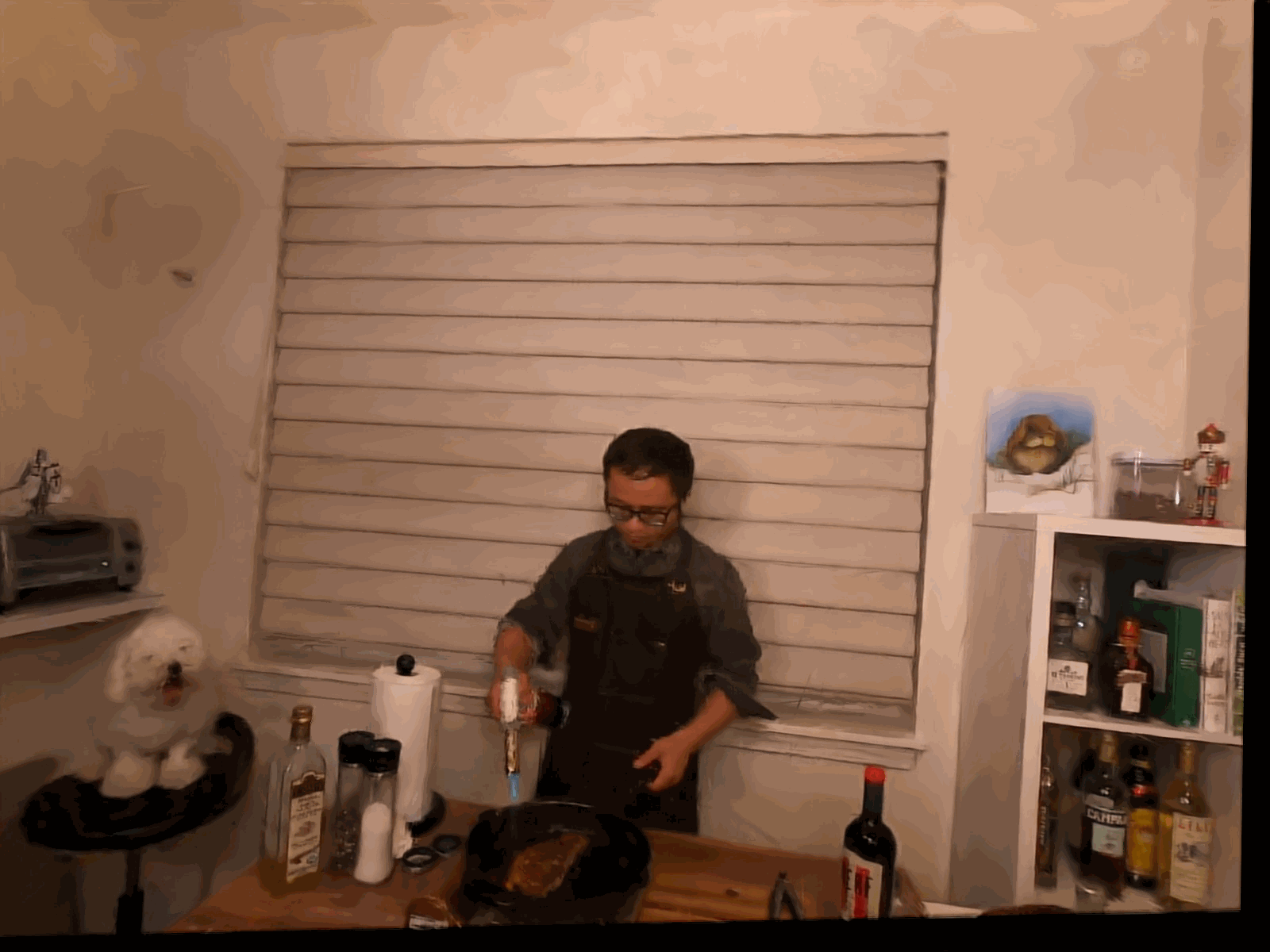}\\[-0.15em]
        \includegraphics[width=\linewidth]{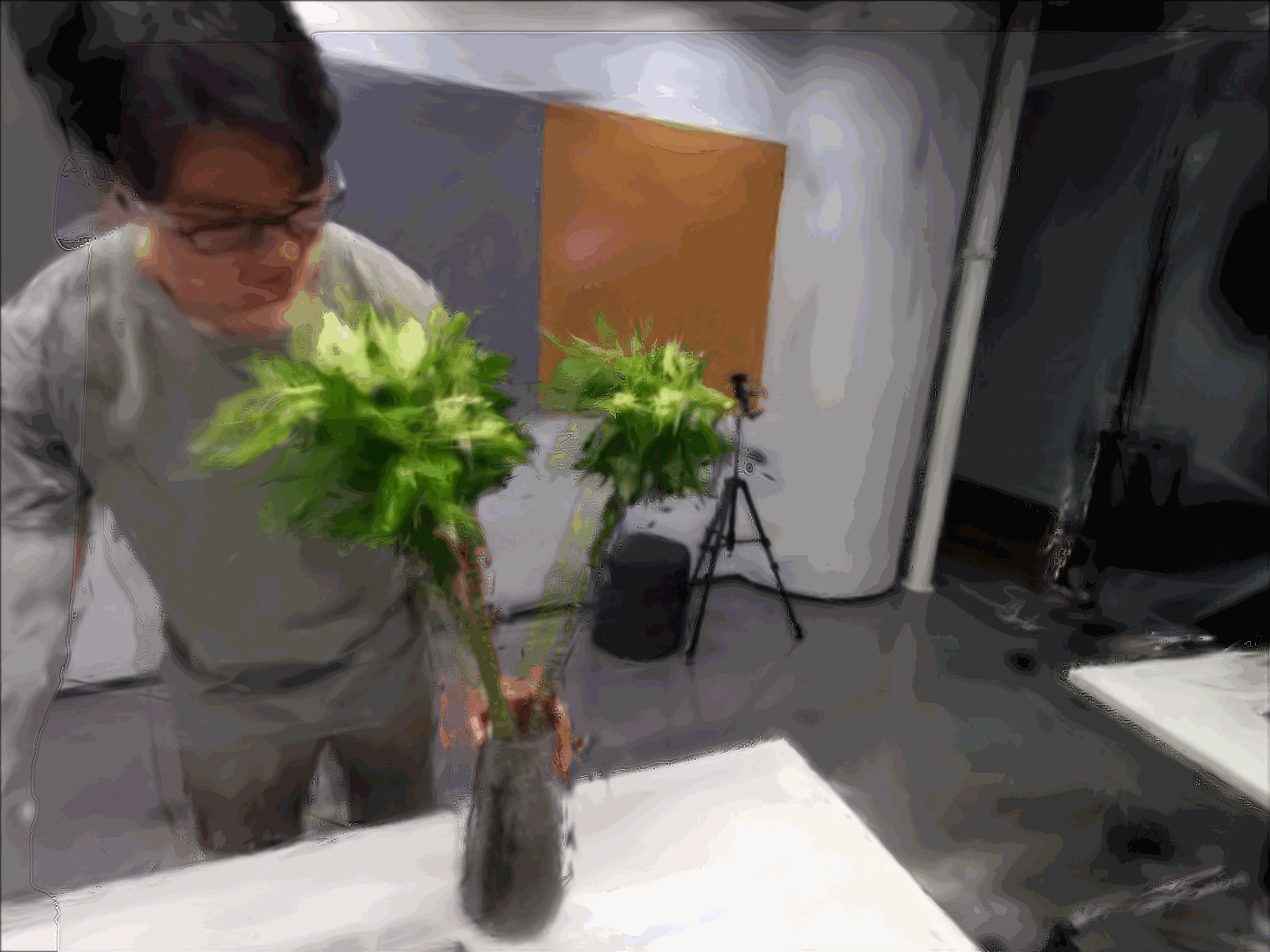}\\[-0.15em]
        \includegraphics[width=\linewidth]{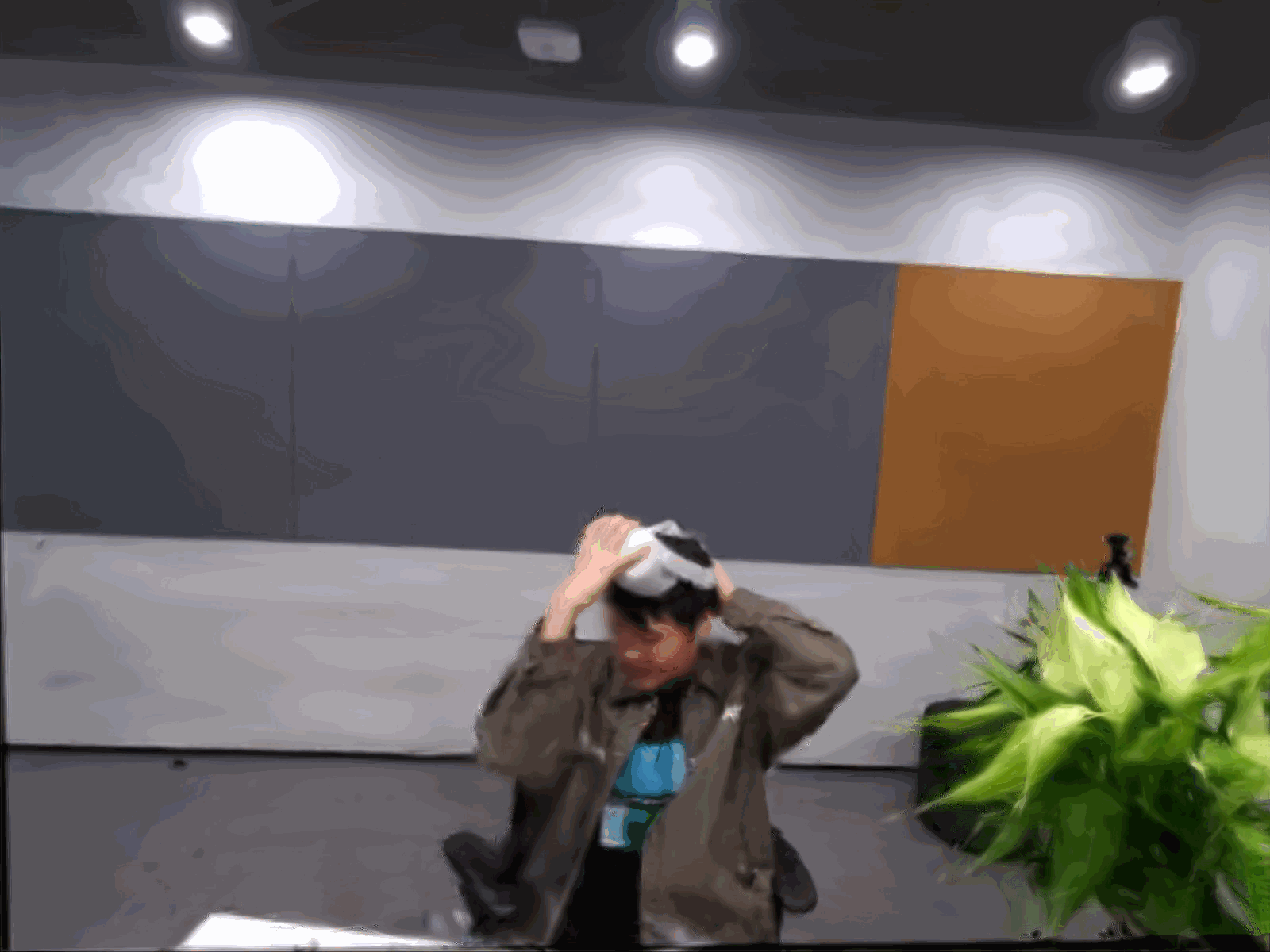}\\[-0.15em]
        \includegraphics[width=\linewidth]{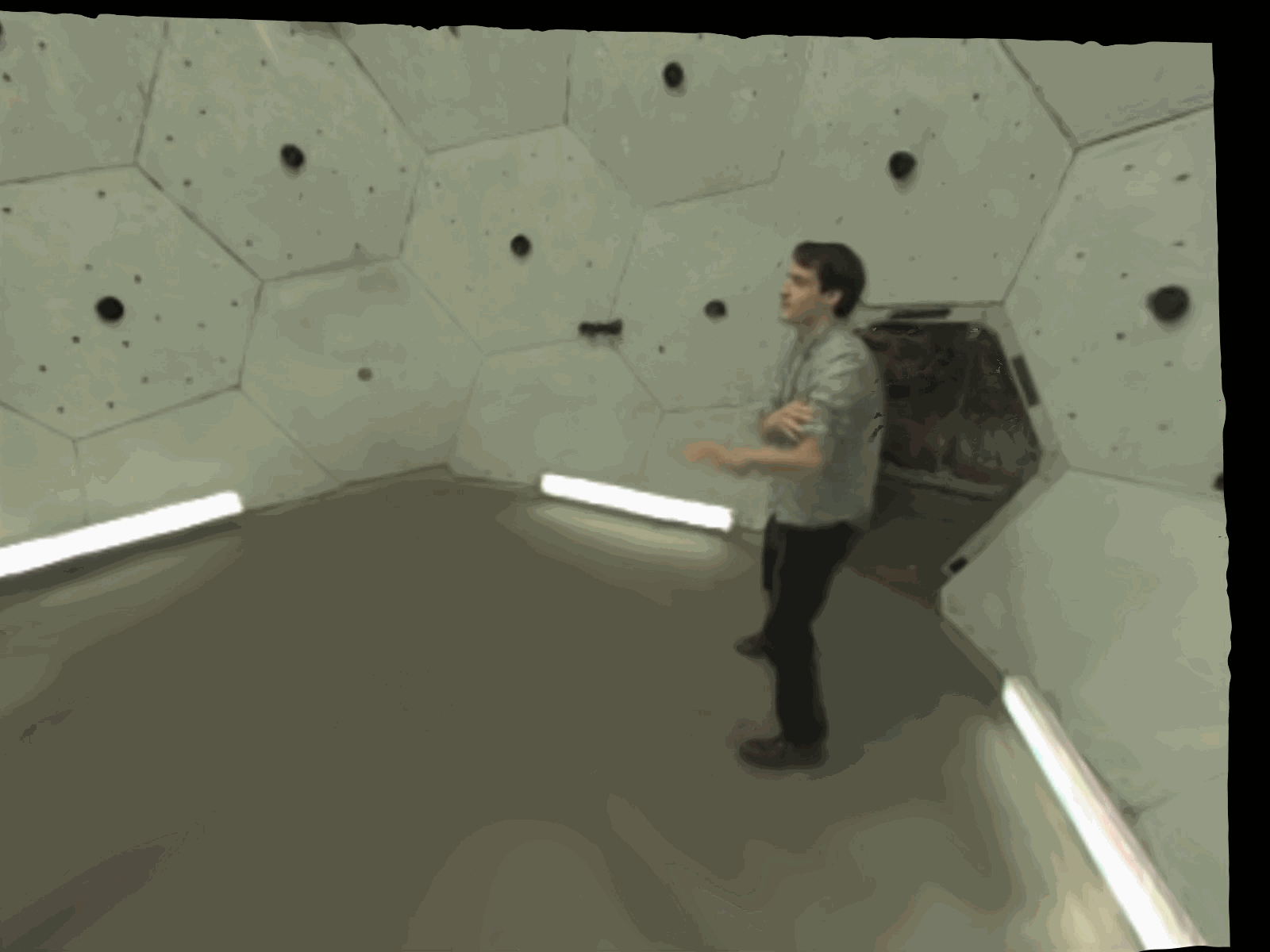}\\[-0.15em]
        \includegraphics[width=\linewidth]{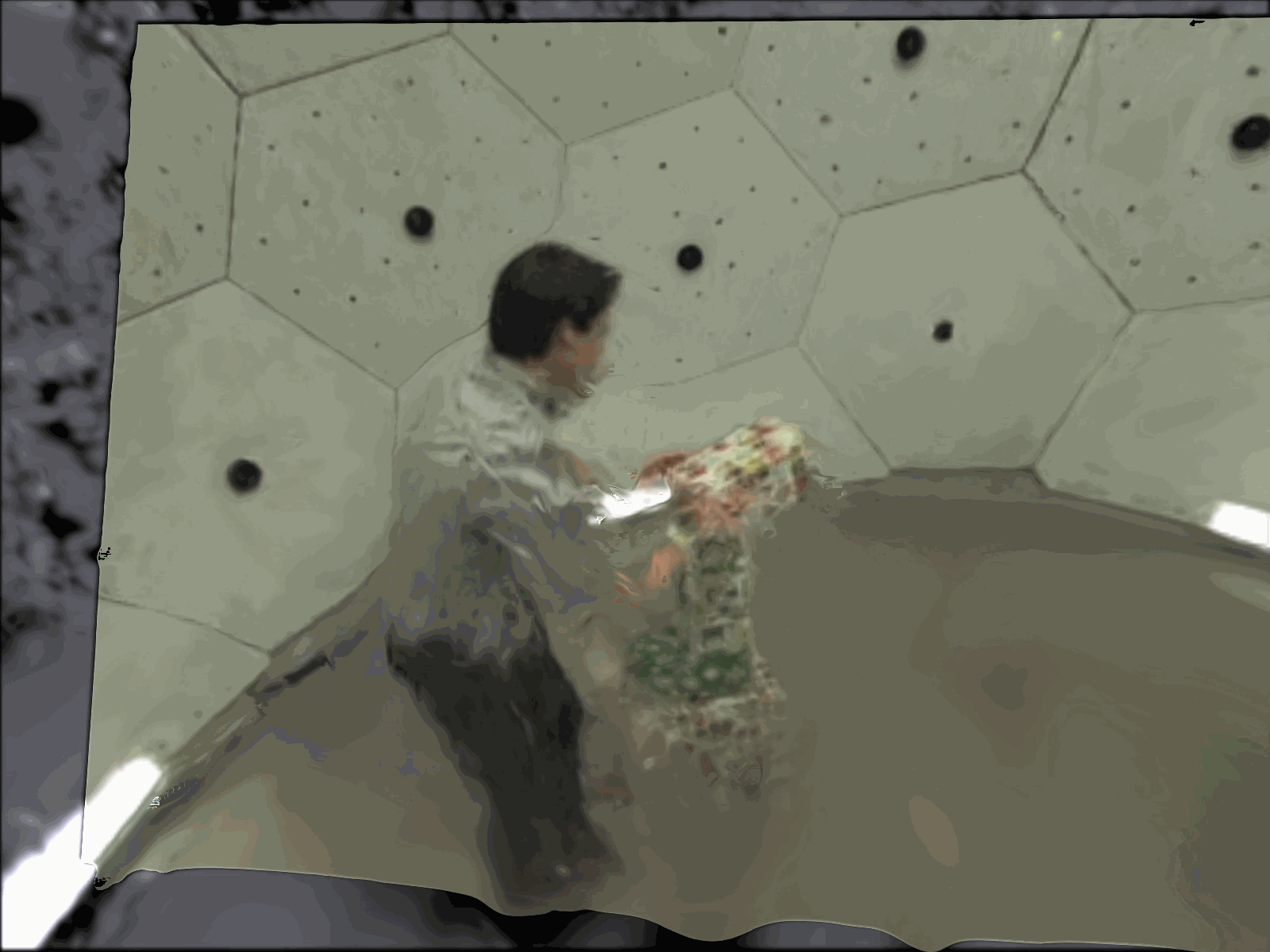}\\[-0.15em]
        \includegraphics[width=\linewidth]{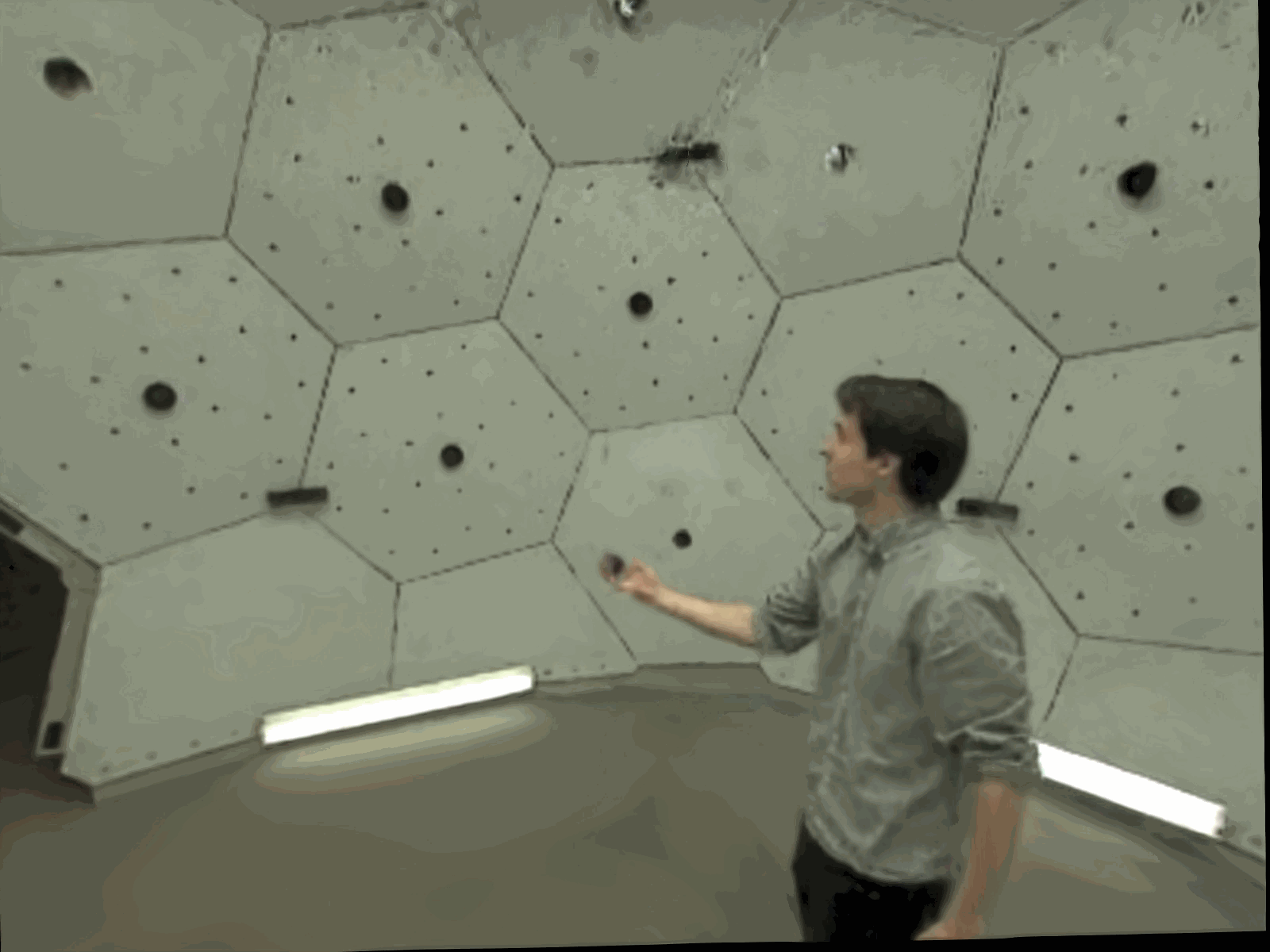}
        \caption{w/o Adaptive FoV}
    \end{subfigure}\begin{subfigure}[t]{0.138\linewidth}
        \centering
        \includegraphics[width=\linewidth]{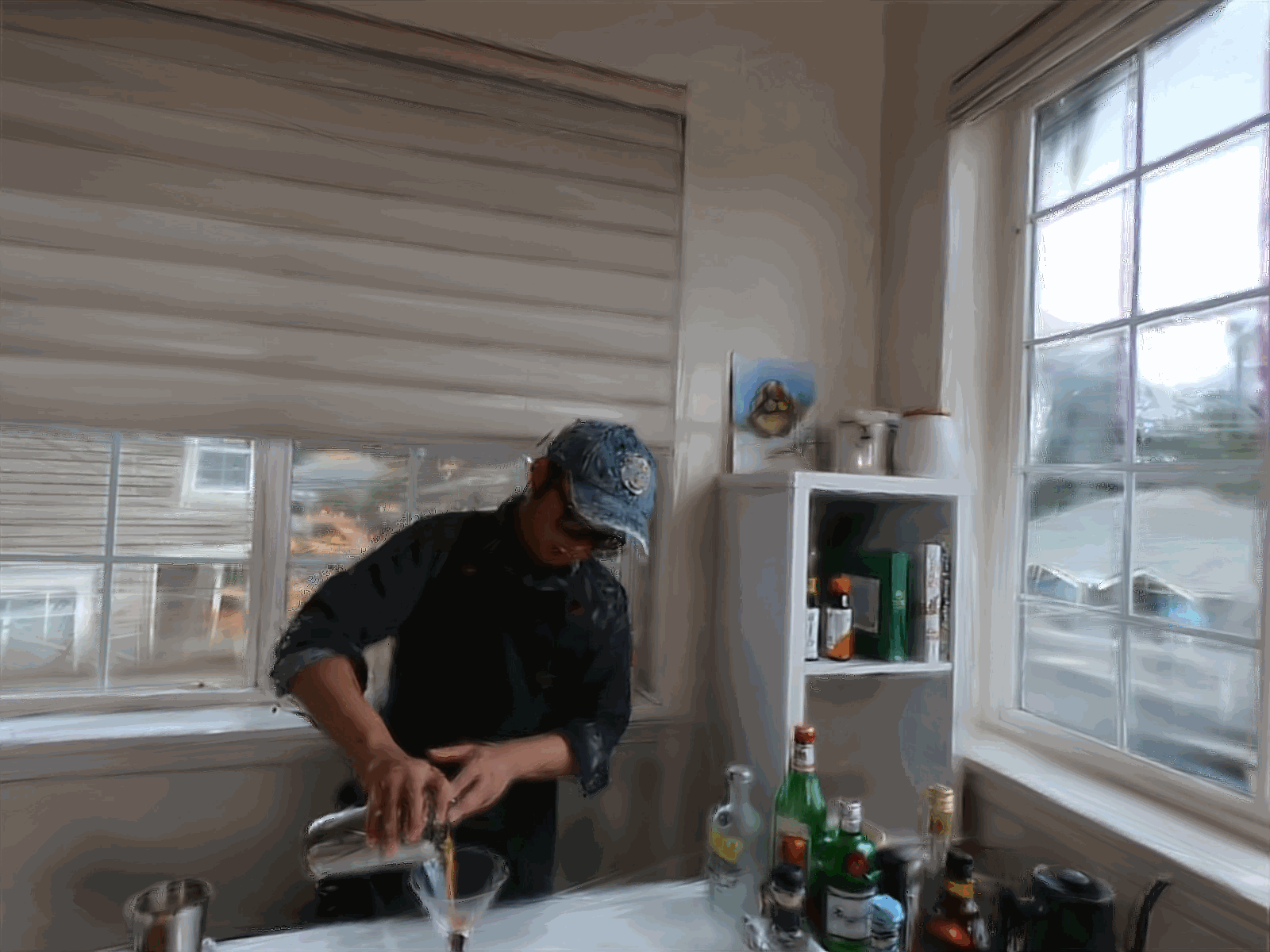}\\[-0.15em]
        \includegraphics[width=\linewidth]{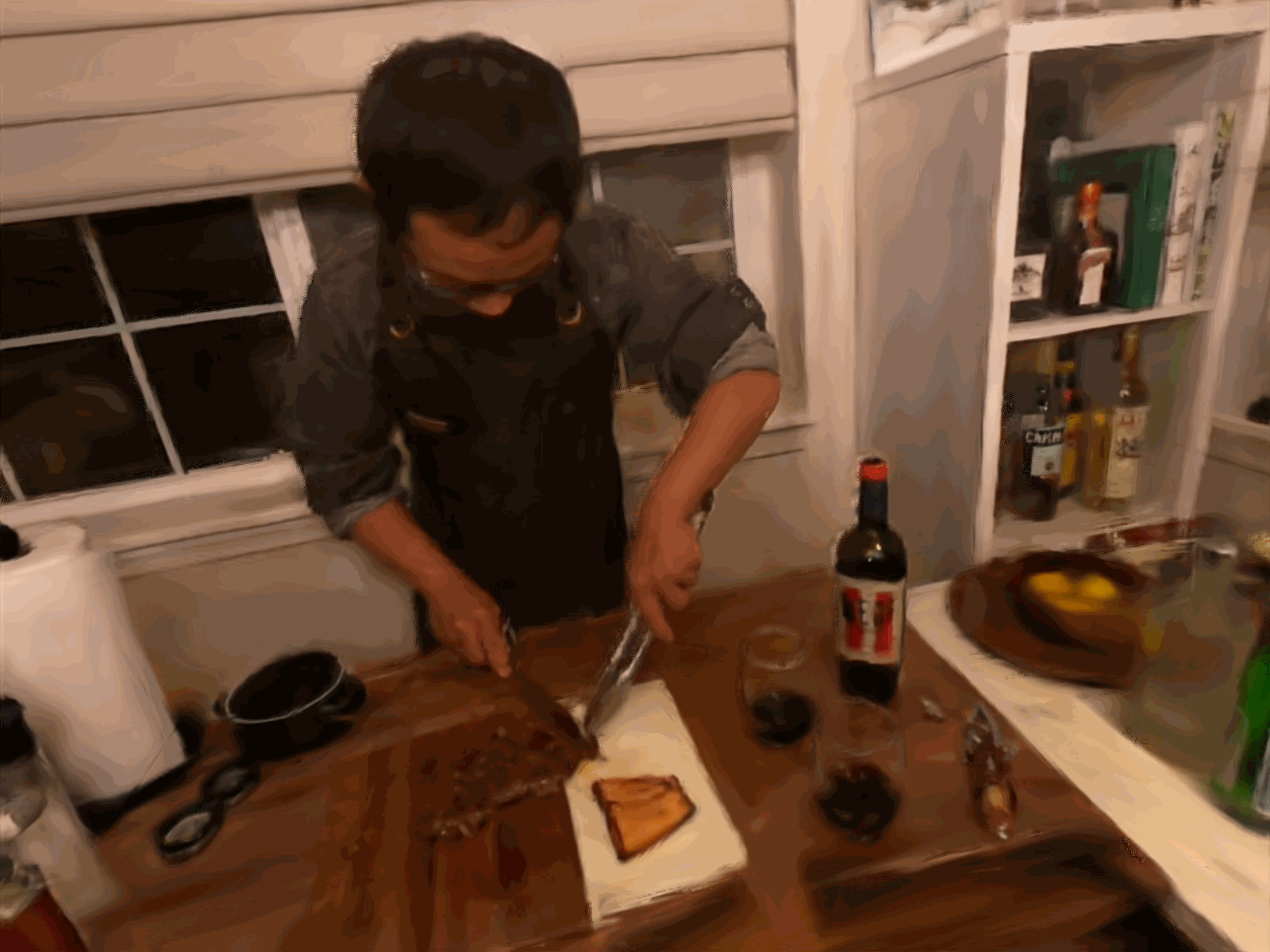}\\[-0.15em]
        \includegraphics[width=\linewidth]{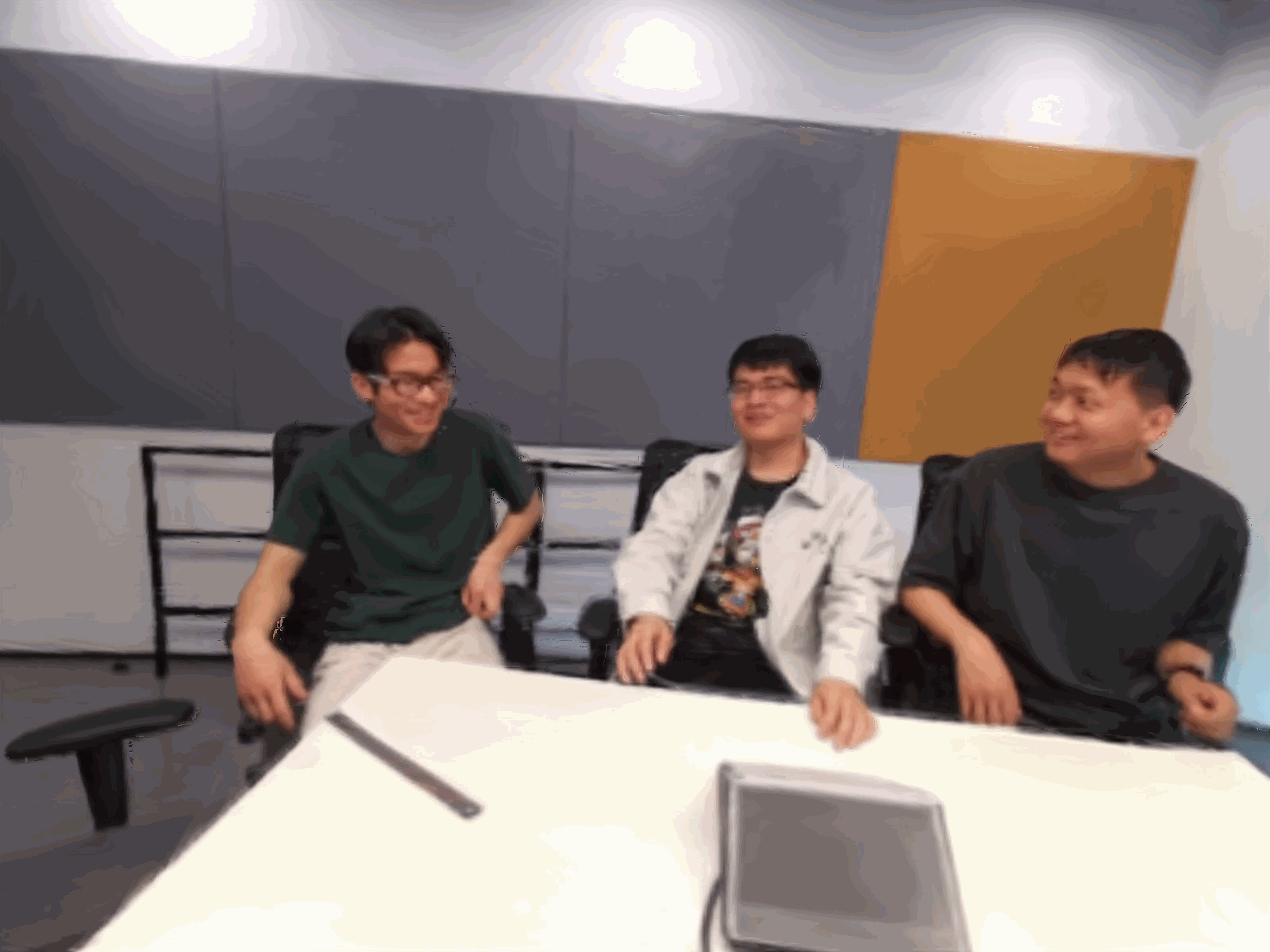}\\[-0.15em]
        \includegraphics[width=\linewidth]{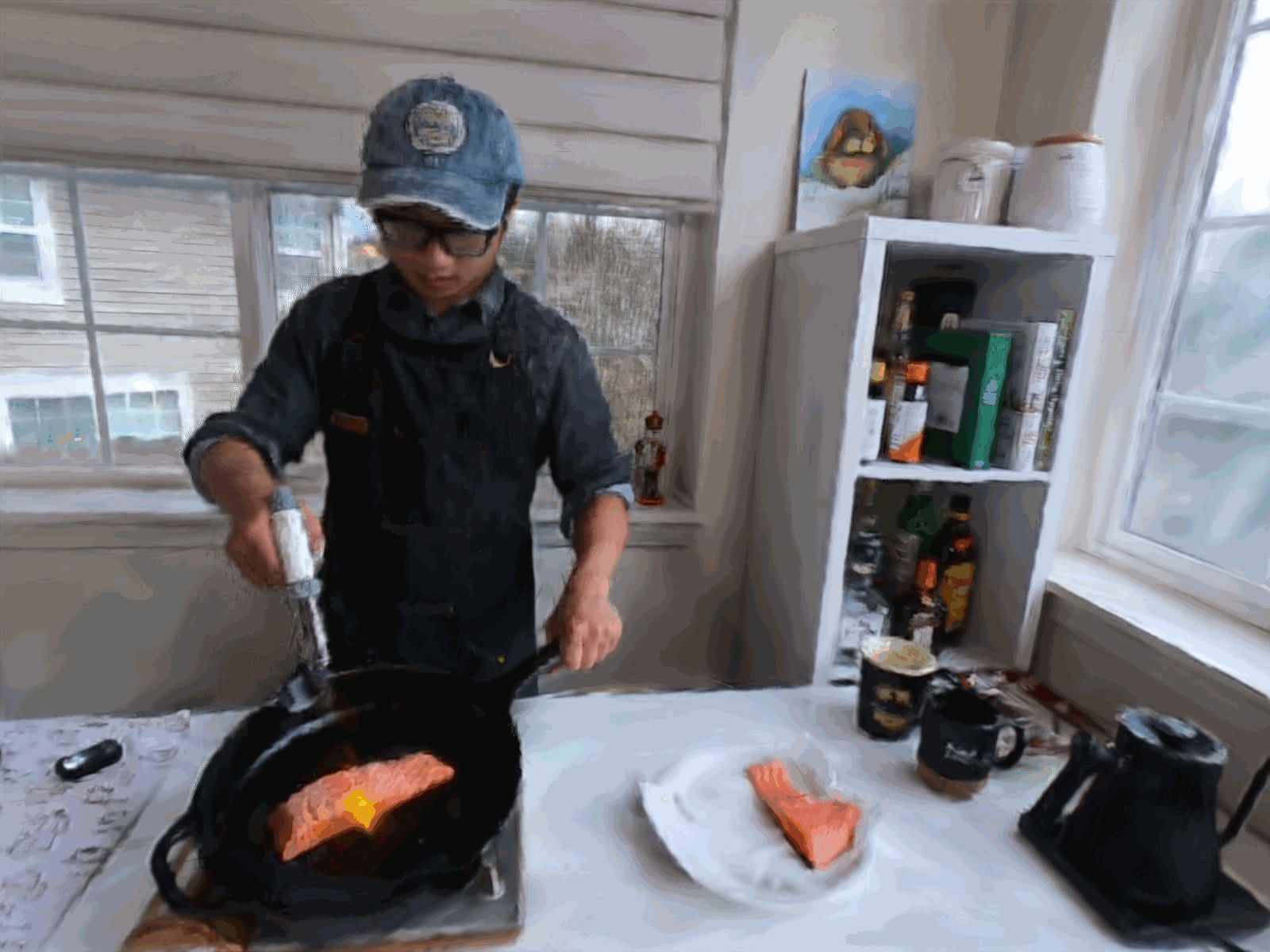}\\[-0.15em]
        \includegraphics[width=\linewidth]{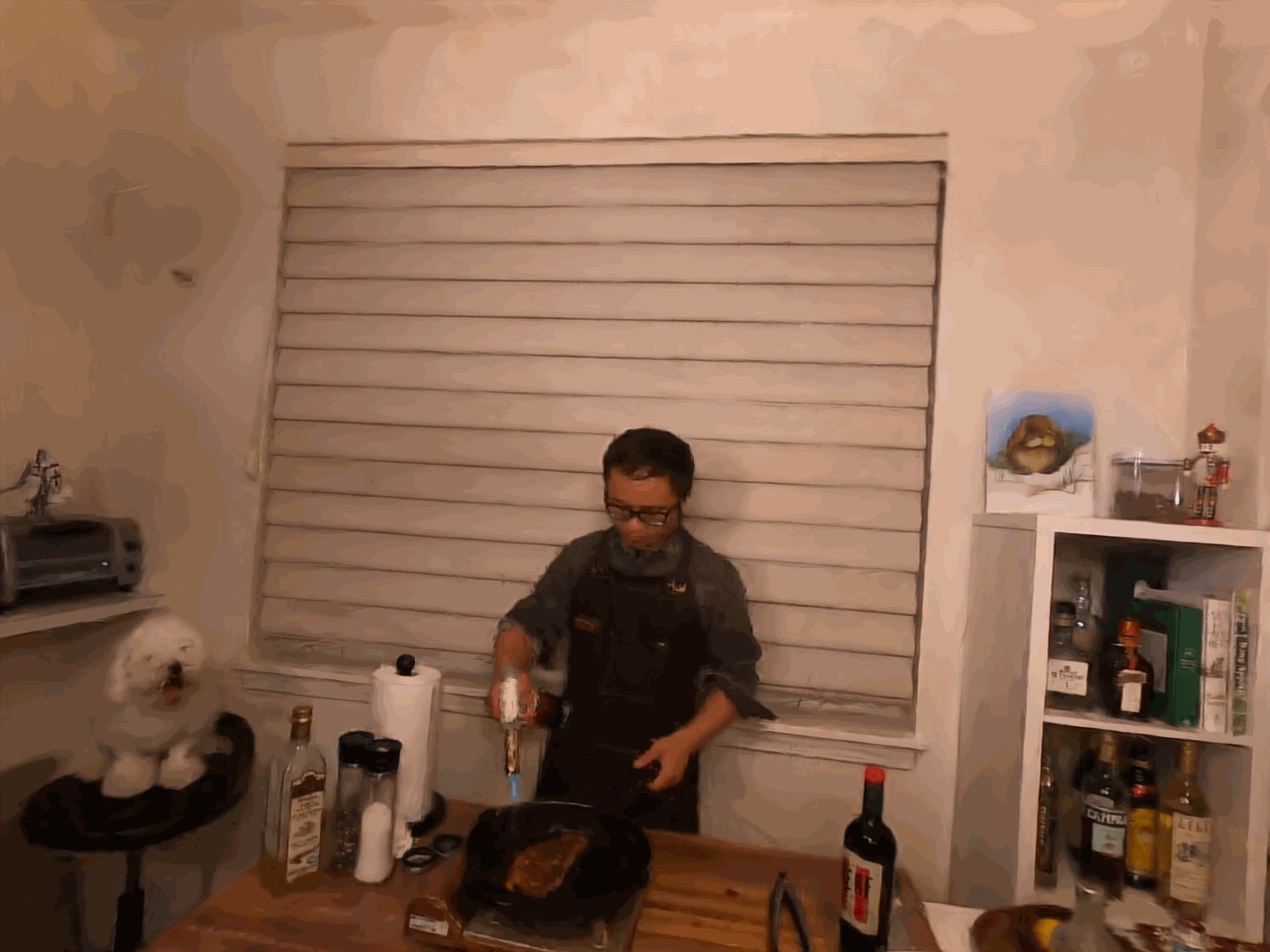}\\[-0.15em]
        \includegraphics[width=\linewidth]{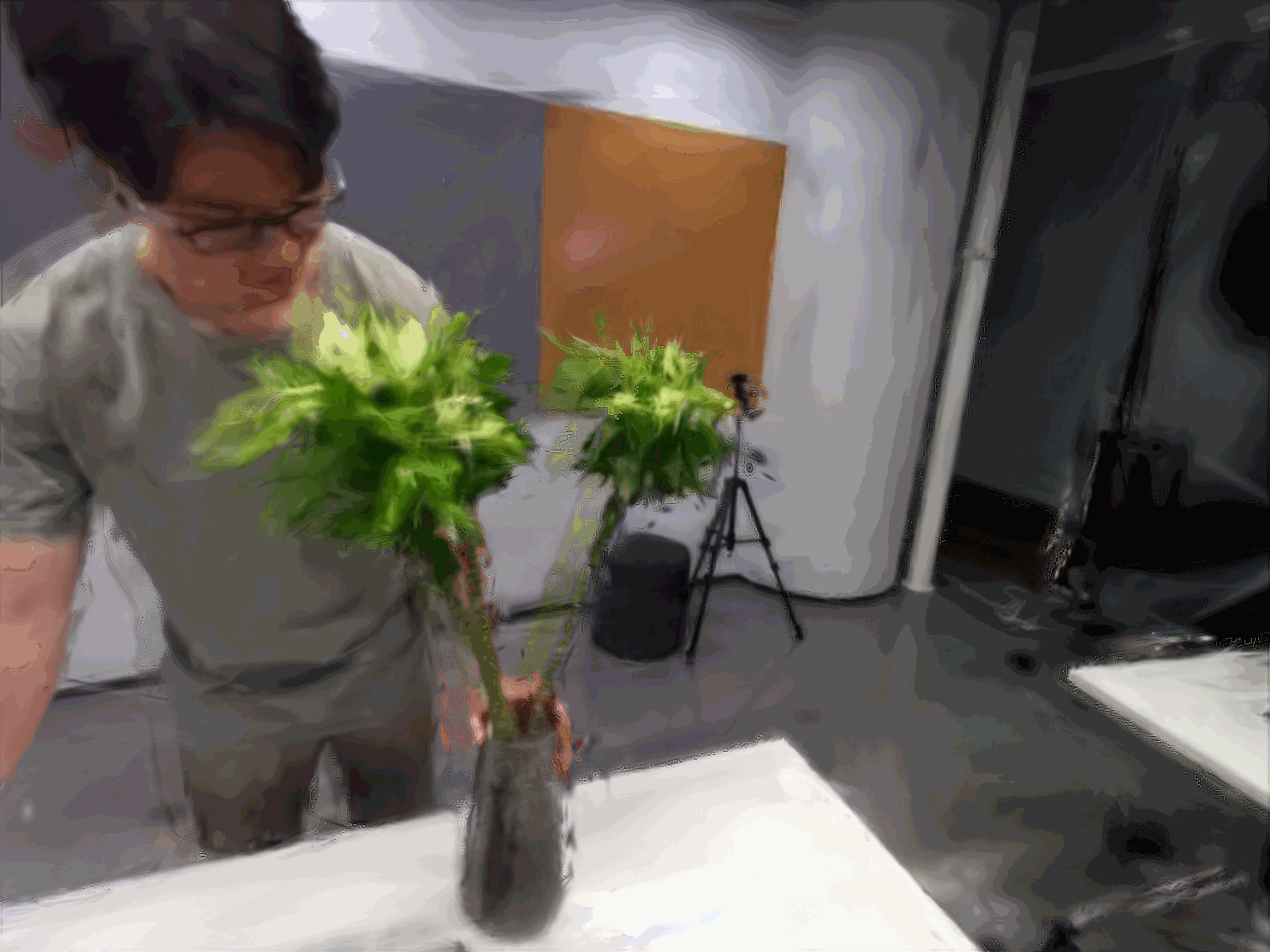}\\[-0.15em]
        \includegraphics[width=\linewidth]{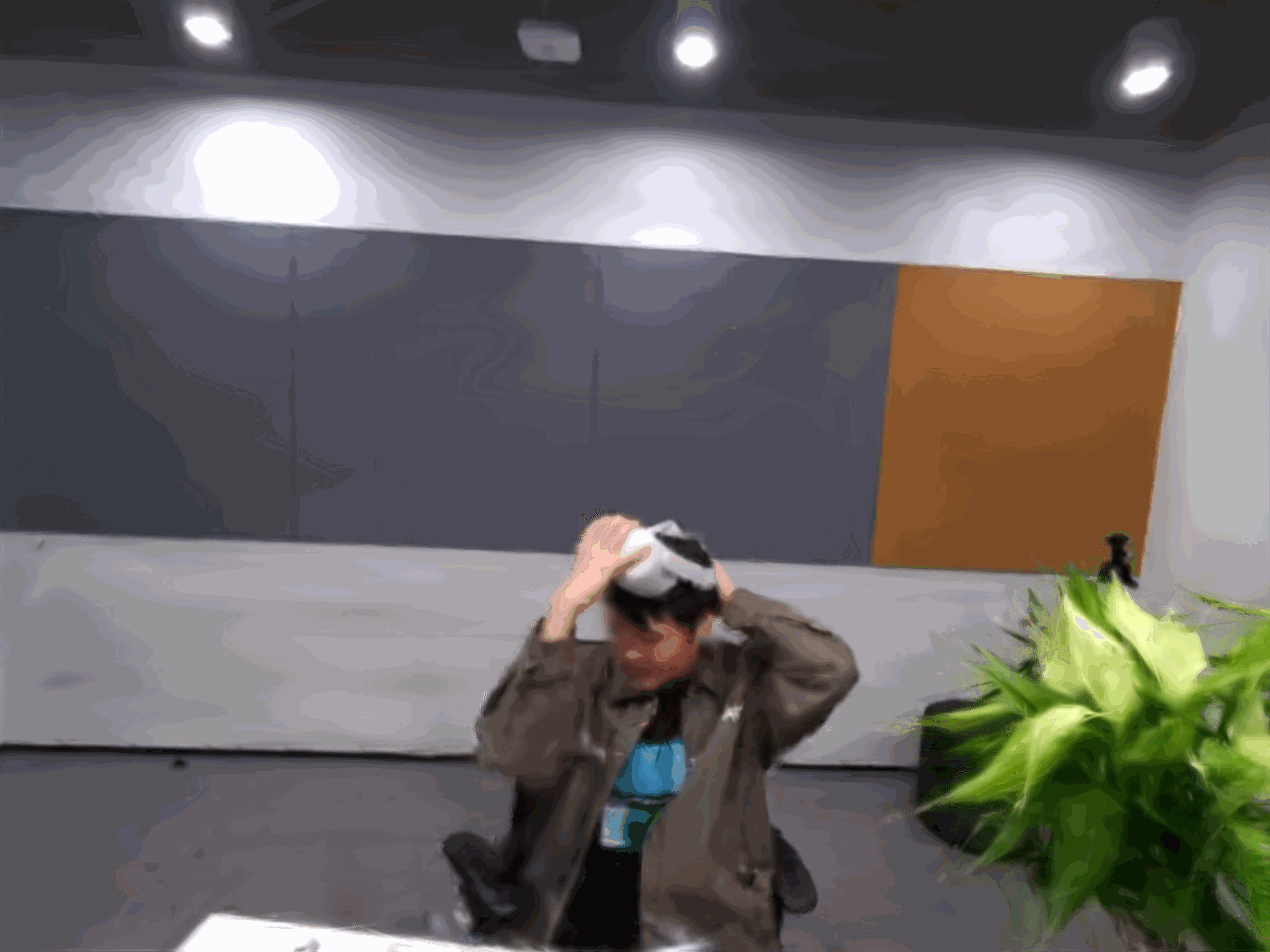}\\[-0.15em]
        \includegraphics[width=\linewidth]{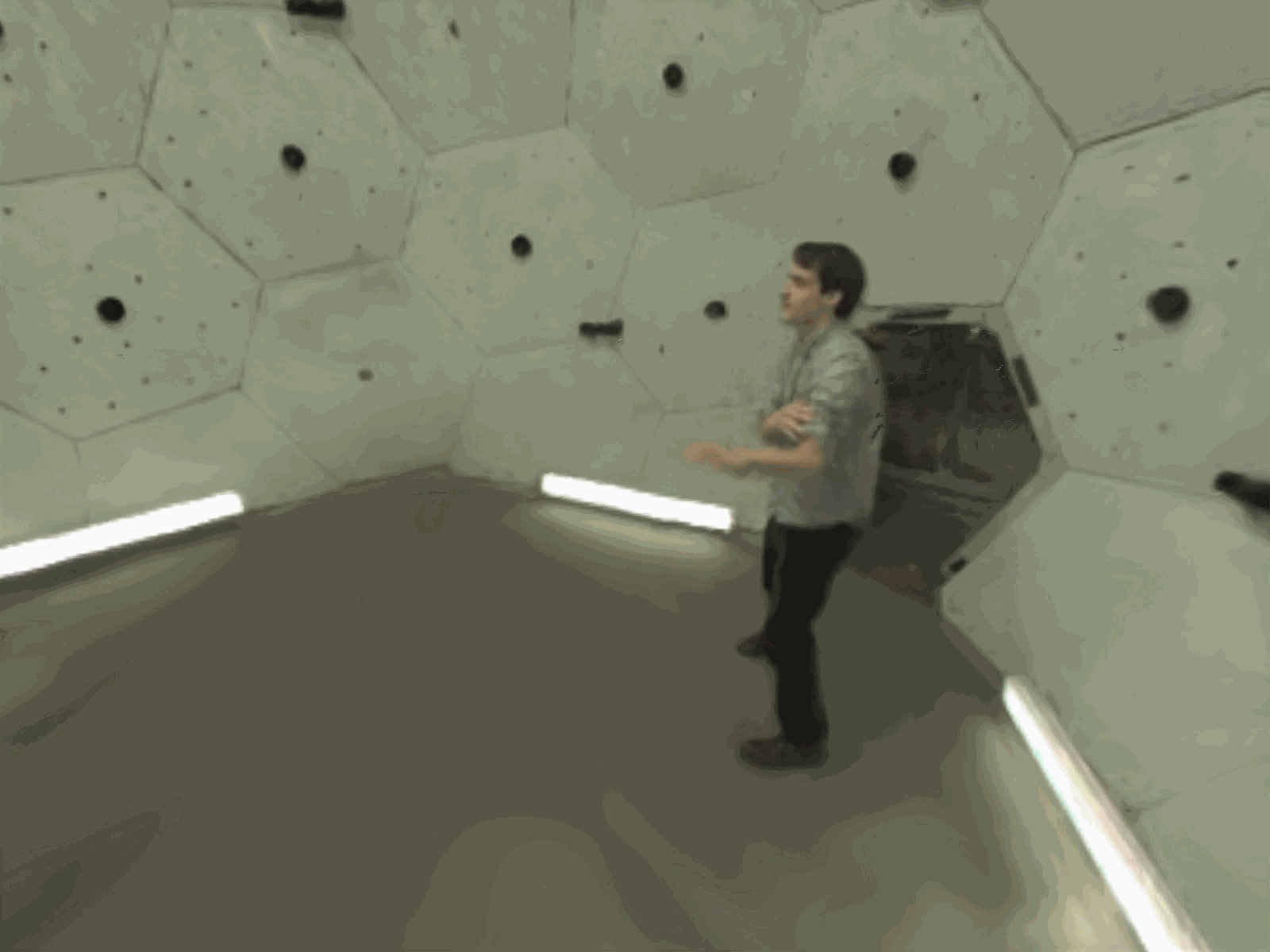}\\[-0.15em]
        \includegraphics[width=\linewidth]{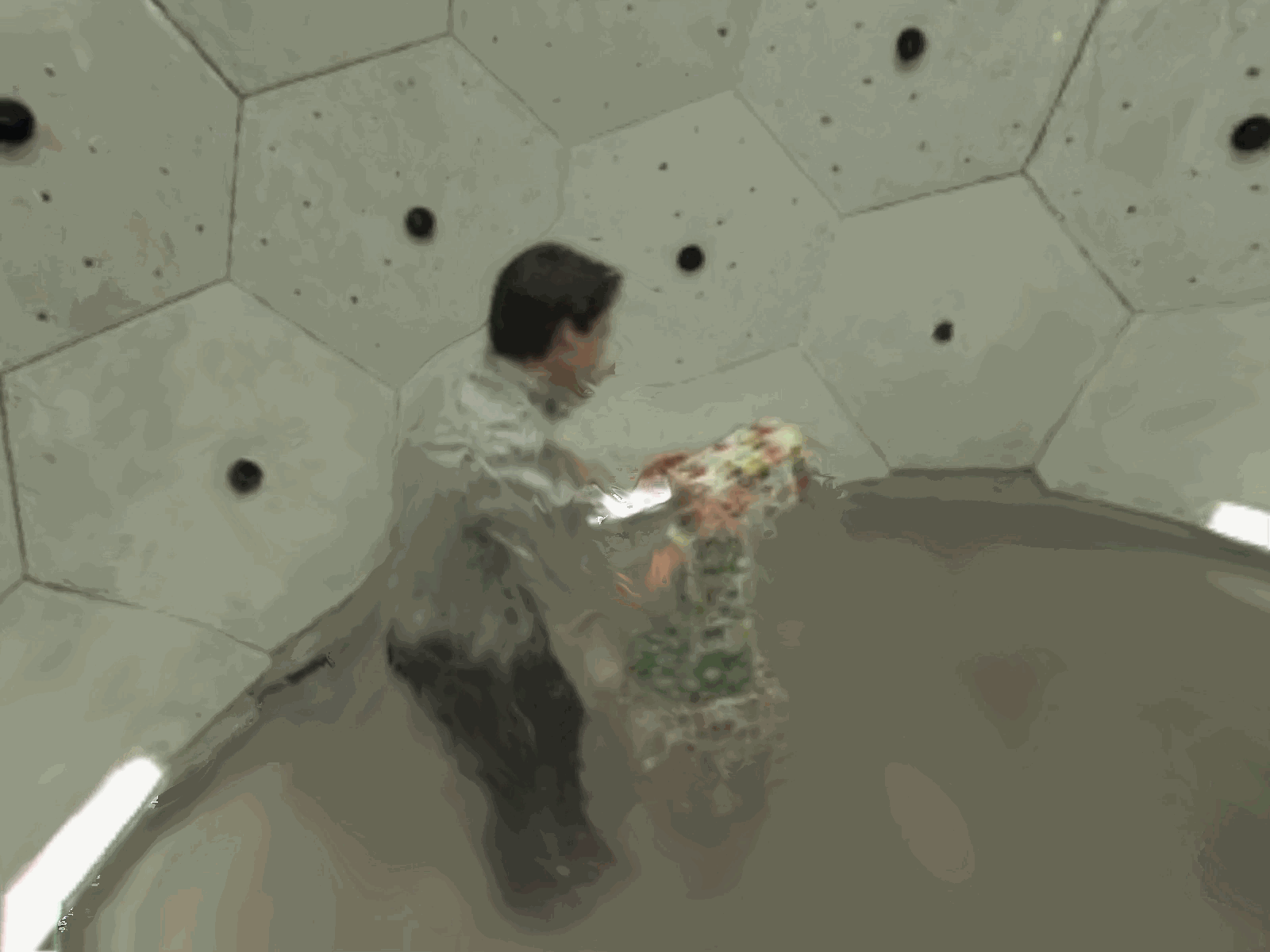}\\[-0.15em]
        \includegraphics[width=\linewidth]{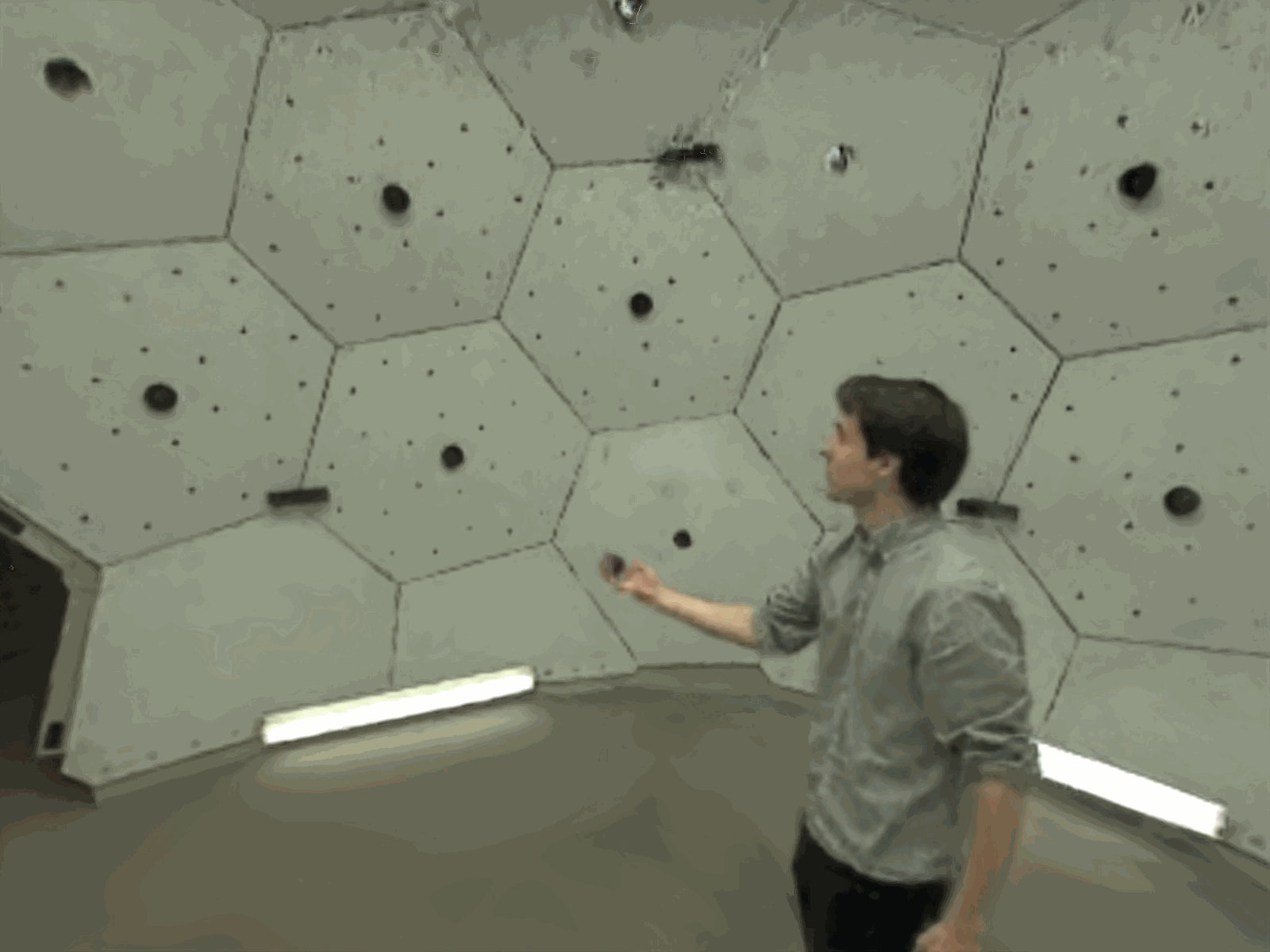}
        \caption{Ours}
    \end{subfigure}\begin{subfigure}[t]{0.138\linewidth}
        \centering
        \includegraphics[width=\linewidth]{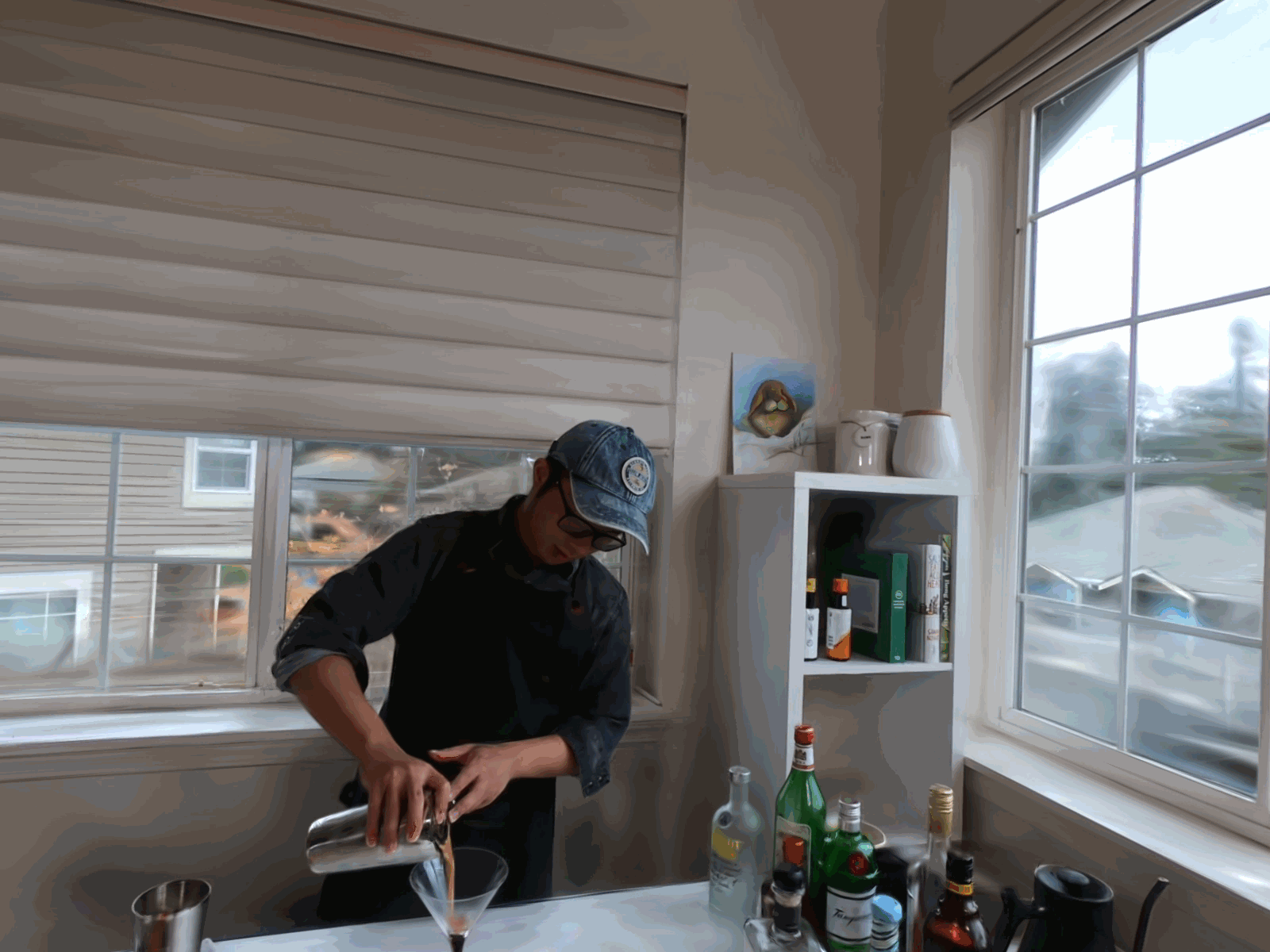}\\[-0.15em]
        \includegraphics[width=\linewidth]{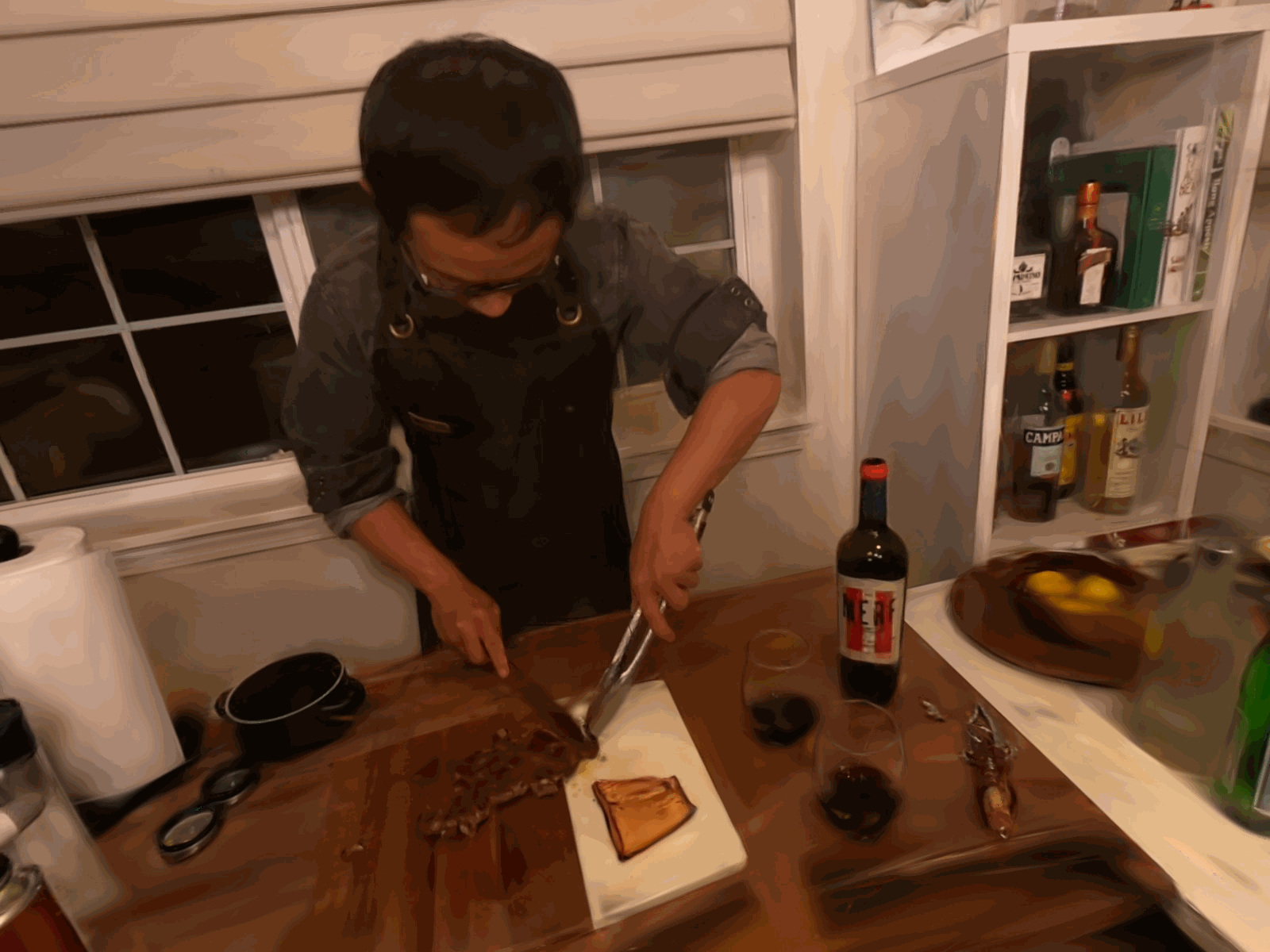}\\[-0.15em]
        \includegraphics[width=\linewidth]{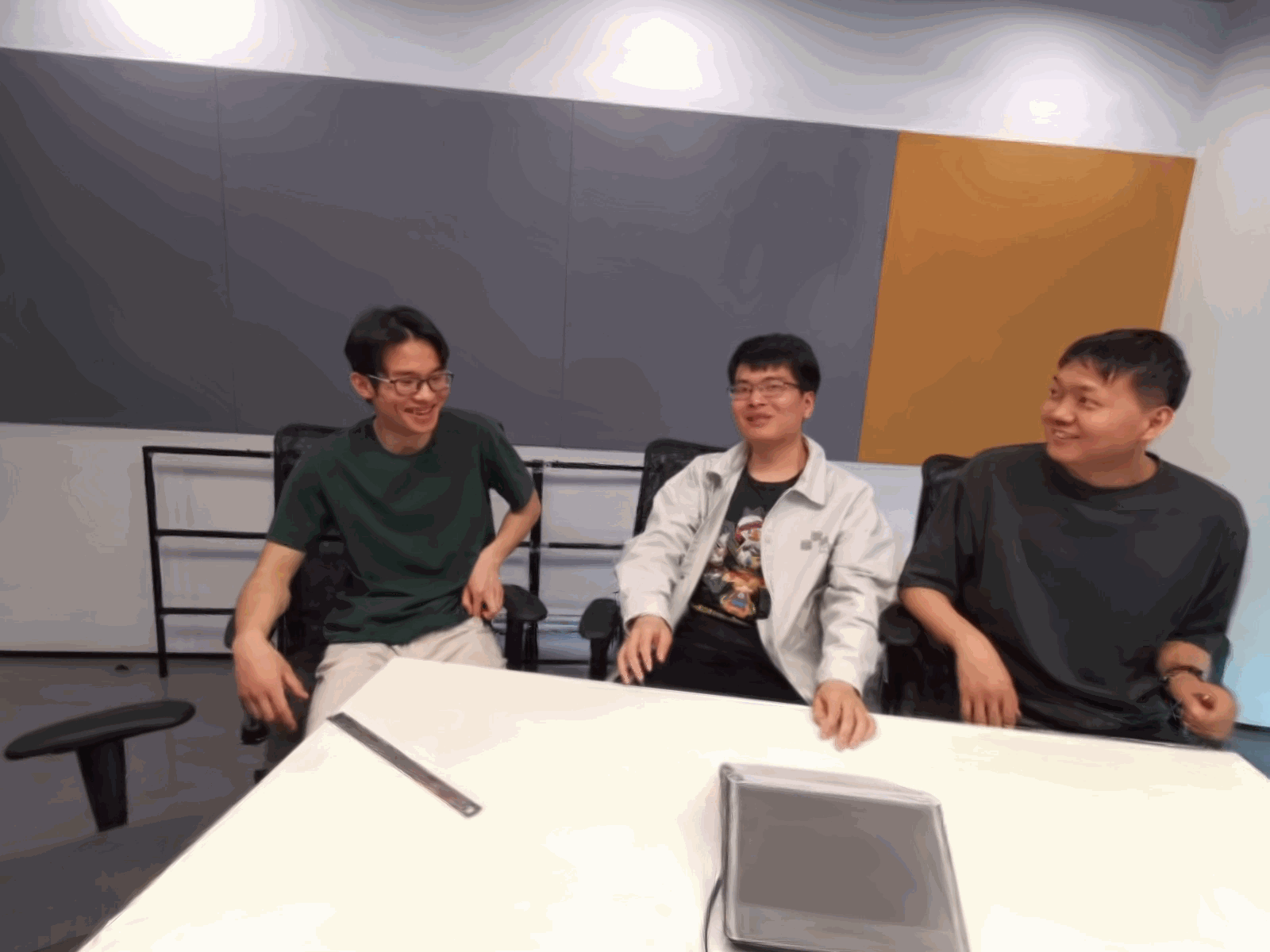}\\[-0.15em]
        \includegraphics[width=\linewidth]{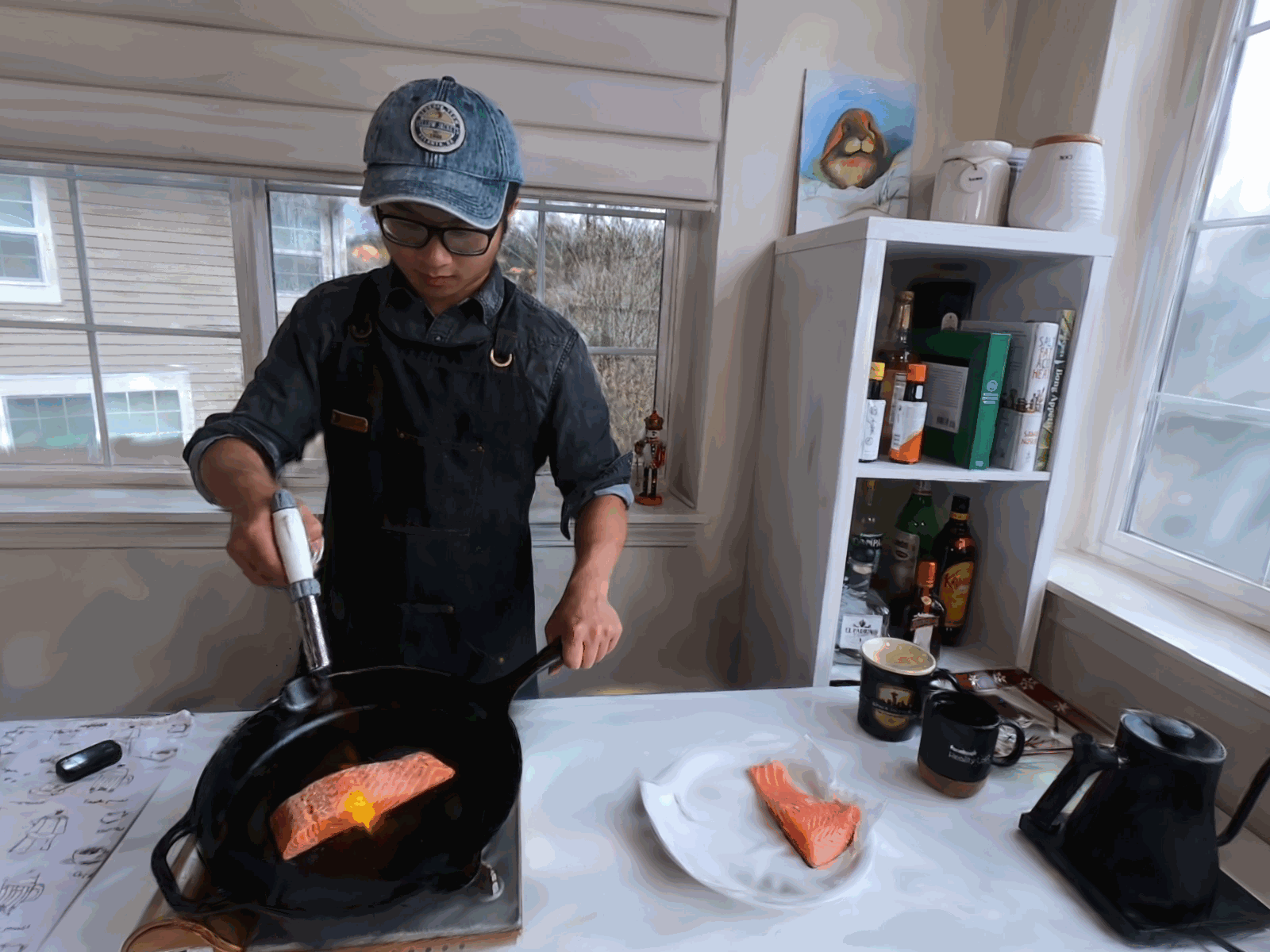}\\[-0.15em]
        \includegraphics[width=\linewidth]{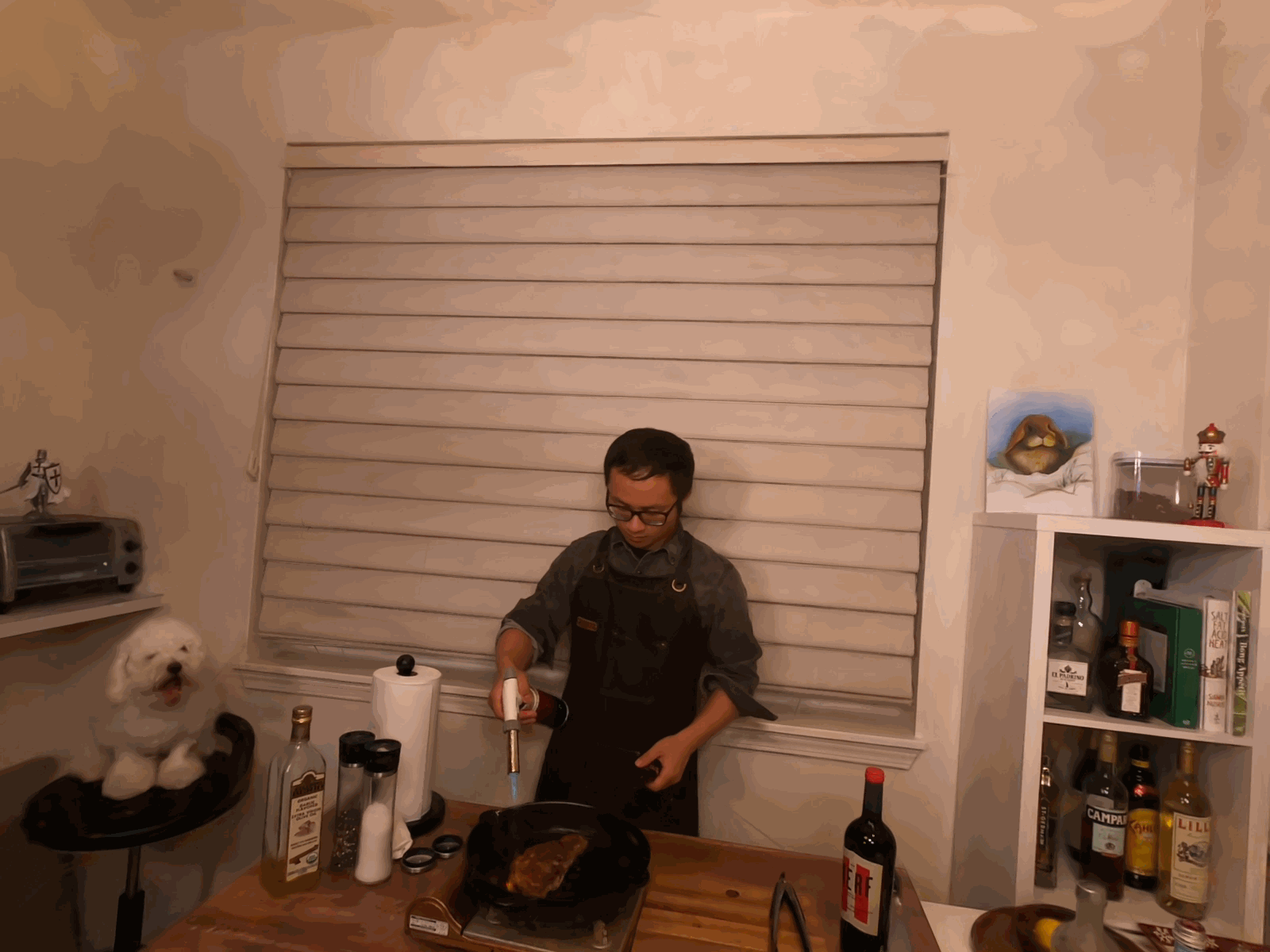}\\[-0.15em]
        \includegraphics[width=\linewidth]{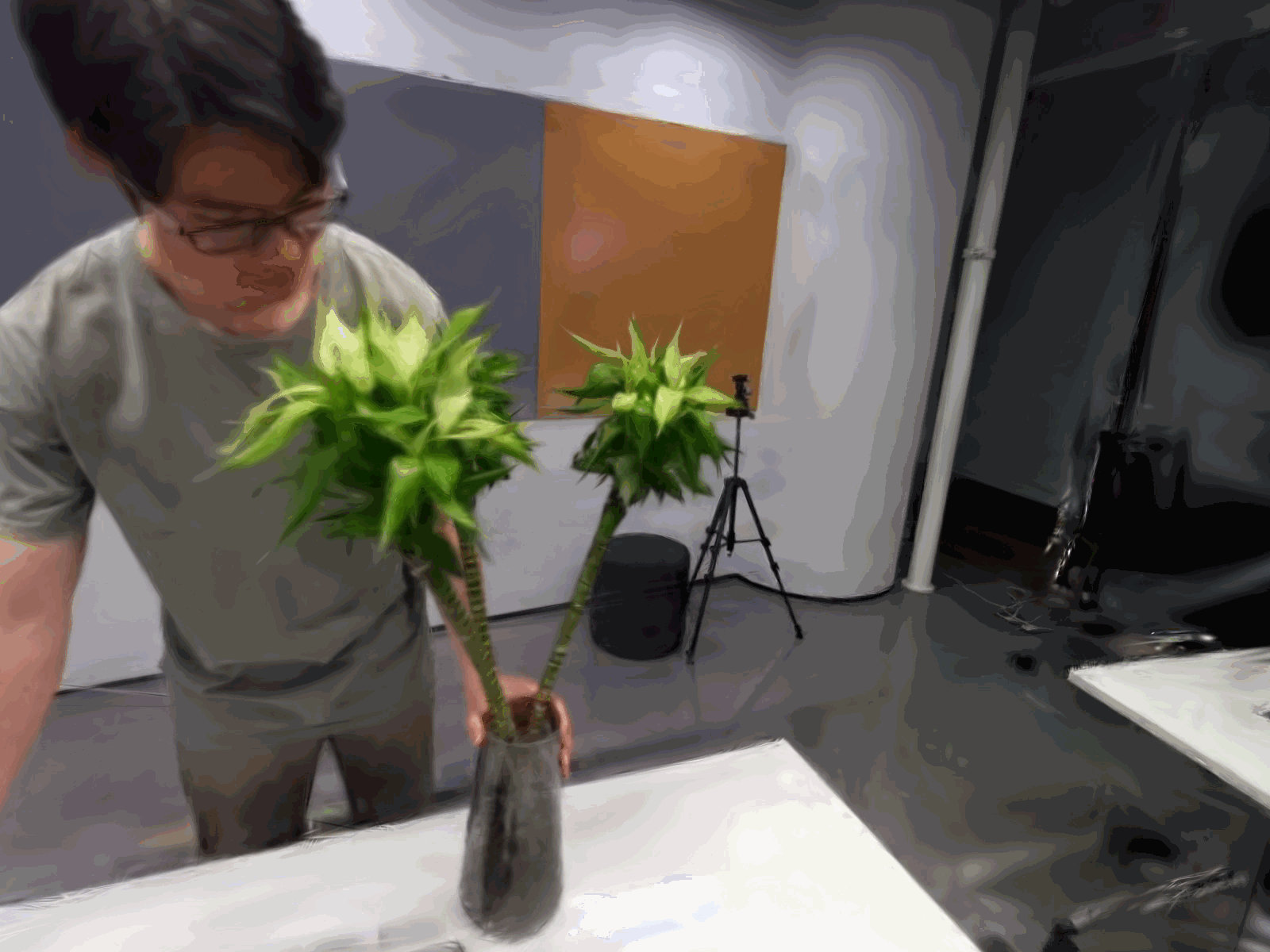}\\[-0.15em]
        \includegraphics[width=\linewidth]{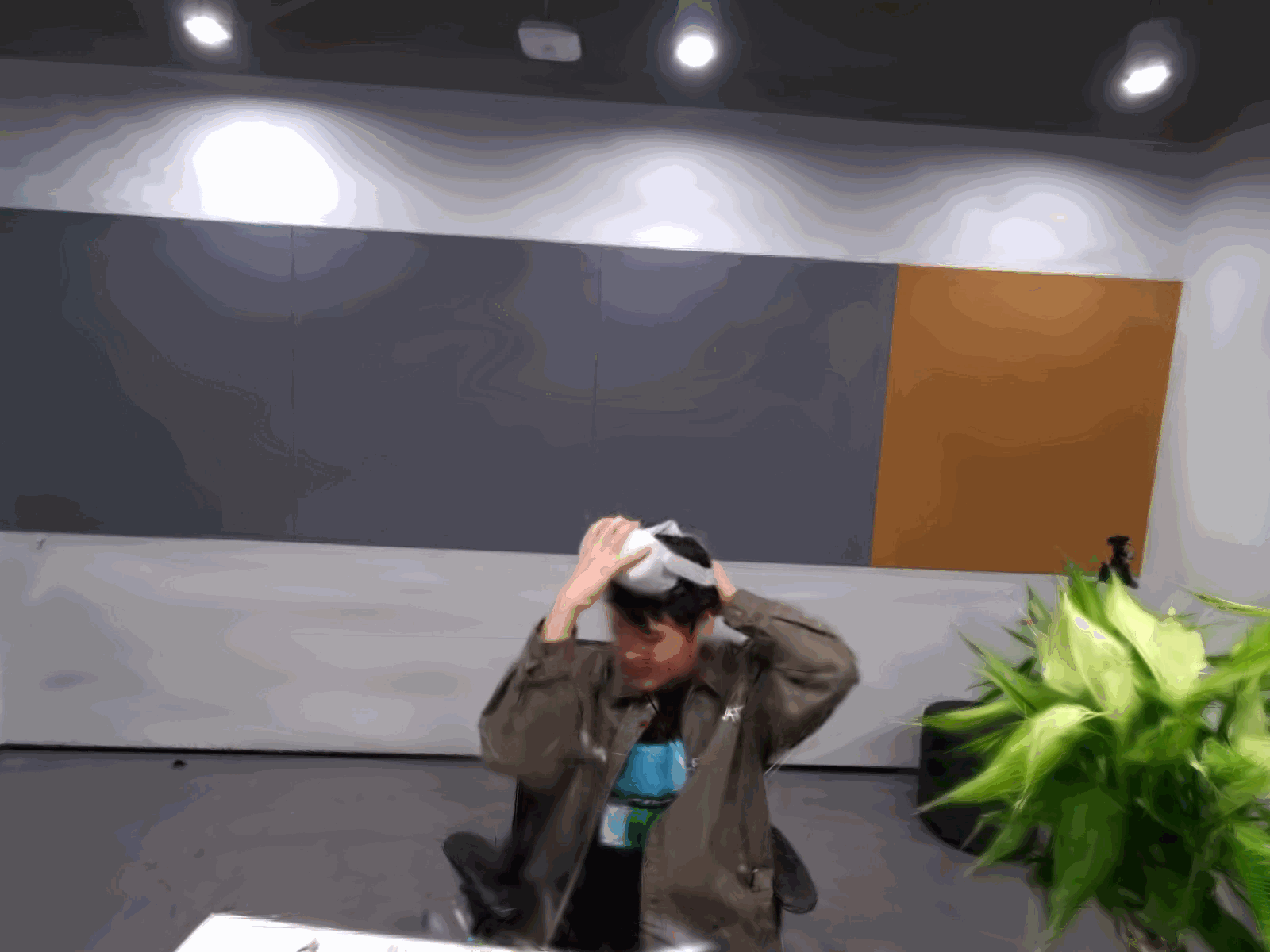}\\[-0.15em]
        \includegraphics[width=\linewidth]{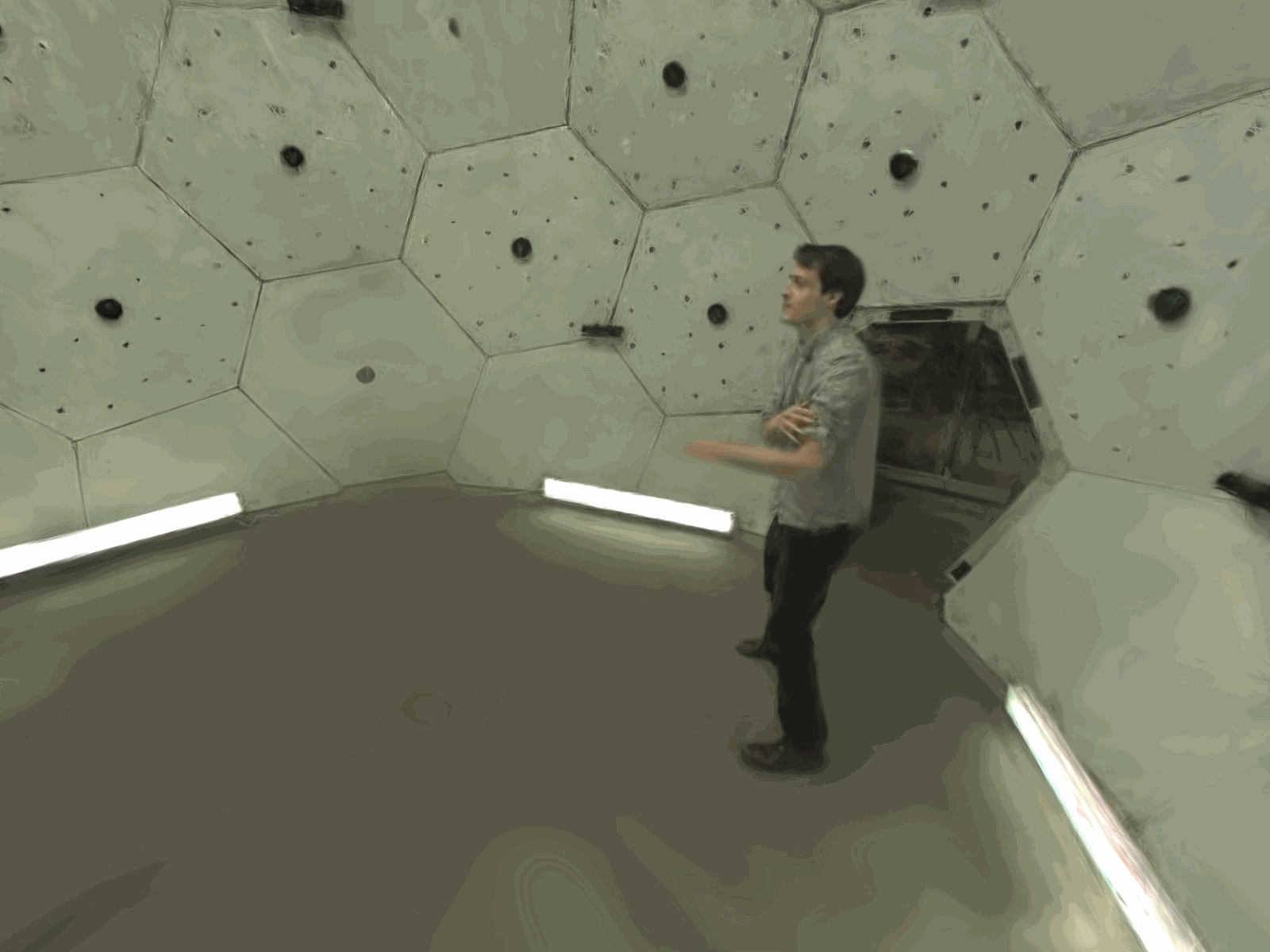}\\[-0.15em]
        \includegraphics[width=\linewidth]{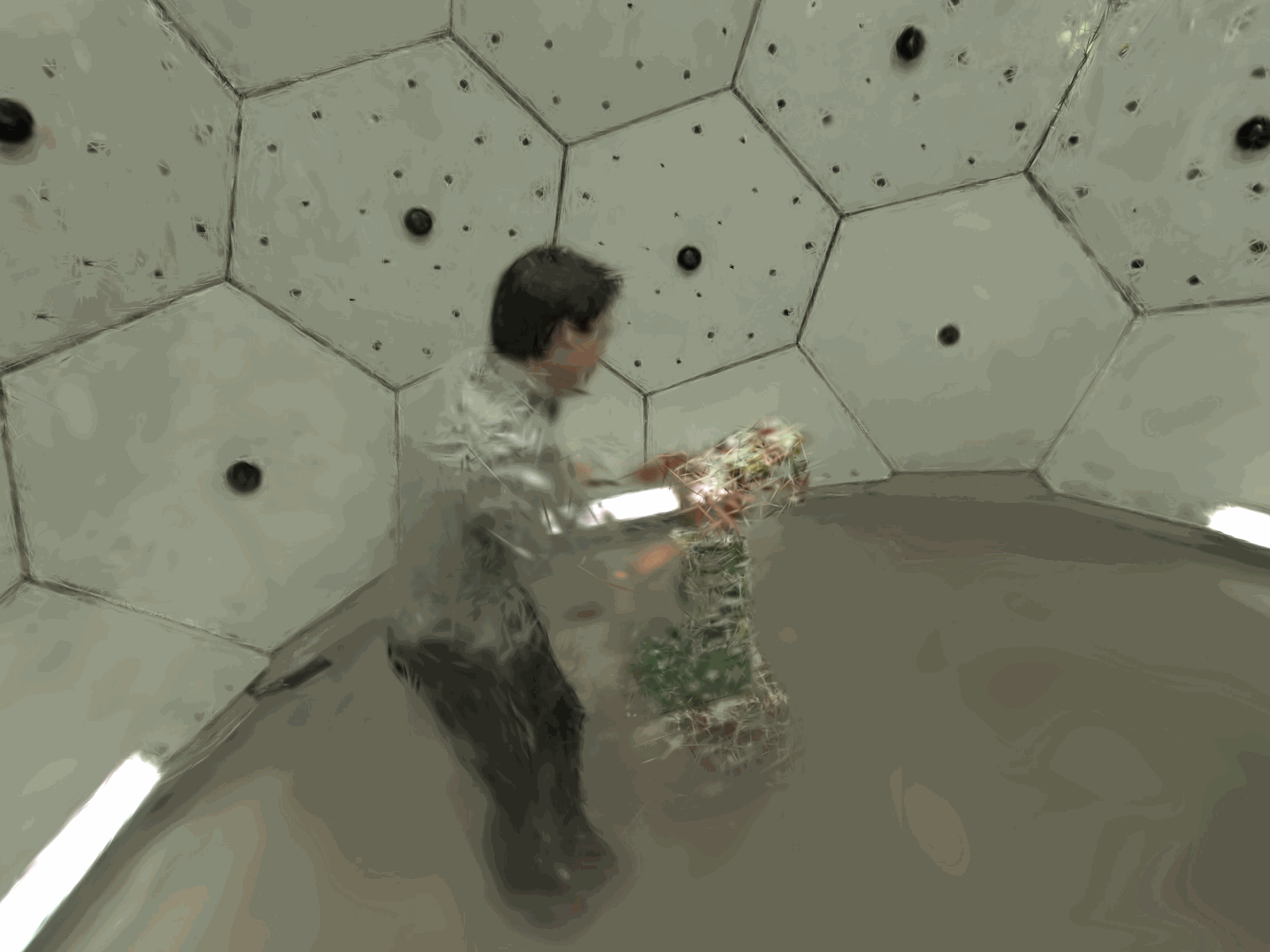}\\[-0.15em]
        \includegraphics[width=\linewidth]{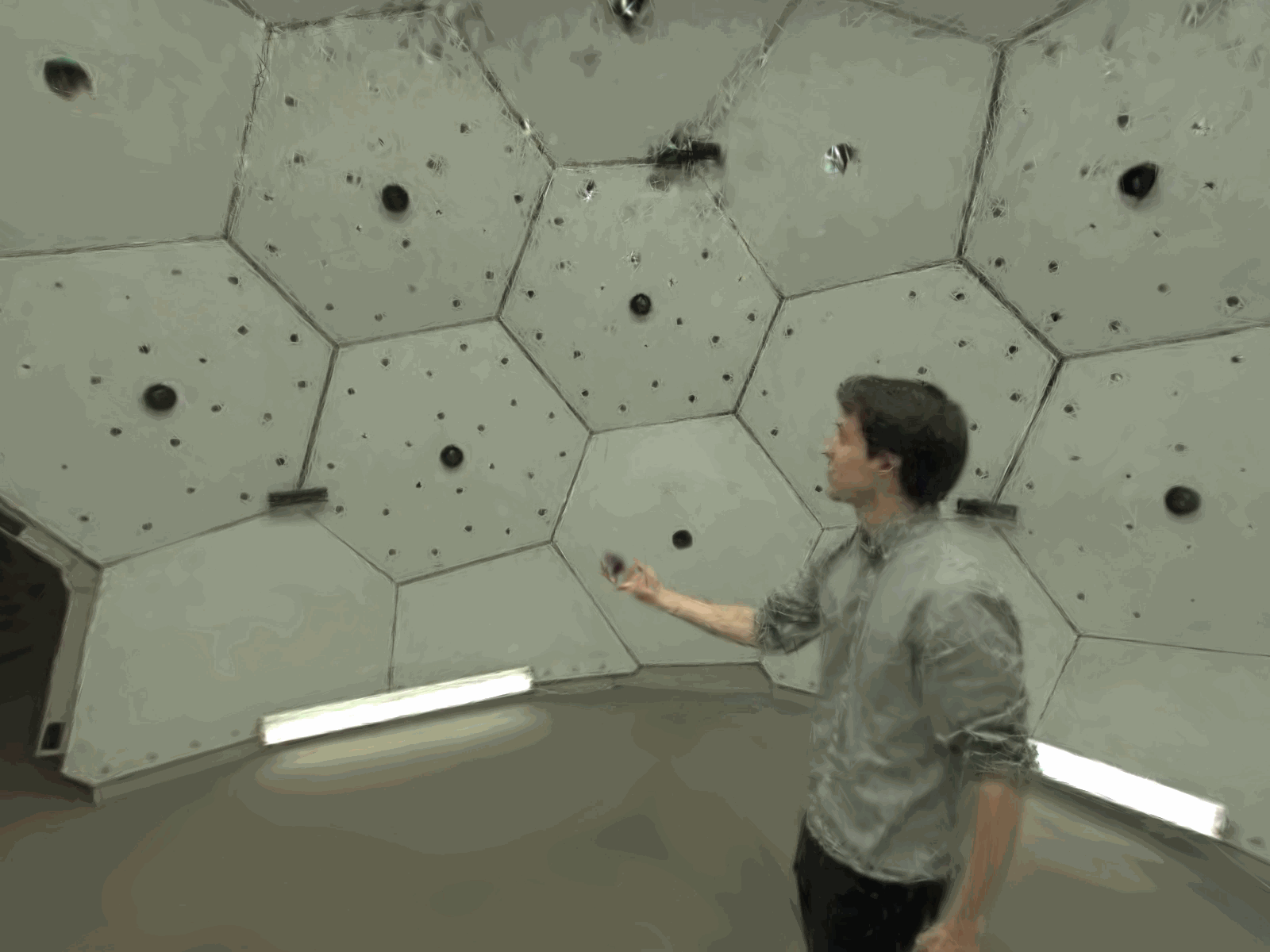}
        \caption{Ground Truth}
    \end{subfigure}
    \caption{
        Visual results of the frame 30 under 30Mbps.
    }\label{fig:visualfull}
\end{figure*}
  
\end{document}